\documentclass[aps,prx,twocolumn,superscriptaddress,nofootinbib]{revtex4-2}
\pdfoutput=1

\usepackage[utf8]{inputenc}
\usepackage[pdfborder={0 0 0}]{hyperref}

\makeatletter
\def\l@subsubsection#1#2{}
\makeatother

\usepackage{graphicx}
\graphicspath{{./figures/}}
\usepackage{dcolumn}
\usepackage{bm}
\usepackage{amsmath}
\usepackage{amssymb}
\usepackage{amsthm}
\usepackage{dsfont}

\usepackage{multirow}
\usepackage[dvipsnames]{xcolor}

\usepackage{physics}
\usepackage{tikz}

\usepackage{layouts}
\usepackage{faktor}

\newtheorem{lemma}{Lemma}
\newtheorem{thm}{Theorem}
\newtheorem{prop}{Proposition}
\newtheorem{cor}{Corollary}

\theoremstyle{definition}
\newtheorem{definition}{Definition}

\newcommand{\mc}[1]{\mathcal{#1}}

\newcommand{\CNOT}{\mathrm{CNOT}}

\newcommand{\NEW}[1]{{#1}}

\newcommand{\redr}{{\color[HTML]{ED598D}r}}
\newcommand{\grg}{{\color[HTML]{A3D671}g}}
\newcommand{\blb}{{\color[HTML]{00D1FA}b}}

\newcommand{\rx}{\texttt{rx}}
\newcommand{\ry}{\texttt{ry}}

\newcommand{\bx}{\texttt{bx}}

\newcommand{\bz}{\texttt{bz}}

\newcommand{\gy}{\texttt{gy}}
\newcommand{\gz}{\texttt{gz}}

\newcommand{\bZ}{\mathbb{Z}}
\newcommand{\bC}{\mathbb{C}}
\newcommand{\bN}{\mathbb{N}}
\newcommand{\cC}{\mathcal{C}}
\newcommand{\cM}{\mathcal{M}}
\newcommand{\cN}{\mathcal{N}}
\newcommand{\cS}{\mathcal{S}}
\newcommand{\cP}{\mathcal{P}}
\newcommand{\cL}{\mathcal{L}}

\newcommand{\cA}{\mathcal{A}}

\DeclareMathOperator{\Aut}{Aut}

\DeclareMathOperator{\sgn}{sgn}

\DeclareMathOperator{\spn}{span}

\newcommand{\isg}[1]{\mc{S}_{#1}}
\newcommand{\lpg}[1]{\mc{N}(\isg{#1})/\isg{#1}}

\makeatletter
\def\l@subsection#1#2{}
\def\l@subsubsection#1#2{}
\makeatother

\makeatletter 
    
\renewcommand\onecolumngrid{
\do@columngrid{one}{\@ne}%
\def\set@footnotewidth{\onecolumngrid}
\def\footnoterule{\kern-6pt\hrule width 1.5in\kern6pt}%
}

\renewcommand\twocolumngrid{
        \def\footnoterule{
        \dimen@\skip\footins\divide\dimen@\thr@@
        \kern-\dimen@\hrule width.5in\kern\dimen@}
        \do@columngrid{mlt}{\tw@}
}%

\makeatother 

\begin{document}

\title{XYZ ruby code: Making a case for a\\ three-colored graphical calculus for quantum error correction in spacetime}

\author{Julio C. Magdalena de la Fuente}
\thanks{these authors contributed equally; \href{mailto:jm@juliomagdalena.de}{jm@juliomagdalena.de}, \href{mailto:j.old@fz-juelich.de}{j.old@fz-juelich.de}}
\affiliation{Dahlem Center for Complex Quantum Systems, Freie Universit\"at Berlin, 14195 Berlin, Germany}

\author{Josias Old}
\thanks{these authors contributed equally; \href{mailto:jm@juliomagdalena.de}{jm@juliomagdalena.de}, \href{mailto:j.old@fz-juelich.de}{j.old@fz-juelich.de}}
\affiliation{Institute for Theoretical Nanoelectronics (PGI-2), Forschungszentrum Jülich, Jülich, Germany}
\affiliation{Institute for Quantum Information, RWTH Aachen University, Aachen, Germany}

\author{Alex Townsend-Teague}
\affiliation{Dahlem Center for Complex Quantum Systems, Freie Universit\"at Berlin, 14195 Berlin, Germany}

\author{Manuel Rispler}
\affiliation{Institute for Theoretical Nanoelectronics (PGI-2), Forschungszentrum Jülich, Jülich, Germany}
\affiliation{Institute for Quantum Information, RWTH Aachen University, Aachen, Germany}

\author{Jens Eisert}
\affiliation{Dahlem Center for Complex Quantum Systems, Freie Universit\"at Berlin, 14195 Berlin, Germany}
\affiliation{Helmholtz-Zentrum Berlin f{\"u}r Materialien und Energie, 14109 Berlin, Germany}

\author{Markus Müller}
\affiliation{Institute for Theoretical Nanoelectronics (PGI-2), Forschungszentrum Jülich, Jülich, Germany}
\affiliation{Institute for Quantum Information, RWTH Aachen University, Aachen, Germany}

\date{\today}

\begin{abstract}
Analyzing and developing new quantum error-correcting schemes is one of the most prominent
tasks in quantum computing research.
In such efforts, introducing time dynamics explicitly in both analysis and design of error-correcting protocols constitutes an important cornerstone.
In this work, we present a graphical formalism based on tensor networks to capture the logical action and error-correcting capabilities of any Clifford circuit with Pauli measurements. We showcase the functioning of the formalism on new Floquet codes derived from topological subsystem codes, which we call \textit{XYZ ruby codes}.
Based on the projective symmetries of the building blocks of the tensor network we develop a framework of \textit{Pauli flows}. Pauli flows allow for a graphical understanding of all quantities entering an error correction analysis of a circuit, including different types of QEC experiments, such as memory and stability experiments.
We lay out how to derive a well-defined decoding problem from the tensor network representation of a protocol and its Pauli flows alone, independent of any stabilizer code or fixed circuit.
Importantly, this framework applies to all Clifford protocols and encompasses both measurement-based and circuit-based approaches to fault tolerance.
We apply our method to our new family of dynamical codes which are in the same topological phase as the $2+1$-dimensional color code, making them a promising candidate for low-overhead logical gates.
In contrast to its static counterpart, the dynamical protocol applies a $\bZ_3$ automorphism to the logical Pauli group every three timesteps. We highlight some of its topological properties and comment on the anyon physics behind a planar layout. Lastly, we benchmark the performance of the XYZ ruby code on a torus by performing both memory and stability experiments and find competitive circuit-level noise thresholds of $\approx 0.18\%$, comparable with other Floquet codes and $2+1$-dimensional color codes.
\end{abstract}

\maketitle

\tableofcontents

\newpage

\section{Introduction}

Reliable quantum computing can be pursued even in the presence of errors: To this aim, the field of \emph{quantum error correction} (QEC) is concerned with designing quantum error-correcting codes, protocols for decoding\NEW{,} and
fault-tolerant schemes for quantum computing~\cite{RevModPhys.87.307}.
With the first QEC codes being proposed in the 1990s~\cite{PhysRevLett.77.793}, only in recent years have steps been taken towards the experimental implementation of large-scale quantum error-correcting codes 
\cite{takita2017experimental,GKPExperiment,krinner2022realizing,zhao2022realization,GoogleQEC,postler2022demonstration,ryan2022implementing,bluvstein2024logical, huang2023comparing,postler2023demonstration}, with remarkable progress. 
Current QEC schemes, however, are still challenging to scale up with current technological capabilities, or require significant resource overhead. Therefore, it remains a fruitful and at the same time crucial endeavour to investigate novel schemes to design and analyze protocols for fault-tolerant storage and computation.

A stabilizer-based error-correcting code is defined to be the joint eigenspace of a group of commuting operators~\cite{gottesman1997stabilizer}. Hence, measuring these operators repeatedly gives rise to an error-correcting protocol where the measurement outcomes are used to infer if a physical error affected the system and potentially correct for that.
Importantly, the order in which the stabilizer operators are measured affects neither the evolution of encoded logical information nor the possibility of inferring a suitable recovery given the measurement outcomes of the protocol.

Recently, novel schemes have been presented that rely on repeatedly measuring non-commuting operators in a specific order. 
While we should not generically expect such a scheme to preserve any information, one can carefully choose the operators that are being measured to design a fault-tolerant \textit{dynamical error-correcting scheme}\,\cite{hastings2021dynamically}.
Removing the commutativity requirements allows one to circumvent many obstacles that appear for na\"ive stabilizer measurements.
Importantly, the weight, i.e.~the physical qubit support of the check operators that are measured throughout a protocol, can be drastically reduced. This constitutes an important avenue of research as high-weight operators usually come along with complicated quantum circuits that induce substantial circuit-level noise in many architectures.
Additionally, the time ordering allows one to protect a qubit with a measurement sequence that without time-ordering (i.e.\ as a subsystem code~\cite{PhysRevLett.94.180501}) would not encode any logical qubits.
Dynamical schemes where the same measurement sequence is repeated periodically over time have been coined \textit{Floquet codes}~\cite{vuillot2022planar, Teague2023Floquetifying} and after a first instance, the honeycomb code, presented in Ref.~\cite{hastings2021dynamically}, the concept has attracted much attention.

For stabilizer-based QEC codes, topological stabilizer codes~\cite{dennis2002topological} have been promising candidates for reliable quantum information storage due to their intrinsic scalability with only local measurements.
Similarly, in the context of dynamical codes, it has taken little time for constructions to arise based on models for topologically ordered phases~\cite{kesselring2022anyon, davydova2023quantum, ellison2023floquet, dua2023engineering, zhang2022xcube, aasen2022adiabatic}.
In turns out that in topological protocols defined in $2+1$ spacetime dimensions, where time-ordering is important, the effective logical degrees of freedom can be understood in terms of the data of an underlying topological phase, characterized by an \textit{Abelian anyon theory} by considering associated 1-form symmetries~\cite{bauer2024topological}.
Importantly, a spacetime perspective is of essence and the operators carrying the encoded information cannot be understood from a static, spatial perspective alone in order to argue about fault-tolerance in the presence of circuit-level noise.

In this work, we contribute to a more thorough understanding of QEC in spacetime by extending existing graphical formalisms for QEC to a tensor network based formalism that captures all protocols defined from Clifford operations. 
On a higher level, our aims are two-fold: We extend the common ZX-calculus~\cite{van2020zx} and obtain a graphical representation of protocols that is closer to the actual components of a circuit involving $Y$ measurements.
Tracking Pauli operators though these networks yields a complete description of all algebraic quantities entering a QEC protocol and an associated decoding problem. 
Moreover, we showcase the usefulness of this framework by analyzing a class of dynamical error-correcting protocols involving $Y$ measurements.
Specifically, we focus on topological codes and introduce a new family of topological Floquet codes.
These codes can be constructed from any 2-colex~\cite{Bombin2007exact}, a trivalent and plaquette-tricolorable planar graph and is defined by a measurement sequence on the gauge generators of Bombin's topological subsystem code~\cite{bombin2006topological}.
Within this family, we mainly focus on the codes obtained from a hexagonal lattice, defined on a system of qubits composed of three qubits per vertex of the original lattice.
We refer to this subfamily of protocols as \textit{XYZ ruby codes}.
As a topological code the associated subsystem code is related to an Abelian anyon theory. In this case, it is the 3-fermion theory~\cite{roberts20203, Ellison2023paulitopological}, which is a chiral subtheory of the color code anyon model.
In the dynamical setting we find, however, that the actual logical operators that are protected and corrected for in the protocol are related to the color code anyon model.
As such, our code realizes the color code phase in spacetime and hence offers a richer set of natively implementable fault-tolerant logical gates. For different schedules, this has already been observed in Ref.~\cite{dua2023engineering}.
In contrast to other known schedules though, the qubits themselves are not in a state of the color code phase at \NEW{every timestep of the} protocol.
They undergo an \textit{instantaneous phase transition} during the measurement sequence to a phase that hosts more anyons. To the best of our knowledge this is the first two-dimensional Floquet code where this happens.

In order to showcase the strength of the three-colored graphical calculus for QEC, we go so far as to design, simulate and decode circuits for the XYZ ruby code. 
Concretely, we combine the spacetime tensor network methods for QEC with the understanding of the anyon physics of the ruby code to devise fault-tolerant QEC experiments that benchmark different properties of the code.
In particular, we perform memory experiments and the first stability experiments for a dynamical code, under different error models (phenomenological, EM3 and circuit-level noise). We find logical error rates and thresholds comparable to other Floquet codes encoding the same number of qubits.

\subsection{Guide to the reader}
In Sec.~\ref{sec:prelim}, we give the necessary background on dynamical QEC protocols and two-dimensional topological codes.
\NEW{In Sec.~\ref{sec:rubycode_definition}, we introduce the XYZ ruby code as a Floquet code and analyze the topological phases of the instantaneous states.
These sections can be read independently of 
Sec.~\ref{sec:prelim-3cgc}, which provides a pedagogical introduction into the three-colored graphical calculus for QEC, essential to understand the remainder of this work.
The reader interested in a more rigorous treatment is referred to App.\,\ref{app:rgbtensornetwork} where we also present our main technical mathematical result on how to understand the sequential application of Clifford maps (including measurement).}

In Sec.~\ref{sec:benchmarking}, we start by giving our perspective on a general ``probing experiment'' in spacetime. We then elaborate on the fault-tolerance properties of the protocols and construct and perform numerical experiments.
We present the necessary considerations when designing planar codes, respectively boundaries in spacetime, of a topological protocol defined in the bulk in Sec.~\ref{sec:planar} and sketch a general construction for topological codes.
Lastly, in Sec.~\ref{sec:remarks}, we comment on more general aspects of spacetime QEC, make connections to existing formalisms and comment on emerging perspectives and open questions arising from a tensor network representation.

\section{Preliminaries}\label{sec:prelim}
In this section, we introduce the important concepts that are required for our analysis of the XYZ ruby code.
First, we give a high-level introduction to dynamical codes based on Pauli measurements. This class of codes has been introduced as instantaneous stabilizer group codes, or \textit{ISG codes} in Ref.~\cite{Teague2023Floquetifying}. Second, we introduce graphical tools to analyze such codes in a spacetime tensor-network language inspired by the ZX-calculus\,\cite{van2020zx, bombin2023unifying}.
Importantly, we include tensors to represent $Y$ measurements more directly in our framework.
This renders the analysis of dynamical protocols involving measurements in the $Y$ basis much easier and more efficient.
Lastly, for readers more familiar with algebraic structures of topologically ordered systems, we give an overview on how certain algebraic features of topological stabilizer codes in two spatial dimensions can be described by an Abelian anyon model~\cite{wang2020and}. 
We will focus on the anyon models associated to toric and color codes in which most of the abstract mathematical language simplifies significantly.

Before we go into details, let us fix some notation.
Consider a system of $n$ qubits with a Hilbert space $(\bC^2)^{\otimes n}$.
For the rest of this work, we denote the Pauli matrices on qubit $j$ by
\begin{align}
\begin{split}
    Z_j = \mqty(1 & 0\\ 0& -1)\qcomma X_j = \mqty(0 & 1 \\ 1&0)\\
    \qq{and} Y_j = iXZ = \mqty(0 & -i\\ i&0).
\end{split}
\end{align}
These Pauli matrices together with a phase $i$ generate the \textit{$n$-qubit Pauli group}, denoted by $\mathcal{P}_n$.
Any element $P\in\mathcal{P}_n$ can be given a \textit{weight} $w:\mathcal{P}_n\to \bZ_{\geq 0}$, which we define to be the number of tensor factors (qubits) on which $P$ acts with a Pauli operator that is not proportional to the identity.

We call an Abelian subgroup of the Pauli group a \textit{stabilizer group} and the common $+1$ eigenspace of all elements in that group the associated \textit{stabilizer code}.
Importantly, any state in the stabilizer code is stabilized by any Pauli element in the stabilizer group.
If we want the stabilized subspace to be at least one-dimensional, we require additionally that $-\mathds{1}$ is not contained in that group~\cite{gottesman1997stabilizer, nielsen2001quantum}.

\subsection{ISG and Floquet codes}\label{sec:prelim_ISG}
Projectively measuring a (multi-qubit) Pauli operator $P$ projects a system of qubits onto the eigenspace associated to the measurement outcome. If the system was in a stabilizer state before the measurement, the post-measurement state will again be a stabilizer state whose stabilizer group can be efficiently calculated from the initial stabilizer group $\cS_0$ and the Pauli operator $P$\,\cite{gottesman1998TheHR}.
We call the stabilizer group of the system immediately after a (Pauli) measurement the \textit{instantaneous stabilizer group (ISG)}.
This term has first been introduced in Ref.~\cite{hastings2021dynamically} together with the honeycomb code, the first Floquet code in the literature.
The idea of \textit{ISG} and \textit{Floquet} codes is to use measurements to drive a quantum system between different ISGs while preserving the logical information.
The goal of the following paragraph is to make this more specific.

We define an \textit{ISG code} via a sequence (ordered list) of sets of Pauli operators
\begin{align}\label{eq:measurement_sequence}
    \cM = [\cM_0,\cM_1, \cM_2 , \dots],
\end{align}
such that $\langle \cM_i\rangle$ is a stabilizer group for all $i$~\cite{Teague2023Floquetifying}.
We call elements in $\cM_i$ \textit{checks}.
In principle, one can consider sequences with a (countably) infinite number of elements, although for this work it suffices to consider finite ones.
Since $\langle \cM_i\rangle$ is Abelian, measuring all checks in $\cM_i$ projects the system unambiguously onto the subspace stabilized by $\langle \{(-1)^{m_j}M_{i,j} \;|\; M_{i,j}\in \cM_i\}\rangle$. The binary numbers $\{m_j\in\{0,1\}\;|\; j=1,2,\dots ,|\cM_i|\}$ label the measurement outcomes of each of the measured Pauli operators.
Note that the structure of the stabilized subspace is the same independent of the measurement outcome. In particular, only the signs of the stabilizer generators will differ.
Hence, without restricting generality, we will later often assume that all measurement outcomes are 0 to simplify the analysis of the instantaneous code states.
To avoid notational clutter, we denote the operation of ``measuring all elements in $\cM_i$'' by $\mathfrak{m}_i$.
If the system has initially been stabilized by a stabilizer group $\cS_{i}$, measuring $\cM_i$ drives the system into a new stabilizer group $\cS_{i+1}$ that can be efficiently constructed from $\cS_{i}$ and $\cM_i$,
\begin{align}
    \cS_{i} \stackrel{\mathfrak{m}_i}{\longrightarrow} \cS_{i+1}.
\end{align}
Starting from an initial stabilizer group $\cS_0$, an ISG code hence defines a sequence of ISGs
\begin{align}
    \mc{S}_0 \stackrel{\mathfrak{m}_0}{\longrightarrow} \mc{S}_1 \stackrel{\mathfrak{m}_1}{\longrightarrow} \mc{S}_2 \stackrel{\mathfrak{m}_2}{\longrightarrow}\dots .
\end{align}
In most cases, we will assume $\cS_0=\{\mathds{1}\}$, i.e., 
we start with an arbitrary state. To initialize a particular logical state, however, one might want to fix $\cS_0$ to some other stabilizer group, for example $\langle\{Z_i\}_{i=1}^n\rangle$, stabilizing the product state vector $\ket{0}^{\otimes n}$.

A \textit{Floquet} code is defined by a finite sequence $\cM$.
The length of $\cM$ is referred to as the \textit{period} \NEW{$T\in\bZ_{> 0}$} of the Floquet code and the measurements are repeated periodically after the first round \NEW{$\mathfrak{m}_0, \mathfrak{m}_1, \dots , \mathfrak{m}_{T-1}$}, i.e., we define $\cM_i=\cM_{i \mod T}$.
The number of logical qubits encoded in the subspace stabilized by a stabilizer group $\cS_i$ is captured by its \textit{rank} $r_i$, the minimal number of independent generators.
Specifically, the number of encoded logical qubits is given by $k_i=n-r_i$.
Note that $r_i$ is a non-decreasing function in $i$, where $i$ is now associated to the \textit{timesteps} in the measurement protocol.

In this work, in order to encode a logical qubit in an ISG code, we require that there exists $\hat{i} \in \mathbb{Z}_{\geq 0}$ such that $r_i$ is constant for all $i\geq \hat{i}$.
Following Ref.~\cite{Teague2023Floquetifying}, we say that the ISG code \textit{establishes} $k = n-r_{\hat{i}}$ logical qubits after $\hat{i}$ measurements.
This is closely connected to the fact that all pairs $(\cS_i,\cS_{i+1})$ for $i\geq\hat{i}$ form a \textit{reversible pair} of stabilizer groups in the sense of Ref.~\cite{aasen2023measurement}:
Two stabilizer groups $(\cS_i,\cS_{i+1})$ form a reversible pair if any element $s\in \cS_{i}$ that commutes with $\cS_{i+1}$ is also contained in $\cS_{i+1}$ and, vice versa, and element $s'\in\cS_{i+1}$ that commutes with $\cS_{i}$ is contained in $\cS_{i}$.
Importantly, this condition is not sufficient to have a fault-tolerant ISG code but only guarantees that the logical information is preserved by the measurement that switches from $\cS_i$ to $\cS_{i+1}$.
To incorporate error-correcting abilities, one needs to specify how to decode on the set of measurement outcomes. 
In the next section we illustrate how we use graphical methods to analyze the fault tolerance properties of a spacetime measurement circuit associated to an ISG code.
For more details on formal definitions and basic properties of sequences of ISGs, we refer the interested reader to Refs.\,\cite{Teague2023Floquetifying, aasen2023measurement}.

\subsection{Topological stabilizer codes and their anyon models}\label{sec:prelim-anyons}
In two-dimensional topological stabilizer codes \textit{anyons} appear as local (point-like) violations of stabilizers that cannot be created by any local operator~\cite{Kitaev2006anyons}.
They can be associated to topological (quasiparticle) excitations of an exactly solvable Hamiltonian
\begin{align}
    H = -\sum_i S_i,
\end{align}
defined by a choice $\{S_i\}_i$ of a (local) generating set of the stabilizer group.
If $\langle\{S_i\}\rangle$ fulfills the error detection condition~\cite{nielsen2001quantum} for \textit{any local error}, the above Hamiltonian is topologically ordered~\cite{bravyi2011ashortproof}
and we identify the (degenerate) ground space of that Hamiltonian with the code space of $\langle \{S_i\}\rangle$.

In two spatial dimensions, any such excitations can be created at the endpoints of Pauli operators supported on string-like regions~\cite{bravyi2009no, bombin2014structure, bombin2012universal}.
The associated (stabilizer equivalence classes of) string operators can be labeled by an (Abelian) \textit{fusion group} $A$ and a complex phase $\theta: A\to U(1)$.
The fusion group $A$ relates to how the (equivalence classes of) string operators multiply, while the phase $\theta$ captures the commutativity of a string of type $a$ with itself. 
Specifically, a string operator defined along a self-intersecting path is stabilizer-equivalent to a string operator defined on a non-intersecting path with the same endpoints times a complex phase, given by $\theta$.
In fact, this data $(A,\theta)$ defines an \textit{Abelian anyon model}~\cite{wang2020and}, an Abelian braided fusion category.
Interestingly, any Pauli topological stabilizer code realizes an anyon model that is equivalent to some (finite) number of copies of the toric code anyon model~\cite{bombin2012universal, bombin2014structure}.

Many properties of topological codes become apparent when interpreting them as ground spaces of the topologically ordered Hamiltonians introduced above.
In particular, this perspective is helpful to construct logical operators as non-trivial anyon string operators, to describe native locality-preserving logical gates in terms of symmetries of the anyon model~\cite{Kesselring2018boundariestwist} and to construct boundaries and domain walls in terms of anyon condensation~\cite{kesselring2022anyon}.
In this section, we introduce the toric code and color code anyon model and show the equivalence of the latter to two copies of the former explicitly.

\subsubsection{Toric code anyon model}
\begin{figure}
    \centering
    \includegraphics[width=\linewidth]{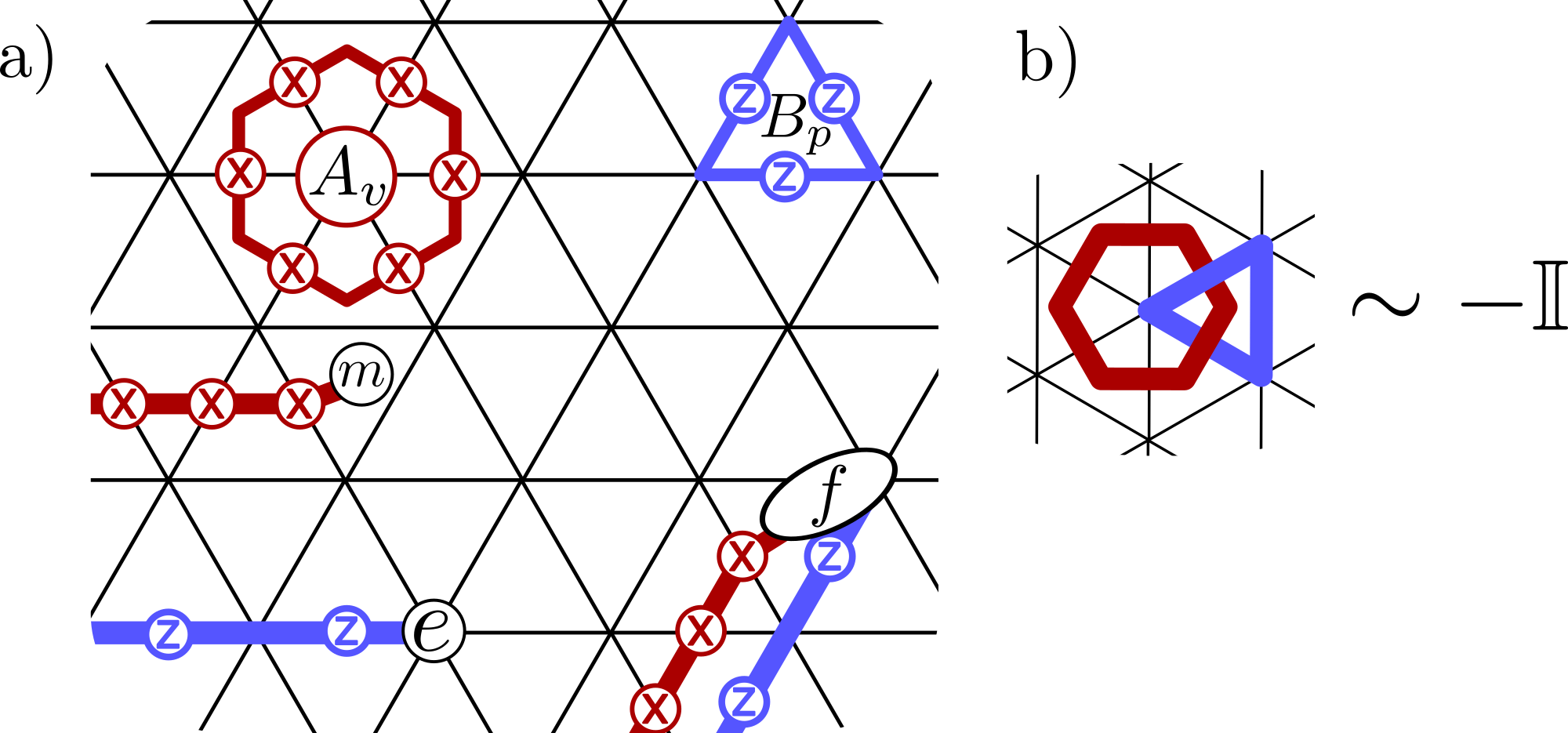}
    \caption{A toric code model can be defined from any planar graph.
    a) The associated codespace is defined by placing a qubit on every edge and adding stabilizer generators on vertices and plaquettes.
    The plaquette stabilizers act with Pauli $Z$ operators around the qubits on the boundary of that plaquette and can be thought of as a (homologically trivial) loop (cycle) of $Z$ operators on the direct lattice.
    The vertex stabilizers act with Pauli $X$ operators on all qubits adjacent to the vertex and can be understood as acting along a (homologically trivial) loop on the dual lattice (cocycle).
    Isolated violations of these stabilizers can be understood as anyons of the toric code anyon model and are created at endpoints of $X$ strings on the dual lattice ($m$ anyons violating a plaquette stabilizer) or $Z$ strings on the direct lattice ($e$ anyons violating a vertex stabilizer).
    Combining these two types of string operators such their endpoints are close to each other creates the fusion product of $e$ and $m$ which is a fermion $f$.
    The fermionic statistics can be seen from the fact that the string operators of $e$ and $m$ anticommute if they intersect an odd number of times.
    b) We depict a small example where two interlinked loops of string operators act like $-\mathds{1}$ on the code space.
    }
    \label{fig:toriccode_anyons}
\end{figure}

The toric code is the most commonly known topological code for quantum error correction.
It is constructed from the \textit{quantum double model} of the Abelian group $\bZ_2$~\cite{kitaev2003fault}.
It can be defined on any planar graph by placing qubits on the edges of that graph.
For each vertex $v$, we introduce a vertex stabilizer $A_v$ acting with $X$ on all qubits on the edges incident to that vertex.
Similarly, we introduce a stabilizer $B_p$ for each plaquette $p$ that acts with $Z$ on every qubit on the edges forming the boundary of that plaquette.
Planarity of the underlying graph guarantees commutativity of both types of stabilizers.
In Fig.~\ref{fig:toriccode_anyons}, we illustrate how anyons appear at the endpoints of the string operators in the toric code defined on a triangular lattice.

The \emph{toric code} hosts four types of anyons, commonly denoted by $A_{TC}= \{1,e,m,f\}$.
Under fusion, they form the group $\bZ_2\times \bZ_2$,
\begin{subequations}
\begin{align}
    a\times 1 = 1\times a = a\qcomma a\times a &= 1,\;\quad\forall a\in A_{TC},\\
    e\times m = m\times e &= f.
\end{align}
\end{subequations}
Additionally, a braiding is defined via
\begin{align}
    \theta_1 = \theta_e = \theta_m = 1\qq{and} \theta_f = -1. 
\end{align}
In analogy with usual particles, we call $1,e$ and $m$ bosons (they have topological spin $1$), and $f$ a fermion (it has topological spin $-1$).
On the lattice, the non-trivial spin of $f$ is captured by the fact that the string operators creating $e$ and $m$ anyons do not commute, see Fig.~\ref{fig:toriccode_anyons}.

\subsubsection{Color code anyon model}
The \emph{color code} has been introduced as a family of topological stabilizer codes defined on any trivalent planar graph with tricolorable plaquettes~\cite{bombin2006topological}.
In this construction, the qubits are placed on the vertices and two stabilizer generators are defined for each plaquette.
One type, $S^X_p$, acts with $X$ on all qubits surrounding plaquette $p$ and the other type, $S^Z_p$, acts with $Z$ on all qubits surrounding $p$.
The trivalency and -colorability ensures commutativity of all the stabilizers on any of these lattices.

\begin{figure}
    \centering
    \includegraphics[width=0.9\linewidth]{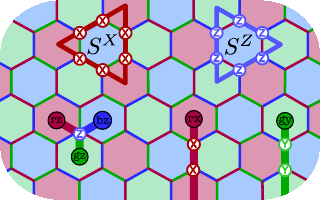}
    \caption{The hexagonal color code is defined on a system of qubits played on vertices of a hexagonal lattice. The stabilizer group of the associated code has two types of generators per plaquette: One acting with a product of Pauli $X$ on each qubits surrounding the plaquette, and one acting with Pauli $Z$ on the same qubits.
    A state violating a set of these generators can be associated with a state with anyonic excitations at the plaquettes whose stabilizers are violated, labeled by the color of the plaquette and the Pauli operator that created them when applied to a codestate.
    A single Pauli error creates a triple of anyons, one at each plaquette touching the qubit. This gives rise to the anyon fusion rule $\texttt{rx}\times\texttt{gx}\times\texttt{bx} = \texttt{1}$.
    An isolated anyon can be created at the endpoints of certain string operators. For example, the boson \texttt{rx} is created at the endpoints of a string of $X$ operators acting on \NEW{pairs of qubits along red} edges, connecting red plaquettes.
    Similarly, one can create the other bosons in the color code anyon model \NEW{by acting with two-qubit Pauli operators of the form $P\otimes P$ on the qubits on edges of a given color}.}
    \label{fig:colorcode_anyons}
\end{figure}

In Fig.~\ref{fig:colorcode_anyons}, we illustrate the color code defined on a hexagonal lattice.
We label each of the violations according to the color and Pauli label of the operators creating it.
For example, a violation of a single $S^X_p$ (that is not simultaneously a violation of $S^Z$) for a red plaquette $p$ is labeled by \texttt{rz} since it is localized at a red plaquette and can be created by a product of Pauli $Z$ operators.
A single violation of that type, labeled by the color $\texttt{c}\in\{\texttt{r},\texttt{g},\texttt{b}\}$ and Pauli label $\texttt{p}\in\{\texttt{x},\texttt{y},\texttt{z}\}$, can only be created by a non-local operator.
For example, when placed on an infinite plane, this could be a half-infinite string operator connecting plaquettes of the same color $\texttt{c}$.
Any color code boson can be created by a product of string operators of definite Pauli and color type, which we label with $\texttt{c}\texttt{p}$.
For now, we will call these string operators \textit{generating string operators} and the associated anyons \textit{generating anyons}.\footnote{This choice of generators is not unique and not even minimal but makes many aspects of the analysis of the anyon model easier.}
Additionally, there are products of string operators labeled by different colors and Paulis that create an inequivalent anyon.
We will see in the next paragraph that these will be the fermions in the color code.

In the color code stabilizer model, we notice that any single-qubit Pauli creates a triple of violations on the plaquettes surrounding the qubit.
On the level of the anyon model this means that any triple of such violations with the same Pauli label corresponds to the trivial anyon,
\begin{align}
    \texttt{rx}\times\texttt{gx}\times\texttt{bx} = \texttt{1},
\end{align}
and analogously for the other Pauli labels.
Similarly, the same holds for a triple of anyons with the same color but different Pauli labels.

It is instructive to order the generating anyons in a $3\times 3$ ``magic square'', where each row corresponds to a Pauli and each column corresponds to a color label~\cite{Kesselring2018boundariestwist},
\begin{align}\label{eq:boson_table}
    \begin{tabular}{c|c|c}
         \texttt{rx} & \texttt{gx} & \texttt{bx}  \\ \hline
         \texttt{ry} & \texttt{gy} & \texttt{by}  \\ \hline
         \texttt{rz} & \texttt{gz} & \texttt{bz}  \\ 
    \end{tabular}.
\end{align}
The square is aligned such that the fusion product of every row and of every column is $\texttt{1}$.
Phrased differently, the fusion product of any two anyons in the same row(column) is the third anyon in the same row(column).
Products of two anyons that do not share a row or a column are anyons of the color code outside of this table.
Taking the constraints above into account, we count 6 additional inequivelant non-trivial anyons.
Looking at the string operators, we notice that each anyon in the square is a boson and the six not directly represented in the square are fermions, i.e., have topological spin $-1$, see Fig.~\ref{fig:colorcode_anyons}.
We conclude that the color code has 16 inequivalent anyons: the trivial anyon \texttt{1}, nine bosons in the table above, and six inequivalent products of these bosons that are fermions.
In the following, we denote the anyon model of the color code with $\cC_{CC}$.

As stated before, any two-dimensional topological Pauli stabilizer code is described by an anyon model that is equivalent to some number of layers of the toric code.
In particular, this is also true for the color code anyon model.
Specifically, there exists an \textit{unfolding}~\cite{kubica2015unfolding} of the color code anyon model into two copies of the toric code.
We demonstrate the unfolding by relabeling a set of generators for the color code anyon model to a set of generators of the anyon model of two decoupled layers of the toric code.
For this we label each anyon of two layers of the toric code by $(a,b)$, where $a,b\in\{1,e,m,f\}$ label the anyon restricted on layer one, respectively two.
For example, we can identify
\begin{align}
    \begin{tabular}{c|c|c}
         \texttt{rx} & \texttt{gx} & \texttt{bx}  \\ \hline
         \texttt{ry} & \texttt{gy} & \texttt{by}  \\ \hline
         \texttt{rz} & \texttt{gz} & \texttt{bz}  \\ 
    \end{tabular} \quad \leftrightarrow\quad 
    \begin{tabular}{c|c|c}
$(e,1)$ & $(e,e)$ & $(1,e)$  \\ \hline
         $(e,m)$ & $(f,f)$ & $(m,e)$  \\ \hline
         $(1,m)$ & $(m,m)$ & $(m,1)$  \\ 
    \end{tabular}.
\end{align}
In this unfolding, the six fermions of the color code are labeled by
\NEW{
\begin{subequations}\label{eq:colorcode_fermions}
\begin{align}
    \texttt{f}_1 = (1,f)\qcomma \texttt{f}_2 = (f,e)\qcomma \texttt{f}_3 = (f,m),\\
    \texttt{f}_4 = (f,1)\qcomma \texttt{f}_5 = (e,f)\qcomma \texttt{f}_6 = (m,f).
\end{align}
\end{subequations}
}
Note that $\texttt{f}_1, \texttt{f}_2, \texttt{f}_3$ as well as $\texttt{f}_4,\texttt{f}_5,\texttt{f}_6$ form a closed subtheory of the color code anyons.
When viewed as an anyon model on its own, each of these subtheories are equivalent to the ``3-fermion" anyon model~\cite{wang2020and, kesselring2022anyon, roberts20203}.
The unfolding into two copies of the toric code anyon model shows that the color code anyon model can be identified as the Abelian group $A_{CC} = \bZ_2^{\times 4}$ 
together with a topological spin $\theta_{CC} = \theta_{TC}\otimes \theta_{TC}$ that acts like the product of toric code spins on the two $\bZ_2\times\bZ_2$-factors of $A_{CC}$.

The symmetries of an anyon model play an important role in the encoding and manipulation of logical information in topological codes.
A symmetry of an anyon model $A$ is a permutation of the anyons that leaves all the anyonic data invariant, i.e., preserves the fusion as well as the braiding properties.
For the two-dimensional color code, the group formed by all symmetries, i.e., 
the \textit{automorphism group} $\Aut(A_{CC})$, can be understood in terms of symmetries of the boson table in Eq.\,\eqref{eq:boson_table}~\cite{Kesselring2018boundariestwist}.
One can permute the rows/columns individually -- permuting color or Pauli labels -- and also mirror along the diagonal -- exchanging color and Pauli label.
This yields an automorphism group of $(S_3\times S_3)\rtimes\bZ_2$.
For a more elaborate treatment of the color code anyon model we refer the reader to Refs.\,\cite{Kesselring2018boundariestwist, kesselring2022anyon}.

\section{The XYZ ruby code}\label{sec:rubycode_definition}

\begin{figure}
    \centering
    \includegraphics[width=\linewidth]{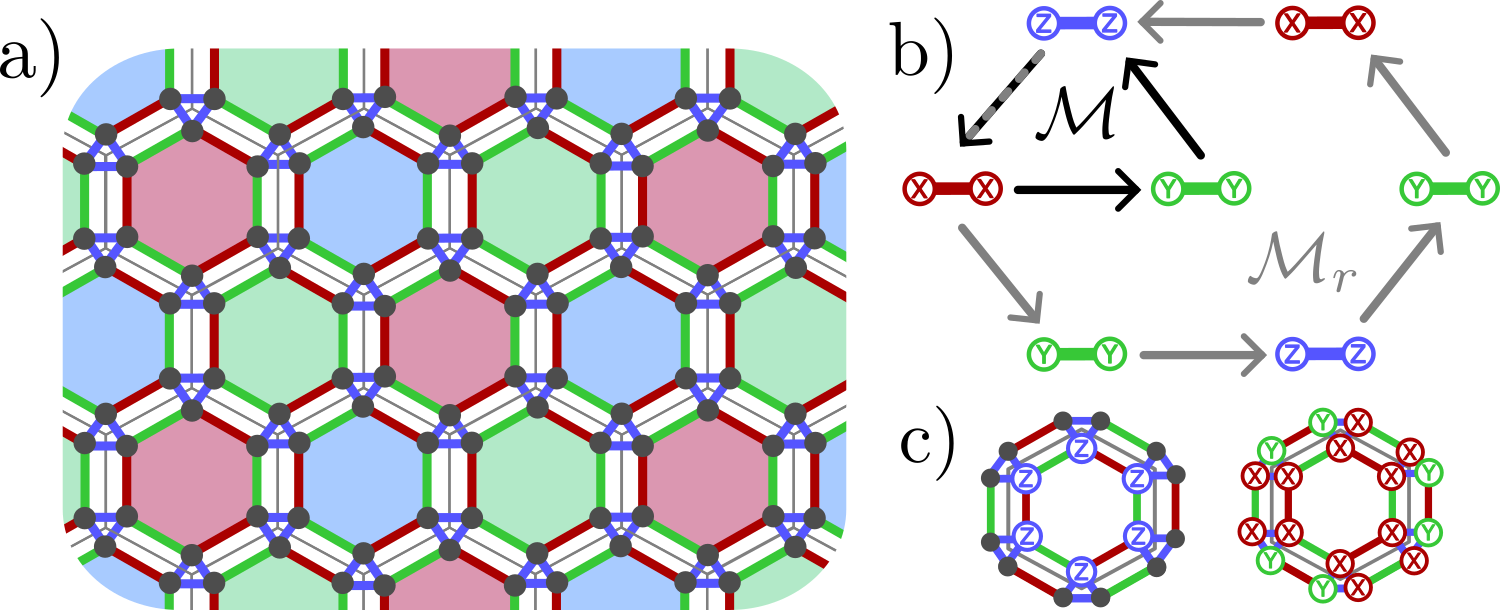}
    \caption{a) The XYZ ruby code is defined on qubits placed on the vertices of the ruby lattice with three-colorable plaquettes.
    The error-correcting protocol only involves weight-2 measurements between neighboring qubits.
    b) The color of the edges indicates the Pauli basis of the measurement.
    Along a blue edge, a $ZZ$ measurement is performed, along a green edge a $YY$ measurement is performed and along a red edge a $XX$ measurement is performed.
    In this work, we consider two different schedules, one with periodicity three (black arrows) and one with periodicity 6 (gray arrows). The period-6 schedule is a rewinding version of the period-3 schedule.
    c) Both schedules are designed to read out the stabilizers of the associated subsystem code~\cite{bombin2006topological}. For each plaque(tte), the subsystem code has two independent stabilizer generators.
    }
    \label{fig:code-definition}
\end{figure}
In this section, we present two variants of the \emph{XYZ ruby code}.
Both are defined on the same lattice of qubits and checks of only weight 2.
We first analyze a period-3 protocol and then a period-6 protocol which can be thought of as a rewinding version of the period-3 protocol (in the sense of Ref.~\cite{dua2023engineering}).
We find that the logicals of the protocol can be described by a color code anyon model, although the instantaneous state is not always in the same phase as a color code state.

The XYZ ruby code is defined on a system of qubits on the vertices of the ruby lattice shown in Fig.~\ref{fig:code-definition}, tessellating a compact, oriented, two-dimensional manifold.
Additionally, we require that we can three-color the interior faces, as indicated in Fig.~\ref{fig:code-definition}.
Later, we refer to these hexagonal faces as \textit{plaquettes} and the rings of 18 qubits around each hexagon as \textit{plaques}.
To each edge, we associate a two-body Pauli check: A $XX$ check to every red edge, a $YY$ check to every green edge and a $ZZ$ check to every blue edge.
The union of these checks generate the gauge group of Bombin's subsystem code~\cite{Bombin2010subsystem}.
As we will see in this section, adding a schedule on the checks reveals richer code properties than just the ones of the subsystem code.
\NEW{We refer the reader interested in the relationship between the XYZ ruby code and the topological subsystem codes from Ref.~\cite{Bombin2010subsystem} to Sec.~\ref{sec:comparison_subsystem}.}

The (period-3) XYZ ruby code is defined by the following sequence of checks
\begin{align}\label{eq:length3schedule}
    \cM = [& \raisebox{-0.3\height}{\includegraphics[height=9pt]{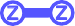}}, \raisebox{-0.3\height}{\includegraphics[height=9pt]{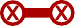}}, \raisebox{-0.3\height}{\includegraphics[height=9pt]{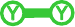}}],
\end{align}
i.e., we define a protocol where we periodically measure the blue, red and green edges in the respective bases.

\subsection{Establishing four logical qubits}\label{sec:establishing}
Having defined the measurement sequence, we now track the ISGs through multiple rounds of measurements following the stabilizer formalism with measurements~\cite{gottesman1997stabilizer, hastings2021dynamically}.
Without loss of generality, we can assume the measurement outcomes are all +1, i.e., that the measurement projects onto the +1 eigenspace of all measurements in that round, see Refs.\,\cite{bombin2023unifying, Teague2023Floquetifying}.
We start at time $t=-4$ with an arbitrary state, stabilized by $\mc{S}_{-4} = \{\mathds{1}\}$.
 
The first round of measurements adds all $ZZ$ terms to the stabilizer group,
\begin{align}
    \mc{S}_{-3} = \left\langle  \raisebox{-0.25\height}{\includegraphics[height=9pt]{ZZedge.pdf}}  \right\rangle,
\end{align}
where the set of generators is understood to be the set of all $ZZ$ operators acting on qubits connected by a blue edge.
Note that at this timestep the stabilized states are superpositions of codestates of 3-qubit repetition codes on the triangles.

The next measurements are $XX$ measurements on the red edges in Fig.~\ref{fig:code-definition}.
None of the $ZZ$ operators measured in the previous timestep commutes with any of the $XX$ operators.
However, there are extensively many elements in $\mc{S}_{-3}$ that do commute with $XX$ on the red edges.
They are Pauli $Z$ operators acting on qubits connected by \NEW{cycles composed of alternating} red and blue edges.
This includes homologically trivial cycles generated by cycles around each plaquette,
\begin{align}
    \raisebox{-0.4\height}{\includegraphics[height=55pt]{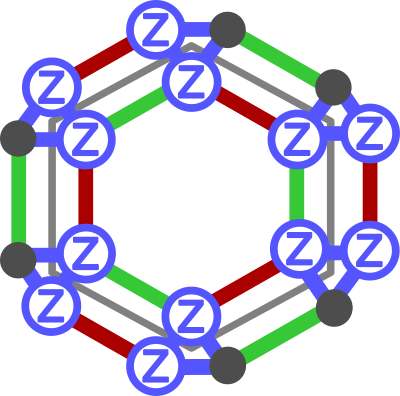}},
\end{align}
and homologically non-trivial cycles around the handles of the ambient manifold,
\begin{align}
    \raisebox{-0.275\height}{\includegraphics[width=0.28\linewidth]{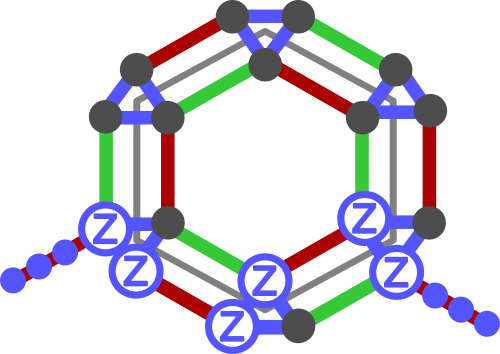}}&, \raisebox{-0.4\height}{\includegraphics[width=0.225\linewidth]{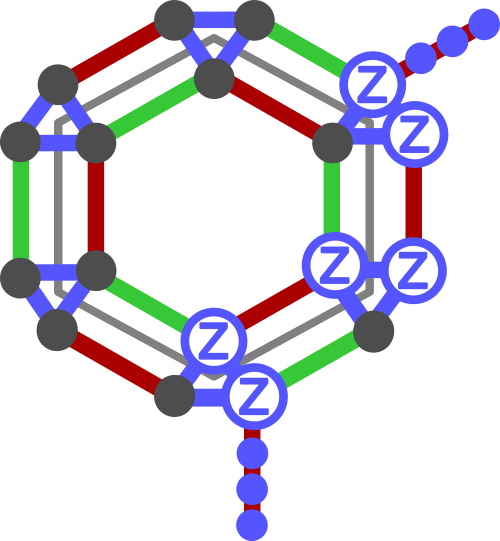}}.
\end{align}
The resulting stabilizer group at $t=-2$ is given by
\begin{align}\label{eq:ISG-2}
\begin{split}
    \mc{S}_{-2} = \Bigg\langle  \raisebox{-0.2\height}{\includegraphics[height=8pt]{XXedge.pdf}} ,  \raisebox{-0.3\height}{\includegraphics[width=0.18\linewidth]{Z_fidget.png}}&,\\
    \raisebox{-0.3\height}{\includegraphics[width=0.25\linewidth]{Z_non-local1.png}}&, \raisebox{-0.425\height}{\includegraphics[width=0.2\linewidth]{Z_non-local2.png}}  \Bigg\rangle.
\end{split}
\end{align}
Note that the generators in the first line are local and of constant weight (2 and 12), whereas the generators in the second line are supported on non-trivial loops around the torus and hence their support grows extensively with the system size.
The path along which the non-local $Z$ stabilizers act can be deformed (locally) by applying the local $Z$ stabilizers.

The next measurements, $YY$ along all green edges, again anticommute with the measurements of the previous timestep.
The product of $XX$ edges within each hexagonal plaquette commutes with all of the $YY$ measurements and hence remains in the stabilizer group.
Additionally, the product of both trivial and non-trivial $Z$-loops with the $XX$ terms along their paths acts only in the $Y$ basis and hence also commutes with the $YY$ measurements.
Taken together, the stabilizer group at $t=-1$ reads
\begin{align}
\begin{split}
    \mc{S}_{-1} = \Bigg\langle  \raisebox{-0.2\height}{\includegraphics[height=8pt]{YYedge.pdf}} ,  \raisebox{-0.4\height}{\includegraphics[width=0.18\linewidth]{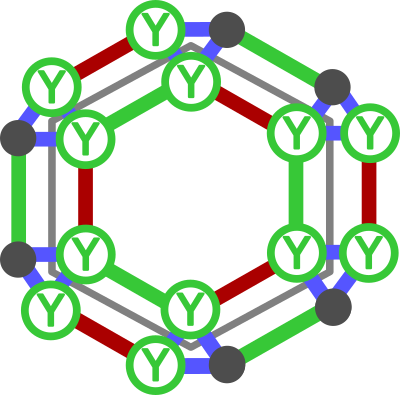}}&,  \raisebox{-0.4\height}{\includegraphics[width=0.18\linewidth]{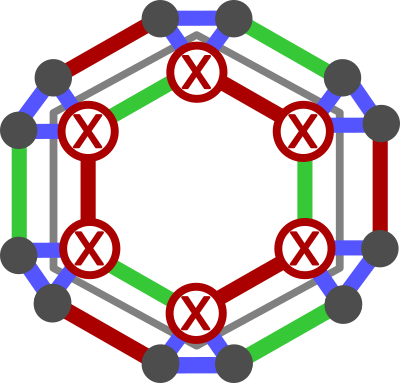}},\\
   \raisebox{-0.3\height}{\includegraphics[width=0.25\linewidth]{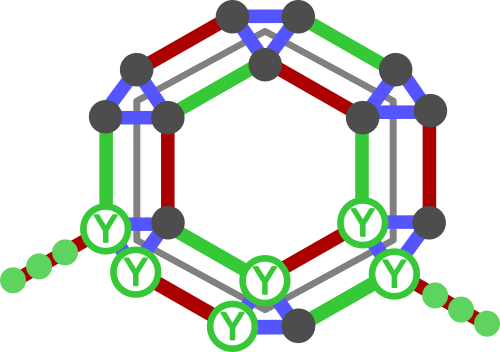}}&, \raisebox{-0.4\height}{\includegraphics[width=0.2\linewidth]{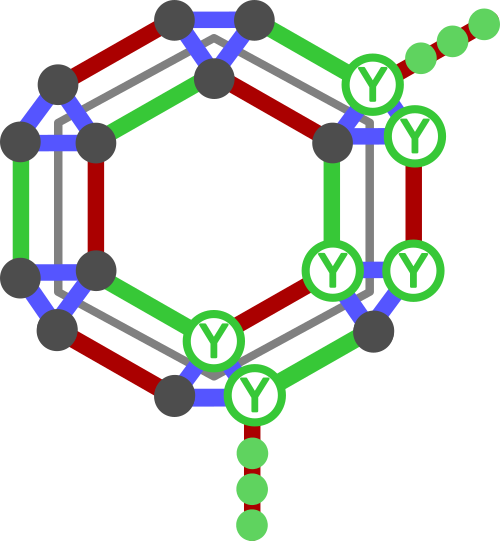}}  \Bigg\rangle.
\end{split}
\end{align}

At $t=0$, we measure $ZZ$ along the blue edges again.
None of the $YY$ generators above commute with all $ZZ$ measurements.
Moreover, none of the non-local stabilizers remain in the stabilizer group after the $ZZ$ measurements.
Both the weight-12 $Y$ loops as well as the weight-6 $X$ plaquettes can be multiplied with $YY$ edges to commute with the just-measured $ZZ$ operators.
Combined, the XYZ ruby code establishes at $t=0$, with the ISG
\begin{align} \label{eq:ISG0}
    \mc{S}_0 = \Bigg\langle \raisebox{-0.25\height}{\includegraphics[height=8pt]{ZZedge.pdf}}, \raisebox{-0.4\height}{\includegraphics[width=0.18\linewidth]{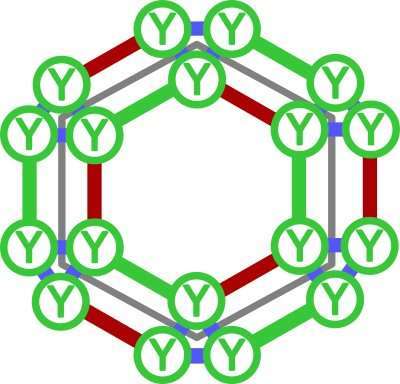}}, \raisebox{-0.4\height}{\includegraphics[width=0.18\linewidth]{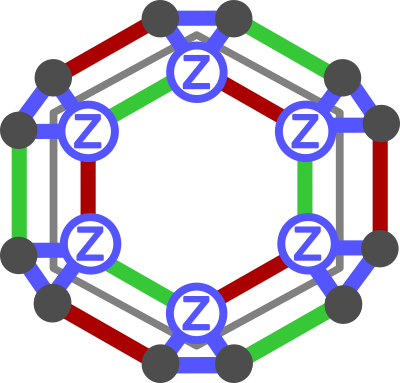}} \Bigg\rangle.
\end{align}
Note that this stabilizer group was discussed in Ref.~\cite{kargarian2010topological} as the group of local integrals of motion of the associated subsystem code Hamiltonian in one corner of its phase diagram. For a longer discussion on the connection of the XYZ ruby code to Bombin's subsystem code we refer to Sec.~\ref{sec:comparison_subsystem}.
The code defined by this stabilizer group hosts four logical qubits per handle of the manifold on which we place the ruby lattice.
To see this, we can count the number of independent generators of the stabilizer group and compare it to the Euler characteristic of the manifold.
Alternatively, we can view the stabilizer group as obtained from a concatenation procedure.
Specifically, the $ZZ$ stabilizers define a 3-qubit repetition code on each of the triangles. 
The effective qubits of the higher level code sit on the vertices of the hexagonal lattice obtained from identifying each blue triangle with a qubit.
On the effective qubits the remaining stabilizers act like the stabilizers of the hexagonal color code\,\cite{bombin2006topological} which encodes exactly four qubits per handle.

\begin{figure}
    \centering
    \includegraphics[width=\linewidth]{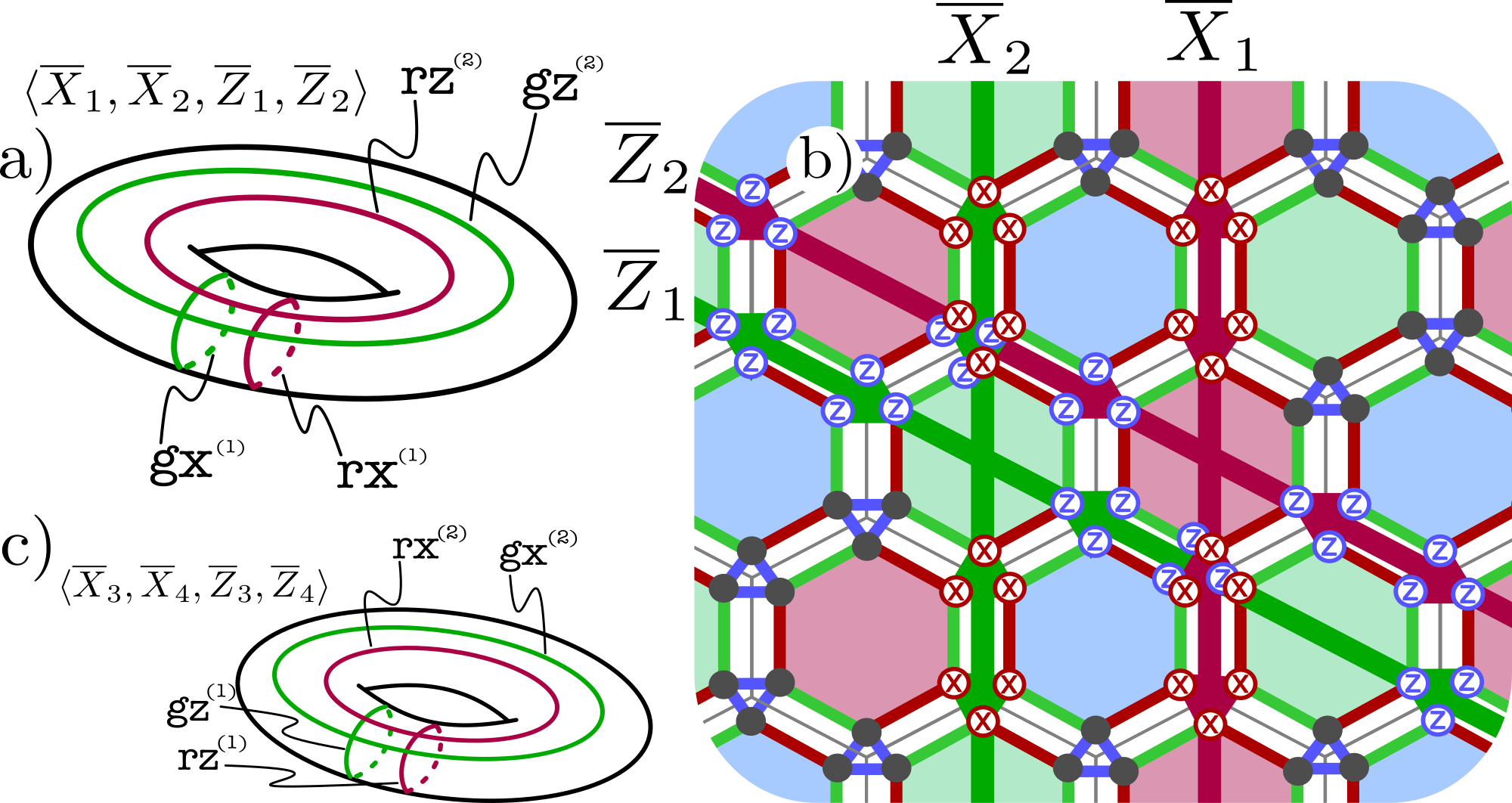}
    \caption{When defined on a torus the XYZ ruby code encodes four logical qubits. In fact, the logical Pauli operators can be understood as color code anyon string operators around homologically non-trivial cycles.
    On the torus, there are two inequivalent cycles which we label with a superscript \texttt{(1)}, respectively \texttt{(2)}, in a) and c).
    In a), we show one choice of generators for the logical Pauli operators on one pair of logical qubits. They are labeled by red and green $X$, respectively $Z$, bosonic string operators. The anyon exchange statistics ensures that they obey the necessary commutation relation \NEW{as can be seen in b) for a set of logical operators at $t=0\mod 3$. We depict that a green $Z$ string overlaps} with one of the red $X$ strings on the three qubits on a single blue triangle.
    The Pauli group on the other two logical qubits are generated by the same type of anyon string operators going around inequivalent cycles, see c).}
    \label{fig:logicals_ISG0}
\end{figure}

\subsection{Reversible stabilizer groups and logical automorphism}\label{sec:automorphism}

In this section, we track the ISGs after having established four logical qubits.
We will see that after $t=0$, the number of logical qubits stays constant even though the ISGs change.
To prove this, we construct logical operators, representing $\lpg{t}$, for any $t\geq 0$.

\begin{figure*}
    \includegraphics[width=\linewidth]{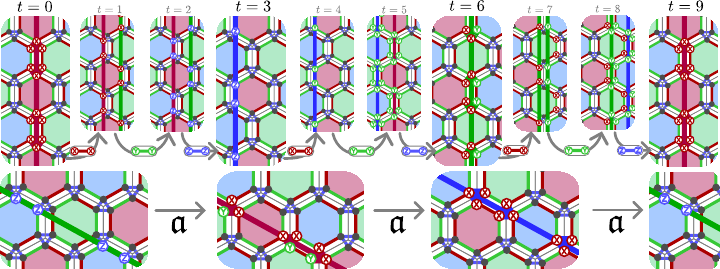}
    \caption{The period-3 XYZ ruby code implements a logical $\bZ_3$ automorphism $\mathfrak{a}$ on the logical Pauli group.
    The figure shows the evolution of a pair of anticommuting logical operators throughout three cycles of the period-3 XYZ ruby code.
    \NEW{At time $t$, for each} logical operator we show a non-trivial representative that is shared among $\cN(\cS_t)$ and $\cN(\cS_{t+1})$.
    Note that to find a logical operator at $t=2\mod 3$ that is preserved by the $ZZ$ measurement the logical operator has to be multiplied by $YY$ edges as well as by non-local $Y$ strings in the the stabilizer group $\cS_2$.
    For the anticommuting logical operator we only show the representative at timesteps $t=0\mod 3$.
    Picking a basis of logical operators (in which the left-most representatives are identified with $\overline{X}_1$ and $\overline{Z}_1$) for the times $t=0\mod 3$ we can understand the action of the automorphism $\mathfrak{a}$ on that pair of logical Pauli operators as $\overline{X}_1\mapsto\overline{Z}_3\overline{Z}_4\mapsto\overline{X}_2\overline{Z}_4\mapsto \overline{X}_1$ and $\overline{Z}_1\mapsto\overline{Z}_2\overline{X}_4\mapsto\overline{X}_3\overline{X}_4\mapsto \overline{Z}_1$.
    Note that we show different representatives for the logical operators at $t=0\mod 3$ compared to Fig.~\ref{fig:logicals_ISG0}.
    How the logical operators transform from one timestep to the next can easily be obtained from the logical flows associated to the logical Paulis, see e.g. Fig.~\ref{fig:logical_flows} for two exemplary logical flows.}
    \label{fig:logical_autmorphism}
\end{figure*}

At $t=0$, after the second $ZZ$ measurement, we can use the logical string operators of the color code to construct representatives of the logical operator classes $\mc{N}(\mc{S}_0)/\mc{S}_0$.
Specifically, the color code admits a basis of logical operators that can be associated to anyonic string operators (cf. Sec.~\ref{sec:prelim-anyons}).
For each handle there are 4 independent pairs of string operators that act like pairs of Pauli $X$ and $Z$ operators on distinct encoded qubits, see Fig.~\ref{fig:logicals_ISG0}.
For example, we can take a non-contractible string of $X$ operators along red edges to represent a logical $X$ operator.
We can represent the corresponding \NEW{anticommuting logical $Z$ operator} by an operator acting with $Z$ operators on green edges \NEW{(connecting green plaquettes) of the effective color code lattice} along a non-contractible loop around the same handle but in the orthogonal direction.
By construction, their support overlaps on an odd number of qubits and the two operators form an anticommuting pair of logical operators that square to one, i.e., generate a logical Pauli group.
Exchanging red and blue in the construction above yields an additional pair of logical Pauli operators.
Finally, we can exchange $X$ and $Z$ in both pairs of operators resulting in a total of four pairs of independent logical Pauli operators per handle of the ambient manifold.
In fact, these are all of the logical Pauli operators for any color code on an orientable manifold (without boundaries).
The physical representatives of the logicals of $\cS_0$ are obtained by replacing each Pauli operator along the non-contractible loop with the logical Pauli operator of the 3-qubit repetition code on the blue triangles.
In Fig.~\ref{fig:logical_autmorphism}, we show a subset of logical representatives, including the ones at $t=0$.

At $t=1$, we measure $XX$ on every red edge.
This updates the stabilizers to form another ISG,
\begin{align}\label{eq:ISG1}
\begin{split}
    \mc{S}_1 = \Bigg\langle \raisebox{-0.25\height}{\includegraphics[height=8pt]{XXedge.pdf}}&, \raisebox{-0.4\height}{\includegraphics[width=0.18\linewidth]{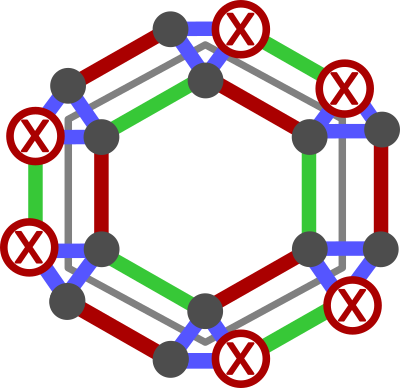}}, \raisebox{-0.4\height}{\includegraphics[width=0.18\linewidth]{Z_plaquette.png}},\\
    \raisebox{-0.3\height}{\includegraphics[width=0.18\linewidth]{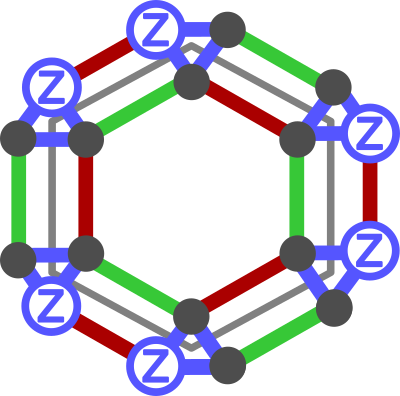}}&,
    \raisebox{-0.3\height}{\includegraphics[width=0.25\linewidth]{Z_non-local1.png}}, \raisebox{-0.425\height}{\includegraphics[width=0.2\linewidth]{Z_non-local2.png}}  \Bigg\rangle,
\end{split}
\end{align}
\NEW{Note that $\cS_1$ needs non-local generators which is not the case for $\cS_0$. This is indicative for a phase transition between states in $\cS_0$ and $\cS_1$. We will later show that, indeed, $\cS_1$ is in the same phase as three toric codes, see Sec.~\ref{sec:phase-transition}.}
$\cS_{1}$ has \NEW{a greater rank} than $\cS_{-2}$ (where we measured $XX$ along red edges the last time) and the same rank as $\cS_0$.
When counting the number of independent generators to determine the rank one has to take into account global relations amongst the generators.
For example, the product of all weight-6 $X$ generators is the same as the product of all $XX$ on red edges.
As a consequence, the logical information encoded in the stabilized subspace of $\cS_0$ is preserved in the subspace stabilized by $\cS_1$.
This can be shown by giving a representative logical Pauli operator that is shared among $\cN(\cS_0)$ and $\cN(\cS_1)$, as shown in Fig.~\ref{fig:logical_autmorphism} for one pair of logical operators.
In fact, we can find such representatives by identifying corresponding logical operators of the lower-level repetition code on the blue triangles.
All the logical $X$ strings, for example, can be represented by $X^{\otimes 3}$ on each triangle which commute trivially with the $XX$ measurements.
For the $Z$ strings, we have to find logical $Z$ representatives of the lower-level code that commute with the $XX$ measurements.
For that, note that all of the $Z$ strings go along edges of a given color (in the higher-level color code).
For each such edge, we can find exactly one red ($XX$) edge in the ruby lattice running perpendicular to it.
It connects two physical qubits of two different triangles. The product of a Pauli $Z$ on both of these qubits implements a logical $Z\otimes Z$ on the logical qubits associated to the two triangles and commutes with the $XX$ measurement along that red edge.
We can combine these $ZZ$ operators along a non-contractible path connecting plaquettes of the same color.
This gives a set of logical operators for both $\cS_0$ and $\cS_1$.
Together with the logicals composed of Pauli $X$ operators, they form a complete set of representatives of the logical Pauli group of both stabilizer groups.
This shows that $\cS_0$ and $\cS_1$ form a \textit{reversible pair} of stabilizer groups.
Importantly, however, since $\cS_1$ needs non-local generators and $\cS_0$ does not, they do not form a \textit{locally} reversible pair.\footnote{Given a pair of locally reversible stabilizer groups, there exists a locality preserving unitary mapping between them, see Ref.~\cite{aasen2023measurement}. Here, we have two reversible stabilizer groups where one is locally generated and the other one is not. Any unitary mapping between them cannot be locality preserving, contradicting local reversibility.}
This manifests itself by the fact that one can view the ISGs as changing their topological phase during the measurement sequence.
We will elaborate on this further in Sec.~\ref{sec:phase-transition}.

At $t=2$, we measure $YY$ along every green edge.
This updates the stabilizer group to
\begin{align}\label{eq:ISG2}
\begin{split}
    \mc{S}_2 = \Bigg\langle \raisebox{-0.25\height}{\includegraphics[height=8pt]{YYedge.pdf}}&, \raisebox{-0.4\height}{\includegraphics[width=0.18\linewidth]{X_windmill.png}}, \raisebox{-0.4\height}{\includegraphics[width=0.18\linewidth]{Z_plaquette.png}},\\
    \raisebox{-0.3\height}{\includegraphics[width=0.18\linewidth]{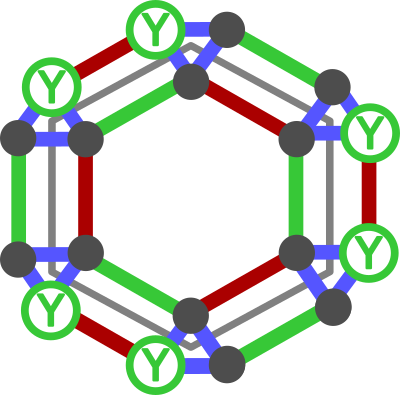}}&,
    \raisebox{-0.3\height}{\includegraphics[width=0.25\linewidth]{Y_non-local1.png}}, \raisebox{-0.4\height}{\includegraphics[width=0.2\linewidth]{Y_non-local2.png}}  \Bigg\rangle.
\end{split}
\end{align}
Again, we can find a logical representative shared amongst $\cN(\cS_1)$ and $\cN(\cS_2)$ for each logical coset.
Starting from the representatives from before, inherited from the $t=0$ round, we have to multiply $XX$ stabilizers along the red edges to obtain the representatives that commute with the $YY$ measurements.
In Fig.~\ref{fig:logical_autmorphism} we show the transformation of a pair of anticommuting logical string operators throughout three measurement cycles.
All other logical string operators are of the same form but act along paths connecting different colors or along a homologically inequivalent loop.
Again, finding a logical representative shared among both $\cN(\cS_1)$ and $\cN(\cS_2)$ for each logical class shows that $\cS_1$ and $\cS_2$ form a reversible pair of stabilizer groups. 

At $t=3$, we measure $ZZ$ again which maps the ISG back to $\mc{S}_0$ and can find logical operators that are shared amongst $\cN(\cS_2)$ and $\cN(\cS_0)$ for each logical coset.
This shows that the ISGs define a reversible sequence of stabilizer groups,
\begin{align}\label{eq:period3-stabilizer-diagram}
\raisebox{-0.5\height}{
\begin{tikzpicture}
    \node (A) at (0,0) {$\mc{S}_0$};
    \node (B) at (-120:2) {$\mc{S}_{1}$};
    \node (C) at (-60:2) {$\mc{S}_{2}$};
    \node[anchor=east] at (-120:1) {$\mathfrak{m}_0$};
    \node[anchor=north] at (-90:1.75) {$\mathfrak{m}_{1}$}; 
    \node[anchor=west] at (-60:1) {$\mathfrak{m}_{2}$}; 
    \draw[->, thick] (A) -- (B);
    \draw[->, thick] (B) -- (C);
    \draw[->, thick] (C) -- (A);
\end{tikzpicture}}.
\end{align}
After one period, the code gets mapped back to itself.
As such, the measurement sequence defines a \textit{logical automorphism} $\mathfrak{a}_t$ on $\cN(\cS_t)/\cS_t$ for each $t$, captured by the following diagram:
\begin{align}
\raisebox{-0.5\height}{
\begin{tikzpicture}
    \node (A) at (2.75,0) {$\mc{N}(\mc{S}_t)/\mc{S}_t$};
    \node (B) at (0.7,-1.5) {$\mc{N}(\mc{S}_{t+1})/\mc{S}_{t+1}$};
    \node (C) at (4.5,-1.5) {$\mc{N}(\mc{S}_{t+2})/\mc{S}_{t+2}$};
    \node[align=center] at (2.75,1.05) {$\mathfrak{a}_t$}; 
    \node[align=center] at (-0.5,-2.4) {$\mathfrak{a}_{t+1}$}; 
    \node[align=center] at (5.7,-2.4) {$\mathfrak{a}_{t+2}$}; 
    \node at (1.25,-0.75) {$\mathfrak{m}_t$};
    \node at (2.75,-1.25) {$\mathfrak{m}_{t+1}$}; 
    \node at (4.25,-0.75) {$\mathfrak{m}_{t+2}$}; 
    \draw[->, thick] (A) -- (B);
    \draw[->, thick] (B) -- (C);
    \draw[->, thick] (C) -- (A);
    \draw[->, thick] (3,0.25) arc (-45:225:0.35);
    \draw[->, thick] (-0.5,-1.5) arc (100:380:0.35);
    \draw[->, thick] (5.35,-1.85) arc (200:470:0.35);
\end{tikzpicture}}.
\end{align}
Each $\mathfrak{a}_t$ itself is a Clifford unitary on the respective codespace that is implemented by a full period of the measurement cycle.
To evaluate the logical automorphism we can pick any reference ISG and track how the logical operators evolve through a cycle of measurements.
Let us pick $\cS_0$ which describes a topological code with an anyon model equivalent to that of the two-dimensional color code.
A single cycle of measurements implements an automorphism on the logical Pauli group of that code which can equivalently be understood as an automorphism of its anyon model.
Ref.~\cite{Kesselring2018boundariestwist} classified all the automorphisms of the color code anyon model.
In particular, any automorphism is uniquely defined by the action it has on the color code boson table, see Eq.\,\eqref{eq:boson_table}.
Tracking the logical string operators through one cycle of measurements we find that the XYZ ruby code implements a shift along the diagonal of that table.
In particular, the automorphism has order 3, which can be seen in Fig.~\ref{fig:logical_autmorphism}, where we show the evolution of one pair of logical string operators through three cycles of measurements.
Specifically, the figure shows the transformation
\begin{subequations}
\begin{align}
\rx\rightarrow \bz\rightarrow \gy\rightarrow \rx\rightarrow \cdots &\qq{and}\\
\gz\rightarrow\ry\rightarrow\bx\rightarrow\gz\rightarrow\cdots&.
\end{align}    
\end{subequations}
Graphically, we depict the automorphism applied on the color code anyons within one period with
\begin{align}\label{eq:automorphism_table}
    \raisebox{-0.45 \height}{\includegraphics[width=0.4\linewidth]{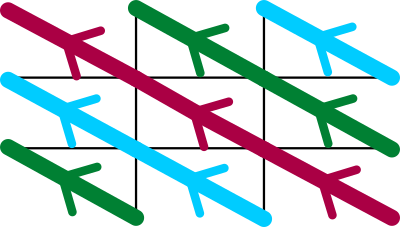}} = \varphi \in \Aut(A_{CC}).
\end{align}
On a subset of logical Pauli operators it acts like
\begin{subequations}
\begin{align}
    \overline{X}_1 &\mapsto\overline{Z}_3\overline{Z}_4 \mapsto\overline{X}_2\overline{Z}_4 \mapsto \overline{X}_1 \mapsto \cdots \qq{and}\\
    \overline{Z}_1 &\mapsto\overline{Z}_2\overline{X}_4 \mapsto\overline{X}_3\overline{X}_4 \mapsto \overline{Z}_1 \mapsto \cdots ,
\end{align}    
\end{subequations}
as we show explicitly in Fig.~\ref{fig:logical_autmorphism}. The explicit labeling of the above logical operators is an arbitrary choice, of course.
However, the automorphism necessarily preserves the commutation relation amongst any pair of logicals, independent of the choice of basis.
\NEW{In a different basis, for example, we would find that some logical operators are ``static'' operators in the protocol.
Consider a logical operator at $t=0\mod 3$ that is associated to moving one of the fermions $\texttt{f}_1$, $\texttt{f}_2$ or $\texttt{f}_3$ as defined in Eq.\,\eqref{eq:colorcode_fermions} around a homologically non-trivial cycle.
The fermions are products of bosons in the table above that lie within the same orbit of the automorphism. In fact, the automorphism simply exchanges the two constitutent bosons and the composite (fermion) is left invariant by the automorphism.
Indeed, we find that the associated logical operators are shared by all of the ISGs.
We discuss this in the context of the associated subsystem code further in Sec.~\ref{sec:comparison_subsystem}.
}
We note that an isomorphic automorphism was found in other schedules on the same set of checks in Ref.~\cite{dua2023engineering}.

\subsection{Instantaneous topological phase transitions}\label{sec:phase-transition}
We have seen that the ISG $\cS_0$ is in the same phase as the two-dimensional color code which helped us significantly in finding the logical operators and their transformations throughout the measurement sequence of the XYZ ruby code.
In this section, we will analyze the other two ISGs, $\cS_1$ and $\cS_2$, with respect to their topological properties.
In particular, we consider the locally generated part of the ISGs as defining the topological phase of the stabilizer group and interpret the non-local stabilizers as additional constraints on the codespace of the topological code defined by the locally generated one.
Interestingly, we find that the code undergoes an \textit{instantaneous phase transition}\footnote{We want to note that the same phenomenon in the 3D X-cube Floquet code was coined ``splitting'' in Ref.~\cite{dua2023engineering}.} from the color code phase into three copies of the toric code phase.
A similar behaviour was observed in Ref.~\cite{zhang2022xcube}, where the ISGs undergo a transition from an X-cube model to layers of the $2+1$-dimensional toric code.
To the best of our knowledge the XYZ ruby code is the first two-dimensional code appearing in the literature that features such an instantaneous phase transition.

\begin{figure}
    \centering
    \includegraphics[width=\linewidth]{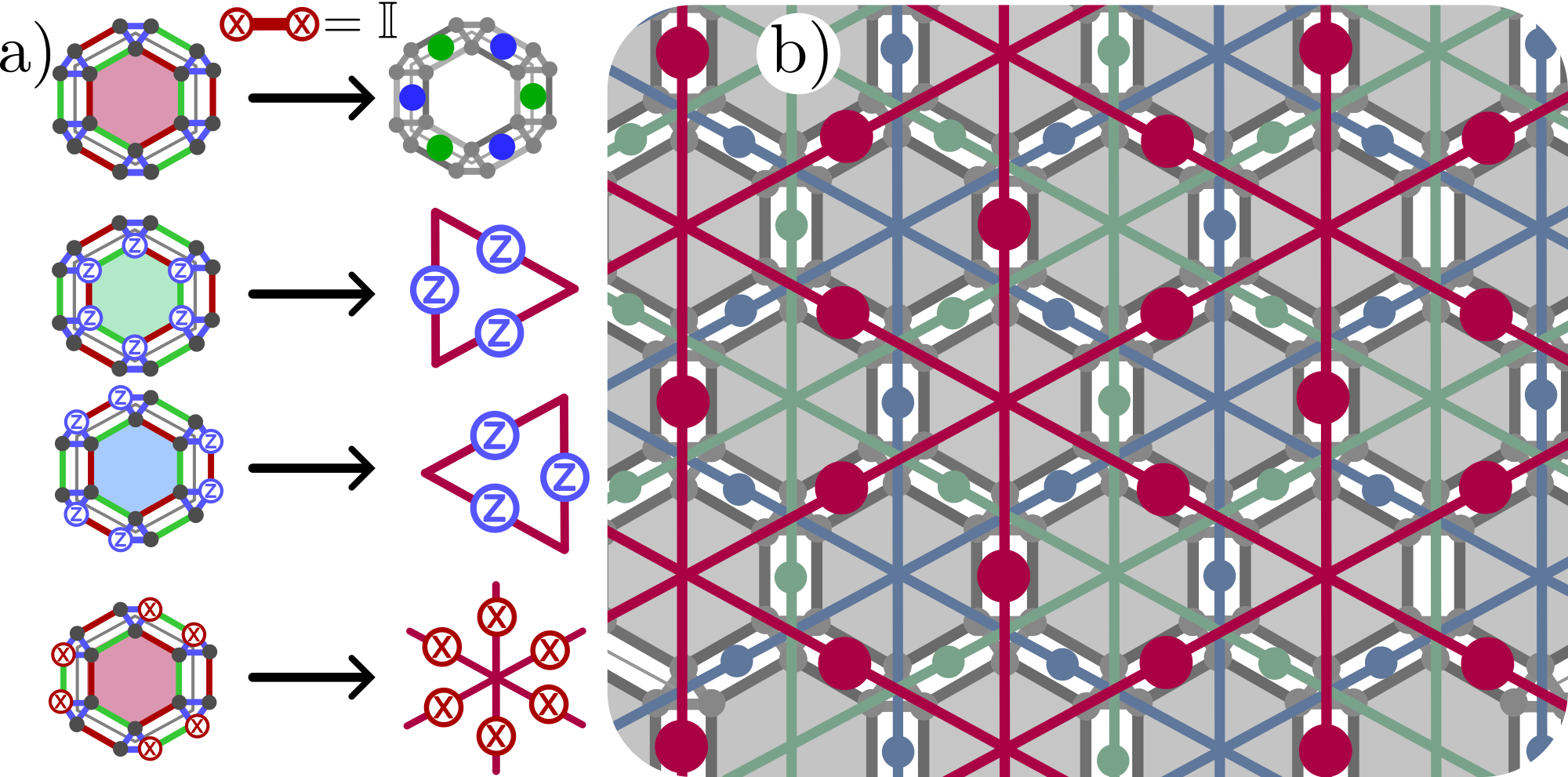}
    \caption{At $t=1\mod 3$ the ISG (after establishing) can be mapped to the one of three copies of toric codes on triangular lattices by a local isometry.
    a) This can be seen explicitly by interpreting the $XX$ stabilizers on the red edges as stabilizers of a lower-level $[2,1,1]$ repetition code.
    \NEW{This defines an effective qubit on each red edge of the ruby lattice which can also be viewed as associating one qubit to each edge of the underlying honeycomb lattice.
    The other stabilizer generators of $\cS_1$ are mapped onto weight-3 $Z$- 
    and weight-6 $X$ stabilizers on the effective qubits. The stabilizer generators can be grouped to act on three non-overlapping subsets of qubits: the ones associated to the red, green and blue edges of the underlying honeycomb lattice.
    We exemplify how the stabilizers act on the effective qubits by showing generators of $\cS_1$ that effectively act only on the qubits associated to red edges.}
    b)
    On each of the three \NEW{subsets}, the effective stabilizers reduce to the ones of a toric code on a triangular lattice, cf. Fig. 1.
    \NEW{We draw the lattices for the effective toric codes overlapping in three different colors.
    For better readability we highlight the lattice on which the effective red toric code is defined.
    The lattices defining the other two toric codes differ only by translation, with vertices placed in the center of plaquettes of a different color.}
    The same mapping can be performed for $\cS_2$, where the basis of the lower-level stabilizers changes from $X$ to $Y$.
    }
    \label{fig:ISG1-concatenation}
\end{figure}

Similar to $\cS_0$, we can view $\cS_1$ and $\cS_2$ as stabilizer groups of a concatenated code, where the $XX$, respectively $YY$, checks are stabilizers of a lower-level $[[2,1,1]]$ code.
Viewed in this way, the other stabilizer generators act as higher-level stabilizers as a product of logical operators on these smaller codes.
For both $\cS_1$ and $\cS_2$ the higher-level code is generated by weight-3 and weight-6 stabilizers, see Fig.~\ref{fig:ISG1-concatenation}.
Additionally, we find that the local part of the stabilizer groups can be decomposed into three factors, each of which acts on a different subset of qubits.
The three subsets can be associated to three colors, red, green and blue, following the colorings of the \NEW{edges of the underlying hexagonal lattice, cf. Fig.~\ref{fig:colorcode_anyons}.}
On each subset, the stabilizer group reduces to the stabilizer 
group of a toric code on a triangular lattice 
(see Fig.~\ref{fig:ISG1-concatenation}).
Since there are no additional local stabilizers, 
we conclude that the topological order of states stabilized by $\cS_1$ and $\cS_2$ is that of three copies of the toric code.
Specifically, the code undergoes an instantaneous phase transition from the color code to three copies of the toric code when the $XX$ or $YY$ checks are measured on a state stabilized by $\cS_0$.
Similarly, the reverse transition is induced by measuring the $ZZ$ checks on $\cS_1$ or $\cS_2$.
The additional non-local stabilizers can be identified with anyon string operators along non-trivial cycles and implement a non-local constraint on the codespace.
In fact, the string operators act non-trivially on all three copies of effective toric codes and can be viewed as moving the electric charges ($e$ anyons) of all three copies jointly around a non-trivial cycle.
As such, the non-local stabilizers can be viewed as logical operators of the three toric codes and it being in the stabilizer group enforces a parity constraint on their logical qubits.\footnote{For one choice of logical basis we can view the non-local string operators as logical operators $\overline{Z}_i\overline{Z}_j\overline{Z}_k$ for some triples of logical qubits $(i,j,k)$.}
This explains how the logical dimension remains constant even though the number of anyons in the code changes periodically in time.

\subsection{Rewinding schedule}
In this section, we explain a \textit{rewinding} schedule based on the schedule discussed so far in Sec.~\ref{sec:rubycode_definition}.
The idea of a rewinding schedule was originally introduced in Ref.~\cite{haah2022boundaries} to define boundaries for the honeycomb code, a Floquet code that applies a non-trivial automorphism after one length-3 period.
Ref.~\cite{dua2023engineering} formalized this concept and applied it to other Floquet codes in two and three spatial dimensions to engineer transitions between desired ISGs.
A rewinding schedule can in general help to construct boundaries, logical gates or other desired features of the ISG.

A Floquet code is considered \textit{rewinding} if the measurement sequence $\cM$ (see Eq.\,\eqref{eq:measurement_sequence}) is symmetric with respect to time-reversal, i.e., is of the form \NEW{
\begin{align}
    [\cM_0,\cM_1,  \dots, \cM_{k-1}, \cM_{k-1}, \dots  ,\cM_1,\cM_0]
\end{align}}
for some \NEW{$k\in\bZ_{> 0}$}~\cite{dua2023engineering}. \NEW{Note that in this definition both $\cM_{k-1}$ and $\cM_0$ are performed twice in a row. In either case, the second of the two measurements will not have any effect on the ISG or logical operators at that time, besides potentially changing signs of some stabilizers if an error occurs. Hence, we also refer to a schedule in which these measurements are not repeated as rewinding.}

For the XYZ ruby code we analyze a rewinding schedule defined by the period-6 measurement sequence
\begin{align}\label{eq:length6schedule}
\begin{split}
    \cM_r = [
    \raisebox{-0.25\height}{\includegraphics[height=8pt]{ZZedge.pdf}}, 
    \raisebox{-0.25\height}{\includegraphics[height=8pt]{XXedge.pdf}},
        \raisebox{-0.25\height}{\includegraphics[height=8pt]{YYedge.pdf}},\\
    \:\raisebox{-0.25\height}{\includegraphics[height=8pt]{ZZedge.pdf}},
    \raisebox{-0.25\height}{\includegraphics[height=8pt]{YYedge.pdf}},
    \raisebox{-0.25\height}{\includegraphics[height=8pt]{XXedge.pdf}}].
\end{split}
\end{align}
Note that in this rewinding schedule there are two types of $Z$ measurements: One sandwiched by $XX$ measurements and one sandwiched by $YY$ measurements.
We pick this rewinding schedule since the second $ZZ$ measurement is needed to establish an ISG in the color code phase, cf. Sec.~\ref{sec:establishing}.

In the following, we describe the ISGs in the rewinding schedule.
Again, consider starting with the first round of $ZZ$ measurements at \NEW{$t=-3$}.
Since the measurement rounds at $t=-3, -2,-1$ and $0$ agree with the non-rewinding schedule (see Eq.\,\eqref{eq:length3schedule}), 
we find that the rewinding schedule also establishes four logical qubits in ISG $\cS_0$ just after the second round of $ZZ$ measurements.
At $t=1$, $YY$ is measured along the green edges which drives the system into a new ISG, which we denote by $\cS_1^r$.
First note that the $Y$ stabilizers all remain in the stabilizer group and so do all the $Z^{\otimes 6}$ stabilizers around the inner plaquettes (see Eq.\,\eqref{eq:ISG0}). Although the individual $ZZ$ stabilizers anticommute with the $YY$ measurement operators, certain combinations commute with them.
Specifically, $ZZ$ operators acting on qubits connected along a loop of green and blue edges remain in the stabilizer group after the $YY$ measurement.
These loops naturally fall into homology classes, so we sort the stabilizers associated to these loops into the same classes.
The homologically trivial stabilizers are all generated from weight-6 operators forming a loop around a single plaquette,
\begin{align}
    \raisebox{-0.4\height}{\includegraphics[width=0.2\linewidth]{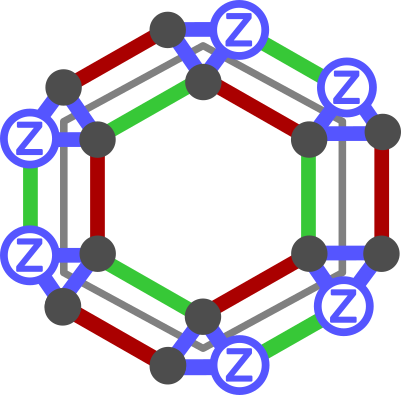}}.
\end{align}
Additionally, there is a stabilizer of extensive weight for each homologically non-trivial loop.
We hence add one generator for each homology class obtained from a product of $ZZ$ operators along a representative loop along green and blue edges.
Taken together, the system is driven into a code stabilized by
\begin{align}\label{eq:ISG_r_1}
\begin{split}
    \cS_1^r = \Bigg\langle \raisebox{-0.25\height}{\includegraphics[height=8pt]{YYedge.pdf}}&, \raisebox{-0.4\height}{\includegraphics[width=0.18\linewidth]{Y_windmill.png}}, 
    \raisebox{-0.4\height}{\includegraphics[width=0.18\linewidth]{Z_windmill_rot.png}},\\
    \raisebox{-0.3\height}{\includegraphics[width=0.18\linewidth]{Z_plaquette.png}}&,
    \raisebox{-0.3\height}{\includegraphics[width=0.25\linewidth]{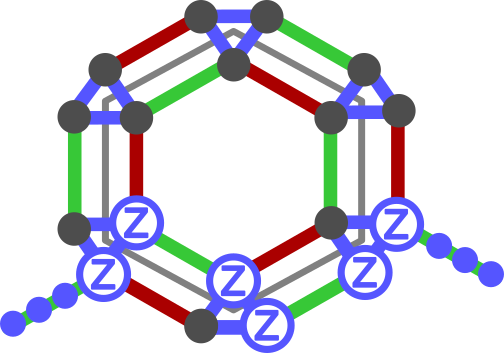}}, \raisebox{-0.4\height}{\includegraphics[width=0.2\linewidth]{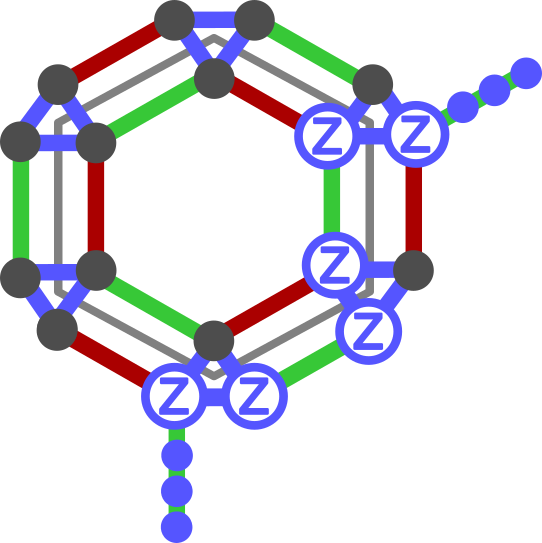}}  \Bigg\rangle.
\end{split}
\end{align}
Note that $\cS_1^r$ is related to $\cS_1$ \NEW{(Eq.\,\ref{eq:ISG1})} by a lattice rotation of $\pi/3$ together with a product of on-site \NEW{unitary} $(X+Y)/\sqrt{2}$\NEW{, swapping $X$ and $Y$ bases}.

At $t=2$, $XX$ operators along red edges are measured.
With similar arguments as for the transition $\cS_1\to\cS_2$ in the non-rewinding schedule we see that the $XX$ measurement at $t=2$ in the rewinding schedule drives the system into a code stabilized by
\begin{align}\label{eq:ISG_r_2}
\begin{split}
    \mc{S}_2^r = \Bigg\langle \raisebox{-0.25\height}{\includegraphics[height=8pt]{XXedge.pdf}}&,
    \raisebox{-0.4\height}{\includegraphics[width=0.18\linewidth]{Y_windmill.png}}, 
    \raisebox{-0.4\height}{\includegraphics[width=0.18\linewidth]{X_windmill.png}},\\
    \raisebox{-0.3\height}{\includegraphics[width=0.18\linewidth]{Z_plaquette.png}}&,
    \raisebox{-0.3\height}{\includegraphics[width=0.25\linewidth]{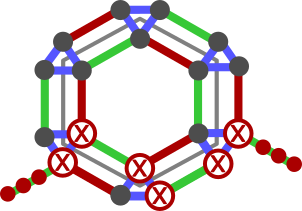}}, \raisebox{-0.4\height}{\includegraphics[width=0.2\linewidth]{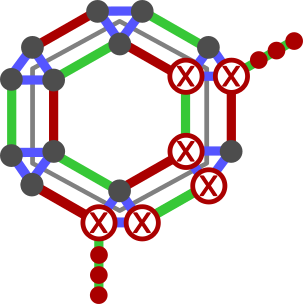}}  \Bigg\rangle.
\end{split}
\end{align}
In particular, the $Y^{\otimes 6}$ generators and the $Z^{\otimes 6}$ generators acting on the inner plaquettes remain in the stabilizer group and the $ZZ$ loops along green and blue edges have to be multiplied with $YY$ stabilizers to commute with the $XX$ measurement operators.
We again find that rotating the lattice by $\pi/3$ and applying $(X+Y)/\sqrt{2}$ to every qubit maps $\cS_2^r$ to $\cS_2$ \NEW{(Eq.\,(\ref{eq:ISG2}))}.

At $t=3$, the $ZZ$ measurements along the blue edges are repeated and we find that the system is driven back to a state stabilized by $\cS_0$.
From there, $XX$ operators along red edges are measured, followed by $YY$ measurements along the green edges, performing the transformation of the non-rewinding schedule until $t=6=0 \mod 6$.
To conclude, after establishing that the stabilizer groups undergo a periodic transformation,
\begin{align}
\raisebox{-0.4\height}{
\begin{tikzpicture}
    \node (A) at ({4/3},0) {$\mc{S}_0$};
    \node (B) at ({2/3},{sqrt(3)*2/3}) {$\mc{S}_{1}$};
    \node (C) at ({-2/3},{sqrt(3)*2/3}) {$\mc{S}_{2}$};
    \node (D) at ({-4/3},0) {$\mc{S}_{0}$};
    \node (E) at ({-2/3},-{sqrt(3)*2/3}) {$\mc{S}_{1}^r$};
    \node (F) at ({2/3},-{sqrt(3)*2/3}) {$\mc{S}_{2}^r$};
    \node[anchor=south east] at (150:1) {$\mathfrak{m}_{2}$}; 
    \node[anchor=north east] at (-150:1) {$\mathfrak{m}_{0}$}; 
    \node[anchor=north] at (-90:1.1) {$\mathfrak{m}_{1}$}; 
    \node[anchor=north west] at (-30:1) {$\mathfrak{m}_{2}$}; 
    \node[anchor=south west] at (30:1) {$\mathfrak{m}_{1}$}; 
    \node[anchor=south] at (90:1.1) {$\mathfrak{m}_{0}$}; 
    \draw[->, thick] (A) -- (B);
    \draw[->, thick] (B) -- (C);
    \draw[->, thick] (C) -- (D);
    \draw[->, thick] (D) -- (E);
    \draw[->, thick] (E) -- (F);
    \draw[->, thick] (F) -- (A);
\end{tikzpicture}},
\end{align}
we see that we encounter a length-6 rewinding schedule.
Note that the same measurements are performed as in the period-3 schedule just in a different order, see Eq.\,\eqref{eq:period3-stabilizer-diagram}.
Since all measurements in the above diagrams are reversible we can infer from the mirror-symmetry of the above diagram that the sequence implements a trivial automorphism on the encoded logical Pauli algebra~\cite{aasen2023measurement, dua2023engineering}.
We can track representatives of the logical operators through a full period of measurements in the rewinding schedule and verify that the cosets of logical operators are mapped back onto themselves, i.e., the rewinding schedule implements a trivial automorphism on the logical Pauli group.
The explicit transformation of logical operators can be deduced from the logical flow as we will see later.

We find that after three timesteps the code is also mapped back onto itself and the logical operators undergo the non-trivial automorphism of the period-3 schedule. Due to the rewinding, the second half of the period implements the inverse of that automorphism rendering the action of the full period of measurements on the logical Pauli group trivial.
For the rest of this work, 
we will investigate the XYZ ruby code with a rewinding schedule.

\section{Three-colored graphical calculus for QEC} \label{sec:prelim-3cgc}

In this section, we introduce the concept of a three-colored graphical calculus for quantum error-correcting protocols defined by circuits composed of Pauli measurements and Clifford operations.
In the following we write ``circuit'' for short since we only consider these types of circuits for now. In principle, one can analyze any such protocol with the usual ZX-calculus, as shown for example in Refs.\,\cite{bombin2023unifying, van2020zx}.
For protocols involving $Y$ measurements, however, the formalism becomes less elegant and the mapping between a ZX-diagram and circuit substantially more tedious.
It is desirable to work with a representation that is closer to the ``native'' operations considered in an QEC protocol.

In the ZX-calculus the elementary building blocks are two types of tensors, also called \textit{spiders}, usually represented by circles of two different colors.
We propose an extension to the graphical representation of circuits with spiders of a third color.
Since tensors of three different colors -- red, green and blue -- will play a central role, we call such a tensor network an \textit{RGB tensor network} (RGB TN).
Additionally, we extend the concept of \textit{Pauli flow}\footnote{A similar concept has been defined as ``stabilizer flow'' in Ref.~\cite{McEwen_2023}, ``Pauli webs'' in 
Ref.~\cite{bombin2023unifying} or identified with stabilizers of tensors in Refs.~\cite{Raussendorf2019Computationally, Cao2022Lego}.} to that setting and rigorously define all algebraic quantities entering a quantum error correction analysis from these flows alone. This gives rise to a purely graphical description of all the quantities needed to perform error correction on a circuit. 

We would like to note that there have been other proposals to add a third color to the ZX-calculus, specifically Refs.\,\cite{yeung2020diagrammatic, Lang_2012}.
From our perspective, both approaches lack some properties that we find desirable for a representation of circuits, independent of their geometric structure.
Ref.~\cite{yeung2020diagrammatic} resorts to a two-dimensional ambient space when defining the spiders.
In particular, the formalism only allows for left- and right-pointing legs attached to individual tensors.
In light of the fact that we want to consider spacetimes of higher dimensions than two and we can interpret the time direction freely in these protocols~\cite{Teague2023Floquetifying}, we want to avoid such choices.
This is resolved in Ref.~\cite{Lang_2012} with a graphical calculus that is defined independently of its ambient space but adds orientations to the edges.
The authors' goal was to obtain a calculus that is fully symmetric among the three colors.
In order to achieve this they have to change one of the two tensors of the ZX-calculus.
In our context this is highly undesirable since simple circuit elements, such as the CNOT gate, are nicely expressed in terms of original ZX-tensors.
We aim to fix both problems with minor additions to the Clifford ZX-calculus by only adding a third type of tensor.
To avoid the dependency on some ambient space, we will allow for edges attaching to that third tensor to be ``flipped" without the need of adding orientations to all edges or tensors.

\subsection{Building blocks}
The idea behind the graphical calculus is to represent the circuit as a tensor network which can be assembled using simple elementary building blocks.
The goal of this section is to first introduce the tensors from which we build up the tensor network representing any circuit.
All tensors will be considered over $\bC^2$, i.e., any ``wire'' will be associated with a two-dimensional complex vector space. As such, it carries a computational basis which we write as $\{\ket{0}, \ket{1}\}$ and identify it with the two orthogonal eigenstates of the Pauli $Z$ matrix.

The key building blocks will be three types of tensors. To each type, we associate a color: blue, red or green. For each color, we define two tensors, labeled by $s\in\{0,1\}$.
In the following, we define each type of tensor in terms of its matrix elements in a computational basis.
We define a \textit{blue $s$-spider} by
    \begin{align}\label{eq:bluetensor}
    \begin{split}
        \raisebox{-0.4\height}{\includegraphics[width=0.2\linewidth]{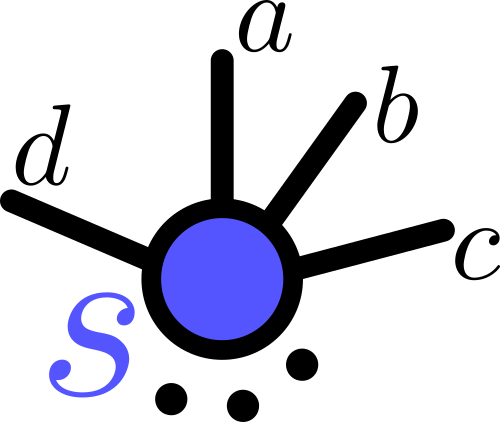}} =& \begin{cases}
        (-1)^{sa} & a=b=c = \dots =d\\
        0 & \text{else}
    \end{cases}\\
    =& (-1)^{as}\delta_{a,b,c, \dots, d},
    \end{split}
    \end{align}
for any number of legs. This tensor implements a constraint on the vector space spanned by all the labelings of the legs, indicated by the three dots next to the tensor.
Specifically, for $s=0$ it is the usual delta tensor enforcing all labels to have the same value.
Similarly, we define a \textit{red $s$-spider} by
    \begin{align}\label{eq:redtensor}
    \begin{split}
    \raisebox{-0.4\height}{\includegraphics[width=0.2\linewidth]{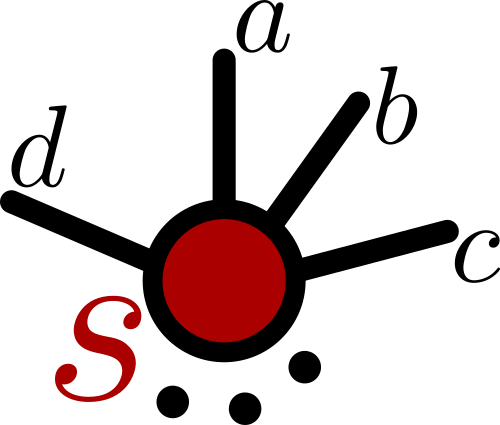}} =& \begin{cases}
        1 & a+b+c+\dots+d=s \mod 2\\
        0 & \text{else}
    \end{cases}\\
    =& \delta_{a\oplus b\oplus c \oplus \dots \oplus d,s}.    
    \end{split}
    \end{align}
It represents a constraint on the joint parity of all the legs of the tensor. Specifically, for $s=0$ it projects onto the even sector whereas for $s=1$ it projects onto the odd sector.
Lastly, we introduce a third type of tensor, namely a \textit{green $s$-spider}, a complex tensor defined by
\begin{align}\label{eq:greentensor}
    \raisebox{-0.4\height}{\includegraphics[width=0.2\linewidth]{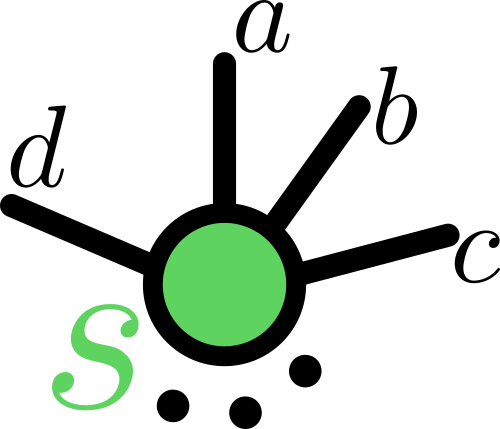}} =& i^{a+b+\dots+c}\delta_{a\oplus b\oplus c ...\oplus d, s}.
\end{align}
The green tensor implements a similar constraint as the red tensor but additionally adds a phase generated by $i$, depending on the explicit configuration on the input legs of the tensor.
For later convenience, we also want to capture tensors that act with a $-i$ on some legs.
Hence, we introduce ``flipping'' of inputs of green tensors, which we denote with a black arrowtip on a leg attached to the tensor.
Specifically, we define
\begin{align}
    \raisebox{-0.4\height}{\includegraphics[width=0.2\linewidth]{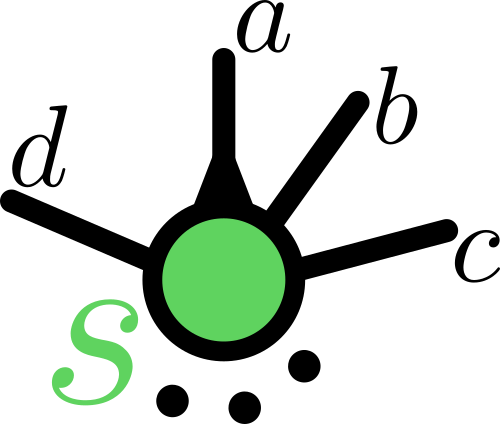}} = i^{-a+b+c+\dots+d} \delta_{a\oplus b\oplus c ...\oplus d, s}
\end{align}    
to denote the tensor with the $a$-leg flipped.
Note that flipping all edges is the same as complex conjugating the tensor,
\begin{align}
    \raisebox{-0.4\height}{\includegraphics[width=0.2\linewidth]{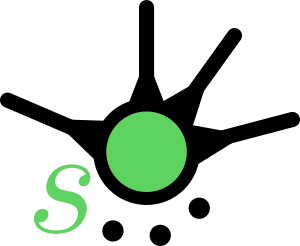}}\quad = \quad \raisebox{-0.4\height}{\includegraphics[width=0.2\linewidth]{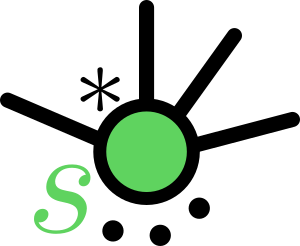}}.
\end{align}    
For the rest of this work, if there is no explicit $s$ label given, we consider the spider to be a $s=0$ spider. Sometimes, we will refer to the $s=1$ spider as a \textit{signed} spider.

Lastly we note that the three spiders are related by basis transformations defined by the  2-legged tensors
\begin{subequations}\label{eq:basistrafo_definition}
    \begin{align}
    \raisebox{-0.4\height}{\includegraphics[width=0.3\linewidth]{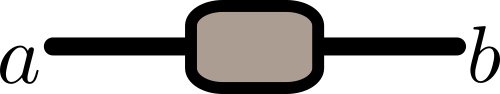}} =& \frac{1}{\sqrt{2}}(-1)^{ab},\\
    \raisebox{-0.4\height}{\includegraphics[width=0.3\linewidth]{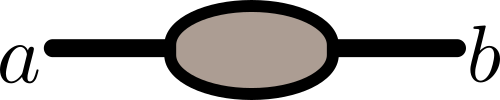}} =& \delta_{a,b} i^a ,\\
    \raisebox{-0.4\height}{\includegraphics[width=0.3\linewidth]{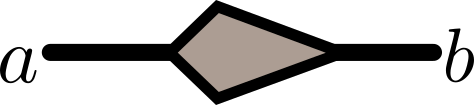}} =& \frac{1}{\sqrt{2}}i^{(1+2a)b}.
\end{align}    
\end{subequations}
For example, the blue spider can be obtained by conjugating either a red or green spider by the above tensors,
\begin{align}\label{eq:blue_basistrafo}
    \raisebox{-0.4\height}{\includegraphics[height=1.3cm]{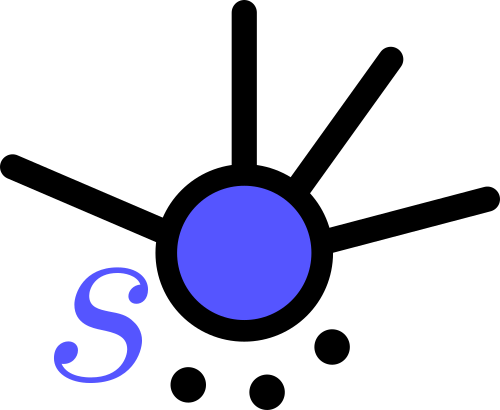}} = \raisebox{-0.4\height}{\includegraphics[height=1.5cm]{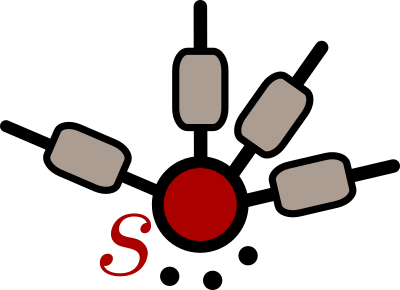}} = \raisebox{-0.4\height}{\includegraphics[height=1.5cm]{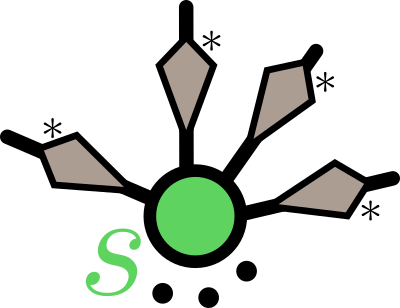}}.
\end{align}
In App.\,\ref{app:rgbtensornetwork}, we give a more complete overview on RGB tensor networks and equivalences thereof.
In particular, we show that relations amongst different tensor networks give rise to a ``calculus'' based on these networks, similar to the ZX-calculus.

\subsection{Graphical representation of circuit elements}\label{sec:circuitelements}
In the following, we show how a generic Clifford circuit with Pauli measurements can be mapped to a tensor network of tensors that were just introduced.

One can think of each of the three colors as corresponding to one of the three Pauli labels, with red corresponding to $X$, green to $Y$ and blue to $Z$.\footnote{Note the different color convention cf.\ ZX-calculus~\cite{van2020zx}.}
Specifically, their two-legged signed versions represent the Pauli matrices
\begin{subequations}
\begin{align}
    \raisebox{-0.25\height}{\includegraphics[width=0.2\linewidth]{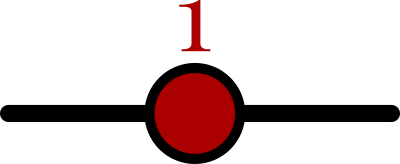}} =& \raisebox{-0.4\height}{  \includegraphics[width=0.2\linewidth]{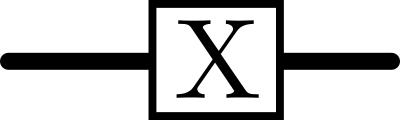}},\\
    \raisebox{-0.25\height}{\includegraphics[width=0.2\linewidth]{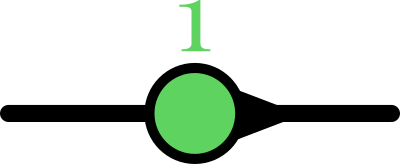}} =& \raisebox{-0.4\height}{  \includegraphics[width=0.2\linewidth]{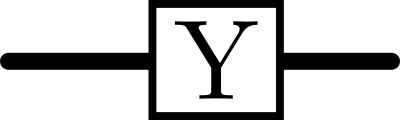}},\\
    \raisebox{-0.25\height}{\includegraphics[width=0.2\linewidth]{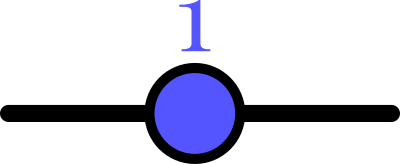}} =& \raisebox{-0.4\height}{  \includegraphics[width=0.2\linewidth]{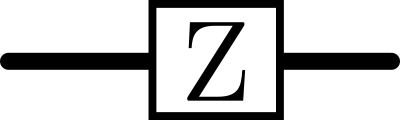}}.
\end{align}    
\end{subequations}
Together with usual Hadamard and $S$ gate, represented by the tensors
\begin{subequations}
\begin{align}
    \raisebox{-0.4\height}{\includegraphics[width=0.2\linewidth]{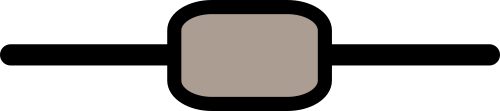}} =& \raisebox{-0.4\height}{  \includegraphics[width=0.2\linewidth]{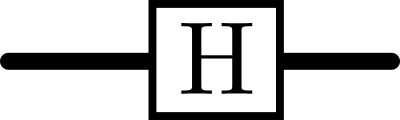}},\\
    \raisebox{-0.4\height}{\includegraphics[width=0.2\linewidth]{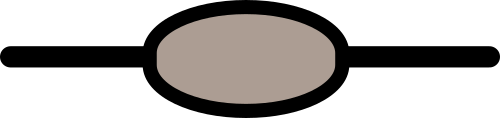}} =& \raisebox{-0.4\height}{  \includegraphics[width=0.2\linewidth]{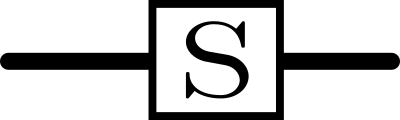}},
\end{align}    
\end{subequations}
the signed blue, green and red spiders generate the single-qubit Clifford group.
To complete the full Clifford group, we need a two-qubit gate, for example the $\CNOT$ gate.
The $\CNOT$ gate can be defined as a matrix with entries
\begin{align}
    \CNOT_{ii',jj'} = \delta_{i,i'}\delta_{j',j\oplus i}
    ,
\end{align}    
which can be represented by the tensor network
\begin{align}
\raisebox{-0.4\height}{
    \includegraphics[width=0.3\linewidth]{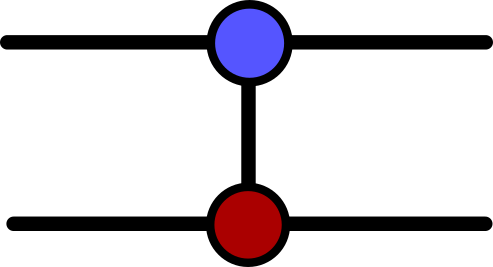}}.
\end{align}
This completes the representation of any Clifford unitary in the three colored calculus.

Additionally, we can represent any multi-qubit Pauli measurement nicely as an RGB tensor network.
More concretely, we represent \textit{projective measurements} in terms of the projectors onto a given measurement outcome.
As such, the measurement outcome appears as a classical label of the network.

Let us start with a simple example of a 2-body measurements.
They constitute the elementary operations of the XYZ ruby code, defined in Sec.~\ref{sec:rubycode_definition}.
We represent the projector onto outcome $m\in\{0,1\}$ of a $Z\otimes Z$ measurement via the following tensor network
\begin{align}
    \raisebox{-0.4\height}{\includegraphics[width=0.3\linewidth]{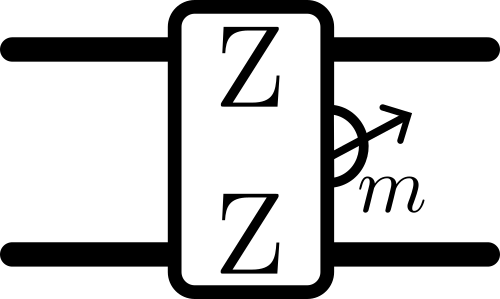}} =
    \raisebox{-0.4\height}{\includegraphics[width=0.3\linewidth]{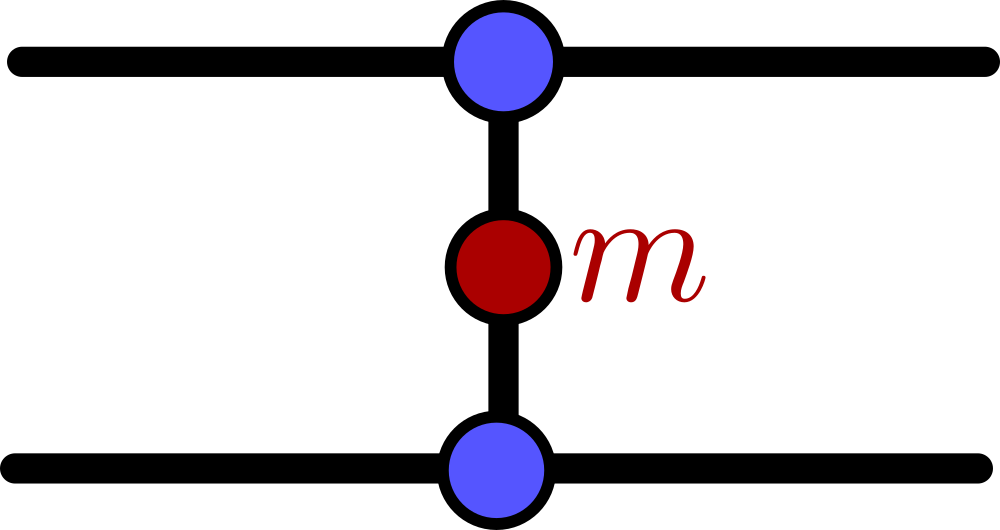}}.
\end{align}
Similarly, an $XX$ measurement can be represented by swapping \includegraphics[height=10pt]{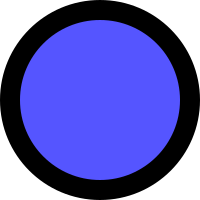} with \includegraphics[height=10pt]{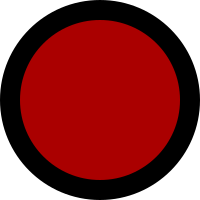} in the above diagram.

We can use only blue and red spiders in combination with Hadamard boxes to also include $Y$ measurements, see for example Fig.~3 in Ref.~\cite{bombin2023unifying}.
However, this comes with a fairly big notational overhead where the components in the tensor network do not resemble the actual building blocks of a circuit implementing that measurement.
Using the green spiders, we can represent measurements involving Pauli $Y$ more directly.
For example, consider a $YY$ measurement.
We find that
\begin{align}
   \raisebox{-0.4\height}{\includegraphics[width=0.3\linewidth]{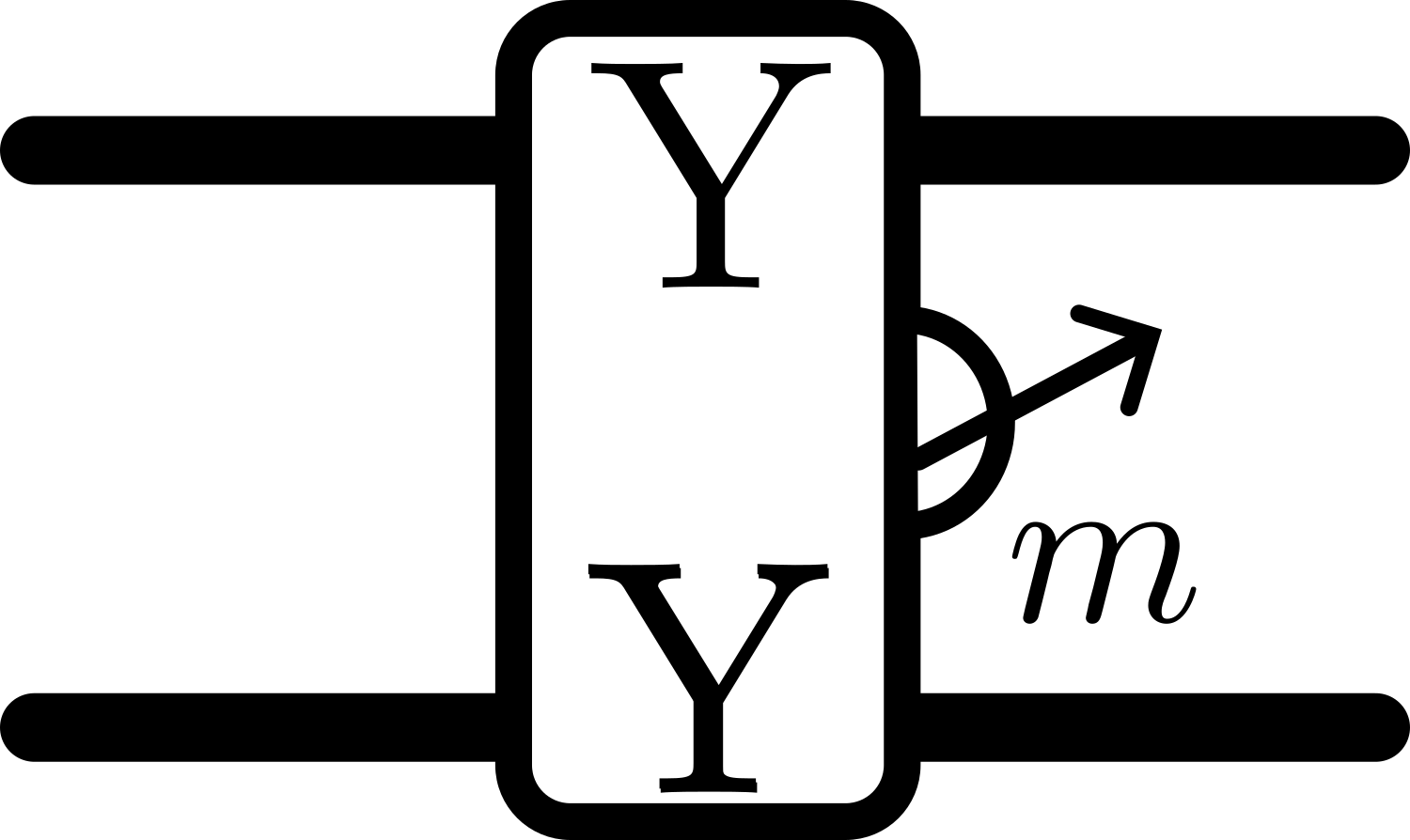}} =
    \raisebox{-0.4\height}{\includegraphics[width=0.3\linewidth]{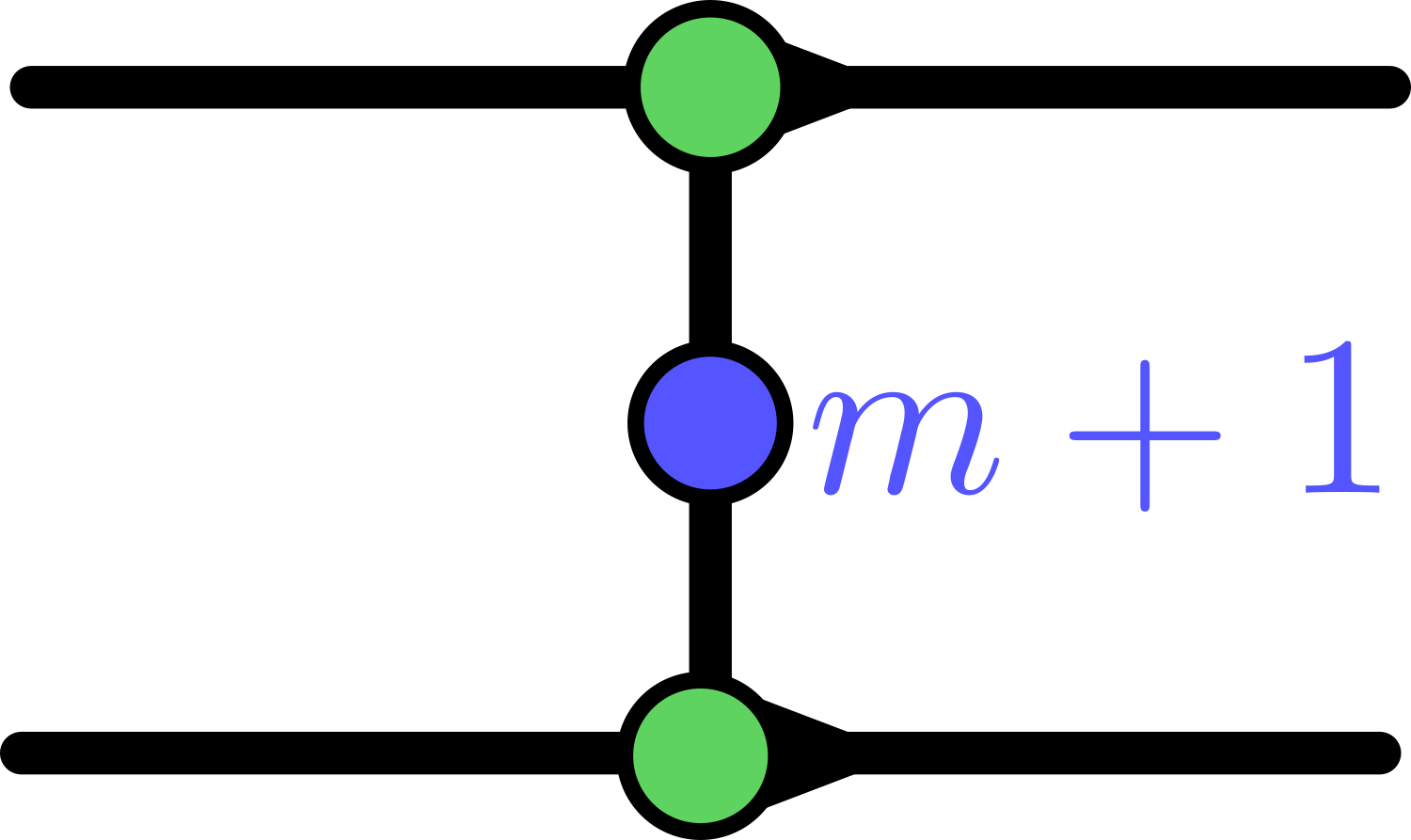}}
\end{align}
represents the projector onto the subspace associated to the $YY$ measurement with outcome $m\in\{0,1\}$.
Note that the sign of the blue spider on the right represents the flipped measurement. This is an artefact on the details of how we defined the green tensor. 
In most cases we do not need to consider the flipped measurement explicitly since the respective stabilizer groups only differ by a sign.

In fact, we can represent any multi-qubit Pauli measurement as a tensor network following a recipe presented in App.\,\ref{app:rgbtensornetwork}.

\subsection{Pauli flow and logical isomorphism}
In this section, we introduce \textit{Pauli flows} in terms of projective symmetries of the building blocks of our tensor network representation of circuits.
We use the term ``projective symmetry'' because the operations we want to consider leave the individual tensors invariant up to a phase.
Specifically, we obtain a graphical rule on how to track the propagation of Pauli operators through the network. 

Let us start with a simple example of the identity tensor represented by a naked wire.
The Pauli matrices behave in the following way:
\begin{align}
    \raisebox{-0.2\height}{\includegraphics[height=16pt]{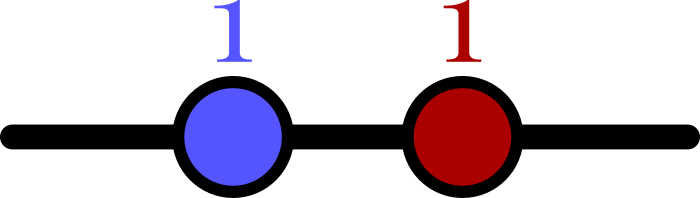}} = -i\; \raisebox{-0.2\height}{\includegraphics[height=16pt]{Ytensor.png}} = - \; \raisebox{-0.2\height}{\includegraphics[height=16pt]{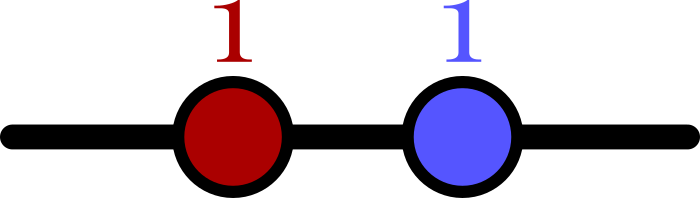}}.
    \label{eq:commutation_pauli_tensors}
\end{align}
We find that, up to phases, they form an Abelian group $\hat{\cP}_1 \simeq \bZ_2\times\bZ_2$.
Since these spiders can be freely moved around a wire and commute up to phases, we only need to keep track of which Pauli acts, not in which order and not which phase factor it carries.
We represent this by highlighting the wire in one of three colors,
\begin{subequations}\label{eq:pauliflowmap}
\begin{align}
    \raisebox{-0.2\height}{\includegraphics[width=0.2\linewidth]{Xtensor.png}} \quad\mapsto&\quad \raisebox{-0.2\height}{\includegraphics[width=0.2\linewidth]{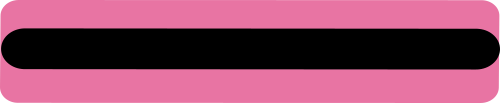}},\\
    \left\{\raisebox{-0.2\height}{\includegraphics[width=0.2\linewidth]{Ytensor.png}},\, \raisebox{-0.2\height}{\includegraphics[width=0.2\linewidth]{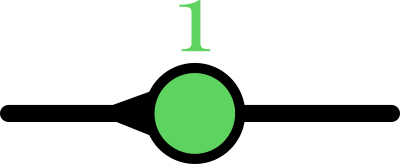}}\right\}  \quad\mapsto&\quad \raisebox{-0.2\height}{\includegraphics[width=0.2\linewidth]{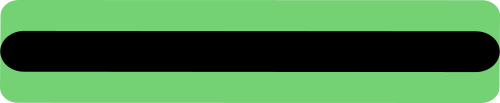}},\\
    \raisebox{-0.2\height}{\includegraphics[width=0.2\linewidth]{Ztensor.png}} \quad\mapsto&\quad \raisebox{-0.2\height}{\includegraphics[width=0.2\linewidth]{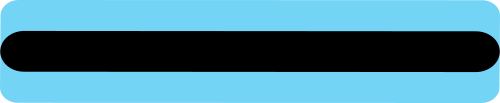}}.
\end{align}
\end{subequations}
Formally, the map depicted by ``$\mapsto$'' can be understood as the map mapping the single-qubit Pauli group to $\bZ_2\times\bZ_2$, whose three non-trivial elements we depict with three different colors.
The commutative addition in $\bZ_2\times\bZ_2$ is represented by the following rules, when ``adding'' two highlights,
\begin{subequations}\label{eq:adding_highlights}
    \begin{align}
\begin{split}
    \raisebox{-0.4\height}{\includegraphics[width=0.175\linewidth]{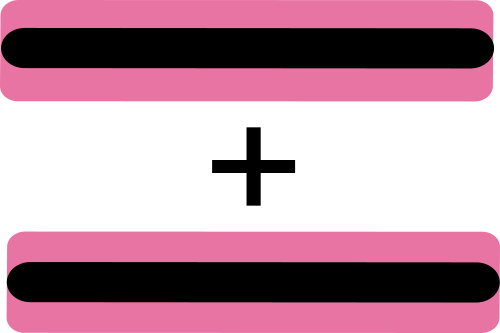}}\; =& \; 
    \raisebox{-0.4\height}{\includegraphics[width=0.175\linewidth]{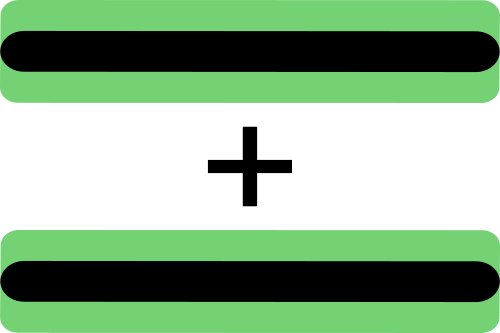}} \;=\;
    \raisebox{-0.4\height}{\includegraphics[width=0.175\linewidth]{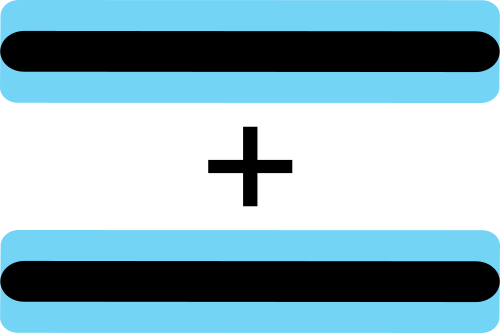}} \\[12pt]
    =&\; \raisebox{-0.2\height}{\includegraphics[width=0.175\linewidth]{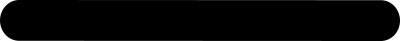}}\qq{and}
\end{split}\\[12pt]
    \raisebox{-0.4\height}{\includegraphics[width=0.175\linewidth]{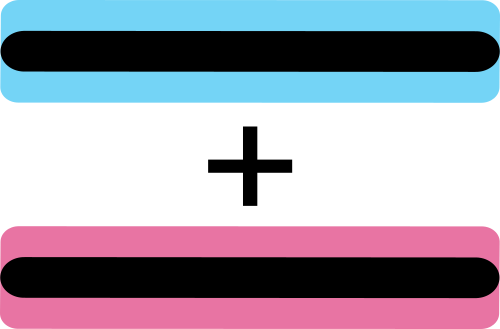}} \;=&\; \raisebox{-0.4\height}{\includegraphics[width=0.175\linewidth]{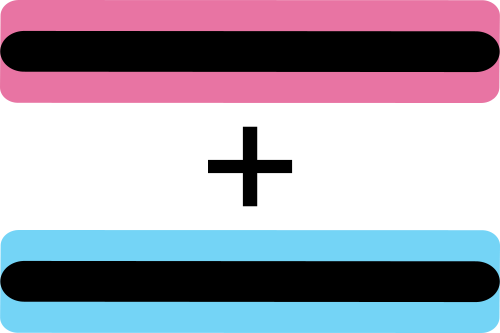}} \;=\; \raisebox{-0.2\height}{\includegraphics[width=0.175\linewidth]{greenflow.png}}\,.
\end{align}
\end{subequations}

In the following, we derive the projective Pauli symmetries of the blue, green and red spider and represent them graphically using the highlights introduced above.
The blue spider obeys the following relations
\begin{subequations}\label{eq:blue_projective_sym}
\begin{align}
    \raisebox{-0.4\height}{\includegraphics[height=37.5pt]{bluespider_nolabel.png}} &= \raisebox{-0.4\height}{\includegraphics[height=37.5pt]{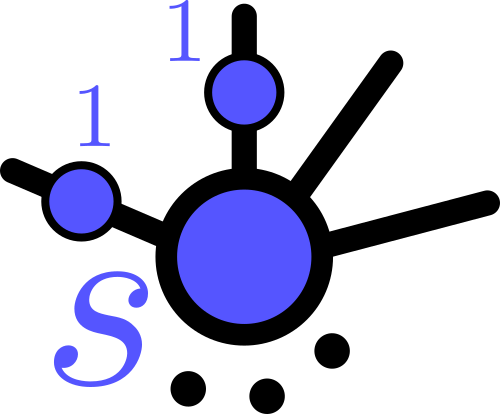}} = (-1)^s \;\raisebox{-0.4\height}{\includegraphics[height=37.5pt]{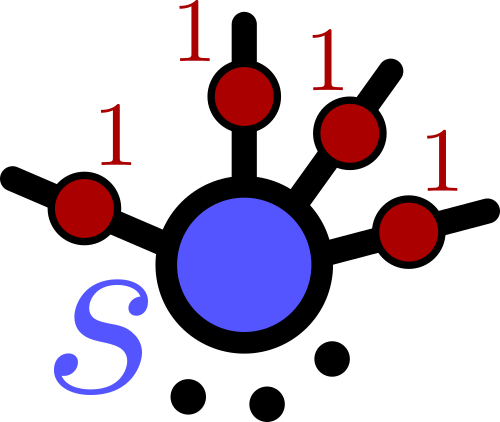}}\\
    &= (-1)^s \;\raisebox{-0.4\height}{\includegraphics[height=37.5pt]{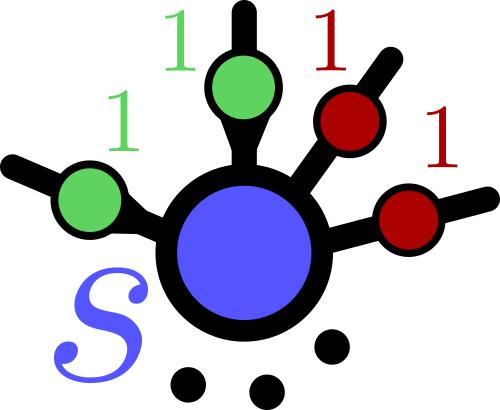}},
\end{align}    
\end{subequations}
that can be understood as its projective Pauli symmetries.
Every symmetry operation defines a valid highlight of the legs around a tensor, which we call \textit{Pauli flows} of that tensor.
Specifically, we highlight a leg in the color of the tensor with which the symmetry acts on that leg.
For a more stringent mathematical treatment we also consider the trivial symmetry operation, acting with an identity tensor on each leg, as a trivial Pauli flow, where no leg is highlighted.
Moreover, we see in Eq.\,\eqref{eq:blue_projective_sym} that the blue tensor acquires a sign when applying a signed red tensor on each of its legs iff $s=1$.
We indicate that by highlighting the (classical) $s$ label of the tensor and say that the tensor is \textit{charged} with respect to that flow.
Together, the set of projective symmetries of the blue tensor, Eq.\,\eqref{eq:blue_projective_sym}, defines the following Pauli flows of the blue tensor,
\begin{align}\label{eq:blue_pauliflows}
\begin{split}
    \raisebox{-0.2\height}{\includegraphics[width=0.2\linewidth]{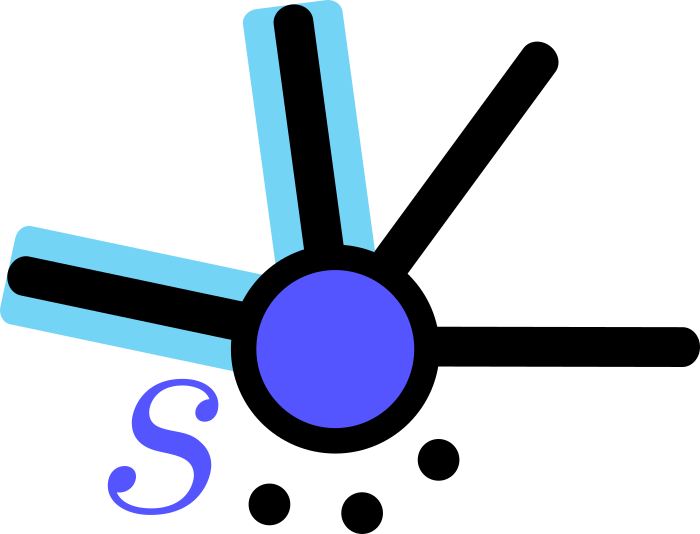}}\qcomma
   \raisebox{-0.2\height}{\includegraphics[width=0.2\linewidth]{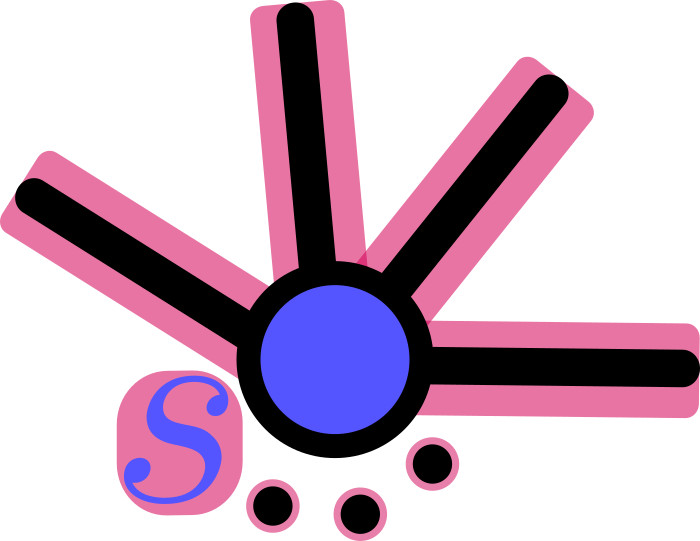}}\qq{and}
   \raisebox{-0.2\height}{\includegraphics[width=0.2\linewidth]{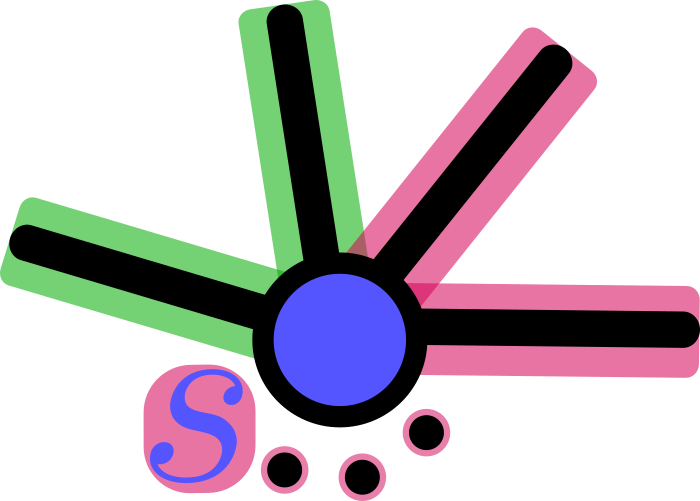}}.
\end{split}
\end{align}
We find that the tensors of the other colors have similar projective symmetries leading to these Pauli flows:
\begin{subequations}
\begin{align}\label{eq:spider_flow}
    \raisebox{-0.2\height}{\includegraphics[width=0.2\linewidth]{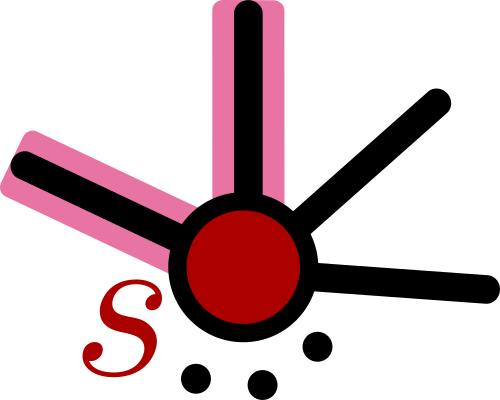}}\qcomma
   \raisebox{-0.2\height}{\includegraphics[width=0.2\linewidth]{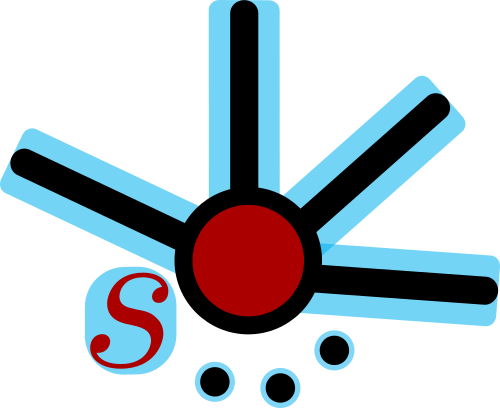}}\qcomma
   \raisebox{-0.2\height}{\includegraphics[width=0.2\linewidth]{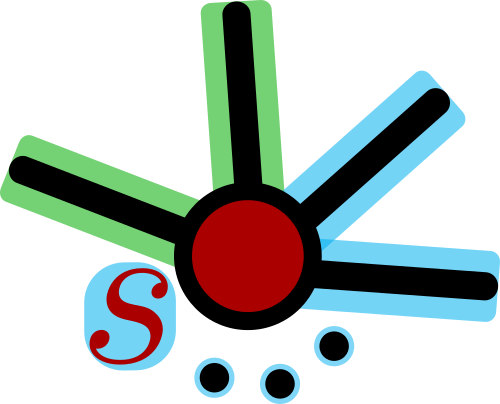}} \\
       \raisebox{-0.2\height}{\includegraphics[width=0.2\linewidth]{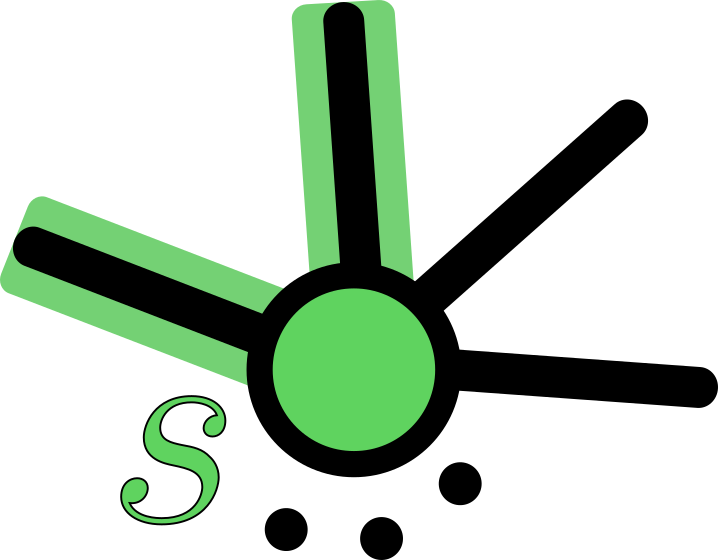}}\qcomma
   \raisebox{-0.2\height}{\includegraphics[width=0.2\linewidth]{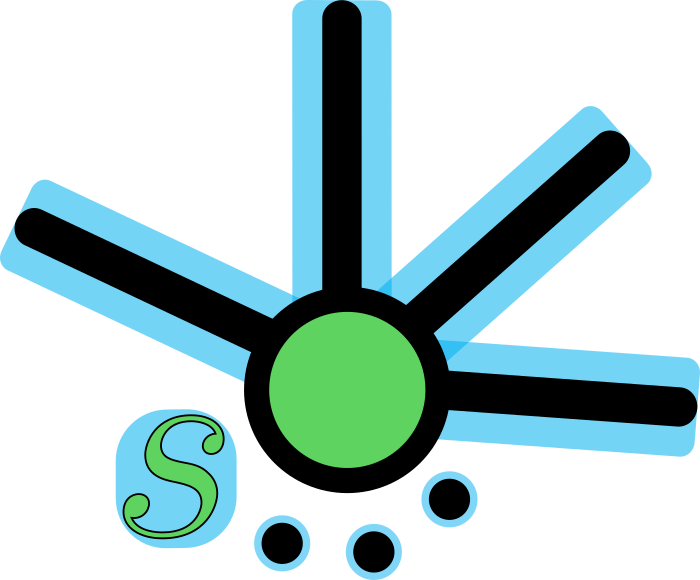}}\qcomma
   \raisebox{-0.2\height}{\includegraphics[width=0.2\linewidth]{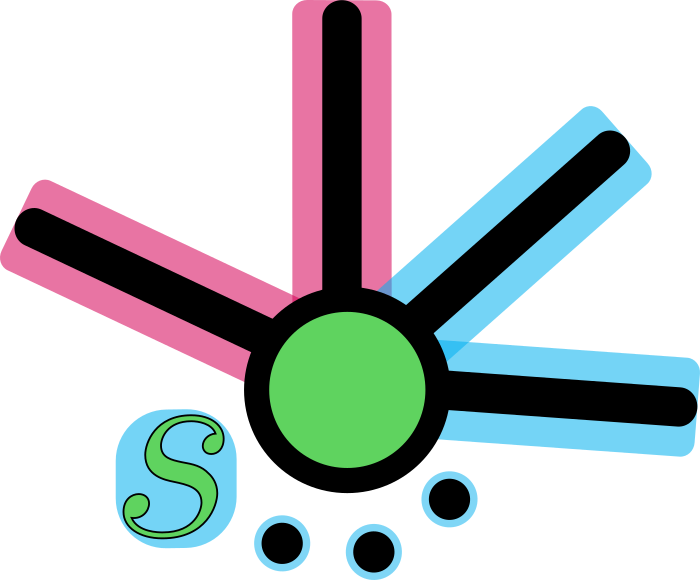}} .
\end{align}
\end{subequations}
It is helpful to think of all allowed Pauli flows around a tensor as being generated by generating flows of two types. The first type highlights an even number of legs in the same color as the tensor.
The second type highlights all legs in one of the two complementary colors, as well as the $s$ label. Adding these highlights gives rise to all Pauli flows of an individual tensor.
Note that single-legged tensors play a special role, since they do not have pairs of legs.
As a result, they only allow for the second type of non-trivial Pauli flow, highlighting the leg and the sign in one complementary color.

From the flows of a single tensor we can construct the flows of any tensor network composed of these tensors.
We define a Pauli flow of a network as a highlight of legs in the network that is a Pauli flow -- as defined above -- when restricted to any of the constituents of the network.
The notion of \textit{charge} carries over directly to a set of tensors: Consider a set of (signed) RGB tensors $T$ in an RGB network\footnote{Note that these tensors do not have to be neighbors in the network but an arbitrary subset of tensors.} and a given flow $F$. $F$ highlights a subset of signs in the network, which we denote by $S$.
The charge of $T$ is defined as the binary sum of highlighted signs of the tensors in $T$, $c_T = \sum_{s\in S\cap T} s \mod 2$.

Ref.~\cite{bombin2023unifying} introduces a similar concept called \textit{Pauli webs}.
There, the authors work with a two-colored (ZX) calculus and two-colored flows.
Implicitly, they include a third highlight as well by allowing a combination of the two.
We take a more direct approach by explicitly working with a third color and adding the third type of tensor.
This makes the flow diagrams more symmetric and easier to work with.
Additionally, we introduce the rigorous notion of tensors being \textit{charged} with respect to a flow which -- to the best of our knowledge -- is a concept newly introduced in our work.
It will be central to our perspective on spacetime quantum error correction (see Sec.~\ref{sec:QEC-spacetime}).
For a rigorous treatment of Pauli flow, we refer to App.\,\ref{app:flows}.

By mapping a given circuit to an RGB tensor network, we can understand its logical action in terms of Pauli flows.
Let us illustrate this with a simple example: a single-qubit teleportation circuit is
\begin{align}\label{eq:teleportation_circuit}
    \raisebox{-0.4\height}{\includegraphics[width=0.8\linewidth]{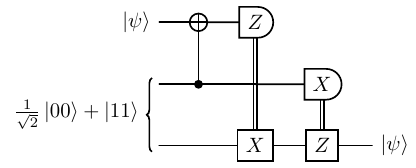}}.
\end{align}
In our tensor network notation, omitting the adaptive Pauli correction, it can be represented by the diagram\footnote{In App.\,\ref{app:sec:rewriterules} we show the steps needed to see that this diagram represents the teleportation circuit starting from a diagram that is closer to the circuit presented above.}
\begin{align}\label{eq:teleportation_network}
    \raisebox{-0.4\height}{\includegraphics[width=0.35\linewidth]{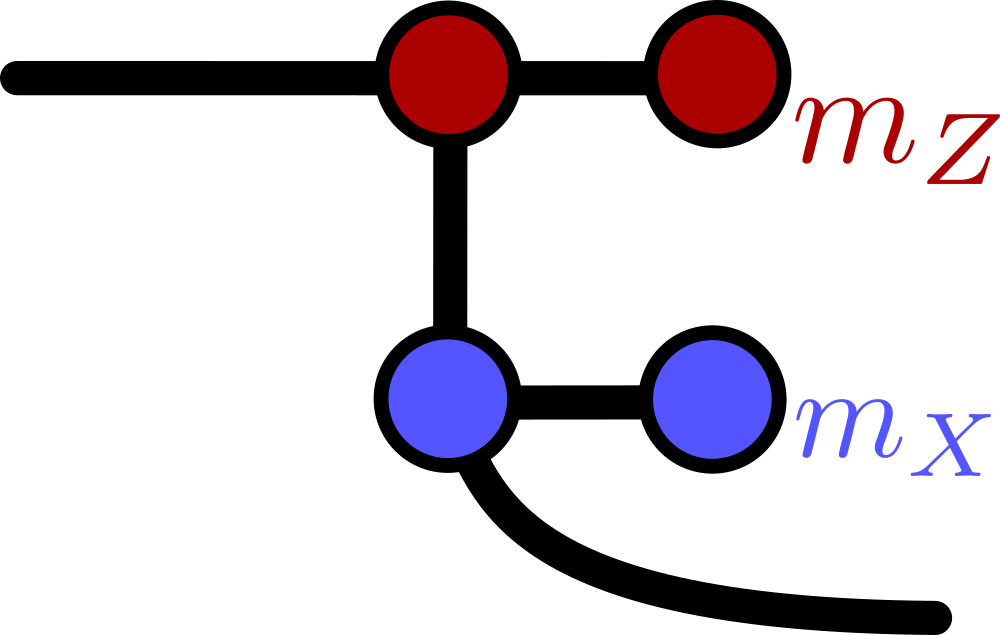}},
\end{align}
where the measurement outcomes $m_Z$ and $m_X$ are represented by signs of two of the tensors.
The flows describe how Pauli operators propagate through the circuit (up to global phases).
In particular, there are flows where the input and output wire of the circuit is highlighted.
Specifically, the network has two independent flows
\begin{align}
    \raisebox{-0.4\height}{\includegraphics[width=0.3\linewidth]{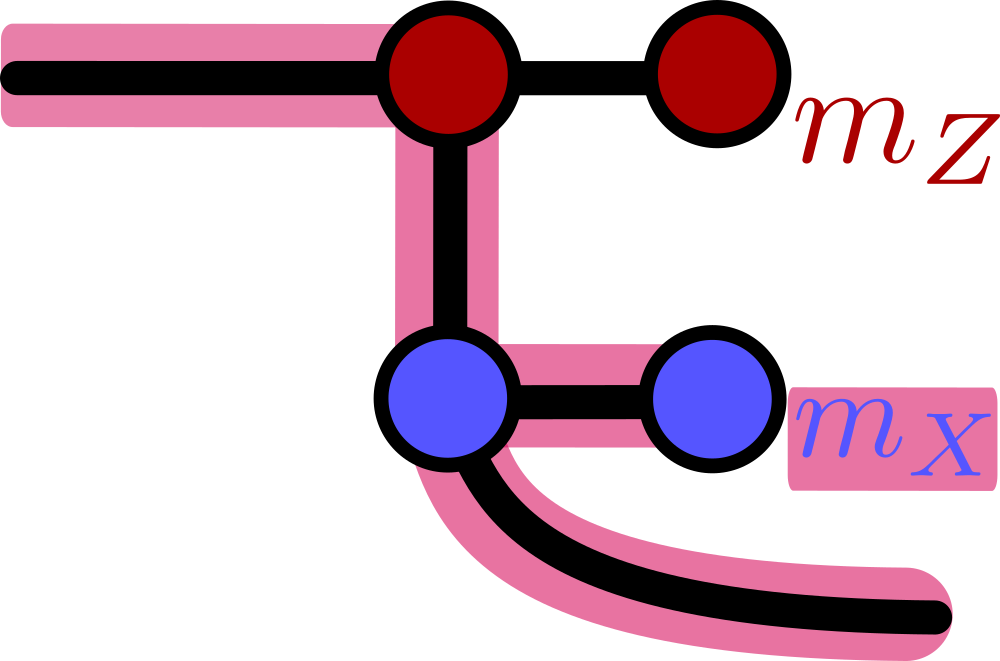}}\qq{and} \raisebox{-0.4\height}{\includegraphics[width=0.3\linewidth]{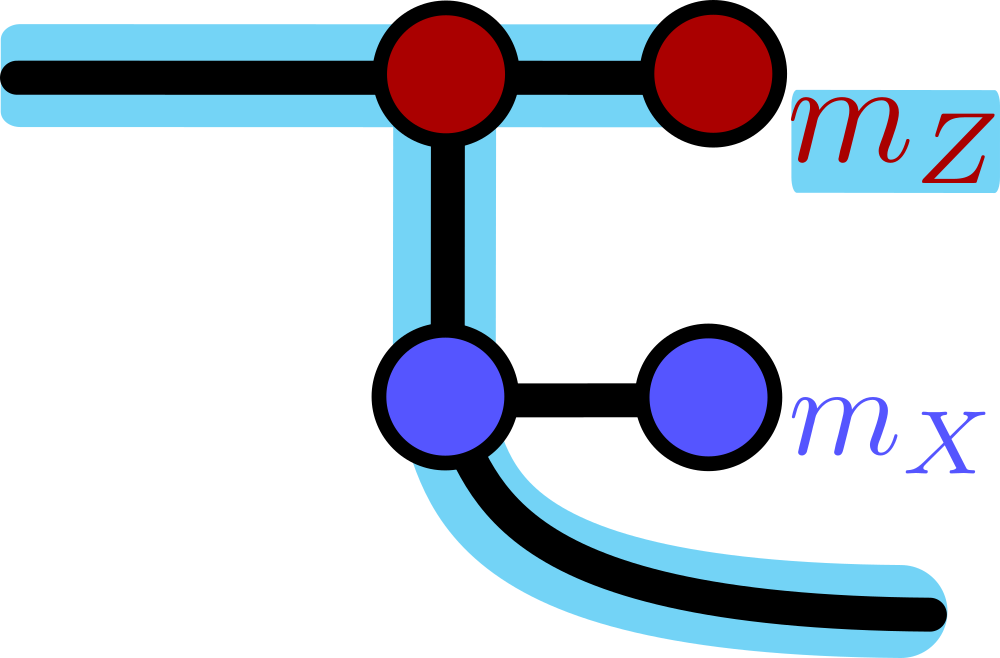}}.
\end{align}
These flows indicate that the circuit maps
\begin{align}\label{eq:teleportation_signed_flows}
    X_1 \mapsto (-1)^{m_X} X_3 \qq{and} Z_1\mapsto (-1)^{m_Z}Z_3,
\end{align}
which fully determines its action since it is a Clifford circuit that acts unitarily on the information encoded on the input qubit.
Moreover, we recover that the circuit acts like $X^{m_Z} Z^{m_X}$ on the logical state, determining the adaptive correction applied in Eq.\,\eqref{eq:teleportation_circuit}.

\subsection{Error correction in spacetime from Pauli flows}\label{sec:QEC-spacetime}
In the previous paragraph we have sketched how the Pauli flows of the teleportation circuit (defined for three physical qubits) give rise to an effective logical (Clifford) circuit on a single qubit.
The concept of a logical isomorphism is central to our analysis of dynamical error-correcting schemes.
A circuit composed of Clifford operators and Pauli measurements implements an isomorphism from an input logical Pauli group to an output logical Pauli group.
We call this isomorphism the \textit{logical isomorphism}.
Both the input logical Pauli group and the output logical Pauli group will be defined in terms of an input stabilizer group  $\cS_{\mathrm{in}}$ and an output stabilizer group $\cS_{\mathrm{out}}$.

In this subsection, we show how the Pauli flows of (a tensor network representation of) a given circuit can be used to understand the logical isomorphism implemented by the circuit.
In particular, we will rigorously define the input and output \textit{stabilizer group}, the \textit{logicals} and \textit{detectors} by their associated Pauli flows.
When introducing errors, the notion of a \textit{charged} tensor will be important to understand the effect of an (Pauli) fault on the syndrome and encoded logical information.
Later in this work, we will use exactly these methods to benchmark the logical isomorphism implemented by the \textit{XYZ ruby code}, see Sec.~\ref{sec:benchmarking}.

Let us consider a circuit from $n$ qubits to $m$ qubits expressed as an RGB tensor network together with all its Pauli flows.
The set of all flows forms a group, which we denote by $F$.
Its identity is given by the trivial flow, not highlighting any edge in the network (non-trivially).
We distinguish between three types of Pauli flows, forming subsets of $F$: \textit{detector flows $D$, stabilizer flows $S$} and \textit{logical flows $L$}.
In the following, we introduce these three types and explain how the Pauli flow gives rise to quantities entering the error correction procedure and illustrate these quantities in a tensor network representing repeated measurements of the stabilizers of a repetition code.
We will see that the notion of tensors being charged with respect to a given flow plays a central role.
For a rigorous definition and treatment of the types of flows and associated quantities in the QEC protocol, we refer to App.\,\ref{app:flows}.

\begin{figure*}
    \centering
    \includegraphics[width=0.75\linewidth]{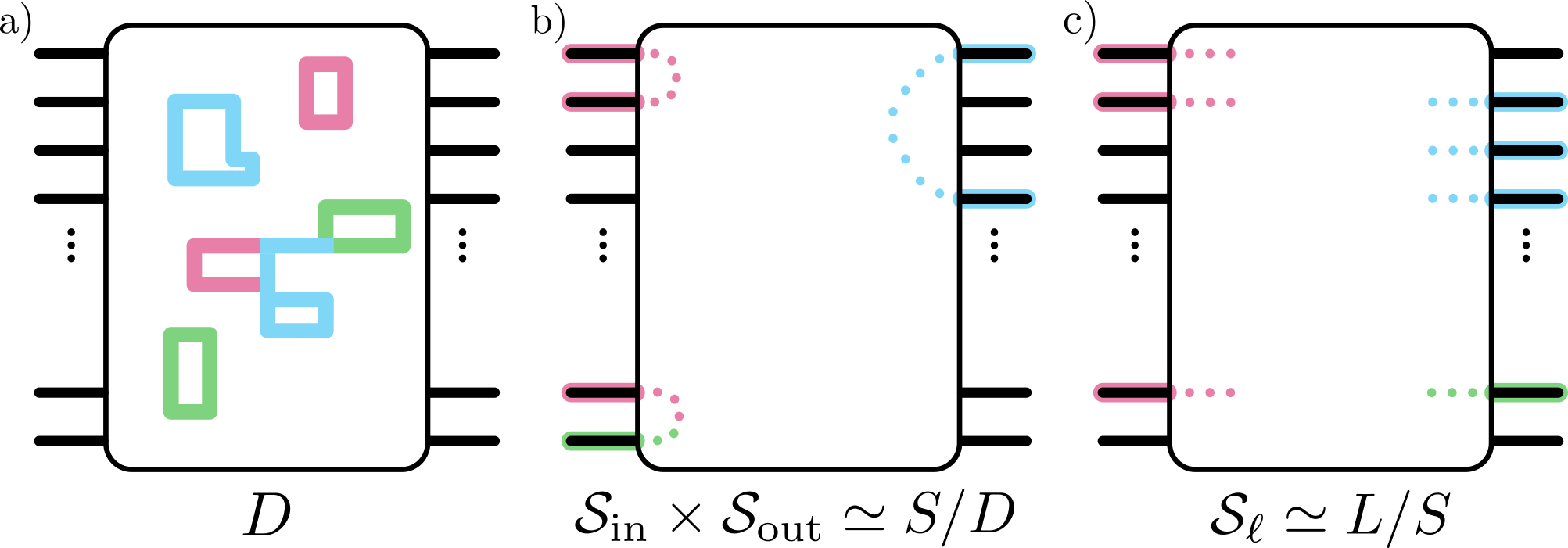}
    \caption{Schematic depiction of the three different types of Pauli flows, defined by how they highlight the input and output legs.
    a) Detector flows are purely internal and give rise to a symmetry group of the RGB tensor network. It determines the classical data that can be used for decoding, see Sec.~\ref{sec:probing} and App.\,\ref{app:flows}.
    b) Flows that are generated by flows that highlight only input or output legs non-trivially can be mapped to an input and output stabilizer groups.
    c) Flows highlighting both input and output legs give rise to a stabilizer group supported on both input and output qubits and together and can define the logical Clifford applied on the logical operators of the input and output logical stabilizer group.}
    \label{fig:Pauli_flows}
\end{figure*}

\begin{enumerate}
    \item \textbf{Detector flows}: Pauli flows that only highlight internal legs non-trivially are called \textit{detector flows}.
    In fact, the set of detector flows form a group $D$ under adding the flows according to Eq.\,\eqref{eq:adding_highlights} and can be considered as a \textit{symmetry group} of the network.\footnote{Note that this is not a symmetry group of operators but a group of sequences of transformations of the network that can be related to inserting certain tensors and propagating them through. For more details, see App.\,\ref{app:detectors}. This is related to the notion of symmetric circuits in Ref.~\cite{bauer2024lowoverhead} but not equivalent in that \NEW{not} every symmetry in the tensor networks of Ref.~\cite{bauer2024lowoverhead} corresponds to a detector flow.}
    We will later use conserved quantities related to that symmetry for decoding.
    
    \item \textbf{Stabilizer flows}: 
    Pauli flows that can be generated by flows that may highlight either input or output legs non-trivially are called \textit{stabilizer flows}.
    We denote the set of all such flows by $S$.
    From $S$, we can in fact derive a stabilizer group
    \begin{align}
    \cS_{\mathrm{in}}\times\cS_{\mathrm{out}}\leq\cP_{n+m}    
    \end{align}
    that factors over input and output Pauli groups.
    Given a stabilizer flow $s$, the associated Pauli word is obtained by taking the highlight of $s$ restricted on the in- and output legs and interpreting it as a Pauli operator that is the product of the Pauli operators associated to each individual highlight on each open leg.
    We identify a trivial highlight with $\mathds{1}$, a red highlight with $X$, a green highlight with $Y$ and a blue highlight with $Z$.
    The sign in front of the Pauli word associated to flow $s$ is obtained from the signs charged with respect to $s$.
    
    In Sec.~\ref{sec:benchmarking}, we consider detectors in the bulk of an RGB tensor network and imagine cutting this network along a set of edges.
    In fact, all the detector flows that had non-trivial highlights on the cut edges will be promoted to non-trivial stabilizer flows of the cut network.
    This perspective helps us to understand slices of detector flows as instantaneous stabilizers.

    \item \textbf{Logical flows}: A Pauli flow that highlights both input and output legs is called a \textit{logical flow}.
    It captures how the network transforms (logical) Pauli operators on the input legs into logical Pauli operators on the output legs.
    Specifically, every logical flow can be associated with a Pauli operator supported on both input and output legs in the same way as discussed for the stabilizer flows.
    Importantly, every Pauli word will carry a sign given by the tensors charged with respect to the corresponding logical flow.
    The set of Pauli words obtained in this way again form a stabilizer group
    \begin{align}
        \cS_{\ell}\leq\cP_{n+m}
    \end{align}
    that commutes with both $\cS_{\mathrm{in}}$ and $\cS_{\mathrm{out}}$.
    Importantly, this group has no element that is only supported on input or output legs that is not contained in $\cS_{\mathrm{in}}$ and $\cS_{\mathrm{out}}$.
    As such, it defines a logically entangled state between the codes of $\cS_{\mathrm{in}}$ and $\cS_{\mathrm{out}}$ and since it is also a Pauli group, it can be identified with a Clifford unitary by identifying one part of the entangled state with its dual and thereby interpreting a stabilizer state on a joint system as a linear map between the two systems. For details on this relation, 
    we refer to App.\,\ref{app:sec:RGB_log_iso}.
    
    As with stabilizer flows, adding a detector flow does not change the Pauli word on the input and output legs assigned to a logical flow.
    In contrast, adding a stabilizer flow does change the Pauli word assigned to the logical flow.
    However, it will not change the logical coset of the Pauli word with respect to $\cS_{\mathrm{in}}\times\cS_{\mathrm{out}}$.
    Hence, we call two logical flows equivalent iff they differ by a detector or a stabilizer flow.
    One can check that this defines a valid equivalence relation, using the Abelian group structure of the Pauli flows of the network and the equivalence classes determine the logical isomorphism applied by the linear operator represented by that network.
\end{enumerate}
Fig.~\ref{fig:Pauli_flows} summarizes the different types of Pauli flows and the associated algebraic structures.
We give a more rigorous construction of all the stabilizer groups defined from Pauli flows in App.\,\ref{app:flows} and prove that the stabilizer group obtained from Pauli flows on the input and output legs of the network fully describe the logical encodings on either side and the explicit isomorphism applied by the network.
We want to remark that the main technical result presented in App.\,\ref{app:sec:RGB_log_iso} applies more generally to any Clifford operations and their composition.

Let us illustrate the concepts introduced above via a guiding example based on a circuit implementing repeated measurements of the stabilizers of a repetition code.
Consider $n$ qubits on a line stabilized by the stabilizer group $\langle \{Z_iZ_j\}_{\langle i,j\rangle}\rangle$, generated by $Z\otimes Z$ on neighboring qubits.
The following tensor network represents two consecutive measurements of a stabilizer generator
\begin{align}
\raisebox{-0.5\height}{\includegraphics[width=0.3\linewidth]{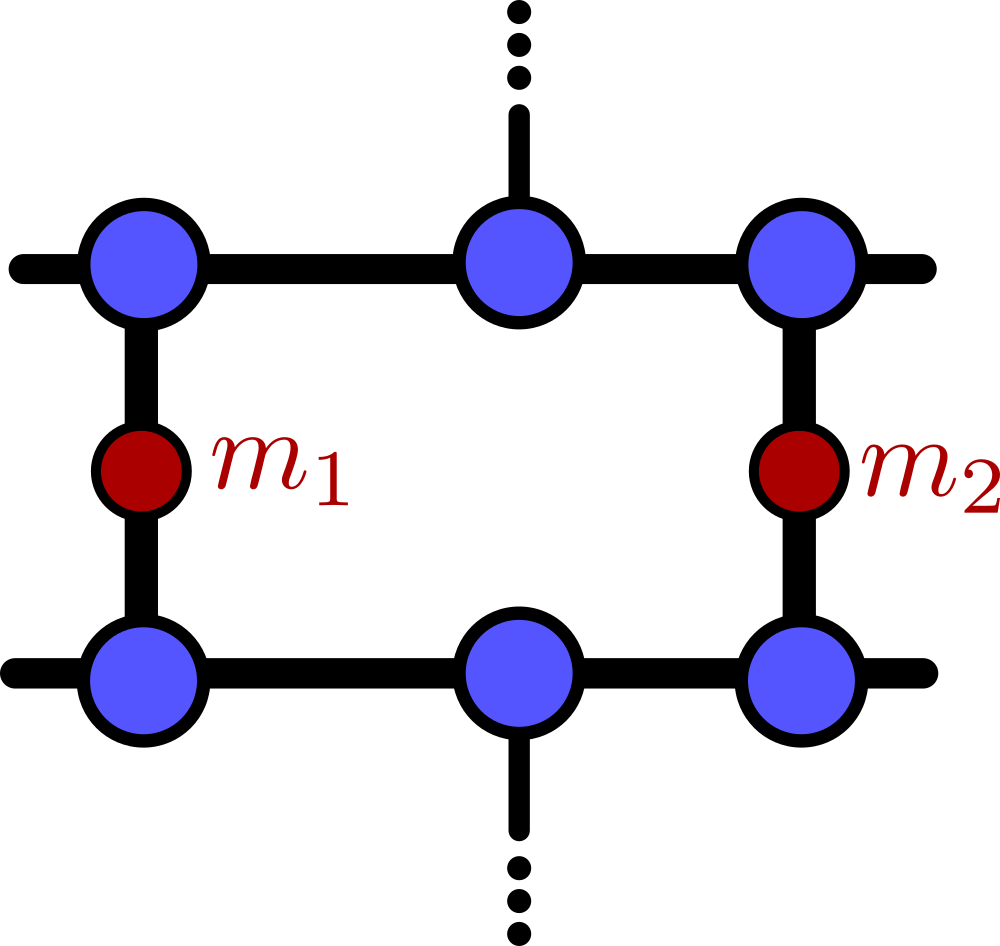}}\;.
\end{align}
Time is understood to go from left to right and the dots indicate the rest of the circuit, performing the same measurements on other pairs.
The labels on the red tensors correspond to the measurement outcomes of the two-body $Z\otimes Z$ measurements.
The detector flows of this network are generated by internal loops of blue flows,
\begin{align}
    \raisebox{-0.5\height}{\includegraphics[width=0.3\linewidth]{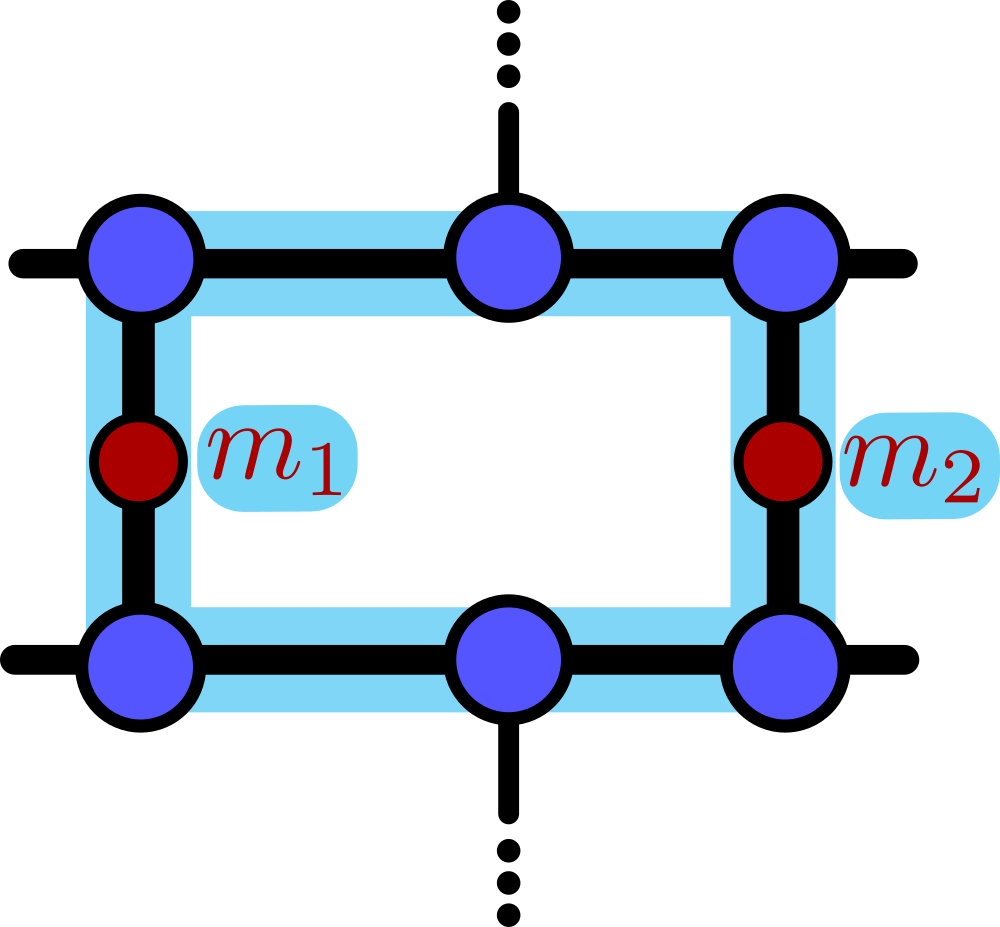}}
\end{align}
with respect to which two red tensors are charged.
This tells us that without noise the measurement outcomes fulfill $m_1+m_2 = 0\mod 2$.
This information can be used for decoding.
Consider a single-qubit $X$ fault happening in between the two measurements.
The tensor network including the fault reads
\begin{align}
    \raisebox{-0.5\height}{\includegraphics[width=0.3\linewidth]{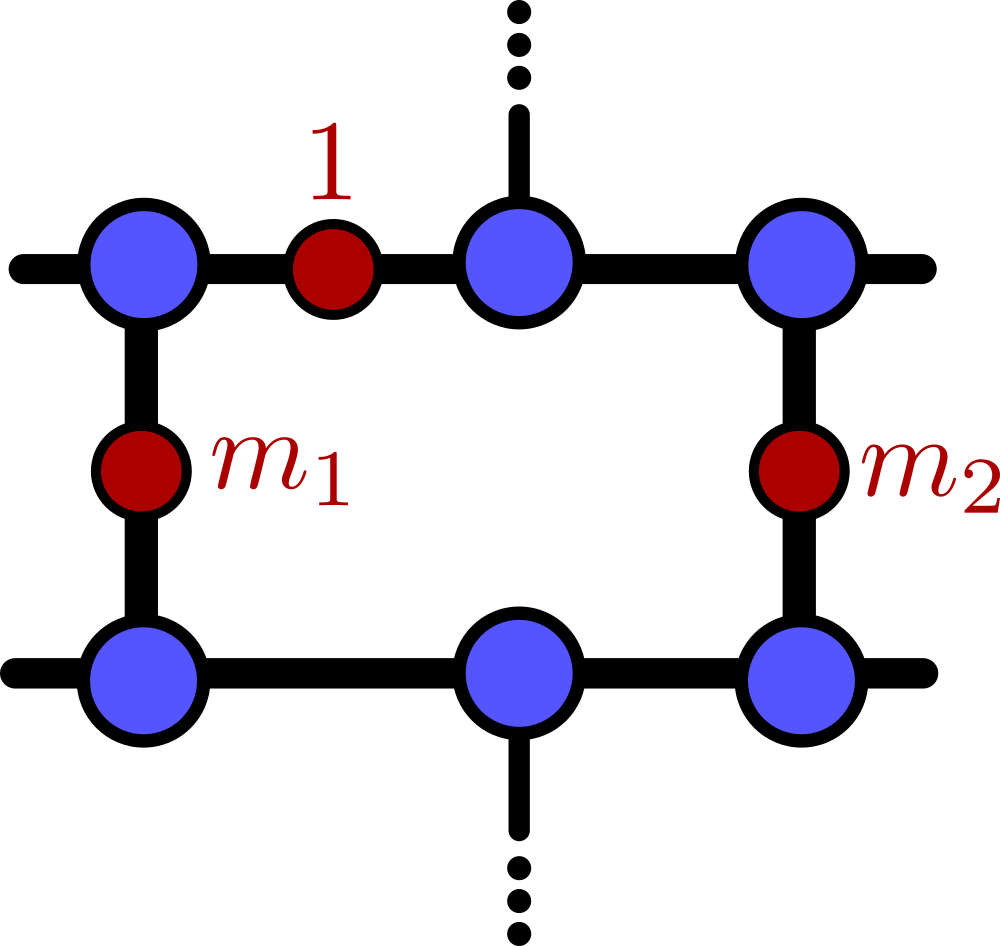}}
\end{align}
and inherits its detector flows from the fault-free network,
\begin{align}
    \raisebox{-0.5\height}{\includegraphics[width=0.3\linewidth]{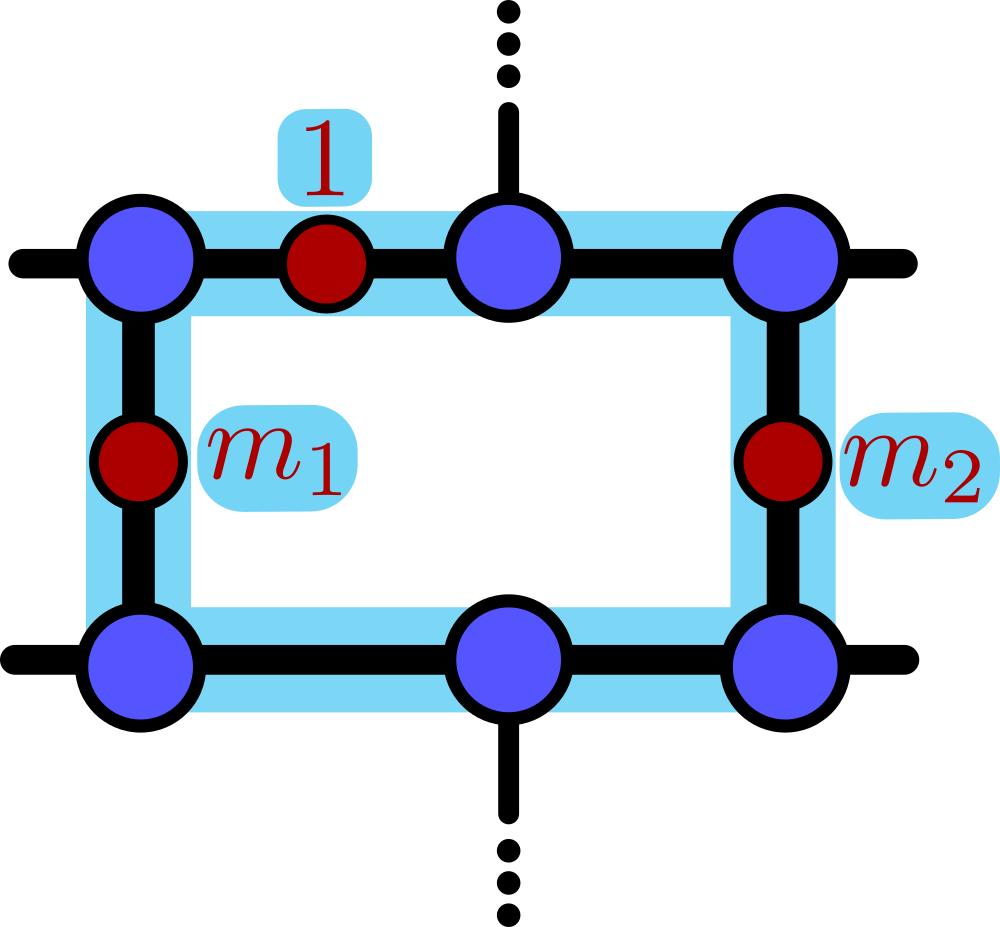}}\;.
\end{align}
We see that the fault is charged with respect to the detector flow and the constraint now reads $m_1+m_2 = 1 \mod 2$.
As such, we can detect this single-qubit Pauli error by calculating the parity of the measurements charged with respect to a detector flow.
Note that this network only has blue detector flows.
Hence, $Z$ faults, represented by blue tensors, can never be charged with respect to any detector flow in this network telling us that they are all undetectable.
To assess the logical effect of an undetectable error, we have to consider the logical flows.
In this simple example there are two independent logical flows, e.g. represented by
\begin{align}
    \raisebox{-0.5\height}{\includegraphics[width=0.3\linewidth]{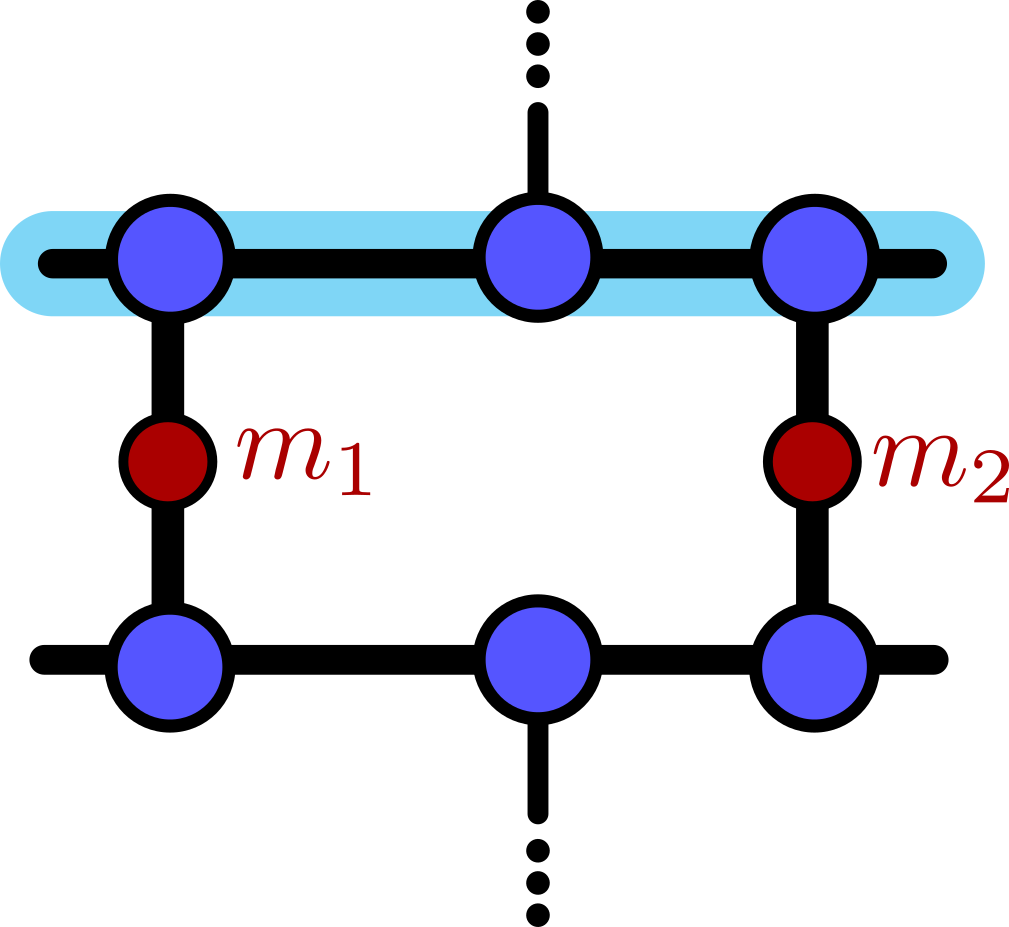}}\qq{and} \raisebox{-0.5\height}{\includegraphics[width=0.3\linewidth]{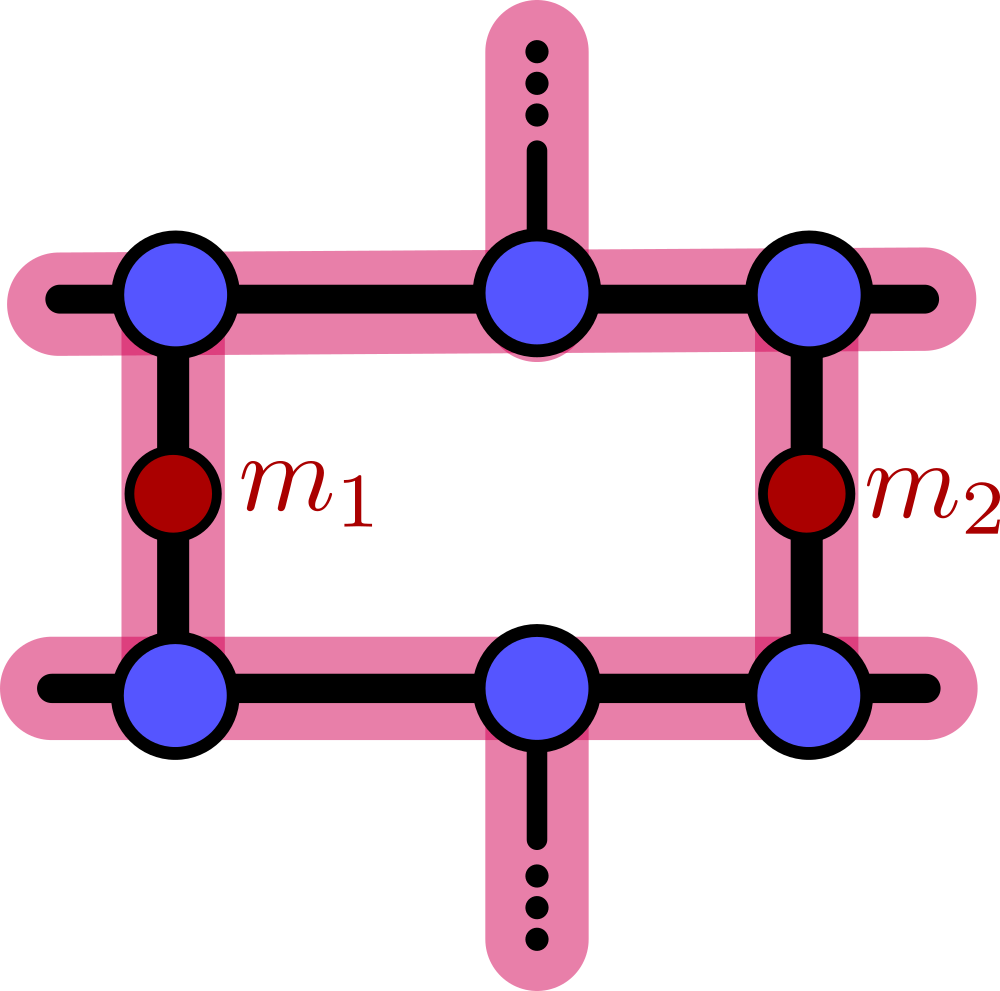}}\;,
\end{align}
showing that a single logical qubit is preserved through the circuit.
Note that when restricted to either input or output legs the highlights of the logical flow give rise to an anti-commuting pair of logical operators.
We find that the undetectable $Z$ fault is charged with respect to the red flow,
\begin{align}
\raisebox{-0.5\height}{\includegraphics[width=0.3\linewidth]{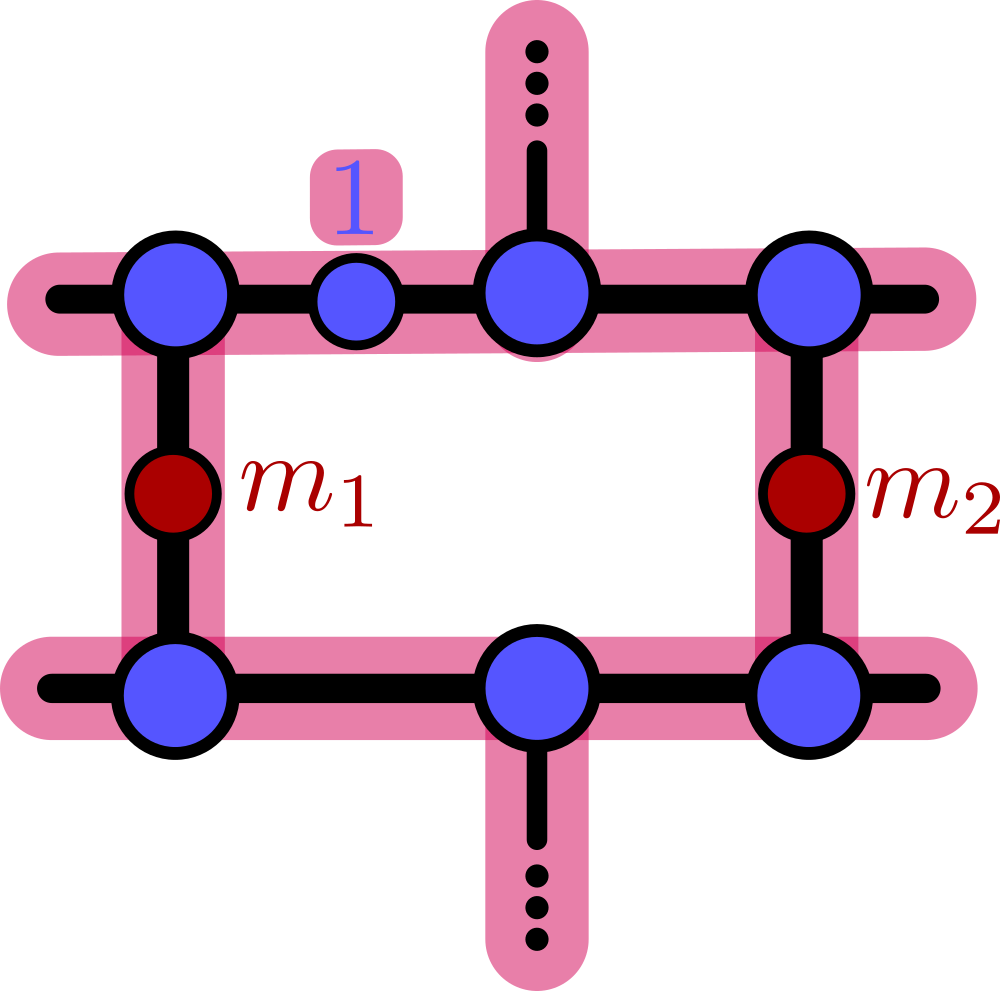}}
\end{align}
and hence corresponds to a non-trivial logical error, i.e. has a \textit{consequence} on an encoded state.\footnote{Concretely, the circuit with the fault maps a logical $X$ operator to a logical $-X$ operator which can also be understood as acting with a logical $Z$ operator.}
Note that this is not the case for every undetectable fault.
For example, consider the two-qubit $Z$ fault,
\begin{align}
    \raisebox{-0.5\height}{\includegraphics[width=0.3\linewidth]{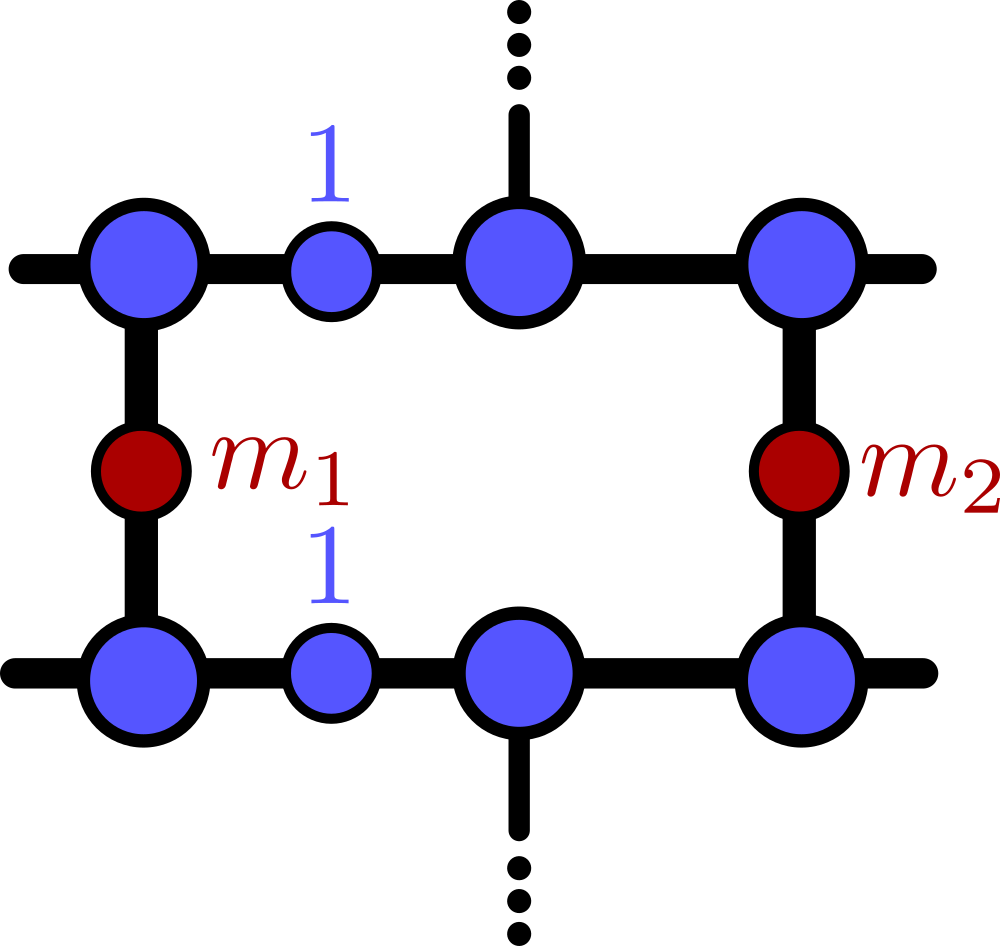}}\;.
\end{align}
It is not charged with respect to any detector flow, i.e. is undetectable.
At the same time, it is not charged with respect to any logical flow (since the charge is defined $\mod$ 2) and hence has trivial action on the encoded qubit. We say such a fault is \textit{inconsequential}.
Taking a step back, this is also expected since at the time when the fault happens the associated Pauli operator $Z_iZ_j$ is an element of the ISG.
With respect to the chosen input (on the left) and output (on the right), the network also has stabilizer flows, e.g.,
\begin{align}
    \raisebox{-0.5\height}{\includegraphics[width=0.3\linewidth]{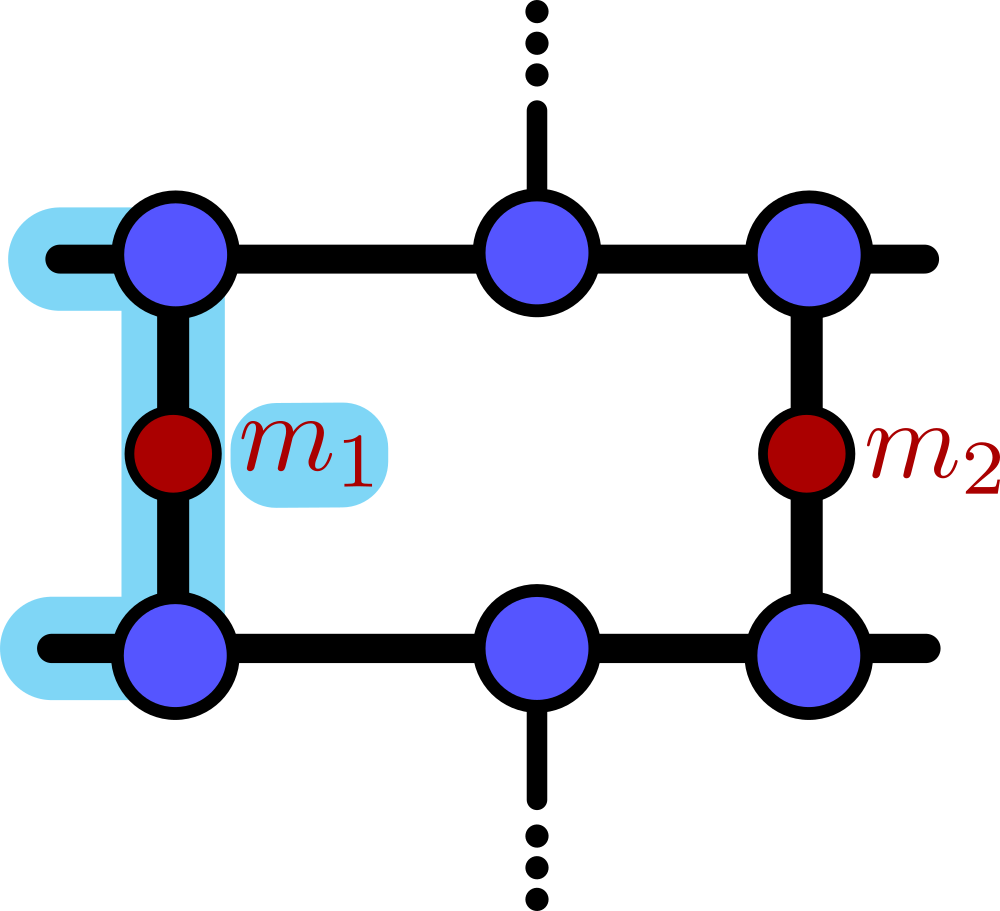}}\;,
\end{align}
which shows that any associated circuit projects any input state into a
$+1$ eigenstate of the stabilizer $(-1)^{m_1}Z_iZ_j$ after the first measurement, up to signs of the stabilizers of the repetition code.
By symmetry of the network we find the same stabilizer flows on the output legs.
Knowing these stabilizer groups is mainly important when designing QEC experiments (see Sec.~\ref{sec:probing}) and interfacing codes (see Sec.~\ref{sec:remarks}).

We \NEW{would like to} highlight that the preceding analysis of the effect of faults was independent of their physical origin.
In particular, we could have placed faults on different edges in the network, possibly representing different types of physical error mechanisms, e.g. measurement errors, and could follow the same arguments.
This naturally leads to the definition of a \textit{fault distance} as the smallest number of (order-$p$) generating faults that need to be combined to form a consequential undetectable fault.
Importantly, this definition  depends on the error model but is applicable to all cases where the noise acts as a stochastic Pauli channel.
Note that the definition of fault distance has appeared similarly in Ref.~\cite{Bombin2023logical}, but we phrase it in terms of the language of RGB tensor networks and Pauli flows.

We also want to note that there is an efficient algorithm to obtain all Pauli flows of a network.
If the network directly comes from a circuit, we can equivalently perform standard stabilizer tableau calculations rendering it unnecessary to describe the algorithm in this work.
There may still be advantages gained in the simulation performance by considering the tensor network representing the whole spacetime evolution instead of time slices through the network which is done in a more conventional simulation with stabilizer tableaus. We leave this analysis to future work.

\section{Fault tolerance in the XYZ ruby code}\label{sec:benchmarking}

In this section, we investigate the fault tolerance properties of the XYZ ruby code.
To set the stage, we first introduce the notion of \textit{probing} an error-correcting protocol.
This will unify various types of error correction experiments\footnote{Note that we mostly think of numerical experiments/simulations, where the true dynamics of the system is known. However, the same formal structure applies to real-life experiments.} and give a constructive framework to assess the quantities that can be probed in a given experiment.
It will also make transparent how initialization and read-out layers have to be designed in order to probe a desired quantity.
We note that our perspective is closely related to the one obtained from the detector error model\,\cite{gidney2021stim,DetectorErrorModels} and Delfosse's spacetime code\,\cite{delfosse2023spacetime} which we comment on in Sec.~\ref{sec:spacetime_code}.

We exemplify our perspective by benchmarking various quantities in the XYZ ruby code \NEW{on a torus}.
To do so, we first identify the Pauli flows of the tensor network of the XYZ ruby code with a rewinding schedule.
We describe the detector, stabilizer and logical flows and give some intuition on how they are affected by Pauli errors within the circuit.
We find that, as expected, the tensor network is truly topological and there is a natural separation between local and non-local flows.
This leads to two types of QEC experiments that we perform: memory and stability experiments.
Both have physical interpretations in terms of the underlying topological phase.
In both experiments, we find competitive thresholds and an exponential suppression of the logical error rate under various noise models for moderate system sizes using a belief propagation decoder enhanced with ordered and localized statistics decoding (BP+OSD/LSD).
Finally, we describe the details of the implementation and the numerical experiments we perform to extract performance indicators.

\subsection{Probing an error-correcting protocol}\label{sec:probing}
In the following, we introduce the concept of \textit{probing} an error-correcting protocol expressed as an RGB tensor network.
Since we live in a classical world, we will never be able to probe the full linear map implemented by an error-correcting protocol in a single experiment.
In practice, ``initialization'' and ``read-out'' circuits have to be chosen, probing only matrix elements.
When probing an RGB tensor network, both initialization and read-out layers can also be expressed as an RGB tensor network, usually of constant depth.
They are networks that only have input, respectively output, legs and their (logical) action is fully described by their stabilizer flows.

We understand a logical probing experiment as the tensor network obtained when terminating the tensor network of the error-correcting protocol with an initialization and read-out layer.
This tensor network does not have any open legs and hence represents a number.
However, this number will not play an important role in the experiment.
More importantly, all Pauli flows of the full experiment are detector flows.
As such, they all give rise to a set of signs of tensors (mostly these will correspond to measurement outcomes) whose sum is $0$ mod $2$ in the absence of noise.
In particular, they define a classical (linear) code.
These will be the only quantities entering the probing experiment.

Every RGB tensor network is labeled by a set of signs of individual tensors.
Some of them are fixed (to 0 or 1), others are open.
One can think of the open ones as being determined by measurement outcomes.
Let $M$ be the $\bZ_2$-vector space spanned by all possible signs in the tensor network.
For convenience, we can disregard the ones being fixed to 0 and assume we investigate a circuit where the remaining signs correspond to measurement outcomes.
The set of detector flows of the experiment forms a group, which we denote by $D$.
It defines a classical linear code $C_D\subseteq M$.
For any probing experiment we pick a subgroup of \emph{decoding flows} $D_{d}\leq D$.
This subgroup also defines a linear classical code, $C_{D}^{d}$ and a quotient group $D/D_{d}$ of \emph{observable flows}.
Importantly, $C_D\subseteq C_{D}^{d}$.
In this context, the decoding problem can be phrased as follows: Given the syndrome information associated to $D_{d}$, respectively $C_D^{d}$, find a prediction for the information associated to $D/D_{d}$.

In practice, we define $D$ and $D_{d}$ in terms of their generators which in turn defines parity check matrices $H_D$ and $H_{D}^{d}$ for both $C_D$ and $C_{D}^{d}$.\footnote{Each generator of of a subgroup of detectors $D$, respectively $D_{d}$, can be associated to a $\bZ_2$ vector in $M$.
The parity check matrix $H_D$, respectively $H_D^{d}$, is the matrix whose rows are the $\bZ_2$ vectors associated to the generators of $D$, respectively $D_{d}$.
This construction leads to $C_D = \ker(H_D)$ and $C_D^{d} = \ker(H_D^{d})$.}
$H_{D}^{d}$ will be a submatrix of $H_{D}$ (obtained by removing rows associated to the constraints $D/D_{d}$) and we denote the complementary submatrix as $H_{D}^{o}$.
Given these definitions, the (numerical) experiment works as follows:
\begin{enumerate}
    \item Sample the full erroneous circuit, resulting in a binary vector of $\bZ_2$ values (e.g.~measurement outcomes) $\vb{m}$
    \item From these infer
    \begin{itemize}
        \item \textit{syndromes} $\vb{s} = H_{D}^{d}\vb{m}$ and
        \item \textit{observations} $\vb{o} = H_{D}^{o}\vb{m}$.
    \end{itemize}
    
    \item Decode the syndrome $\vb{s}$, ideally using information about the error model,\footnote{For example, the detector error model in \texttt{stim} matches the syndrome onto a decoding graph obtained from the error model. Note that this graph can be easily constructed from the tensor network representation of the probing experiment and the noise model.
    For details on this, see Sec.~\ref{sec:circuits}.} to obtain a binary vector $\vb{m'}\in M$ that can be interpreted as a proposed set of measurements that need to be flipped to ``redo'' the effect of the errors in the circuit. In that reading, $\vb{m'}$ has a 1 at the position of each measurement that needs to be flipped and 0 everywhere else.
    
    \item From $\vb{m'}$, calculate a predicted value for the observations $\vb{o'} = \vb{o}\oplus H_{D}^{o}\vb{m'}$.
    \item Declare a failure if $\vb{o'} \neq \vb{0}$.
\end{enumerate}
Of course, only specific choices for $D$ and $D_{d}$ give rise to a meaningful experiment.
Moreover, the choice of initialization and read-out layer highly influences the choices that are even possible.

In practice, however, there are clear guidelines in designing both the initialization and read-out layer as well as for the choice of $D_{d}$.
For the main part of this section we focus on topological codes where there are two main classes of experiments that differ by the choice of observables.
In both cases, the detector group $D$ splits into a locally generated part $D_{\mathrm{loc}}$ and a part which cannot be created locally.
We pick $D_{d} = D_{\mathrm{loc}}$ and identify $D/D_{\mathrm{loc}}$ with the set of observables of the experiment.

\begin{figure*}
    \centering
    \includegraphics[width=0.7\linewidth]{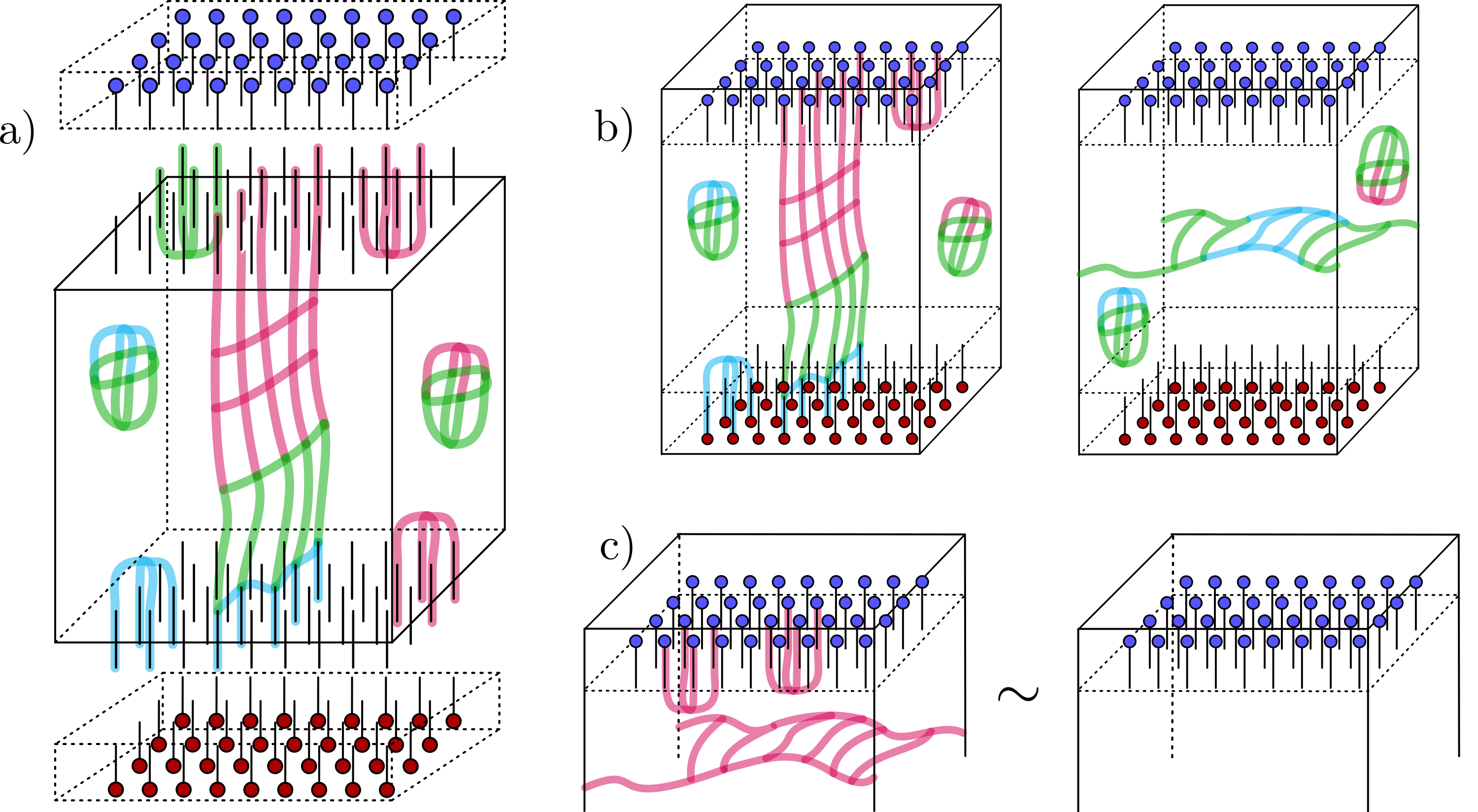}
    \caption{
    Schematic depiction of the process of designing and performing an error correction experiment to probe (non-trivial) observables in a protocol. We think of the left and the right of the cubes being identified, as well as the front and the back, as the code is defined on a torus, and time going upwards. a) Given the bulk of a protocol and its flows, input and output layers need to be chosen to preserve the desired set of stabilizer and logical flows.
    b) After closing off the input and output, the experiment is described by the detector flows of the resulting network. In topological protocols the homologically non-trivial flows are probed, using the local detector flows as input to the decoder. On the left, we depict the flow probed in a memory experiment, on the \NEW{right} the flow that is probed in a stability experiment.
    In general, they can be performed within the same experiment.
    c) The choice of temporal boundaries determines which flows are probed in the experiment. A read-out layer can render a flow that was non-trivial in the bulk  of the protocol trivial in the actual experiment.
    Here, we illustrate this for a red-colored flow that one would na\"ively consider a non-trivial observable for a stability experiment.}
    \label{fig:probing_QEC_protocol}
\end{figure*}

\subsubsection{Memory experiments} 
\NEW{A} memory experiment \NEW{probes} the capability of a circuit to preserve logical information over time~\cite{dennis2002topological}.
The observables are defined by flows that connect the two time-like boundaries of the experiment in a non-trivial way, i.e., are non-trivial elements in $D/D_{\mathrm{loc}}$.
In particular, they are formed by logical flows of the circuit that are ``closed off'' by stabilizer flows of the initialization and read-out layer. Additionally, some of the stabilizer flows of the circuit are promoted to local boundary detector flows. Together with the local detector flows of the circuit itself they generate the group $D_d$ and with that define the syndromes for decoding. We depict the different classes of Pauli flows in a memory experiment in Fig.~\ref{fig:probing_QEC_protocol}.

For example, to probe a logical flow that has only blue highlights on the output legs, we want to choose an initialization and read-out layer that only have blue stabilizer flows. Moreover, it should allow for an attachment of the blue logical flow.
This is achieved, for example, by a single-qubit initialization in the $Z$-basis represented by single-legged red spiders.

Importantly, the choice of logical flow that can be attached to the initialization and read-out layer automatically rules out other logical flows being probed and boundary detector flows being formed.
For example, probing a logical $Z$-type flow rules out a probe of an anticommuting $X$-type flow.
Additionally, the local detector flows at the boundary are also constrained.
As a consequence, close to the boundary, there will be Pauli errors that are not detectable in the experiment.
However, since the logicals that would be flipped by that error are not probed by the experiment, any errors of that form are inconsequential and hence do not lower the distance observed in these experiments.

\subsubsection{Stability experiments}
The concept of a \textit{stability experiment} has been  introduced in Ref.~\cite{gidney2022stability} as a numerical experiment that probes how well a code can perform lattice surgery-like operations.
More specifically, it allows us to give a better estimate on how many times measurement rounds have to be repeated to suppress undetectable error-chains in the time direction.

In this subsection, we focus on stability experiments in the context of $2+1$-dimensional topological codes.
The observables in a stability experiment are defined by non-trivial detector flows that connect spatially separated boundaries without connecting any of the temporal boundaries.
Again, we understand non-trivial as being in a non-trivial equivalence class in $D/D_{\mathrm{loc}}$.
Note that these flows only exist for certain spatial boundary configurations.
Specifically, they are supported on a spacelike region in which the topological code has a property called \textit{topological charge conservation} that guarantees a given parity on the stabilizers supported in that region~\cite{brown2023conservation}.
For example, the stabilizers of the toric code (placed on, e.g., a surface without a boundary) multiply to the identity operator.
In general, one can think of these types of non-local flows as giving rise to a global constraint on the stabilizers of an ISG. 
Any topological code has this property when placed on a closed, compact manifold, such as a torus or a surface with a single connected boundary component.

\subsubsection{Interplay of stability and memory experiment}
As stated above, it might seem that a stability experiment and memory experiment can be designed independently of one another but performed simultaneously.
However, this is not true for topological codes.
In order for it to be a non-trivial stability experiment we want the detector flow of the observable to be non-trivial in $D/D_{\mathrm{loc}}$.
In particular, this depends on the local detectors close to the boundaries, including the temporal ones.
In order to perform a stability experiment we have to choose initialization and read-out layers that do not close off the local flows that could move the stability observable into the boundary and thereby \NEW{make} it trivial.
This restricts the type of stability experiments that can be performed simultaneously to a memory experiment.
For example, we cannot probe an $X$-type memory observable together with an $X$-type stability observable, even if the circuit itself has both types of flows to begin with.
Either the stability observable will be trivial or the $X$-type logical flow will be trivial in the experiment, i.e., can be written as a sum of local detector flows.

We illustrate this phenomenon in Fig.~\ref{fig:probing_QEC_protocol}.
Such trade-offs exists in any topological code and are related to the fact that all boundaries in the experiment have to be topologically protected boundaries in order to have a protocol with macroscopic fault distance.
In $2+1$ dimensions, these are fully determined by so-called \textit{Lagrangian subgroups} which determine what type of anyon worldlines can be terminated at that boundary.
These in turn define which detector flows exist close to the boundary (the ones detecting the confined anyons) and which logical flows can terminate at the boundary (the ones that correspond to the worldsheet of the logical operators associated to the anyons in the Lagrangian subgroup).

\subsection{Pauli flows of the XYZ ruby code} \label{sec:pauliflows}

In the following, we identify Pauli flows of the XYZ ruby code protocol defined by the rewinding schedule. 
As described in the section above generically for topological protocols, we partition the Pauli flows into \emph{local} and \emph{non-local} flows.

\subsubsection{Decoding flows}
We find detector flows of the rewinding protocol by first considering space-like symmetries and then consecutively highlighting edges of the XYZ tensor network until we find a spatially and temporally confined Pauli flow, so we will use them as decoding flows in probing experiments. In the following, we will refer to them as \emph{detectors}.
Consider one unit cell of the ruby lattice with a few timesteps before and after a $ZZ$-measurement.
Inserting a generator of the ISG,
\begin{align} \label{eq:ISG0_2}
    \mc{S}_0 = \Bigg\langle \raisebox{-0.25\height}{\includegraphics[height=8pt]{ZZedge.pdf}}, \raisebox{-0.4\height}{\includegraphics[width=0.18\linewidth]{Z_plaquette.png}}, \raisebox{-0.4\height}{\includegraphics[width=0.18\linewidth]{Y_ring.png}} \Bigg\rangle,
\end{align}
as faults at that timestep have, by definition, trivial action on the circuit. In the three-colored graphical calculus, this corresponds to placing \NEW{two-legged} spiders for each qubit a generator acts on  which leaves the RGB tensor network invariant.
Mapping these to highlights (cf. Eqs.\,\eqref{eq:pauliflowmap}) does not give a valid detector flow yet. For this, we have to construct highlights of adjacent edges, such that at each tensor, the highlights correspond to a Pauli flow, and non-trivial highlights form a spatially and temporally compact set.
We can assume that the highlights emanating from the ISG generator are part of a detector flow, because the previous measurement rounds have initialized this generator. Owing to the mirror symmetry of the network, we can expect that the corresponding measurements of the following rounds will contribute in forming a detector.
We therefore also only need to highlight edges in one time direction, e.g. in time forward after the $ZZ$-measurement, and the other half of the highlights are their mirror. 

\begin{figure*}
    \centering
    \includegraphics[width=0.9\linewidth]{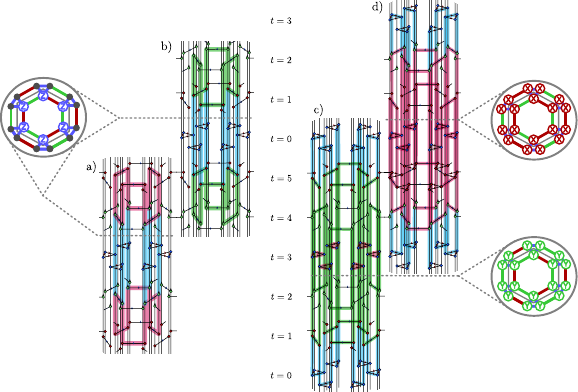}
    \caption{
    Detectors of the XYZ ruby code in the rewinding schedule, with time going upwards. For each detector flow, there exists an equivalent one shifted by $6$ timesteps, corresponding to the period of the schedule and equivalent ones centered around every plaque(tte).  
    Whenever a measurement spider is charged with respect to the flow (i.e. it is highlighted by a different color), it contributes to the set of measurements that are constrained to be deterministic in the error free case. 
    \NEW{Cutting detector flows gives rise to instantaneous stabilizers.}
    a) $\redr \blb \redr$  and b)  $\grg \blb \grg$ plaquette detector flows. These flows combine $12$ measurements that correspond to reading out the eigenvalue of the $6$-body $Z$-plaquette twice by combining $XX$- and $YY$-measurements.
    c) $\blb \grg \blb$  and d) $\blb \redr \blb$ plaque flows. They correspond to measuring a $18$-body $Y$- or $X$- plaque operator two times.
    Taken together, all detector flows densely cover the RGB tensor network.
    }
    \label{fig:detectors}
\end{figure*}

We start with the $Z$-plaquette generator and highlight the corresponding inner\footnote{When we talk about the 18-site plaque, we use the notion of \emph{inner} to refer to edges related to the six qubits (or edges) forming the hexagon of the plaquette and \emph{outer} regarding the twelve qubits (or edges) around the outside of the plaque.} time-like edges after the $ZZ$-measurement at $t = 3$ in blue (cf. Fig.~\ref{fig:detectors} a)).  
At the green spider in the next timestep $t=4$, valid highlights of legs are all blue with an even number of red highlights, cf. Eq.\eqref{eq:spider_flow}.
Having the next $XX$-measurement in mind we choose a pair of red highlights on the horizontal and vertical upwards outgoing legs of the spider.
At the red spiders ($t=5$), we choose to highlight the horizontal legs in red, leading to a (half) closed flow. 
A reflection about the $ZZ$-measurement at $t=3$ generates a closed Pauli flow in one unit cell from times $t=1 \mod 6$ to $t=5 \mod 6$ and therefore a detector flow.

Analogously, detectors can be formed by starting with blue highlights after the $ZZ$-measurement at $t=0$. 
After the next (and before the previous) $XX$-measurement, the color of the detector flow should be green in order to be closed at the $YY$-measurement in $t = 2$ and $t = 4$, see Fig.~\ref{fig:detectors} b).
These detectors exist for every unit cell, and we call them $\redr \blb \redr$- and $\grg \blb \grg$ \emph{plaquette detectors}.

Also from an operational perspective these are reasonable detectors: measuring $XX$ and $YY$ once gives the eigenvalue of the $Z$-plaquette as a product of the outcomes around the plaquette. 
This operator trivially commutes with the following $ZZ$ measurement. In the rewinding schedule, now measuring $YY$ and $XX$ reads out the plaquette a second time 
without randomizing its eigenvalue. 

The $Y$- and $X$-plaque stabilizer generators give rise to a second set of detectors, shown in Fig.~\ref{fig:detectors} c) and d). 
These \emph{plaque detectors} start and end in a $ZZ$-measurement and have again two different colors. Depending on the measurements around the middle $ZZ$ step, we call them $\blb \grg \blb$ and $\blb \redr \blb$ (right) for $YY$- and $XX$-measurements. 

Another set of local detectors is purely space-like and formed by the \emph{triangles} of the $ZZ$-measurements highlighted in blue. They correspond to the constraint that the value of the third $ZZ$-measurement is already determined by the sum of the first two.
 
This completes the set of generating detectors for the rewinding schedule. In the period-3 schedule, we find analogous plaquette and plaque detectors. Note that they do not exhibit the mirror symmetry about the $ZZ$-measurement due to the missing symmetry in the schedule.

\subsubsection{Observable flows}
Here, we show two types of flows that we will use as  observables in memory and stability experiments.

\begin{figure}
    \centering
    \includegraphics[width=\linewidth]{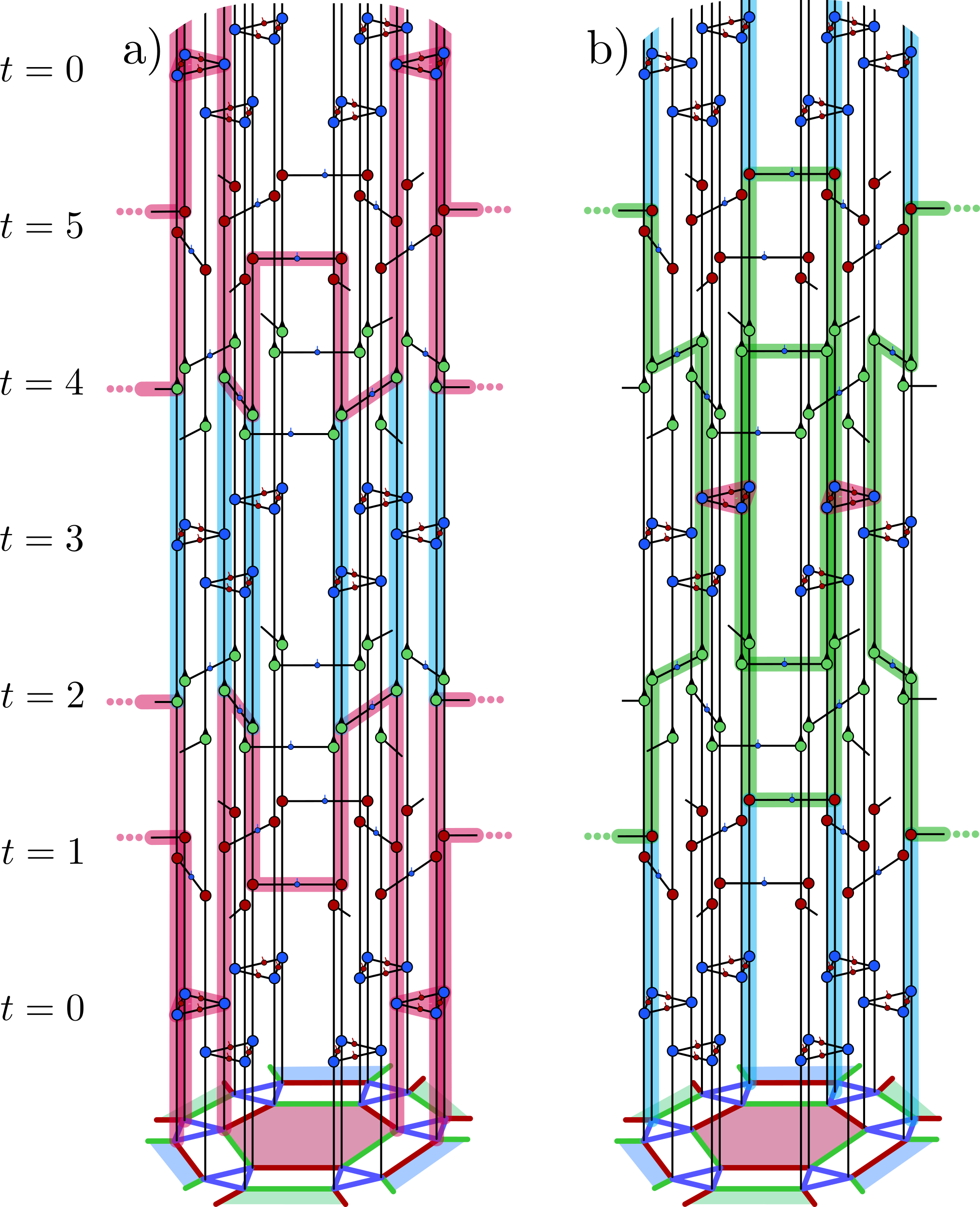}
    \caption{We show parts of two logical flows in the bulk of the XYZ ruby code with the rewinding schedule.
    When cutting the network along a plane perpendicular to the time direction, i.e., at some time $t_0$, the flows give rise to logical operators of the ISG at $t_0$.
    As in any $2+1$-dimensional topological protocol, the logical flows are membrane like, indicated with the colored dots on the left and the right of both flows.
    Each non-trivial logical flow is supported along one homologically non-trivial membrane.
    In a), we show an $\redr \blb$ ($\mathcal{L}_X$)  logical flow that when cut after $t=0$ gives rise to a \texttt{rx} string operator going through the red plaquette shown at the bottom, cf. $\overline{X}_1$ of Fig.~\ref{fig:logicals_ISG0}. Cut after $t=3$, the string operator has evolved to a \texttt{bz} operator through the blue plaquettes adjacent to the blue plaquette. 
    Similarly, the $\grg \blb$ ($\mathcal{L}_Y$) logical flow in b) on the right can be associated with a \texttt{rz} $\times$ \texttt{gz} $=$ \texttt{bz} (after $t=0$) and a \texttt{gy} (after $t=3$) string, as expected from the automorphism applied in these three timesteps, cf. Eq.\,\eqref{eq:automorphism_table}.
    }
    \label{fig:logical_flows}
\end{figure}

The first are the logical flows of the QEC protocol that is implemented by the RGB TN observable flows. They are directly related to logical operators of the ISGs, cf. Fig.~\ref{fig:logicals_ISG0}. 
In fact, using the intuition developed in the construction of detectors, we can find the measurements required to properly transform the logical operator from one timestep to the next, cf. Fig.~\ref{fig:logical_autmorphism}. 
In Fig.~\ref{fig:logical_flows}, we depict two (Pauli-)inequivalent logical flows, which we will refer to as $\redr \blb$ (or $\mathcal{L}_X$) and as $\grg \blb$ (or $\mathcal{L}_Y$) logical flows. They are supported along homologically non-trivial membranes as in any $2+1$-dimensional protocol. 
When constructing a memory experiment, we will use thin layers that are compatible with one type of these and declare them as observables of the memory experiment.

\begin{figure}
    \centering
    \includegraphics[width=0.75\linewidth]{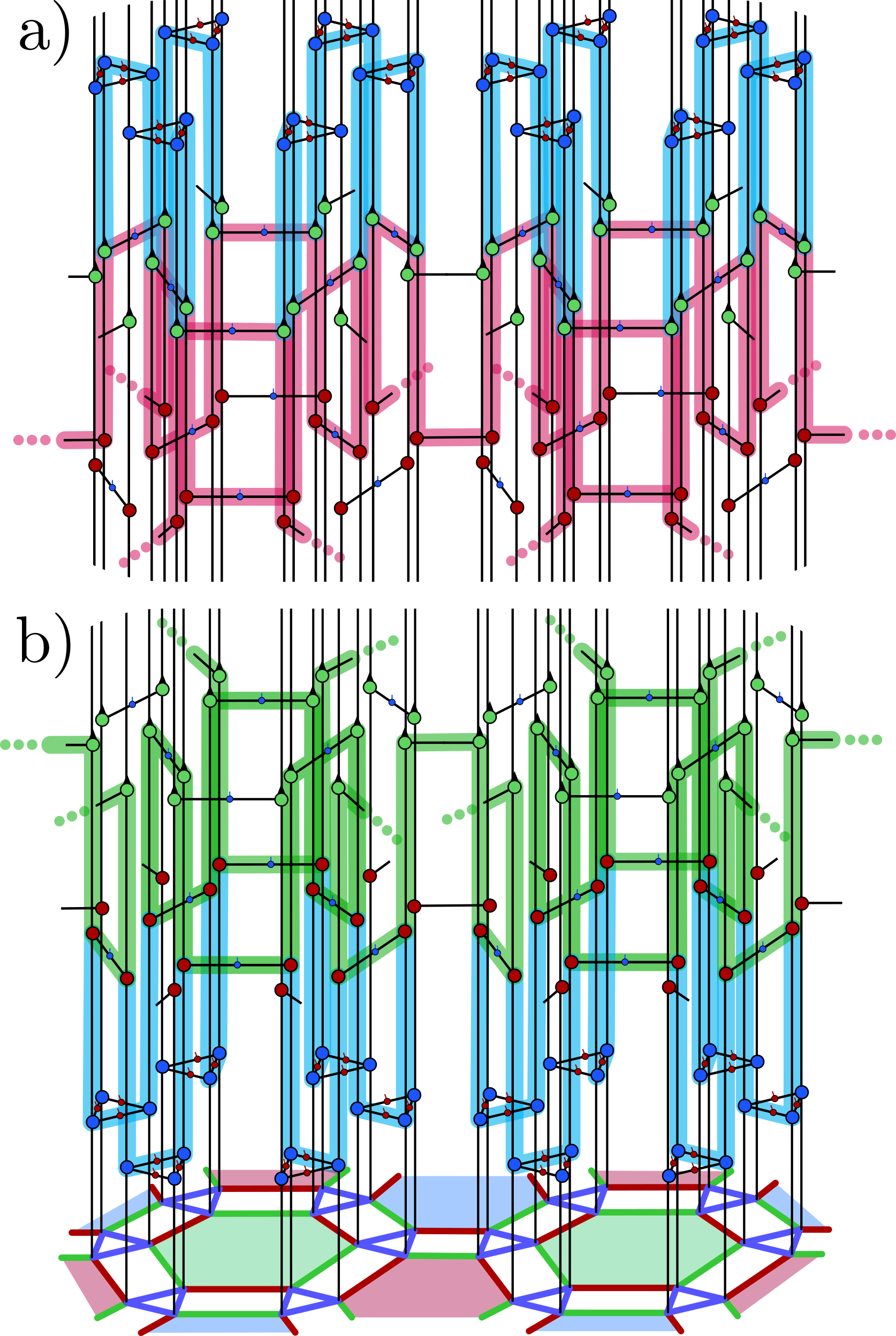}
    \caption{Observable flows related to global constraints of the ISGs, i.e., products of ISG elements. There are two independent constraints related to which color of plaquettes are chosen in space (see also main text). Additionally, an offset in time gives two additional independent flows to get a total of 4 as expected by topological charge conservation in the color code phase.  In a), we show the $\redr \blb$ closing off blue highlights on $ZZ$- edges around a red plaque. In b), we show a time offset $\grg \blb$ flow, starting with blue highlight around a green plaque.}
    \label{fig:global_constraints}
\end{figure}

The second type are detector flows that cannot be generated by above introduced local detector flows. We employ these global detector flows as observables in a stability experiment.
From topological charge conservation, we expect the existence of $4$ independent global constraints, one for each independent anyon in the color code\,\cite{kitaev2003fault}. These give rise to two types of logical flows that are closely related to conservation laws in the (static) subsystem code. 
There, a global constraint is the product over all small string operators of one color around all vertices that evaluates to identity (cf. Eq.\ (5) of Ref.~\cite{Bombin2010subsystem}). 
To obtain those, we highlight the blue edges around all plaques of one color, e.g. red as shown in Fig.~\ref{fig:global_constraints} a). Adding highlights backward in the time direction, there is a choice of consistent highlight to apply at the next ($YY$) step. We change the color to red such that at the next $XX$ measurement, the highlights can be chosen to close the flow obtaining an $\redr \blb$  observable flow.
Choosing a second color of plaques for highlighting the blue edges at the $ZZ$-measurement timestep, we get an independent observable flow. 
The third color, however, is the product of the flows of the other two colors, up to local triangle detectors. This gives a total $2$ independent observable flows.
A second set of two independent flows can be  equivalently obtained starting at the $ZZ$-measurement step at $t = 0 \mod 6$, sandwiched by two $XX$-measurements, which we show for highlight around green plaques in Fig.~\ref{fig:global_constraints} b). Again, the product of two such flows of two colors yields the flow of the third color. We therefore get two more independent ($\grg \blb$) observable flows.
It is apparent in the structure of both observable flows that they cannot be generated by products of local detectors which only deform the observable flows in spacetime.

Already at that point, even before constructing the actual probing experiment, we can make statements on the fault tolerance of the protocol, based on decoding and observable flows as we show in the next subsection.

\subsection{Phenomenological fault tolerance} \label{sec:phenomFT}
To investigate the fault-tolerance of the protocol, we consider a simplified but complete error model based on the observations\footnote{A similar approach was used to argue about fault-tolerance in the honeycomb code\,\cite{hastings2021dynamically}.} that
\begin{enumerate}
    \item a measurement error during measurement $M_{PP}$ is equivalent to a Pauli error $P' \neq P$ just before and after a noiseless measurement,
    \item any Pauli $P$ is proportional to a product of two other Paulis,  $P \propto P' P''$, and it is therefore enough to consider a basis of two Pauli operators after each measurement,
    \item a Pauli fault $P$ can be commuted through a measurement $M_{PP}$. 
    \item after a measurement $M_{P_i P_j}$, the Pauli operator $P_i P_j$ is part of the ISG, such that the errors $P_i$ and $P_j$ are stabilizer-equivalent $P_i \sim P_j$. 
\end{enumerate}
The last property is a consequence of the just-measured two-body operators on edges being inconsequential faults or fault equivalences.\,\footnote{In the literature, minimum fault configurations that do not violate any detectors are also called elementary equivalences, 
see, e.g., Ref.~\cite{andrist2010tricolored}.}

This leaves a restricted error model where we place single-qubit $X,Y$ and $Z$ faults after the $XX$, $YY$ and $ZZ$ measurements respectively. Any other Pauli fault (of arbitrary weight involving any spatial or temporal location) can be generated by this set. We show some of these properties in the RGB TN in App.\ref{app:noise}, Fig.~\ref{fig:noise_model_reduction}.
Since any single-qubit fault on the other qubit(s) of the just-measured edge (triangle) is equivalent, we can also think of the red, green edges and blue triangles as the elementary faults.

All the elementary faults of this error model violate three detectors.
This is similar to the action of elementary faults in a standard color code, where $X$- and $Z$- faults violate the three adjacent plaquette detectors, $Y \propto ZX$ and $c_Y = c_Z + c_X \mod 2$ for $X$- and $Z$-faults inserted in the network at the same location. Here $c_P$ is the charge of the corresponding $P$-spiders with respect to the Pauli flows of the network, or equivalently, the syndrome.
We find, however, that other properties are different from $2+1$-dimensional color code decoding graphs.

Firstly, the XYZ ruby code is not CSS and a $Y$-elementary fault is independent of $X$- and $Z$- elementary faults.
Secondly, the elements of the ISG of the current timestep represent inconsequential errors at that timestep, i.e., any fault can be multiplied by an element of the ISG to get an equivalent fault with the same action on the codespace. 
As shown in Sec.~\ref{sec:automorphism}, the elements of the ISG can be regarded as stabilizers of a color code for $\isg{0}$ and three copies of toric codes for $\isg{1}^{(r)}$ and $\isg{2}^{(r)}$. This phase transition is reflected in the behavior of faults at different timesteps, for which we would like to develop a physical intuition.

\begin{figure*}
    \centering
    \includegraphics[width=\linewidth]{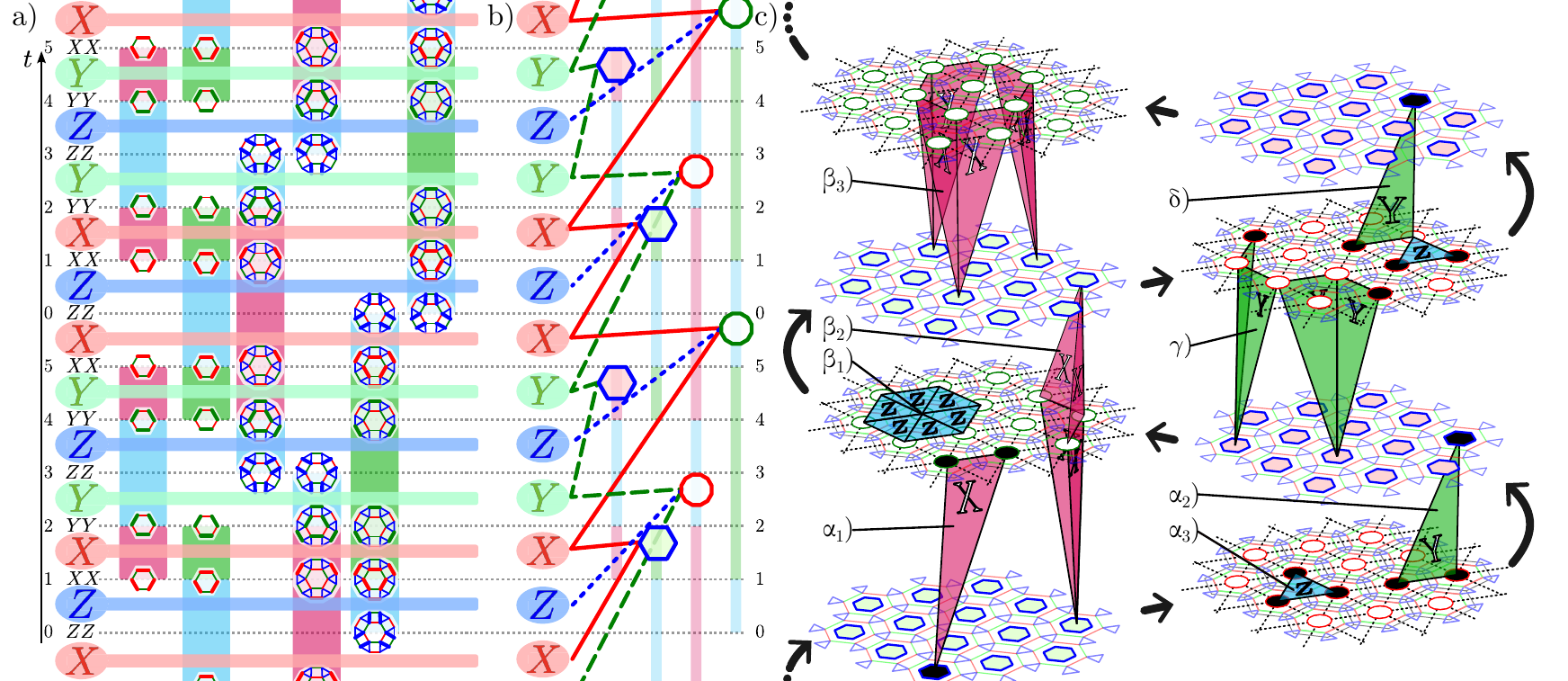}
    \caption{a) Action of elementary faults on detectors. We place single-qubit $X$-, $Y$- and $Z$- faults after $XX$-, $YY$-, and $ZZ$ measurements respectively and represent them by horizontal shades. We draw the temporal extent of the detectors vertically and use a shade corresponding to the highlights of the Pauli flows. We also indicate the edges contributing to the value of the detector. Whenever the colors of single-qubit faults and detectors differ, the corresponding detector is violated. b) \NEW{This allows one to draw a time-slice of the decoding graph, a bipartite graph with elementary faults connected to the detectors they trigger.  We represent detectors by hexagons and dodecagons attached to the last time step of the corresponding shade.} For the $X$- and $Y$- elementary faults, \NEW{the decoding graph} corresponds to two Tanner graphs of (classical) repetition codes. $Z$-elementary faults violate \NEW{plaques of a single time-step}.
    c) \NEW{For every time-step where detectors close, we draw a plane associating them to their spatial location. Since elementary faults flip three detectors, we now represent them as triangles. We draw a subset of faults of the resulting} space-time decoding graph. Time follows the arrows, elementary faults are red ($X$), green ($Y$) and blue ($Z$) triangles. \NEW{Elementary faults occurred on qubits that are supported on a red (green) edge cut by the short side of the $X$- ($Y$-) triangle, or on one of the three qubits covered by the $Z$-triangle.} A detector that is violated is filled. 
    $\alpha$\textsubscript{1-3}) Elementary faults violate three detectors. $\beta$\textsubscript{1-3}) We show some faults that do not flip any detectors. They are typically related to elements of the ISG ($\beta$\textsubscript{1,3})), but can also involve faults of different time steps. $\gamma$) Pure $Y$ (and $X$) faults always violate two plaque stabilizers on the endpoints of a string. $\delta$) Combination of $Y$ and $Z$ faults.  }
    \label{fig:fault_tolerance_detector_graph_time}
\end{figure*}

To this end, suppose we place the elementary faults on a single qubit and investigate how these violate the detectors.
In Fig.~\ref{fig:fault_tolerance_detector_graph_time} a), we show detectors in time, represented as vertical shades in the color of the respective highlights of the flow. Horizontal lines represent single-qubit $X$-, $Y$- and $Z$-faults after the corresponding 2-body measurements. 
Whenever crossing horizontal and vertical lines have different colors,  the fault is charged with respect to the corresponding decoding flows and triggers the associated 
check in the decoding graph.
Note that elementary $X$-faults only violate $\grg$-type detectors (i.e., $\blb \grg \blb$ plaques and $\grg \blb \grg$ plaquettes) and elementary $Y$-faults only violate $\redr$-type detectors (i.e., $\blb \redr \blb$ plaques and $\redr \blb \redr$ plaquettes). 
The $Z$- faults violate plaques only.
To describe how syndromes of elementary faults combine, we draw decoding graphs that have vertices for faults and detectors and edges between fault $E_i$ and detector $D_j$ if $E_i$ triggers $D_j$. 
In Fig.~\ref{fig:fault_tolerance_detector_graph_time} b), we show a decoding graph abstracting away the spatial locations of faults and detectors, focusing on their action in time. 
The $X$- and $Y$-faults generate Tanner graphs of classical repetition-codes, similar to how measurement errors trigger two temporally spaced detectors of a surface code experiment. 
The $Z$-faults violate plaques of a unique timestep, in a similar way to data-qubit errors of a color code experiment.

In Fig.~\ref{fig:fault_tolerance_detector_graph_time} c), we extend that picture to include faults at different spatial locations.
First, the temporal behavior induces a natural bipartition of the detectors into those violated by $X$ and $Z$ faults after $t=0 \mod 6$, and those by $Y$ and $Z$ faults after $t=3 \mod 6$. 
For every detector we place vertices on the ruby lattice in the center of the plaque(ttes).
Because the elementary faults always trigger three detectors, faults are extended to hyperedges which we draw as triangles in Fig.~\ref{fig:fault_tolerance_detector_graph_time} c).
Whenever an odd number of a triangle's corners touch a detector, it is violated. 

We draw subsets of fault configurations, starting with the elementary faults in Fig.~\ref{fig:fault_tolerance_detector_graph_time} c) $\alpha$).   
The $Y$- and $X$-faults violate a single plaquette and two plaque detectors spanning over different timesteps. 
The $Z$ faults violate three adjacent plaques in a single timestep. Superimposing a triangular lattice on the ruby lattice with triangles in the faces, this is equivalent to how $Z$-faults behave in a regular color code on a hexagonal tiling.

In Fig.~\ref{fig:fault_tolerance_detector_graph_time} c) $\beta$), we show some other inconsequential errors. In $\beta$\textsubscript{1}), $6$ $Z$-faults in a loop surrounding six triangles, such that each involved plaquette detector is flipped twice, are undetectable. These fault configurations correspond to the stabilizer of the ISG at that timestep, cf. Eq.\,\eqref{eq:ISG_r_1}, and have a resemblance to the smallest homologically trivial loops in hexagonal color codes.
In $\beta$\textsubscript{2}) we show two $X$-faults on one plaquette, both repeated at consecutive timesteps as an inconsequential error that spans space and time.
The weight $6$ $X$-fault shown in $\beta$\textsubscript{3}) is another fault related to ISG elements, this time a stabilizer of one of the three toric codes of $\isg{1}$, cf. Fig.~\ref{fig:ISG1-concatenation}. 

The relation to the decoding problem of toric codes becomes apparent in Fig.~\ref{fig:fault_tolerance_detector_graph_time} c) $\gamma$). If we restrict the $Y$ ($X$)-faults to the subset of plaque detectors, they always trigger them in pairs at the endpoint of a string of faults. We believe that this can be used to design a matching-based decoder, similar to how matching decoders are adapted to color codes\,\cite{stephens2014efficient,gidney2023new,lee2024color}. 

Having established that the distance of the XYZ ruby code is $\mc{O}(s)$ (where $s$ is the 1-dimensional system size) and considering that any single-qubit fault can be uniquely identified, we conjecture that the XYZ ruby code also exhibits a threshold. This is supported by the numerical results of the following sections.
A mapping to a random bond Ising model that incorporates the above mentioned equivalences can provide a lower bound on a maximum likelihood decoder but is left for future work (cf. e.g.~\cite{andrist2010tricolored,kovalev2013fault}). We would like to note that we expect an interesting model, considering that the decoding graph (of the restricted noise model) is that of interleaved color code and toric code decoding graphs.

\subsection{Circuits for memory and stability experiments and noise models}\label{sec:circuits}
\begin{figure}
    \centering
    \includegraphics[width=\linewidth]{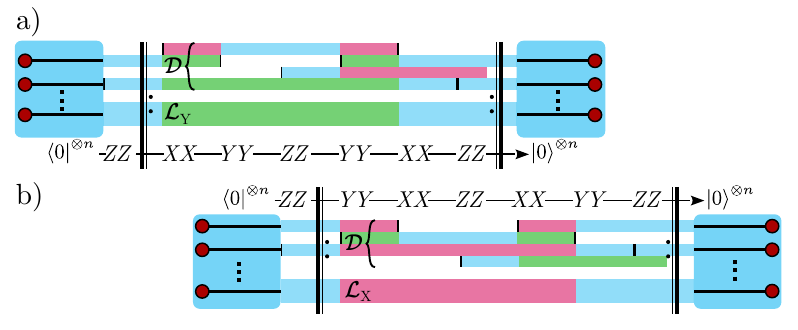}
    \caption{Memory experiments. We show thin $Z$-basis initialization and measurement slices that sandwich the bulk protocol. The upper stripes of the (potentially repeated) bulk sequence indicated by $\mathcal{D}$ corresponds to decoding flows (cf. Fig.~\ref{fig:fault_tolerance_detector_graph_time} and the lower to observable flows (cf. Fig.~\ref{fig:logical_flows}). In a), we cut the network after measuring $XX$- and before measuring $ZZ$ to attach the initialization spiders. Similarly, we cut after the $ZZ$ (and before an $XX$)  measurement to attach the read-out spiders.
    The (blue) Pauli flows of the single leg spiders match the blue highlight of the $\grg \blb$ logical flows; this choice of cut therefore initializes and reads out these logical operators. Additionally, $\grg\blb\grg$-plaquette stabilizer flows can be closed to generate detectors at the boundaries. In b), cuts shifted by one half-cycle achieve the same for the $\mathcal{L}_X$ logical flow.}
    \label{fig:memory_probing}
\end{figure}

Knowing the colors of the highlights of decoding and observable flows, we can design memory and stability experiments by choosing compatible input and output layers corresponding to single-qubit initialization and measurement. Take, e.g., an initialization in the $n$-qubit product state vector $\ket{0}^{\otimes n}$. This corresponds to a collection of single-legged red tensors as already mentioned above. 
These have blue Pauli flows and can therefore attach to any other blue Pauli flow. Both the $\mathcal{L}_X$ and $\mathcal{L}_Y$ memory logicals have blue flows around one of the $ZZ$ measurements. 
Cutting a bulk protocol in between $ZZ$ and $YY$ or $ZZ$ and $XX$-measurement respectively therefore allows us to initialize the $\mathcal{L}_X$ and $\mathcal{L}_Y$ logicals.
Crucially, these cuts allow us to \emph{fault-tolerantly} initialize the corresponding logical, because we can form enough detectors by closing stabilizer Pauli flows. In particular, the first $ZZ$ measurements are deterministic, as are the Pauli flows that are half of the plaquette detectors (i.e., the product of subsequent $XX$ and $YY$ measurements, or vice versa). 
We show the memory experiment protocols graphically in Fig.~\ref{fig:memory_probing}.

\begin{figure}
    \centering
    \includegraphics[width=\linewidth]{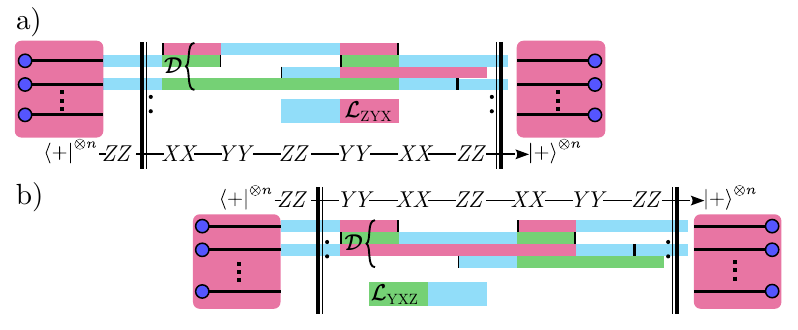}
    \caption{Stability experiments. We show thin $X$-basis initialization and measurement slices that sandwich the bulk protocol. The upper stripes of the (potentially repeated) bulk sequence indicated by $\mathcal{D}$ corresponds to decoding flows (cf. Fig.~\ref{fig:fault_tolerance_detector_graph_time} and the lower to observable flows (cf. Fig.~\ref{fig:global_constraints}). In a), we cut the network after measuring $XX$- and before measuring $ZZ$ to attach the initialization spiders. Similarly, we cut after the $ZZ$ (and before an $XX$  measurement to attach the measurements.
    The (blue) Pauli flows of the single leg spiders do not match the blue highlight of the $ZYX$- like stability flows, this choice of cut therefore leads to random initialization. Additionally, $\grg\blb\grg$-plaquette stabilizer flows do not close to generate detectors at the boundaries. In b), cuts shifted by 1 half-cycle achieve the same for the $\mathcal{L}_{YXZ}$ stability flow.}
    \label{fig:stability_probing}
\end{figure}

For stability experiments the relevant observables are initially random, so we initialize in a different basis and declare the global constraints of Fig.~\ref{fig:global_constraints} as observables, see Fig.~\ref{fig:stability_probing}.

We benchmark two different circuit implementations with different noise models. 
First, we assume that direct 2-body Pauli measurements are available. When measuring the three edges of a triangle one after another, a cycle ($XX \to YY \to ZZ$) takes $5$ steps.
Unlike for the honeycomb code, some qubits will therefore be idling during the $ZZ$-measurements. 
For this circuit, we employ two noise models. The first is a \emph{phenomenological noise model} where we insert single-qubit depolarizing channels $\mathcal{D}^{\otimes 1}(p)$ in between the check measurements. Additionally, every measurement outcome is flipped with the same probability $p$.  
The second is the \emph{EM3} noise model~\cite{gidney2021faulttolerant} that is commonly used in the Floquet code literature.
Here, single-qubit depolarizing channels are replaced with two-qubit depolarizing channels $\mathcal{D}^{\otimes 2}(p)$ after a check measurement. 
This is motivated by the expected behaviour of hardware noise on potential Majorana based quantum computing platforms~\cite{knapp2018modeling}.
We expect the EM3 noise model to have a reduced distance due to 2-qubit faults that already appear at first order. A detailed analysis of the noise channels of the elementary building blocks is deferred to the Appendix App.~\ref{app:noise}.

\begin{figure}
    \centering
    \includegraphics[width=\linewidth]{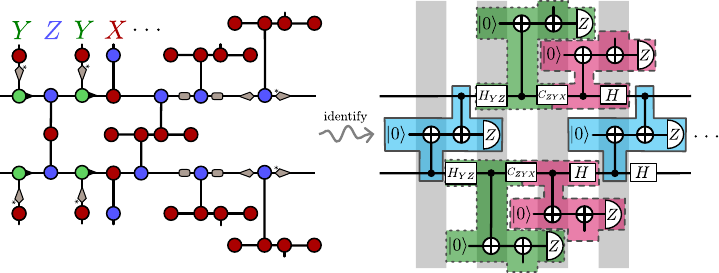}
    \caption{Translating the XYZ ruby code tensor network to a circuit. We first use basis changes to express the network with blue and red spiders. We then split the red measurement spiders to $Z$-basis initialization, two CNOTs and a $Z$-basis measurement. The circuit can be scheduled such that in the bulk of the protocol no qubit is ever idle. Here, $C_{ZYX} = H_{YZ} H$ permutes the bases anti-cyclically. Note that any circuit obtained from rewrite rules performs the readout.}
    \label{fig:measurement_circuit_aux}
\end{figure}

Additionally, we construct circuits with single-qubit initialization and measurement in the $Z$-basis, single-qubit Clifford rotations on the data qubits, and entangling $\CNOT$ gates. We place one auxiliary qubit on each edge to facilitate the two-body measurements and schedule the gates as shown in Fig.~\ref{fig:measurement_circuit_aux}. With the exception of the first three layers in this circuit, no data or auxiliary qubits are idling. We use a uniform circuit-level noise model, with noisy operations specified in Tab.~\ref{tab:noise_models}.      

\begin{table}
    \caption{Noise Models used in bench-marking the circuits in this work. The noisy multi-Pauli measurement gate $\text{M}_{PP}(p)$ has two different realizations. In the phenomenological model,  $\text{M}_{PP}(p)$ corresponds to a single-qubit depolarizing channel on data qubits of its support and a noisy measurement outcome. In the EM3 model, the single-qubit depolarizing channel is replaced by a two-qubit depolarizing channel on data qubits of its support and a noisy measurement outcome.}
    \label{tab:noise_models}
    \begin{ruledtabular}
    \begin{tabular}{lll}
       Noise Model   & Phenomenological, EM3   & Circuit-Level \\
        \hline
        \multirow{5}{*}{Noisy Gateset}  & $\text{Init}_Z(p)$ & $\text{Init}_Z(p)$    \\
                                        & $\text{M}_{PP}(p)$ & $\text{M}_Z(p)$       \\
                                        &                    & $\text{C}_1(p)$       \\
                                        &                    & $\CNOT(p)$      \\
                                        & $\text{Idle}(p)$   & $\text{Idle}(p)$      \\
    \end{tabular}
    \end{ruledtabular}
\end{table}

\subsection{Sampling, decoding and notes on implementation}\label{sec:sampling}

We construct the circuits in \texttt{stim}~\cite{gidney2021stim} and annotate the measurements that contribute to detectors and observables, as well as error channels. 
We generate a \emph{detector error model} by evaluating with respect to which  elementary faults defined by the error channels the detector and observable flows are charged\footnote{This is done in \texttt{stim} using \texttt{circuit.detector\_error\_model()}}. The detector error model is a list of elementary faults with their probability and the detectors and logical observables this fault flips. 
We can directly sample  detection events (syndromes)  $\mathbf{s} \in \mathbb{F}_2^{n_{\mathrm{d}}}$ and logical flips (observations $\mathbf{o} \in \mathbb{F}_2^{k}$)\footnote{Using \texttt{stim}'s \texttt{CompiledDetectorSampler()}.}.
This completes the first two steps of probing the error-correcting protocol, cf. Sec.~\ref{sec:probing}.
To decode this syndrome, we convert the detector error model into a \emph{detector matrix} $\mathbf{M}_d \in \mathbb{F}_2^{n_{\mathrm{d}} \times n_{\mathrm{e}}}$, a \emph{logical matrix} $\mathbf{M}_l \in \mathbb{F}_2^{k \times n_{\mathrm{e}}}$ and a \emph{prior vector} $\mathbf{p} \in \mathbb{R}^{n_{\mathrm{e}}}$. Here $n_{\mathrm{d}}$, $ n_{\mathrm{e}}$ and $k$ are the number of detectors, error mechanisms and logical observables respectively.

The task of a decoder is to find a set of error mechanisms $\mathbf{e} \in \mathbb{F}_2^{n_e}$ that fulfills $\mathbf{M}_d \mathbf{e} = \mathbf{s}$, i.e., a set of error mechanisms that produces the observed syndrome.
To verify whether a correction would result in a logical error, the prediction $\mathbf{o}^\star = \mathbf{M}_l \mathbf{e}$ has to equal the observed logical flips, i.e., $ \mathbf{o} \oplus \mathbf{o}^\star \overset{!}{=} \mathbf{0}$.
Note that we decode directly for error mechanisms and not for flipped measurements, therefore adapting the protocol compared to step 4 and 5 of Sec.\ref{sec:probing} to make use of information on the error model. Using the faults or the measurement outcomes is equivalent for getting the predicted values of the observables. However, including information on the probability of faults can only improve the accuracy of decoding.

Based on our discussion above, the effective error model derived from the circuit is not trivially decomposable into errors with at most two syndromes. This renders a non-adapted matching-based decoder such as pymatching not suitable for this problem. We therefore employ the general purpose two-stage decoders 
\emph{belief propagation + ordered statistics decoding} (BP+OSD) and the recently introduce \emph{belief propagation + localized statistics decoding} (BP+LSD) ~\cite{panteleev2021degenerate,roffe2022ldpc,hillmann2024localized}.
The first stage uses belief propagation on the Tanner graph defined by the detector matrix $\mathbf{M}_d$ to approximate the single error mechanisms' marginal probabilities based on the observed syndrome $\mathbf{s}$ and the a priori probabilities of elementary error mechanisms $\mathbf{p}$. 
Typically -- and in particular if the Tanner graph contains many small loops -- BP is inconclusive in its output, returning an error guess that does not match the observed syndrome. However, it usually points to a subset of likely erroneous qubits; OSD then brute-forces the solution of the decoding problem on that subset via matrix inversion. LSD provides a speed-accuracy trade-off by doing inversion on localized cluster, which we exploit to generate more data for larger codes and smaller physical error rates. We give details on the implementation of BP+O(L)SD and the used parameters in the appendix App.\,\ref{app:bp}. 

\subsection{Results}\label{sec:results} 

In all following results, we take a conservative approach and report the logical failure probability as the probability that at least one of the $4$ independent (logical) observables probed in the experiment is predicted wrongly by the decoder. 

\subsubsection{Memory experiment}
\begin{figure*}
    \centering
    \includegraphics[width=\linewidth]{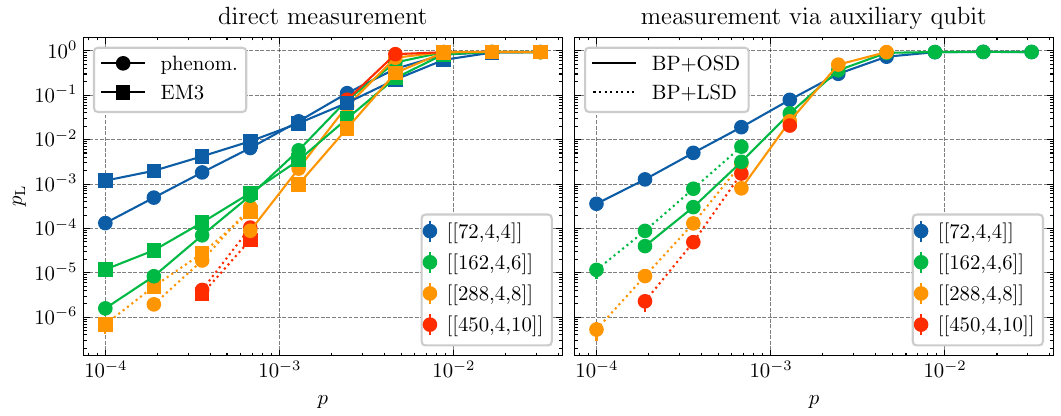}
    \caption{Memory experiment for the $\mathcal{L}_Y$ logical observable using direct $2$-body measurements and measurements using an auxiliary qubit. 
    We decode using BP+OSD (solid lines) and BP+LSD (dotted lines) for low physical error rate. BP+LSD has a higher logical error rate, but decodes orders of magnitude faster. 
    For the direct measurements, we compare a phenomenological and an EM3 noise model with the latter showing a slightly higher threshold ($p^{\mathrm{em3}}_{\mathrm{th}} \approx 0.36 \pm 0.04 \%$ vs. $ p^{\mathrm{ph}}_{\mathrm{th}} \approx 0.28 \pm 0.02 \%$) but a worse scaling for small physical error probabilities $p$. Measuring using an auxiliary qubit and simulating full circuit-level noise shows a crossing at $p^{\mathrm{cl}}_{\mathrm{th}} \approx 0.18 \pm 0.01 \%$. Here and in all other plots, error bars are standard Monte Carlo sampling errors and are smaller than the symbol.}
    \label{fig:plot_pL_d}
\end{figure*}

In Fig.~\ref{fig:plot_pL_d}, we show logical error rates  for the $\mathcal{L}_Y$ logical observable with increasing distance $d$ and $d$ cycles for the phenomenological, EM3 and circuit-level noise model respectively. 
We find that for all three circuits and noise models, the logical error rate curves cross, indicating the existence of a threshold. Errors are obtained from a finite size scaling analysis, which we detail in App.\,\ref{app:fss}. 
For the phenomenological noise model we find
\begin{align}
    p^{\mathrm{ph}}_{\mathrm{th}} \approx 0.28 \pm 0.02 \%.
\end{align}
For the EM3 noise model we find
\begin{align}
    p^{\mathrm{em3}}_{\mathrm{th}} \approx 0.36 \pm 0.04 \%.
\end{align}
For the circuit-level noise model we find
\begin{align}
    p^{\mathrm{cl}}_{\mathrm{th}} \approx 0.18 \pm 0.01 \%.
\end{align}

The scaling of the logical error rate in the low $p$ regime is $\propto p^s = p^{\frac{d}{2}} $ for the phenomenological and the circuit-level noise model, as expected. Because the distance of the codes on the torus is even, no failure occurs for faults with weight less than $\frac{d}{2}$. This confirms that these circuits and noise model are fault-tolerant, as the (dynamic) fault distance coincides with the static distance of the codes defined by the ISGs. As expected, the fault distance is halved when using the EM3 noise-model. This is reflected in a flattening of the curves at low physical error rate $p$. 

The standard depolarizing threshold for $d$ rounds of the (toric) Honeycomb Floquet code is estimated in Refs.\,\cite{gidney2021faulttolerant, gidney2022benchmarking}  to lie within $0.2-0.3\%$, which is not significantly larger than our $\approx 0.18 \%$. Note, however, that for the toric honeycomb code, they report the failure rate of one of the logical qubits to better compare to planar surface codes. Asymptotically, the threshold for failure of one or any of the logical qubits is the same. 
Due to the considered system sizes and the sub-optimality of the decoder used, we consider our numbers a conservative estimate of the asymptotic threshold.
We observe an improved threshold for the EM3 noise model, however only by a factor of $\approx 2$ compared to  a factor of $\approx 5$ for the (toric) Honeycomb Floquet code.
One reason for the difference is that we measure the triangles in three separate $2$-body measurements. 

\begin{figure*}
    \centering
    \includegraphics[width=.9\linewidth]{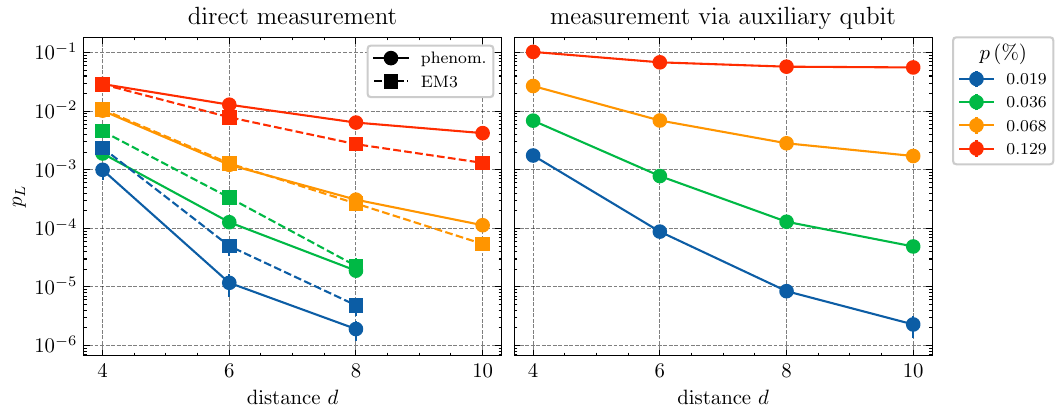}
    \caption{Cuts for the $\mathcal{L}_Y$ memory experiment decoded with BP+LSD at different physical error rates $p$ showing the suppression of logical error rate with code distance. Well below threshold, we observe an exponential suppression for small distances. We attribute the flattening of the curves, in particular for circuits with auxiliary qubits, to non-optimal decoding of zeroth order BP+LSD, see main text and appendix\,\ref{app:bp}.}
    \label{fig:plot_suppression_distance}
\end{figure*}

Constructing and benchmarking fault-tolerant circuits for stabilizer readout of planar static color codes with triangular boundaries encoding a single logical qubit has seen progress in circuit-level noise thresholds from $0.1 \%$\,\cite{wang2010graphical}  to $0.143 \pm 0.001 \%$~\cite{stephens2014efficient} and $0.2\%$\,\cite{chamberland2020triangular}. These improvements arise on the one hand by constructing more involved circuits for syndrome measurements using cat states or flag constructions, and on the other hand designing better decoders. 
The circuits to measure the two-body stabilizers in our protocols are inherently simple and we achieve thresholds that compare well to values from the literature, despite reporting the probability that any of the $4$ logical qubits fails.
Additionally, a comparison is difficult because of our usage of BP+O(L)SD decoders, which is a fundamentally non-optimal approach. Moreover,  BP+O(L)SD has a reduced accuracy for large decoding problems, i.e., many qubits or a lot of rounds, see, e.g., Ref.~\cite{higgott2023improved}. 
This can be observed in Fig.~\ref{fig:plot_suppression_distance}, where we show the suppression of logical error rate with growing size in the sub-threshold regime. The suppression is exponential for low error rates and small distances. In particular for the circuit-level noise, the suppression deviates from exponential for larger distances. 
It is left for future work to construct a decoder that is efficient and accurate enough to perform reasonable simulations for larger decoding problems. This can involve designing a new decoder based on the considerations in the previous section, or decoding with a sliding-window decoder\,\cite{tan2023scalable,ParallelWindow,PhysRevLett.128.080505,sahay2022decoder,gidney2023new}. 

\subsubsection{Stability experiment}
\begin{figure*}
    \centering
    \includegraphics[width=.9\linewidth]{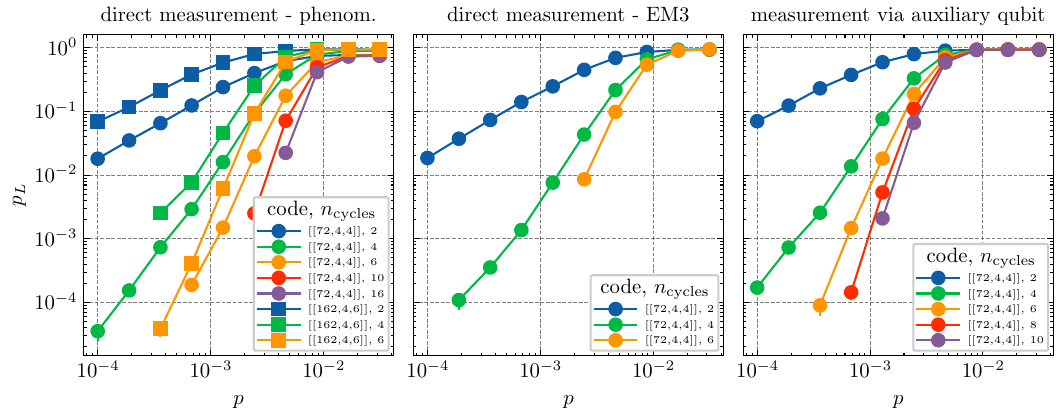}
    \caption{Stability experiment of the $\mathcal{L}_Y$ stability observable for direct measurements using phenomenological and EM3 noise model, as well as measurements using an auxiliary qubit. In all noise models, the logical error rate scales $\propto p^{n_{\mathrm{cycles}}/2}$ for low physical error rates $p$. This implies that this time-like distance of the stability experiment coincides with the space-like fault distance for $n_{\mathrm{cycles}} = 2s$.}
    \label{fig:plot_pL_stability}
\end{figure*}

We also perform stability experiments and show results  in  Fig.~\ref{fig:plot_pL_stability}.
In the phenomenological noise model, we simulate sizes $2$ and $3$. For size $s=2$, we simulate $2,4,6,10$ and $16$ measurement cycles.
In all noise models, the logical error rate scales $\propto p^{n_{\mathrm{cycles}}/2}$ for low physical error rates $p$. This implies that this time-like distance of the stability experiment coincides with the space-like fault distance for $n_{\mathrm{cycles}} = 2s$.
This means concretely, that we can read off these experiments how many rounds of measurements are needed to reach a desired distance for the global observable. Or, more practically, how many rounds are needed to obtain a target logical error rate given a physical error rate. 
The global observables are typically measured in space-like parts of dynamical protocols, like moving a logical qubit or performing lattice surgery for entangling operations like logical CNOTs\,\cite{horsman2012surface}.
The stability experiments also allow us to explore how different error rates for data qubit and measurement errors influence the performance of the error correction protocol, which we consider an interesting avenue for future research.   

\section{Towards planar codes}\label{sec:planar}

So far, we have considered topological codes with logical qubits coming from the non-trivial topology of the ambient manifold on which the model is defined.
It is also possible to use \textit{stable boundaries} to encode logical information robustly in a topological code.
From the anyon perspective a stable boundary type is defined by a Lagrangian subgroup of anyons $L$, determining which anyons can condense at the boundary~\cite{kapustin2011topological, Levin2013Protected}.
On the level of the stabilizer group and logical operators the Lagrangian subgroup determines which string operators can be terminated at the boundary without leaving the codespace.
In general, devising models for boundaries of a well-defined type is a non-trivial task.
Specifically, for ISG codes, if a bulk measurement sequence transforms an anyon string operator $s$ at time $t$ into a string operator $s'$ at time $t+1$, a boundary where $s$ can terminate at $t$ has to transform into a boundary where $s'$ can terminate at $t+1$.
At the same time, the measurement is not allowed to read-out the associated string operators.
Moreover, a Floquet code can apply a non-trivial automorphism to the logical operators after multiple rounds of measurements.
As a consequence, (discrete) time-translation symmetry has to be broken at the boundary into a coarser symmetry.
This leads to an enlargement of the periodicity of the schedule close to the boundary.
If the automorphism applied after one cycle has finite order $o$, the period at the boundary has to be enlarged from $T$ to $oT$.

For example, the honeycomb code is in a toric code phase throughout the whole evolution.
As such, it can have two boundary types, usually referred to as \textit{rough} and \textit{smooth}, condensing $e$ and $m$ anyons, respectively.
Since it applies a $\bZ_2$ automorphism after one length-3 cycle, exchanging the electric and the magnetic anyons of the underlying toric code, the boundaries have to be transformed into each after one cycle of measurements which leads to a period doubling in the presence of boundaries~\cite{haah2022boundaries}.

Additionally to transforming according to the bulk evolution of the ISGs\NEW{,} a spacetime boundary has to be fault-tolerant.
In the case of topological protocols this means that there are no constant-sized fault configurations that lead to a logical error.
So far, boundary constructions have been understood well for some specific examples, with the most general framework that uses dynamical anyon condensation to understand $2+1$-dimensional Floquet codes~\cite{haah2022boundaries, kesselring2022anyon, davydova2023quantum}.

In the following, we will present two equivalent perspective on boundaries in the XYZ ruby code.
First, we give a macroscopic description of how the ISGs have to transform in the XYZ ruby code in order to preserve the encoded logical information.
Specifically, we pick $\cS_0$ as our reference ISG and use the automorphism applied after one cycle of $XX$-$YY$-$ZZ$ measurements to argue about how each of the six possible boundaries have to transform.
Second, we sketch an explicit method to construct boundaries for topological protocols using the spacetime tensor network representation.
A subset of boundaries obtained in this way for the XYZ ruby code can be found in App.\,\ref{app:boundaries} and are by construction fault-tolerant.
We want to note that this construction, in principle, works for any topological Clifford protocol, that is, for any protocol whose bulk is defined by local quantum operations and has a macroscopic fault distance with only local detector flows.

\subsection{Tracking Lagrangian subgroups}
For our first analysis, we pick $\cS_0$ as a reference ISG and assume we devised a schedule that has established a topological boundary at $t=0$.
Since $\cS_0$ is in the phase of the two-dimensional color code (see Sec.~\ref{sec:establishing}) there are only six possible topological boundaries~\cite{Kesselring2018boundariestwist}, three different ``color'' boundaries, where bosons of a given color in \{\texttt{r}, \texttt{b}, \texttt{g}\} can condense, and three different ``Pauli'' boundaries, where bosons of a given Pauli label in \{\texttt{x}, \texttt{y}, \texttt{z}\} can condense.

Assume the ISG at $t=0$ realizes a boundary of type $a$ described by a Lagrangian subgroup $L_a$. 
After three rounds of measurements, the anyon permuting automorphism $\varphi$ from Eq.\,\eqref{eq:automorphism_table} is applied.
This does not change the bulk stabilizer group (cf. Eq.\,\eqref{eq:period3-stabilizer-diagram}) but acts non-trivially on the stabilizer generators close to the boundary.
On the macroscopic level, this changes the boundary type.
To obtain the new boundary type we apply the automorphism to the Lagrangian subgroup and get a possibly changed boundary described by $L_{a'} = \varphi(L_a)$, where the function is understood to be applied element-wise.
Since $\varphi$ preserves the braiding data of the anyon model, $L_{a'}$ is again a Lagrangian subgroup but potentially of different type.
We find that there is no boundary that is invariant under $\varphi$.
There are, however, two transitive subsets of boundaries: Color boundaries get mapped to color boundaries and Pauli boundaries get mapped to Pauli boundaries.
For example, $\varphi$ maps a $\texttt{r}$-boundary to a $\texttt{b}$-boundary and the $\texttt{b}$-boundary to a $\texttt{g}$-boundary.
Following the notation introduced in Ref.~\cite{Kesselring2018boundariestwist} for the different boundary types as rows, respectively columns, of the color code boson table (Eq.\,\eqref{eq:boson_table}) we can for example track the transformation of the color boundaries in the following way:
\begin{align}
\begin{split}
\begin{tikzpicture}
    \node (A) at (-150:1.5) {\includegraphics[height=1cm]{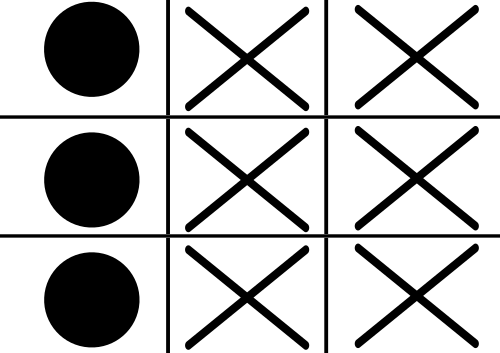}};
    \node (B) at (90:1.5) {\includegraphics[height=1cm]{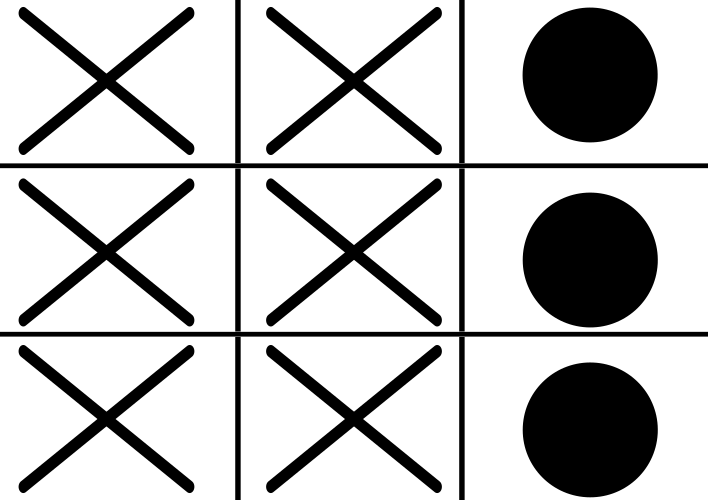}};
    \node (C) at (-30:1.5) {\includegraphics[height=1cm]{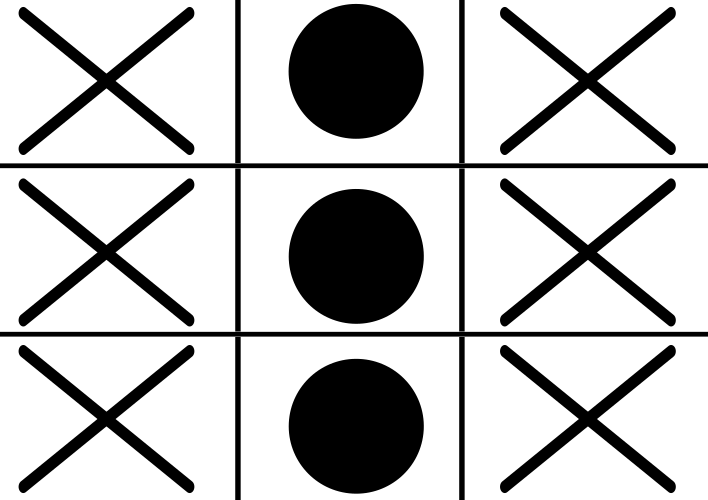}};
    \draw[->, thick] (A) -- (B);
    \draw[->, thick] (B) -- (C);
    \draw[->, thick] (C) -- (A);
    \node[anchor=east] at (150:0.75) {$\varphi$};
    \node[anchor=west] at (30:0.75) {$\varphi$};
    \node[anchor=north] at (-90:0.75) {$\varphi$};
\end{tikzpicture}
\end{split}
\end{align}
Similarly, the Pauli boundaries need to transform in the following way:
\begin{align}
\begin{split}
\begin{tikzpicture}
    \node (A) at (-150:1.5) {\includegraphics[height=1cm]{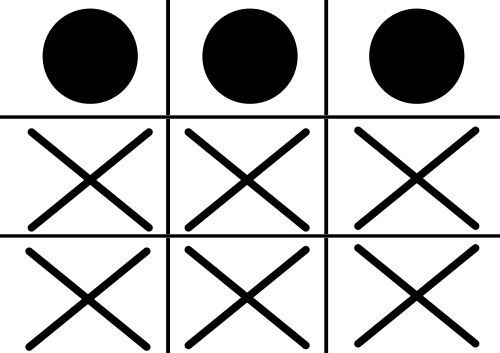}};
    \node (B) at (90:1.5) {\includegraphics[height=1cm]{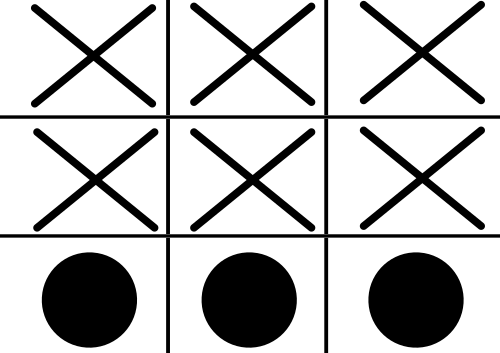}};
    \node (C) at (-30:1.5) {\includegraphics[height=1cm]{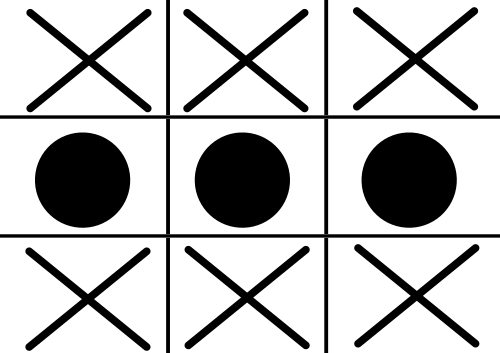}};
    \draw[->, thick] (A) -- (B);
    \draw[->, thick] (B) -- (C);
    \draw[->, thick] (C) -- (A);
    \node[anchor=east] at (150:0.75) {$\varphi$};
    \node[anchor=west] at (30:0.75) {$\varphi$};
    \node[anchor=north] at (-90:0.75) {$\varphi$};
\end{tikzpicture}
\end{split}
\end{align}
In the rewinding schedule, the automorphism is applied after three rounds -- the first half the period -- and then inverted throughout the second half.

\subsection{Constructing boundaries in spacetime}
Given the rather intricate transitions that the instantaneous boundaries need to undergo throughout one cycle of the rewinding schedule, a systematic construction of boundaries for topological protocols might seem out of reach.
We find that this can be achieved using the RGB tensor network representation of the protocol together with the Pauli flows.
Specifically, we can construct a boundary of a given type by cutting the RGB network describing the bulk of the protocol along an arbitrary plane and consider the stabilizer state defined by the Pauli flows of the bulk.
In this perspective, a boundary corresponds to terminating the tensor network along the cut with a (thin) RGB tensor network that attaches to the cut legs but has no additional open legs.
The network that is attached along the cut has to be chosen such that it preserves the desired set of (logical) flows.
Additionally, the stabilizers mapping between the logicals representatives have to be ``read-out'' along the boundary to form the detector flows close to the boundary.
In App.\,\ref{app:boundaries}, we present three distinct types of boundaries that can be combined to form a triangular code.\footnote{Note that in order for the resulting protocol to be fault tolerant \textit{corners}, the interfaces between the boundaries (one-dimensional objects in spacetime) have to be designed accordingly.}
For the instantaneous color codes (at $t=0\mod 3$), these boundaries correspond to the three Pauli types discussed in the previous section.
A more elaborate treatment of the construction is left for future work.

\section{Remarks}\label{sec:remarks}
In this section, we want to remark on some aspects of the XYZ ruby code and spacetime fault tolerance not touched on in the main part of this work.
First, we compare the Floquet code to its associated subsystem code, introduced by Bombin in Ref.~\cite{bombin2006topological}, and other topological subsystem codes.
Secondly, we comment on how fault tolerance beyond topological codes can fit into the framework of Pauli flows and highlight some potentially interesting aspects of protocols defined via quantum operations of constant weight.
Moreover, we comment on how the concept of Pauli flows of a protocol and the design principles for a probing experiment discussed in Sec.~\ref{sec:probing} are related to the recently introduced spacetime code construction for Clifford circuits~\cite{delfosse2023spacetime}.
Lastly, we comment on how rewrite rules of RGB tensor networks naturally lead to an equivalence class of protocols and highlight some interesting equivalences between protocols defined in different spatial dimensions.
We argue that Pauli flows could play a central role in a topological classification of these equivalence classes and how such a classification, also beyond the Clifford case, could help in the construction of composable fault tolerant protocols.

\subsection{Comparison to topological subsystem codes}\label{sec:comparison_subsystem}
The group generated by all check operators of a Floquet code can be thought of as a gauge group of a subsystem code.
We call this subsystem code the \textit{associated subsystem code}~\cite{Teague2023Floquetifying}.
One can understand the bare logical operators of a two-dimensional topological subsystem code as generating a 1-form symmetry associated to an anyon model~\cite{Ellison2023paulitopological}.

The associated subsystem code of the XYZ ruby code is Bombin's topological subsystem code constructed from a hexagonal lattice~\cite{bombin2006topological}.
Its anyon model is the 3-fermion model~\cite{bombin2006topological, roberts20203}, a chiral anyon model.
It attracted some attention in the community of topological phases since there cannot exist a gapped, commuting projector Hamiltonian on finite dimensional degrees of freedom -- and with that no stabilizer code -- whose ground states represent a chiral anyon theory~\cite{bombin2012universal, haah2021classification, Kapustin2019, Kitaev2006anyons}.

In contrast to the anyon model of the subsystem code, we find that the XYZ ruby code as a three-dimensional protocol, in fact, is described by the color code anyon model, which is equivalent to two copies of the 3-fermion anyon model and non-chiral.
One can think of the logical qubits of the subsystem code as ``static'' and the additional logicals as ``dynamical'' qubits of the XYZ ruby code.
For the subsystem code these dynamically generated qubits are gauge qubits.
These gauge qubits, however, have logical operators of extensive size.
We have chosen the measurement schedule such that the extensive gauge qubits are never fixed which renders them logical qubits of the protocol.
In fact, any measurement schedule on the generators of the gauge group that reads out the stabilizers of the subsystem code that we investigated gives a protocol associated to the color code anyon model.\footnote{Note that we have not rigorously defined yet how to make sense of an ``anyon model'' or an anomalous 1-form symmetry of a three-dimensional topological protocol.
For the tensor networks constructed in Ref.~\cite{bauer2024lowoverhead} the anyon model describing the protocol follows from the construction via (twisted) Abelian gauge theories.
We aim to provide a constructive framework to identify the anyon model represented by more generic (families of) topological protocols in future work.}

In that sense, the anyon model of the subsystem code does not characterize the topological phase of the measurement protocol, as already popularized by the honeycomb code and its CSS counterpart~\cite{hastings2021dynamically, Davydova2023Floquet, kesselring2022anyon}, where the anyon theory of the associated subsystem code, respectively of its stabilizer group, is a proper subtheory of the toric code anyon model.
It would be interesting to investigate the phases arising in protocols derived from other subsystem codes representing chiral anyon theories, for example chiral subsystem codes from Ref.~\cite{Ellison2023paulitopological}.

From the QEC perspective it might also be interesting to investigate how different measurement schedules on the same set of checks affect the properties of the arising protocol.
For example, the instantaneous topological phase transition present in our schedule seems like a fairly unique feature compared to other schedules, compare, e.g., with Ref.~\cite{dua2023engineering}.
It is an open question if this has implications on the error-correcting capabilities of the schedule.

\subsection{Fault tolerance in non-topological protocols}
In the sections above we have mainly discussed topological protocols as they are central to our analysis.
However, both memory and stability experiments can be thought of in the context of non-topological protocols.
Here, we define topological protocols as error-correcting protocols that admit a complete set of local detector flows.
That means we consider a family of QEC circuits to be topological if they give rise to a family of locally generated detector flows such that the fault distance 
-- with respect to that set of detectors -- grows with the number of physical qubits.
In these cases, there is a natural partition of detector flows of any probing experiment into local and non-local flows, as described in the subsections above.
Moreover, for topological protocols there has been established a thorough understanding of how to understand fault tolerance with respect to local noise~\cite{dennis2002topological, roberts2017symmetry, Iverson2020Coherence, bauer2024lowoverhead}.

One might ask in which sense fault tolerance can be understood generically in the case of non-topological protocols.
i.e., given an experiment with its flows, can one make rigorous statements about the fault tolerance with respect to a given error model?
And more importantly, can we argue about fault tolerance based on the structure of the detector flows of the protocol, say, e.g., 
their weight distribution and a guarantee on the fault distance?
The natural generalization of ``locality'' of detector flows is obtained by ignoring their geometric locality but considering only the scaling of the number of edges that are non-trivially highlighted by the flows within the a family of protocols.
If one can choose a set of detector flows of bounded weight to define syndromes and correct all observables (with extensive weight), we propose to call the corresponding protocol ``spacetime LDPC''.\footnote{Note that Ref.~\cite{delfosse2023spacetime} introduces the notion of an LDPC spacetime code. Although it refers to a stabilizer code derived from detector flows, the notion of LDPC is the same: What we call a spacetime LDPC protocol will give rise to a LDPC spacetime code.}
As far as we know there is no good reason to believe that this notion of spacetime LDPC is similarly important to fault tolerance as it is believed to be for an instantaneous stabilizer code.
Importantly, quantum operations of constant weight can lead to protocols that are not LDPC.
An interesting example in that regard is a recently proposed (modified) schedule for the Bacon-Shor code that is believed to have a threshold but does not fall into that category, see Ref.~\cite{gidney2023less}.
Even though most detector flows used for decoding have low weight, the fractal structure of the protocol creates some high-weight detector flows and hence a decoding graph with unbounded check weight.
In light of the fact that the previously considered (unmodified) schedule associated to the Bacon-Shor code only has detectors of extensive weight and is not thresholded we consider the question if there has to exist a minimum amount of constant-weight detector flows in order to be fault tolerant an interesting question for further research.
It would also be interesting to investigate the structure of the detector flows in protocols based on fracton codes~\cite{Haah2011local, Brown2020parallelized} to inspire new and in particular efficient decoding strategies for type-II fracton models.
The more exotic structure of the resulting decoding graph could reveal insights into the origin of fault tolerance beyond more well-behaved topological codes.
The example above indicates that detector flows do not need to be of strictly bounded weight in order to achieve fault tolerance with a protocol based on bounded-weight quantum operations.

\subsection{Connection of Pauli flow to outcome and spacetime codes}\label{sec:spacetime_code}
In Ref.~\cite{delfosse2023spacetime} Delfosse and Paetznick introduce the \textit{outcome code} of a Clifford circuit.
It is a classical affine linear code on the space spanned by all measurement outcomes.
In fact, we recover similar constraints on the set of signs in the network based on detector flows.
In our formulation, we end up with a linear code because we include signs of tensors that might flip measurements also without noise (e.g. Pauli unitaries).
The way that the outcome code is translated to form a linear code in Ref.~\cite{delfosse2023spacetime} is equivalent to identifying these operations and flipping the measurements accordingly, i.e., inverting the affine translation with which a linear code is translated to obtain the outcome code.
In App.\,\ref{app:detectors}, we give a rigorous derivation of the linear code obtained from detector flows.

Additionally, the authors of Ref.~\cite{delfosse2023spacetime} introduce the \textit{spacetime code}, a quantum code derived from the circuit and the outcome code.
One can understand the stabilizers of the spacetime code directly from the tensor network picture.
The spacetime code is defined with respect to a time direction, i.e., an interpretation of the tensor network as a sequential application of stabilizer quantum instruments on a set of qubits.
Given that interpretation, we can identify each detector flow with a stabilizer of the spacetime code.
Specifically, each stabilizer is the product of Pauli operators associated to the highlights of the detector flow on the edges associated to qubit worldlines.\footnote{For a rigorous mapping, we refer to the map $\rho$ and $\Tilde{p}$ in App.\,\ref{app:sec:flows_logical_iso}.}

For a fixed circuit and error model, the spacetime code gives a faithful description of the behaviour of faults in the QEC protocol.
The commutation relation of a fault with a stabilizer of the spacetime code are exactly given by the charge of the fault with respect to the associated detector flow.

More generally, Pauli flows carry more information.
Since they are preserved by all rewrite rules of the network, they are independent of the interpretation of the network.
For example, consider two circuits $A$ and $B$ that implement the same stabilizer measurement.
In the RGB tensor network formulation any such pair of equivalent circuits is related by a rewrite rule presented in App.\,\ref{app:sec:rewriterules}.
Pauli flows are a preserved quantity in the sense that we can update the flows locally with each rewrite without computing the full detector flow and the associated constraints on the signs explicitly, see App.\,\ref{app:sec:flows_rewrite}.
In that way, there is a direct translation between these quantities in the RGB tensor network formulation.
Note that this includes mappings between protocols based on static stabilizer codes, dynamical Floquet-style readout schemes as well as measurement-based computation schemes.
It would be interesting to see if one can extend the formalism in Ref.~\cite{delfosse2023spacetime} to construct direct mappings between the outcome and spacetime codes of circuits related to the same underlying tensor network.

In a fault-tolerant architecture it makes sense to think about logical blocks, specific building blocks from which any logical operation is built out of.
We think of them as being expressed in terms of protocols on the physical level that have an effective logical action.
When composing logical Clifford blocks it is natural to understand the Pauli flows of the combined block in terms of the Pauli flows of the individual blocks.
In the spacetime code, it is not that obvious how to combine the spacetime codes of two logical blocks (see also Lem.\,\ref{app:lem:composition}).
We see the reason for that in the fact that there is no direct analogue of stabilizer and logical flows in the spacetime code. They can be mapped to a certain subset of logical operators of the spacetime code but there are many more logical operators that have no direct interpretation in the composition of logical blocks as they act purely internal to the network, respectively circuit.
It would be interesting to see a more direct way to incorporate the composition of logical blocks into the spacetime code formalism.

\subsection{Equivalence classes of protocols and associated invariants}
We want to finish on a more general thought arising from considering Pauli flows in RGB tensor networks. 

One can use certain rewrite rules of RGB tensor networks, presented in App.\,\ref{app:sec:rewriterules}, to define a sensible equivalence relation between RGB tensor networks.
Under that equivalence relation we find that many protocols are equivalent even though they live in different spatial dimensions.
A well-studied example of that is the equivalence between the fault-tolerant identity gate in \emph{measurement-based quantum computation} (MBQC) with the three-dimensional cluster state and the $2+1$-dimensional protocol based on the surface code on a square lattice as illustrated in Ref.~\cite{bombin2012universal}.

One can make this equivalence much more explicit and rigorous and concrete by 
viewing a MBQC scheme as a full protocol where qubits are initialized in a product state, a 
constant depth Clifford circuit is applied to prepare the resource state and single-qubit 
measurements are performed.
If the preparation circuit is local, and constant depth, in some dimension $D$, the TN of that protocol is equivalent to a TN of a protocol over time composed of operations that are local in $D-1$ dimensions by treating one spatial direction as the time direction and reinterpreting qubit worldlines accordingly.
Vice versa, any protocol defined by a tensor network that is local in some spacetime dimension $D$ can be rewritten and reinterpreted as a constant-shot (MBQC) protocol.
A similar connection was made in Ref.~\cite{bauer2024lowoverhead} and also applies to other three-dimensional stabilizer states such as states of the three-dimensional toric code~\cite{Hamma2005String}, 3D color code~\cite{Bombin2007exact} or topological single-shot protocols~\cite{Bombin2015Single, Bridgeman2024lifting} that can, in principle, be identified with $2+1$-dimensional protocols.

The striking feature of Pauli flows is that one can think of them as being preserved 
throughout the process, giving rise to an invariant of an equivalence class of protocols.
A direct consequence of that is that the logical isomorphism applied by a protocol as well as the in- and output stabilizer groups are preserved by all rewrites.
Finding an (algebraic) invariant that captures the fault tolerance properties of a protocol, possibly with respect to some ``unit volume'' would constitute a big step in the understanding of fault tolerance beyond individual (classes of) protocols and in defining a ``phase'' of a stabilizer tensor network.

These invariants would also be important when combining logical blocks.
Given one block with invariants $A$ and one block with invariants $B$, can one make statements about the invariant of the composed block?
So far, we only know how to explain this for Clifford protocols in terms of Pauli flows and techniques presented in App.\,\ref{app:rgbtensornetwork}.
A more general algebraic treatment of the Clifford case could inspire new perspectives on the understanding of fault-tolerance for non-Clifford operations.

\section{Conclusion and outlook}

In this work, we have established a Pauli flow formalism which unifies an error-correction analysis for protocols based on Clifford circuits and Pauli measurements.
The formulation in terms of a tensor network leads to a well-defined equivalence between different protocols and circuits that share their fault-tolerance properties and also provides a systematic procedure to design various implementations of the same logical operation with a clear handle on the error-correcting properties.
The concept of Pauli flow emerges from combining projective symmetries of the elementary building blocks of an RGB tensor network consistently to obtain symmetries of the full network.
Defining the notion of tensors being charged with respect to such a symmetry allows us to tie together the design and fault tolerance analysis of a protocol directly in spacetime.
Specifically, we can rigorously derive a unified perspective on properties of an error-correcting protocol from first principles within the tensor network representation.
This goes beyond the specific representation of a protocol in terms of the RGB tensors and applies to all Clifford protocols, independent of the underlying paradigm of quantum computation.
With this, we tie together existing notions on error correction in spacetime~\cite{delfosse2023spacetime, gidney2021stim, bombin2023unifying} that have taken different approaches, where either a family of a topological protocol or explicit circuits take center stage.
We do this by proving very general algebraic properties of stabilizer tensor networks that are not tied to specific implementations or even error-correcting operations.
Although a tensor network representation is very natural for topological protocols our analysis on how to think about QEC in spacetime does not require any (spatial) locality.
Hence, we believe it is helpful to investigate any active QEC protocol through the lens of a tensor network representation of the full spacetime evolution of the system.
Formulating the tensor network in terms of RGB tensors allows for additional flexibility as local rewrite rules can be used to turn non-fault-tolerant schemes into fault-tolerant ones.
This makes this method appealing for both code design and fault-tolerant circuit compilation.

To make this case, we have constructed a new dynamical error correcting code based on repeated measurements of two-body operators, which we call the XYZ ruby code. 
We have analyzed the dynamic protocol using `traditional' methods and have shown how the graphical calculus allows us to think of the protocol directly in spacetime and devise the algebraic object required for error correction and fault-tolerance properties. 
We have further used the RGB tensor networks to construct fault-tolerant protocols and circuits to probe the memory and stability qualities of the XYZ ruby code under different noise models. We have explained how to generate and decode a decoding graph from the detector flows and have developed a thorough physical understanding of faults and their action on the detectors.
We have performed numerical experiments, decoded using general-purpose decoders, and have found that the XYZ ruby codes show a threshold in the vicinity of $0.18 \pm 0.01 \%$ for full circuit-level noise. 
We believe that this could be further improved upon by designing decoders specifically tailored for the code using e.g. restriction or matching-based approaches.
Even as such, the thresholds are comparable to and competitive with other Floquet and non-Floquet codes in the literature. In particular, static color codes require contrived syndrome readout constructions using e.g. flag qubits, whereas our dynamical version relies on $2$-body nearest neighbor measurements only.

The locality and performance of XYZ ruby codes places them as promising candidates for near-term fault tolerance.
Additionally, the XYZ ruby code allows for more transversal gates than Floquet codes realizing toric codes\,\cite{dua2023engineering}, because it is in the topological phase of a color code.
Logical gates implemented transversally are naturally fault-tolerant and introduce a very limited amount of noise in the protocol\,\cite{gottesman1997stabilizer, Bombin2007exact,zhou2024algorithmic}. 
Similarly, the small weight of measurements in Floquet codes limits the spreading of faults.
With properly introduced boundaries, one can design logical gates based on lattice surgery protocols, which rely on fault-tolerant joint logical measurements between different code patches\,\cite{horsman2012surface,cohen2022low,thomsen2022low}. 
For the honeycomb code, one way to perform these joint measurements was shown in Ref.~\cite{haah2022boundaries}.
We have sketched how boundaries could be introduced in the XYZ ruby code. 
Important next steps include the microscopic realization of these boundaries and design of measurement sequences that join multiple copies of the code along those boundaries to perform said joint logical measurements.

More generally, Pauli flows offer a versatile tool to design interfaces between logical blocks in spacetime without resorting to a specific code on which the protocol could be based on.
In particular, Pauli flows allow one to systematically join microscopically different blocks with little overhead.
For topological protocols these interfaces can be described abstractly as topological defects in a topological spacetime.
Analyzing how other types of defects, such as twist-defects, can be described locally in terms of Pauli flows constitutes an interesting avenue for further research.
We believe that this connects to formulations of fault-tolerant logical gates within a measurement-based computation scheme based on Walker-Wang models enriched with external symmetry defects~\cite{Walker2011, williamson20211-form}.
We hope that, once we understand how to construct these defects explicitly within the bulk of a tensor network, we can use this insight to design similar defects in more general classes of protocols that involve non-local operations.
This would extend our understanding of how to implement fault-tolerant logical operations to protocols based on QLDPC codes that offer better distance and encoding rates.

To achieve computational universality, fault-tolerant logical non-Clifford operations are needed. Towards a more general understanding of those, we think it is important to investigate non-Pauli symmetries in our tensor network formulation. 

We hope that Pauli flows in RGB tensor networks and the new competitive dynamical QEC code will contribute to inspire further research and drive the progress towards scalable fault-tolerant quantum computing.

\section{Author contributions}
JM and JO led this project. They jointly analyzed the XYZ ruby code as a Floquet code and as an RGB tensor network and devised the probing experiments.
JM had the idea to formalize Pauli flows based on projective symmetries and devised the proofs in the appendix.
JM, JO and ATT jointly worked out the details of the formalism.
ATT contributed to the design of the QEC experiments, the boundaries presented and gave important feedback on the draft at many stages.
JO performed all the numerical experiments, designed the circuits and implemented the code, with feedback from MR.
MM and JE contributed to shaping the direction of the project and contributed in the writing of the paper.

\section{Acknowledgements}
We want to thank M.\ Kesselring for comments and ideas on constructions of Floquet codes in the color code phase.
JM wants to thank A. Bauer for fruitful discussions on tensor network representations of topological protocols and D.\ Williamson, T.\ Ellison, J.\ Conrad, 
T.\ Scruby, S.\ Roberts and T.\ Hillmann for 
stimulating conversations.
Moreover, JM wants to thank R. Raussendorf for bringing up the usage of Pauli symmetries in Ref.~\cite{Raussendorf2019Computationally} to our attention.
We gratefully acknowledge support by the European Union’s Horizon Europe research and innovation programme under Grant Agreement No.\ 101114305 (“MILLENION-SGA1” EU Project). This research is also part of the Munich Quantum Valley (K-8), which is supported by the Bavarian state government with funds from the Hightech Agenda Bayern Plus. We additionally acknowledge support by the BMBF project MUNIQC-ATOMS (Grant No.\ 13N16070). We also acknowledge support for the research that was sponsored by IARPA and the Army Research Office, under the Entangled Logical Qubits program through Cooperative Agreement Number W911NF-23-2-0216. The views and conclusions contained in this document are those of the authors and should not be interpreted as representing the official policies, either expressed or implied, of IARPA, the Army Research Office, or the U.S. Government. 
The U.S. Government is authorized to reproduce and distribute reprints for Government purposes notwithstanding any copyright notation herein.
JM and JE acknowledge \NEW{additional} support from the DFG (CRC 183), the ERC (DebuQC), the BMBF (QSolid, RealistiQ), the Quantum Flagship (PasQuans2, \NEW{Millenion}), \NEW{and Berlin Quantum}.
MM, JO and MR ackknowledge support by the Deutsche Forschungsgemeinschaft (DFG, German Research Foundation) under Germany’s Excellence Strategy “Cluster of Excellence Matter and Light for Quantum Computing (ML4Q) EXC 2004/1” 390534769 and the ERC Starting Grant QNets through Grant No. 804247. 
The authors gratefully acknowledge the computing time provided to them at the NHR Center NHR4CES at RWTH Aachen University (Project No. p0020074). This is funded by the Federal Ministry of Education and Research and the state governments participating on the basis of the resolutions of the GWK for national high performance computing at universities.

\clearpage

\onecolumngrid
\begin{appendix}

\section{RGB tensor networks and Pauli flow}\label{app:rgbtensornetwork}

In this section, we give some additional properties and relations for \emph{RGB tensor networks} (RGB TNs).
We will use the definitions introduced in Sec.~\ref{sec:prelim-3cgc}.
Eq.\,\eqref{eq:blue_basistrafo} shows how the blue tensor can be seen as the red, respectively green, tensor in a different local basis.
Using unitarity of the basis transformations,
\begin{align}\label{eq:basistrafos_unitarity}
\raisebox{-0.3\height}{\includegraphics[width=0.15\linewidth]{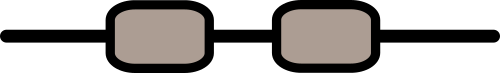}} = \raisebox{-0.3\height}{\includegraphics[width=0.15\linewidth]{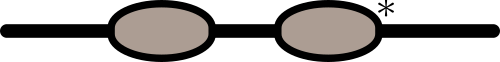}} = \raisebox{-0.3\height}{\includegraphics[width=0.15\linewidth]{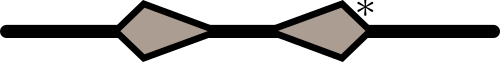}} = \raisebox{-0.2\height}{\includegraphics[width=0.1\linewidth]{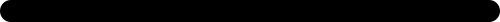}}
\end{align}
we find similar relations for red and green tensors,
\begin{align}\label{app:eq:red-green-basistrafo}
    \raisebox{-0.4\height}{\includegraphics[width=0.1\linewidth]{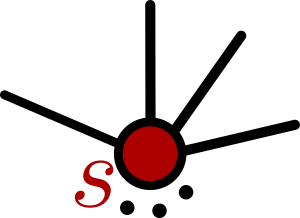}} =\raisebox{-0.4\height}{\includegraphics[width=0.1\linewidth]{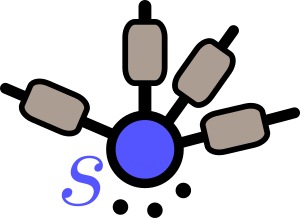}} = \raisebox{-0.4\height}{\includegraphics[width=0.1\linewidth]{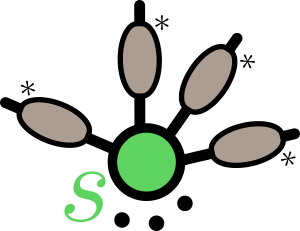}}
    \qq{and}
    \raisebox{-0.4\height}{\includegraphics[width=0.1\linewidth]{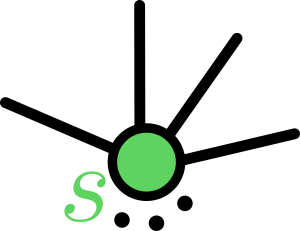}} =\raisebox{-0.4\height}{\includegraphics[width=0.1\linewidth]{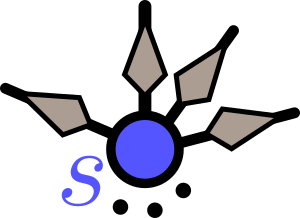}} = \raisebox{-0.4\height}{\includegraphics[width=0.1\linewidth]{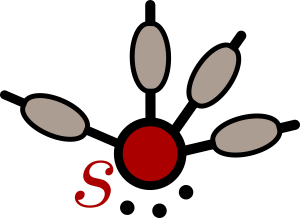}}.
\end{align}

\subsection{Representing any Pauli measurement as an RGB TN}\label{app:stablizerwithmeasurement}
Consider a system of $n$ qubits.
Projective measurements of an operator $O$ project the system into a definite eigenspace of $O$, associated with the image of a measurement outcome labeled by a projector $P_m$.
Specifically, starting with $\ket{\psi}$, the post-measurement state, knowing the 
outcome $m$, is proportional to $P_m\ket{\psi}$.
Within our graphical formalism we represent any Pauli measurement with a binary outcome $m\in\{0,1\}$ with the associated projector.
For a generic $n$-qubit Pauli operator $P_1\otimes P_2\otimes \dots  \otimes P_n$ this projector reads
\begin{align}
    P_m = \frac{1}{2}\left(\mathds{1} + (-1)^m P_1\otimes P_2\otimes \dots  \otimes P_n\right).
\end{align}
In the following, we describe a recipe of how to construct the graphical representation of that projector with an RGB TN.
\begin{enumerate}
    \item Each qubit comes with its own worldline, which we represent by a line,
    \begin{align}
    \raisebox{-0.4\height}{
        \includegraphics[width=0.15\linewidth]{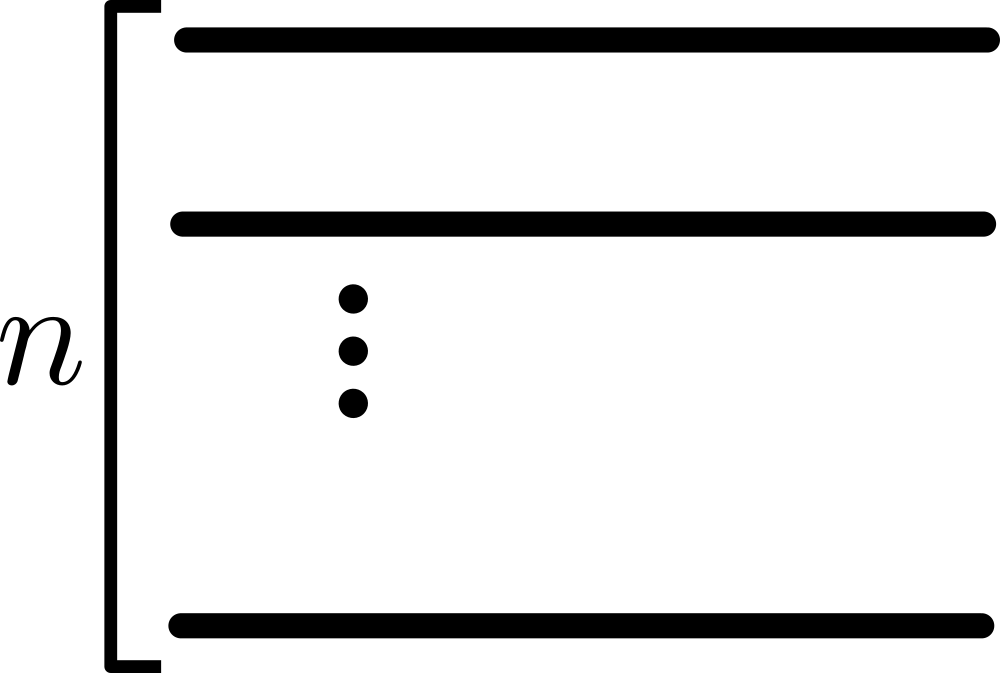}}
    \end{align}
    We identify the left hand side with the ``input'' and the right hand side with the ``output''.
    This can be motivated by interpreting that diagram as representing an identity operator on $(\bC^{2})^n$, mapping from the left to the right.
    
    \item Introduce a red $m$-tensor for the measurement,
    \begin{align}
    \raisebox{-0.4\height}{
        \includegraphics[width=0.15\linewidth]{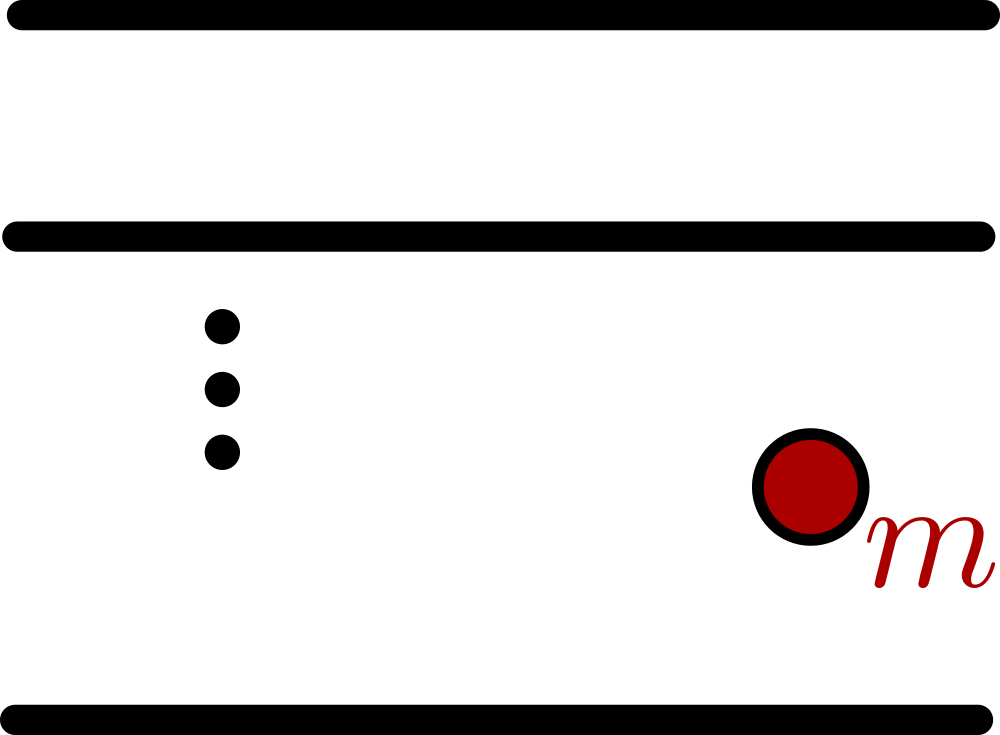}}
    \end{align}
    
    \item Place an $s=0$-tensor on each qubit wire taking part in the measurement. The color of the tensor is determined by the Pauli type of the measurement operator on that qubit ($X$: red, $Y$: green and $Z$: blue). For the green tensor we have to ``flip'' the right-pointing output to represent the correct measurement.
    For example, for a $X\otimes Y\otimes \cdots \otimes Z$ measurement, we place the following tensors:
    \begin{align}
    \raisebox{-0.4\height}{
        \includegraphics[width=0.15\linewidth]{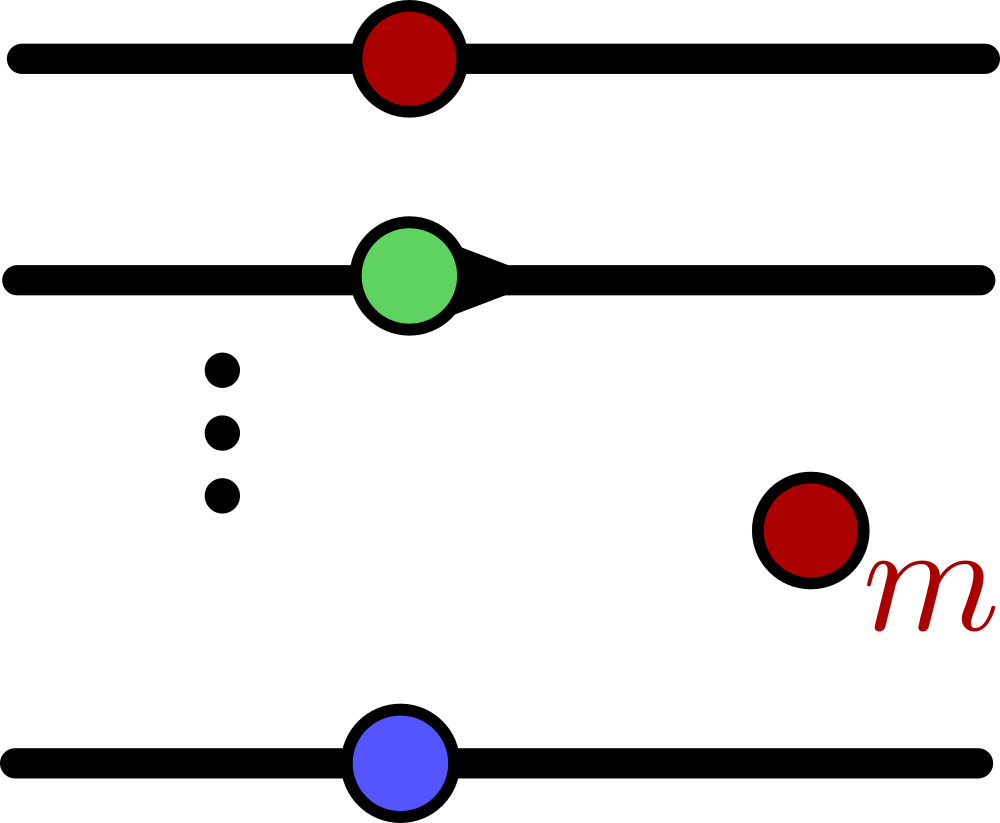}}
    \end{align}
    By flipping the input of the green tensors where the wire attaches from the right we have chosen to interpret the resulting diagram as a map from the open wires on the left to open wires on the right.
    Specifically, the projector of a measurement of a Pauli operator that includes a $Y$ operator is not generically symmetric (as a linear operator) so the diagram cannot be mirror symmetric either.

    \item Finally, we connect the newly introduced tensors to the measurement tensor. Depending on the colors of the tensors that are connected, we place an additional basis transformation on the connecting wire.
    A blue tensor is connected to the measurement tensor with a simple wire.
    A red tensor is connected via the basis transforming tensor that maps it to the blue tensor (in this case, the Hadamard box).
    Similarly, the green tensor is connected via the basis transforming tensor that maps it to the blue tensor.
    For the $X\otimes Y\otimes \cdots \otimes Z$ example above, this yields
    \begin{align}
    \raisebox{-0.4\height}{
        \includegraphics[width=0.15\linewidth]{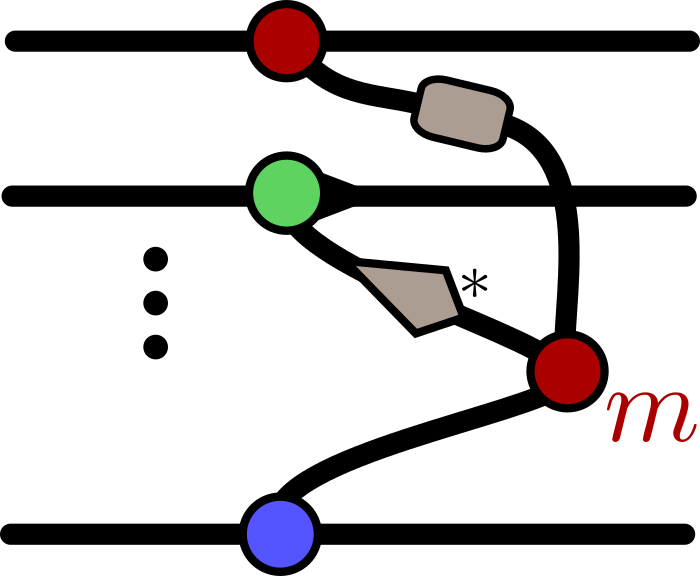}}
    \end{align}
\end{enumerate}

Single-qubit measurements can be viewed as projections onto eigenstates of single-qubit Pauli operators, i.e., initializations of single-qubit Pauli eigenstates.
As such it is helpful to directly represent these as single-legged tensors,
\begin{align}
    \ket{s}\sim\quad \raisebox{-0.4\height}{\includegraphics[height=16pt]{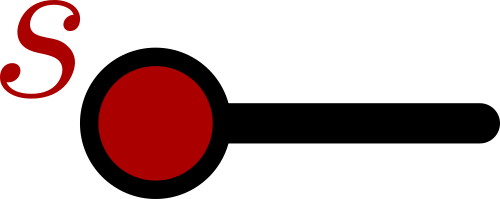}}\qcomma \ket{(-1)^s}\sim\quad \raisebox{-0.4\height}{\includegraphics[height=16pt]{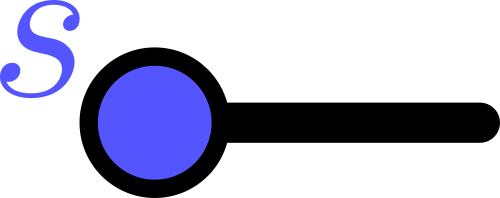}} \qq{and} \ket{(-1)^s i}\sim\quad \raisebox{-0.4\height}{\includegraphics[height=16pt]{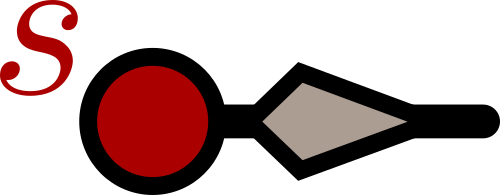}}.
\end{align}

\subsection{Putting the calculus onto the three colors}\label{app:sec:rewriterules}
In the main text, we have focused on constructing an RGB TN given an explicit circuit, i.e., sequence of gates and then using this representation to analyze the QEC properties of that circuit.
However, we find that RGB TNs have \textit{local rewrite rules} that allow one to locally rewrite a given RGB TN without changing the global action, i.e., the linear operator it represents.
These types of rewrite rules induce an equivalence relation amongst RGB tensor networks.
These equivalences are a core part of the ZX-calculus~\cite{van2020zx}, where the central objects are equivalence classes of diagrams and not individual diagrams.
One could say that these equivalences constitute the ``calculus'' part of the ZX-calculus.
Ref.~\cite{bombin2023unifying} showed that (the bulk of) a fault-tolerant measurement-based quantum computation protocol, the CSS Floquet honeycomb code~\cite{Davydova2023Floquet, kesselring2022anyon} and the two-dimensional surface code are equivalent in exactly that sense. 
In Ref.~\cite{Teague2023Floquetifying}, these rewrite rules were used to ``Floquetify'' the two-dimensional color code, i.e., construct a Floquet code that only uses one- and two-body measurements that is equivalent to the protocol where the stabilizer generators of the hexagonal color code are periodically measured.
These equivalences are also useful when considering individual circuit components.
For example, Ref.~\cite{McEwen_2023} used these equivalences to improve read-out circuits for the square surface code.
All these examples show that such equivalences are a desirable feature for a graphical representation of any circuit.

Since the ZX-calculus is complete~\cite{van2020zx}, we could express our green tensor in terms of ZX-tensors and use the ZX rewrite rules.
We argue, however, that working with more direct relations involving only green tensors themselves is more helpful when actually working with RGB networks.
In the following, we present these relations for all three types of tensors\footnote{The relations on the red and blue tensors are special cases of the ZX-calculus rules. For completeness, we still include them here.}.
Let us start with the blue tensors.
A straightforward tensor network calculation shows that they obey the following relations
\begin{align}\label{app:eq:blue_split}
    \raisebox{-0.4\height}{\includegraphics[width=0.3\linewidth]{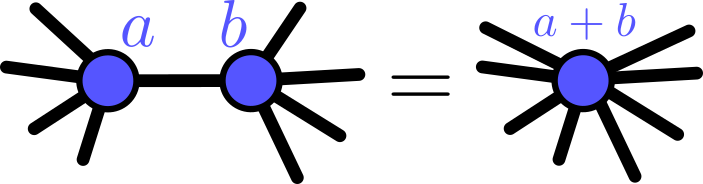}}
\end{align}
Using the basis transformations from Eq.\,\eqref{app:eq:red-green-basistrafo} and unitarity, Eq.\,\eqref{eq:basistrafos_unitarity}, this implies that the tensors of the other two colors fulfill the following relations
\begin{align}\label{app:eq:redgreen_split}
    \raisebox{-0.4\height}{\includegraphics[width=0.3\linewidth]{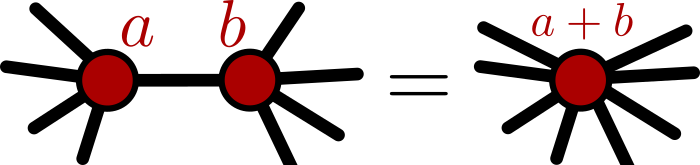}}\qq{and}
    \raisebox{-0.4\height}{\includegraphics[width=0.3\linewidth]{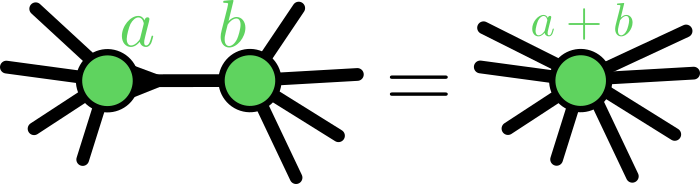}}.
\end{align}
Note that when splitting/fusing the green tensors, the orientation of the input/output is important.
These relations hold for any number of legs.

Using Eq.\,\eqref{app:eq:red-green-basistrafo} and the (projective) symmetries of the blue tensor, shown in Eq.\,\eqref{eq:blue_projective_sym}, we obtain similar symmetries for the tensor of the other two colors,
\begin{subequations}\label{app:eq:red_green_sym}
\begin{align}
    \raisebox{-0.3\height}{\includegraphics[width=0.08\linewidth]{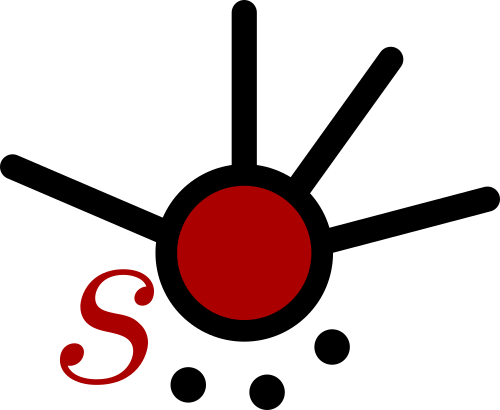}} =& \raisebox{-0.3\height}{\includegraphics[width=0.08\linewidth]{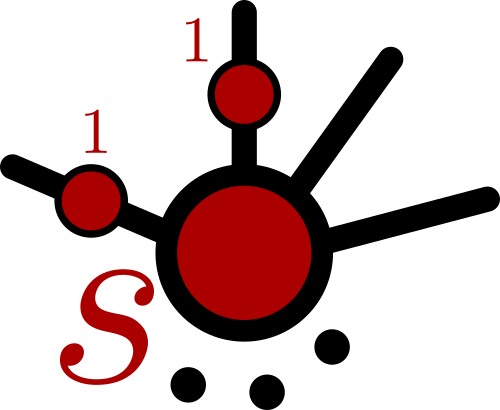}} = (-1)^s \raisebox{-0.3\height}{\includegraphics[width=0.08\linewidth]{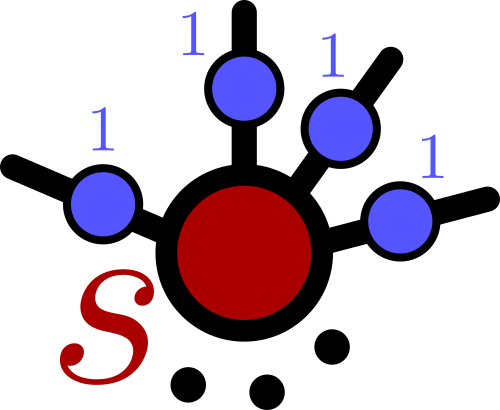}}\qq{and}\\
    \raisebox{-0.3\height}{\includegraphics[width=0.08\linewidth]{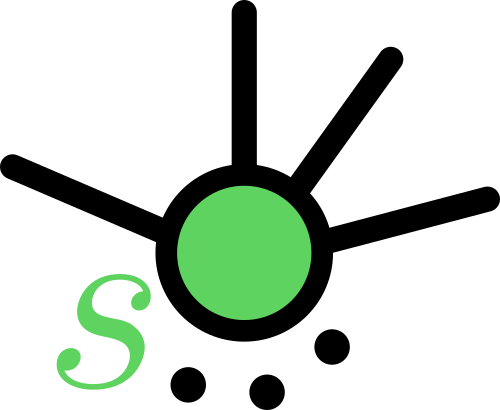}} =& \raisebox{-0.3\height}{\includegraphics[width=0.08\linewidth]{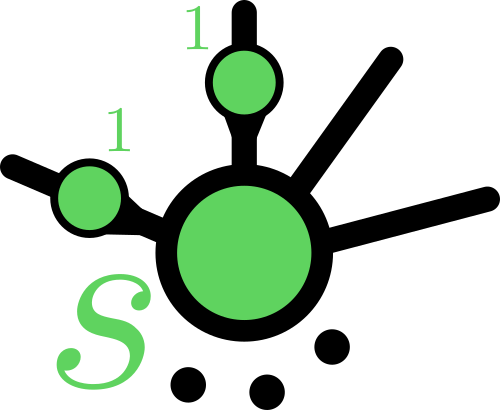}} = (-1)^s\raisebox{-0.3\height}{\includegraphics[width=0.08\linewidth]{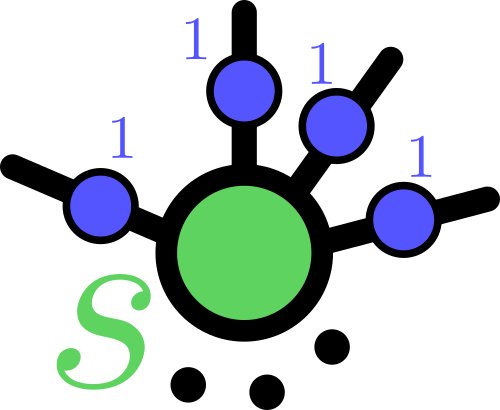}}.
\end{align}    
\end{subequations}

Additionally, they fulfill a \textit{generalized bialgebra rule} for the $Z$ and $X$ tensors\footnote{Note that this is just a usual ZX rule~\cite{van2020zx}.},
\begin{align}
    \raisebox{-0.4\height}{\includegraphics[width=0.2\linewidth]{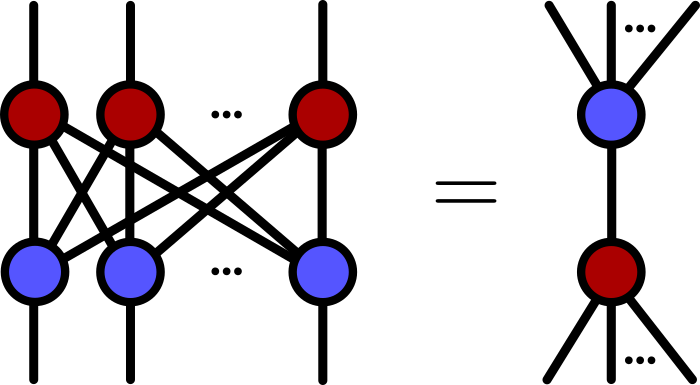}}
\end{align}
Applying the basis transformation rules \eqref{app:eq:red-green-basistrafo} we can derive similar rules involving the green ($Y$) tensor.
Importantly, this particular rule is not fully symmetric with respect to permuting the colors.

There is a second set of rewrite rules of the form
\begin{align}\label{app:eq:loop_split}
    \raisebox{-0.4\height}{\includegraphics[width=0.08\linewidth]{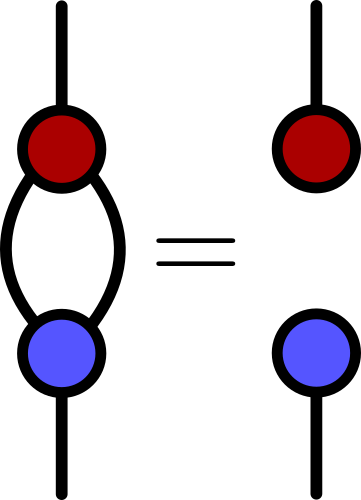}}\; \qcomma
    \raisebox{-0.4\height}{\includegraphics[width=0.15\linewidth]{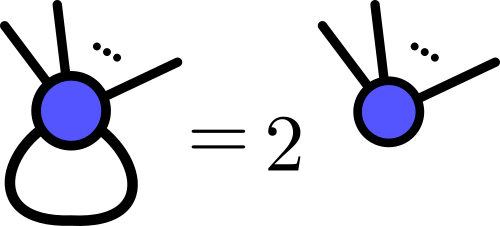}} \qcomma \raisebox{-0.4\height}{\includegraphics[width=0.15\linewidth]{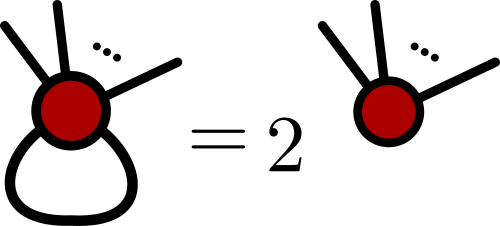}} \qq{and} \raisebox{-0.4\height}{\includegraphics[width=0.15\linewidth]{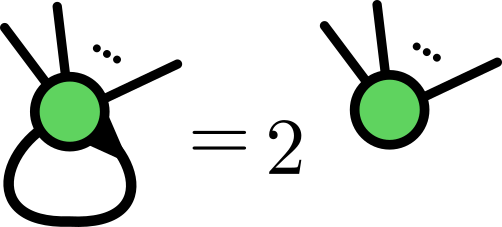}}
\end{align}
which allow one to remove or add loops into the network.
Often, we do not care about constants appearing via such contractions and think about a normalized version of each network.
Note that for the first rule shown above we can use Eqs.\,\eqref{eq:blue_basistrafo} and \eqref{app:eq:red-green-basistrafo} to obtain a similar a relation involving a green tensor.
This will, however, have some basis transformations on contracted legs so the rule is not fully symmetric among the three colors.

Another straightforward calculation shows how we can graphically see that single-qubit Pauli measurements initialize a qubit in a given Pauli eigenstate,
\begin{align}
    \raisebox{-0.4\height}{\includegraphics[height=1.2cm]{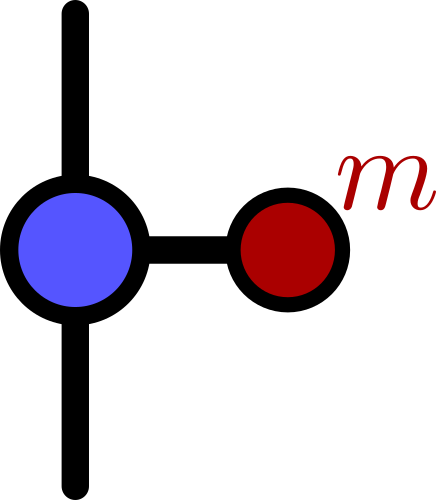}} = \raisebox{-0.4\height}{\includegraphics[height=1.2cm]{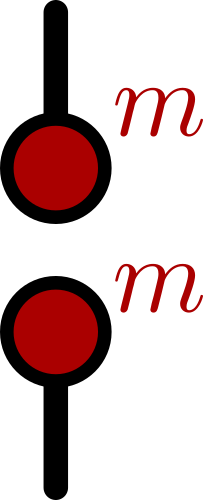}}\qcomma 
    \raisebox{-0.4\height}{\includegraphics[height=1.2cm]{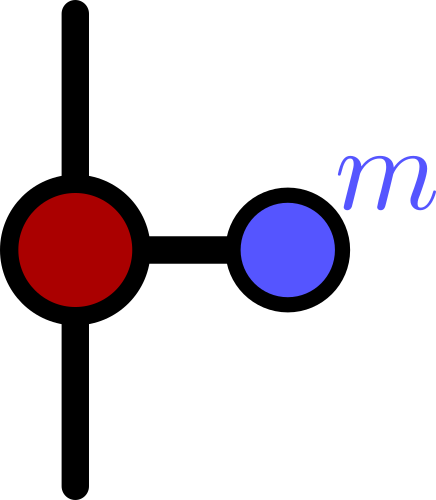}} = \raisebox{-0.4\height}{\includegraphics[height=1.2cm]{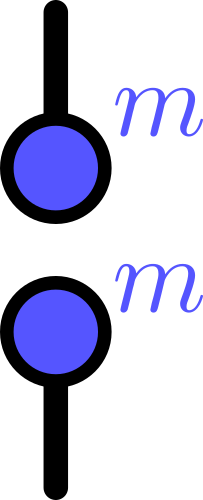}}\qq{and}
    \raisebox{-0.4\height}{\includegraphics[height=1.2cm]{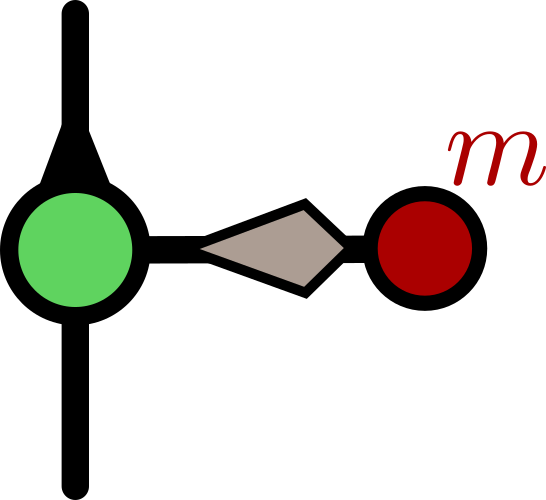}} = \raisebox{-0.4\height}{\includegraphics[height=1.5cm]{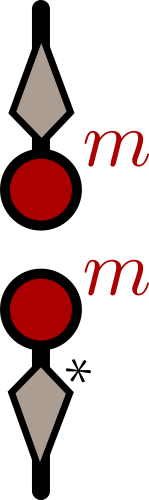}}.
\end{align}

Note that in this section we do not give a minimal nor a complete set of graphical rules but we solely aim to gather the most useful rules when considering analyzing and working with active QEC schemes.

\subsubsection{Example: Teleportation circuit}
Here, we show how a na\"ive representation of the teleportation circuit shown in Eq.\,\eqref{eq:teleportation_circuit} is rewrite-equivalent to
\begin{align}\label{app:eq:teleportation_network}
    \raisebox{-0.4\height}{\includegraphics[width=0.2\linewidth]{teleportation_tensor.png}}.
\end{align}
Including a preparation circuit of the Bell pair on the left we start with the following network representing the teleportation circuit which we can rewrite in two steps in to the diagram above:
\begin{align}
    \raisebox{-0.4\height}{\includegraphics[height=2cm]{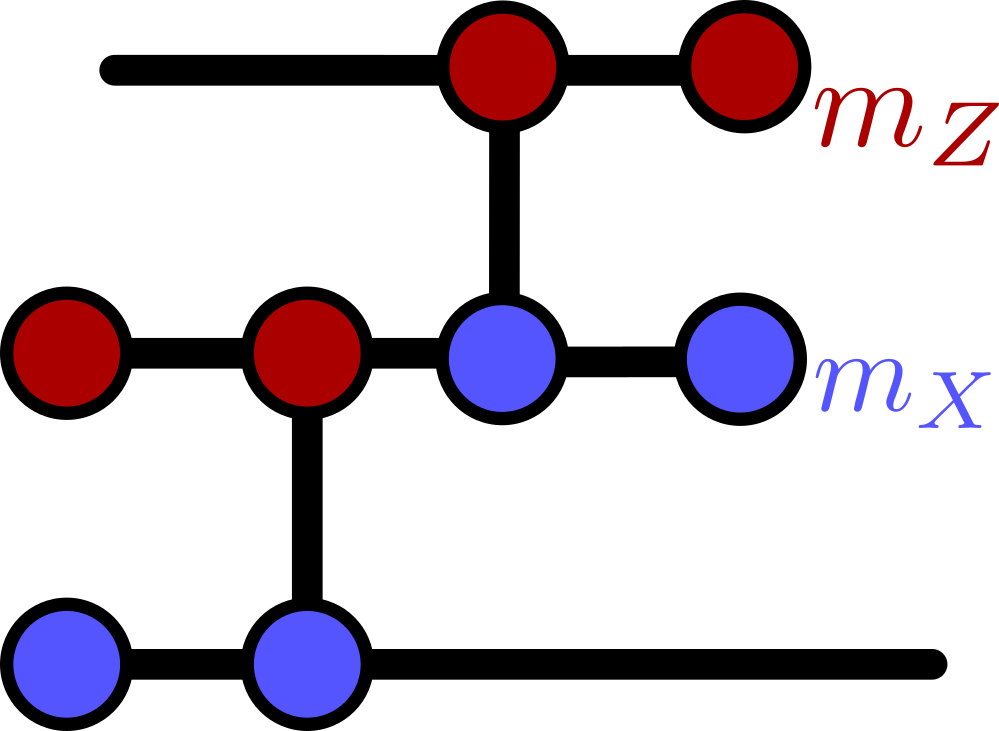}} = \raisebox{-0.4\height}{\includegraphics[height=2cm]{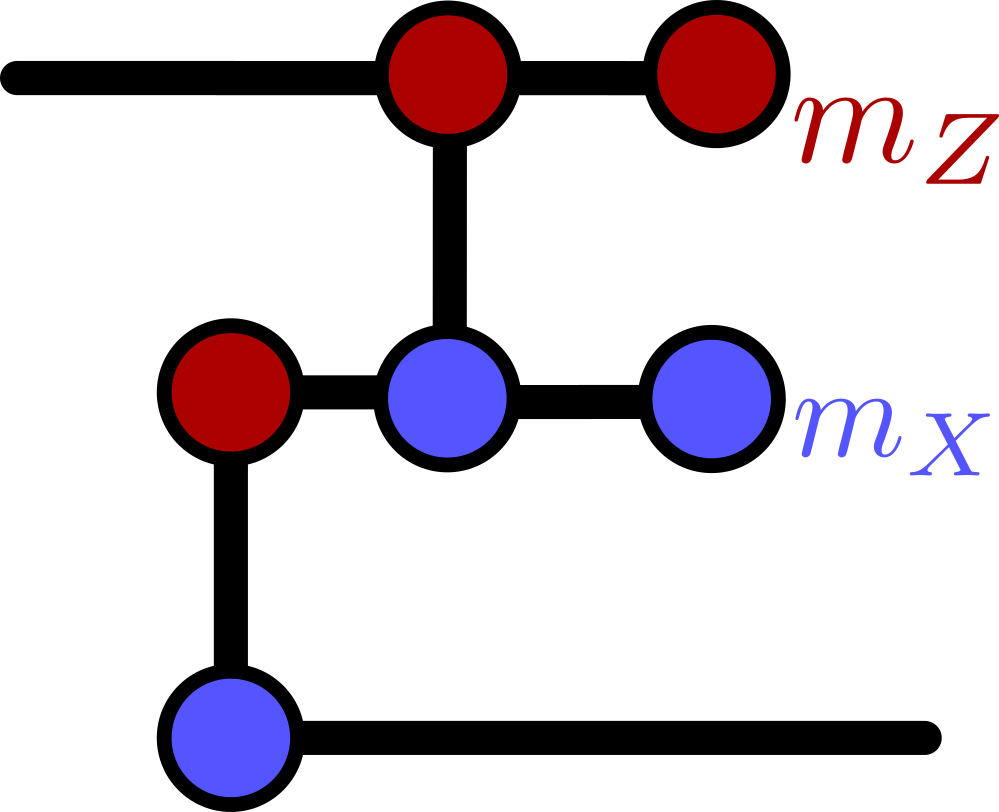}} = \raisebox{-0.4\height}{\includegraphics[height=2cm]{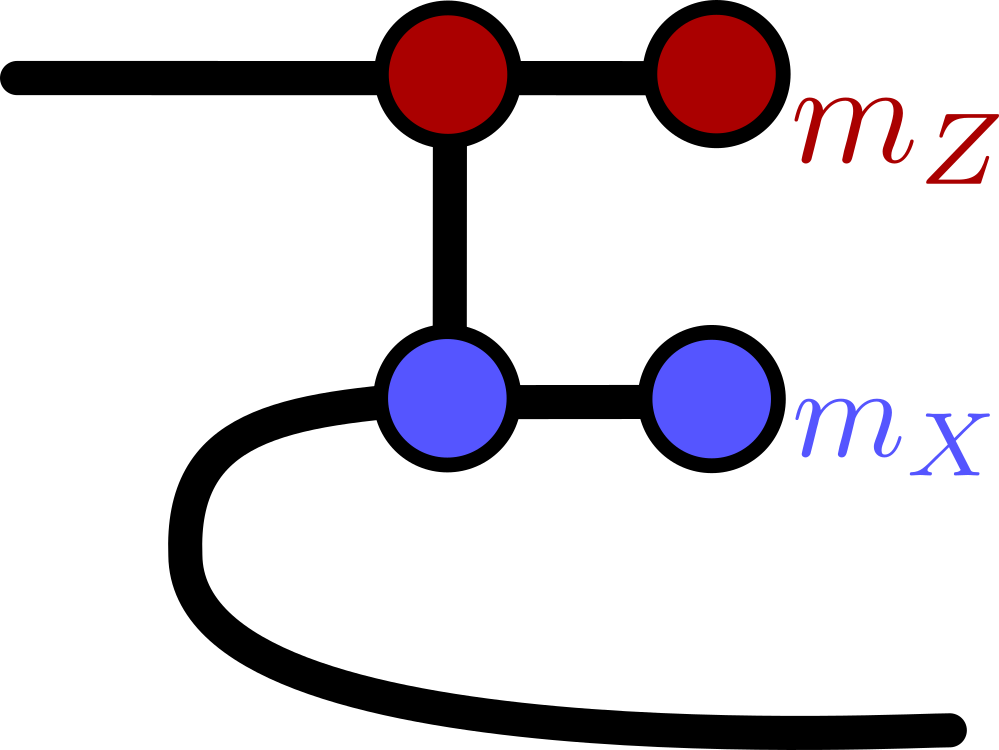}},
\end{align}
where we first used the splitting rules and then used that for each type of tensor, its two-legged version with $s=0$ corresponds to the identity matrix, and, can just be removed.
The diagram on the right is exactly the same as the one in Eq.\,\eqref{app:eq:teleportation_network} by construction as a tensor network which is independent on how the lines are drawn on the paper.

\subsection{Physical meaning of Pauli flows: a logical isomorphism protected by a classical code}\label{app:flows}
In the main text, we introduced a set of colorings of edges of an RGB tensor network that represents how Pauli operators can propagate through the network.
Here, we want to make the meaning of these highlights more precise in terms of their associated constraints and stabilizer groups on the in- and output qubits.

First, we want to complete the rules of highlighting introduced in the main text (see Eq.\,\eqref{eq:spider_flow}) to all components of an RGB network.
Specifically, we have not yet defined a highlight for the basis transformations, defined in Eq.\,\eqref{eq:basistrafo_definition}.
Calculating how these tensors map between different Pauli types,
\begin{subequations}\label{app:eq:basistrafo_sym}
\begin{align}\label{eq:basis_trafo_sym}
\raisebox{-0.4\height}{\includegraphics[width=0.1\linewidth]{Htensor.png}} = \raisebox{-0.4\height}{\includegraphics[width=0.1\linewidth]{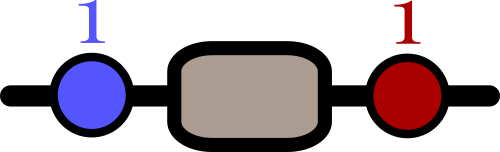}} = \raisebox{-0.4\height}{\includegraphics[width=0.1\linewidth]{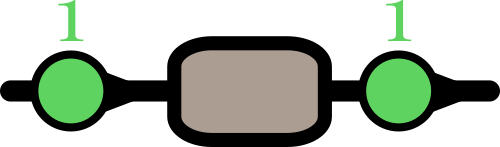}},\\
\raisebox{-0.4\height}{\includegraphics[width=0.1\linewidth]{Stensor.png}} = \raisebox{-0.4\height}{\includegraphics[width=0.1\linewidth]{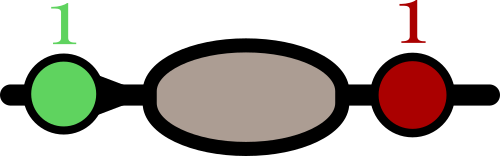}} = \raisebox{-0.4\height}{\includegraphics[width=0.1\linewidth]{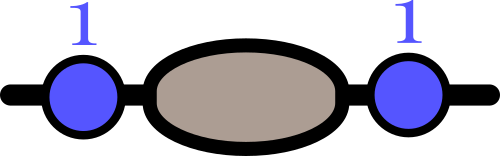}} ,\\
\raisebox{-0.4\height}{\includegraphics[width=0.1\linewidth]{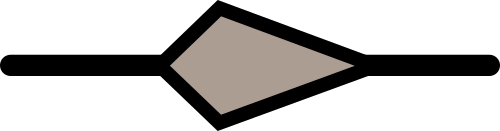}} = \raisebox{-0.4\height}{\includegraphics[width=0.1\linewidth]{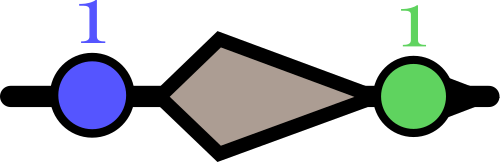}} = \raisebox{-0.4\height}{\includegraphics[width=0.1\linewidth]{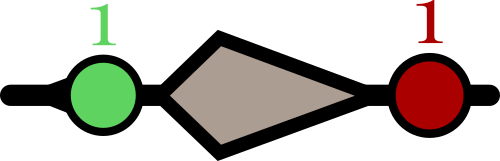}}
\end{align}
\end{subequations}
we obtain the following generators of ``valid'' highlights for these elements
\begin{align}\label{eq:basis_trafo_flows}
    \raisebox{-0.4\height}{\includegraphics[width=0.25\linewidth]{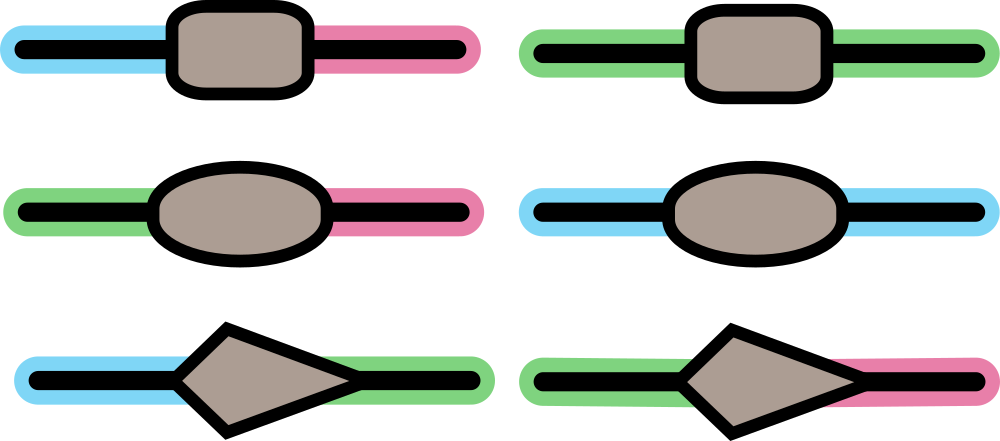}}.
\end{align}

\subsubsection{Setting the stage}

Consider an $n$-qubit Pauli group $\cP_n$, $n\in\bN$, and a bipartition of $n$ into $n_A$ and $n_B=n-n_A$ qubits.
We can write $\cP_n = \cP_{n_A}\otimes \cP_{n_B}$.
In the following, we show that a stabilizer group $S\leq\cP_n$ of rank $n$ defines a unique isomoprhism between a logical Pauli group embedded into $\cP_{n_A}$ and $\cP_{n_B}$.
If clear from context, we denote a single-qubit Pauli operator $P\in\cP_1$ on qubit $i$ with $P_i$.
We denote the center of a group $G\leq \cP_n$ by $Z(G) = \{x\in G\;|\; xg=gx\,\forall g\in G\}$ and its centralizer over the Pauli group by $C(G) = \{p\in \cP_n\;|\; pg=gp\,\forall g\in G\}$. Note that we omit the subscript in the centralizer whenever we consider centralizers over the Pauli group.
Sometimes we will refer to a ``real Pauli group''.
With that, we denote the subgroup of the Pauli group that only has real entries, generated by all $X$ and $Z$ operators. Note that $Y$ is not directly included in that group but only $\pm i Y$.

\begin{lemma}\label{app:lem:stab-bipartition_isomorphism}
    Consider the space of $n$ qubits with a bipartition $\bC^{2^n} = A\otimes B$ with $A\simeq\bC^{2^{n_A}}$ and $B\simeq\bC^{2^{n_B}}$.
    Let $S\leq \cP_n$ be a rank-$n$ stabilizer group on $A\otimes B$.
    Let $S_A=\{a\;|\; a\otimes \mathds{1}\in S\}$ be a group isomorphic to the maximal subgroup of $S$ solely supported on $A$ and analogously $S_B$, the maximal subgroup of $S$ supported on $B$.
 Then
    $S$ together with a bipartition $(A,B)$ defines a group isomorphism
    \begin{align}
        \varphi_S(A,B): \faktor{N(S_A)}{S_A} \to \faktor{N(S_B)}{S_B}\;,
    \end{align}
    where $S$ gives rise to a complete set of representatives of both $N(S_A)$ and $N(S_B)$.
    Vice versa, any such isomorphism defines a stabilizer group on $A\otimes B$ of maximal rank.
\end{lemma}

We remark that this holds for any bipartition and holds due to the maximality of the rank of $S$.

\begin{proof}
    The main object in our proof will be the group generated from truncated stabilizers, such as $S|_A = \langle\{a\;|\; a\otimes b\in S\}\rangle$ and $S|_B =\langle \{b\;|\; a\otimes b\in S\}\rangle$.
    By including all operators generated from the truncated stabilizer group we might add $-\mathds{1}$ to the group due to non-commutativity of restricted stabilizers.
    Let $S_A = \{a\;|\; a\otimes\mathds{1}\in S\}$ be a group of operators on $A$ that is isomorphic to the maximal subgroup of $S$ supported on $A$ and similarly $S_B$ for $B$.
    Since $S$ is a stabilizer group, $S_A$ is in the center of $S|_A$.
    Hence, we can decompose $S|_A$ as~\cite{gorenstein2007finite}
    \begin{align}
        S|_A \simeq S_A\times \faktor{S|_A}{S_A}\; .
    \end{align}
    In fact, $S_A$ is the center of $S|_A$ up to operators proportional to the identity, $\langle i\mathds{1}\rangle$, due to the maximality of the rank of $S$.
    We can see this by taking an element $a'\in Z(S|_A) = \{x\in S|_A\;|\; xs=sx\;\forall s\in S|_A\}$.
    It commutes with $a\in S|_A$, hence, defines an element $a'\otimes \mathds{1}$ that commutes with $S$ and hence, due to its maximality, must be included in $S$.
    By definition of $S_A$, $a'\in S_A$.

    Hence, $Z(S|_A/S_A) \subseteq \langle -\mathds{1}S_A\rangle$.
    In fact, this inclusion is an equality iff $S|_A$ includes anti-commuting Pauli operators, i.e., $S_A$ has non-maximal rank (on $A$).
    
    $S|_A/S_A$ admits a generating set of pair-wise anti-commuting operators.\footnote{This can be proven using the fact that it is isomorphic to a group generated by products of $X$ and $Z$ operators and there is no Pauli operator besides potentially $-\mathds{1}$ that commutes with all. This allows us to iteratively construct a set of mutually anti-commuting Pauli operators.}
    As such, it is isomorphic to the group of logical Pauli operators with real entries, i.e., $S|_A \simeq N(S_A)\cap \langle \{X_i,Z_i\}_{i=0}^n\rangle $.
    The same applies for $S|_B$, the stabilizer group truncated to the qubits in $B$.

    \paragraph*{Construction of isomorphism}
    Let $S_A\otimes S_B = \{a\otimes b\;|\; a\in S_A,b\in S_B\}$.
    Consider the quotient group $S/(S_A\otimes S_B) = \{xS_A\otimes yS_B\;|\; x\otimes y\in S\}$.
    Note that we can write any element in $S/(S_A\otimes S_B)$ as $xS_A\otimes yS_B$ for $x\in S|_A$ and $y\in S|_B$.
    From this we define the map
    \begin{align}
    \begin{split}
       \varphi_S(A,B): \faktor{S|_A}{S_A} &\to \faktor{S|_B}{S_B} \;,\\
       xS_A &\mapsto yS_B\qq{for} x S_A\otimes y S_B\in \faktor{S}{S_A\otimes S_B} \;.
    \end{split}
    \end{align}
    In the following, we show that this is a well-defined group isomorphism.

    First, we show that it is well-defined, i.e., there exists no two cosets $yS_B\neq y'S_B$ such that $x S_A\otimes yS_B, xS_A\otimes y'S_B\in S/(S_A\otimes S_B)$ for some $x\in S|_A$.
    Assume $\exists y,y'\in S|_B$ s.t. $yS_B\neq y'S_B$.
    Since $S_A\otimes S_B$ is central in $S$ the cosets $S/(S_A\otimes S_B)$ admit a group structure.
    This implies that $S_A\otimes yy'^{-1}S_B \in S/(S_A\otimes S_B)$, i.e., $\mathds{1}\otimes yy'^{-1}\in S$.
    By definition of $S_B$, $yy'^{-1}\in S_B$ and we obtain $yS_B = y'S_B$, contradicting our assumption.
    This shows that $\varphi$ is well defined.

    Let us tweak the argument slightly to see that $\varphi$ is in fact injective.
    Assume it wasn't. This means $\exists x\neq x' \in S|_A: xS_A\otimes yS_B, x'S_A\otimes yS_B \in S/S_A\otimes S_B$.
    Again by the group structure of the cosets, this implies $xx'^{-1}S_A\otimes S_B\in S/S_A\otimes S_B$ and $xx'^{-1}\in S_A$.
    Which contradicts the assumption $xS_A\neq x'S_A$.
    Surjectivity of $\varphi$ can be shown similarly. 
    By construction via a quotient group, for any $yS_B\in S|_B/S_B$ there exists a coset $xS_A\otimes yS_B\in S/S_A\otimes S_B$ for some $x\in S|_A$ and with that $yS_B\in \Im(\varphi_S)$.
    This shows that $\varphi$ is a bijection.

    It remains to be shown that $\varphi_S$ is a group homomorphism.
    This is straightforward to calculate using the centralness of $S_A$ and $S_B$ in $S|_A$ and $S|_B$.
    Let $x\otimes y,x'\otimes y'\in S$.
    Then
    \begin{align}
        \varphi_S(xS_A)\varphi_S(x'S_A) = yS_B y'S_B = yy'S_B = \varphi_S(xx'S_A),
    \end{align}
    where we used centrality of $S_B$ and the group structure on $S/S_A\otimes S_B$.

    We can extend $\varphi_S$ to act on a full logical Pauli group $N(S_A)/S_A$ by letting it act like the identity on all phases $\langle i\rangle$, the factors by which $S|_A$ differs from $N(S_A)$, i.e., we define $\varphi(i^m xS_A) = i^m\varphi(xS_A)$ for any $m\in\bZ, x\in S|_A$.
    This completes the proof that $\varphi$ is an isomorphism.

    \paragraph*{Construction of stabilizer group}
    To construct a maximal rank stabilizer group from a given isomorphism, we reverse the construction above.
    Consider isomorphism $\varphi'$ between logical Pauli groups, on given codes, defined by stabilizer groups $S_A$ and $S_B$ of rank $n_A-k$ and $n_B-k$, respectively.
    Consider the subgroup of the domain of $\varphi'$ that is generated by cosets that act like $X$ and $Z$ operators, i.e., it does not contain the logical $-\mathds{1}$ and any imaginary logical Pauli and denote it with $\Tilde{P}_{\varphi'}$.
    We use $\varphi'$ to construct cosets $xS_A\otimes \varphi'(xS_A)$, for $xS_A\in \Tilde{P}_{\varphi'}$.
    Take a complete set of representatives $R_{\varphi'} = \{x_i\otimes y_i\}_{i=1}^{2k}$ for these cosets.
    Since $\varphi'$ is a group isomorphism, they generate a stabilizer group $S_{AB} = \langle R_{\varphi'}\rangle$ on $A\otimes B$ that is in the centralizer of both $S_A\otimes \mathds{1}$ and $\mathds{1}\otimes S_B$.
    We define the stabilizer group associated to $\varphi'$ as
    \begin{align}
        S_{\varphi'} = \langle S_A\otimes\mathds{1}, S_B\otimes \mathds{1}, S_{AB}\rangle \simeq  S_A\times S_B \times S_{AB}\;.
    \end{align}
    It remains to be shown that it is of maximal rank.
    Since every factor above is independent, the ranks add up and we obtain
    \begin{align}\label{app:proof:eq:rank}
        \rank(S_{\varphi'}) = \rank(S_A) + \rank(S_B) + \rank(S_{AB}).
    \end{align}
    Let $n_A$ be the number of qubits in $A$ and $n_B$ be the number of qubits in $B$.
    We can use the fact that the isomorphism is on $k$ logical qubits to obtain $\rank(S_A) = n_A- k,\rank(S_B) = n_B-k, \rank(S_{AB})  = 2k$.
    Plugging this into Eq.\,\eqref{app:proof:eq:rank} yields $\rank(S_{\varphi}) = n_A-k+n_B-k+2k = n_A+n_B$ which is the maximal rank of a stabilizer group on a system of $n_A+n_B$ qubits.
\end{proof}
Note that this can be generalized straightforwardly to Pauli stabilizer groups on qudits.
In that case, one should take the adjoint on the $B$ factor to not overcomplicate things in the interpretation of the resulting stabilizer group.
Besides that, the proof goes through in the same way, as long the qudit dimension is finite.

\paragraph*{An algebraic perspective}
Before getting into more details on how to identify an RGB tensor network with an isomorphism above, we want to comment on how to understand the above isomorphism analogously on the level of a full ``logical algebra'', the algebra of matrices acting within the logical subspace.
This is not too important for any logical Clifford operation but should be considered when generalizing logical operations beyond Clifford operations in a unified manner.
Once a non-Clifford operation is applied, the full evolution of the system cannot be described by any Pauli group isomorphism anymore but should be described by a $^\ast$-isomorphism on a logical algebra.
In the following, we want to sketch how the Clifford case and a logical group isomorphism fits into this algebraic perspective.

Consider a Pauli stabilizer group $S\leq \cP_n$ and its normalizer group $N(S)=\{p\in\cP_n\;|\; pS=Sp\}$. Both $S$ and $N(S)$ are sets of linear operators on a finite-dimensional vector space that are closed under multiplication.
As such, they span a $^\ast$-algebra where the matrix adjoint $^\dagger$, transposing and complex conjugating, plays the role of the $^\ast$-operation.
Let $\cS = \spn_{\bC}(s \in S)$ be the commutative $^\ast$-algebra spanned by the stabilizer group.
Since for Pauli stabilizer groups, the normalizer equals the centralizer, $N(S) = C(S) = \{p\in\cP_n\;|\; ps = sp\,\forall s\in S\}$, it spans the commutant
\begin{align}
    \cC(\cS) = \spn_{\bC}(p\in C(S)).
\end{align}
Note that any element in that algebra acts the same within one ``syndrome sector'', a subspace that transforms irreducibly under the action of $\cS$\footnote{These subspaces can be identified with the images of minimal central idempotents in $\cS$.}.
Within the trivial syndrome sector, the space of states\footnote{Note that this can directly be generalized to density matrices by considering the set of square matrices containing a representation of the algebra above where the algebra acts via conjugation, not via left-multiplication as for state vectors.} in the image of
\begin{align}
    P_0 = \frac{1}{\dim(\cS)}\sum_{s\in S} s\in \cS,
\end{align}
the logical action of elements in $\cC(\cS)$ is formally characterized by $\cA_S = \spn_{\bC}(C(S)/S)$ which again is a $^\ast$-algebra.
In fact, since $C(S)/S$ is isomorphic to a Pauli group, it spans the full matrix algebra on the logical subspace, $\cA_S\simeq \cM_{2^k}$, where $k = \log_4(\dim(\cA))$ denotes the number of logical qubits on which $\cA$ acts.

The isomorphism $\varphi_S(A,B)$ obtained in Lem.\,\ref{app:lem:stab-bipartition_isomorphism} can be lifted to a $^\ast$-isomorphism
\begin{align}
    \Phi_S(A,B): \cA_{S_A}\to \cA_{S_B}
\end{align}
on two (formal) logical algebras.
Importantly, a logical Clifford isomorphism is an isomorphism of that kind that happens to be fully described by an isomorphism on a particular basis of $\cA_{S_A}$ and $\cA_{S_B}$, the Pauli basis.
A non-Clifford operation will not be described by such a group isomorphism but is intrinsically an algebra isomorphism that can map single Paulis to sums of Paulis.

\subsubsection{The Logical isomorphism implemented by an RGB tensor network}\label{app:sec:RGB_log_iso}

In this section, we want to show that any RGB \emph{tensor network} (TN) implements a logical isomorphism on some logical Pauli algebra.
As a byproduct, we show in the proof of Lem.\,\ref{app:lem:composition}, how to think of the composition of (logical) Clifford blocks in a unified manner, independent of a specific encoding.
This applies to any Pauli stabilizer tensor network and gives a complementary perspective on how to view the stabilizer formalism including measurements and may give insight onto how to better understand architectures for quantum computations on the logical level.

\begin{prop}\label{app:prop:clifford}
    Let $T$ be an RGB tensor network composed of red, green and blue tensors, introduced in Eqs.\,\eqref{eq:bluetensor}, \eqref{eq:redtensor} and \eqref{eq:greentensor}, as well as the basis transformations defined in Eq.\,\eqref{eq:basistrafo_definition},
    with $n\in\bZ_{\geq 0}$ open legs.
    Consider a bipartition of the open legs into two subsets $A$ and $B$. Equivalently, we denote the vector space associated to these legs with $A$ and $B$.
    $T$ can be interpreted as a linear map $A\to B$, considered as the vector spaces assigned to the input and output legs,
    and can act via conjugation as a map $B(A)\to B(B)$ mapping linear operators on $A$ to linear operators in $B$.
    As such, it implements an isomorphism
    \begin{align}
        \varphi_T(A,B): \faktor{N(S_A)}{S_A} \to \faktor{N(S_B)}{S_B} \;,
    \end{align}
    for some stabilizer groups $S_{A}$ on $A$ and $S_{B}$ on $B$.
\end{prop}

Given Lem.\,\ref{app:lem:stab-bipartition_isomorphism}, this reduces to showing that any RGB TN admits a stabilizer group of maximal rank, i.e., when it is interpreted as a state, it is a unique stabilizer state.
We proceed step by step and first show that this structure holds for any of the constituents of an RGB TN and then, in Lem.\,\ref{app:lem:composition}, show how to construct the stabilizer group of a contracted network, knowing the stabilizer groups of the constituents.
This Lemma is independent of the specific RGB TN representation and applies to all stabilizer tensor networks and with that to any linear map composed of Clifford operations.

In the following, the red, green and blue spider, defined in Eqs.\,\eqref{eq:redtensor}, \eqref{eq:greentensor}, \eqref{eq:bluetensor}, as well as the basis transformations, defined in Eq.\,\eqref{eq:basistrafo_definition}, are considered ``elementary building blocks'' of an RGB TN.

\begin{lemma}\label{app:lem:stabilizer-elem}
Let $t$ be an elementary building block of an RGB TN with $n$ legs.
As a state, it is stabilized by a stabilizer group of rank $n$.
\end{lemma}

\begin{proof}  
Let us go through each building block.
We will map the projective symmetries of the tensors to a stabilizer group of maximal rank.
Each of the symmetries of the tensors considered in this work are Pauli symmetries, i.e., they can be viewed as multiplying the tensor with a tensor product of Paulis on the vector space associated to the open legs of a tensor.

We find that the Pauli operators associated to the symmetries of the basis transformations in Eq.\,\eqref{app:eq:basistrafo_sym} indeed commute and generate a rank-2 stabilizer group.
For example, the first tensor in Eq.\,\eqref{app:eq:basistrafo_sym}, admits two generators $X\otimes Z$ and $Y\otimes Y^T$.
Since they have two legs, this is a maximal stabilizer group and fully defines the tensor.

For the red, green and blue spiders, we consider the symmetries defined in Eqs.\,\eqref{eq:blue_projective_sym} and \eqref{app:eq:red_green_sym}.
We find that their symmetry groups are related by permuting the color (i.e., Pauli type), such that it suffices to consider a single color.
Take the blue tensor with $n$ open legs.
Its symmetries are shown in Eq.\,\eqref{eq:blue_projective_sym}.
It has one $X$-type symmetry with which we associate the stabilizer $(-1)^s X^{\otimes n}$, where $s$ is the value of the sign of the tensor, and a $Z$-type symmetry for each pair of legs $(i,j)$  associated to the stabilizer $Z_iZ_j$.
A simple counting argument shows that the group of these Pauli operators is indeed a stabilizer group of rank $n$.
One can view this stabilizer group as the stabilizer group of a GHZ-like state.
Similarly, for the other colors there are two types of stabilizer generators, one that acts on all input legs and might carry a sign of $-1$ and a set of operators that act on pairs of legs with a different Pauli type.

This completes the proof for each of the elementary building blocks.
\end{proof}

Given that the elementary building blocks are defined by a stabilizer group of maximal rank, we want to proceed by showing how to obtain the stabilizer group of a tensor network composed of these building blocks.
Specifically, the following Lemma and its proof give an explicit construction of the resulting stabilizer group when contracting two stabilizer TNs along an arbitrary subset of legs.

\begin{lemma}\label{app:lem:composition}
    Let $t$ and $t'$ be two tensor networks with $n$ and $n'$ open legs that admit stabilizer groups $S^{(t)}$ and $S^{(t')}$, both with maximal rank ($n$ and $n'$).
    Consider an ordered subset $\ell = \{ l_1,l_2,...,l_c\}$ of $c$ legs of $t$, with $c\leq \min(n,n')$, and identify them with another ordered subset of legs $\ell'$ of $t'$ of the same size.
    We can define a unique contraction of $t$ and $t'$ by joining $\ell$ with $\ell'$ according to their order and denote the contracted network with $t'\circ_{\ell} t$.

    The contracted network $t'\circ_\ell t$, admits a stabilizer group of maximal rank that is uniquely determined from the stabilizer groups of $t$ and $t'$, respectively.
\end{lemma}

\begin{figure}
    \centering
    \includegraphics[width=0.6\linewidth]{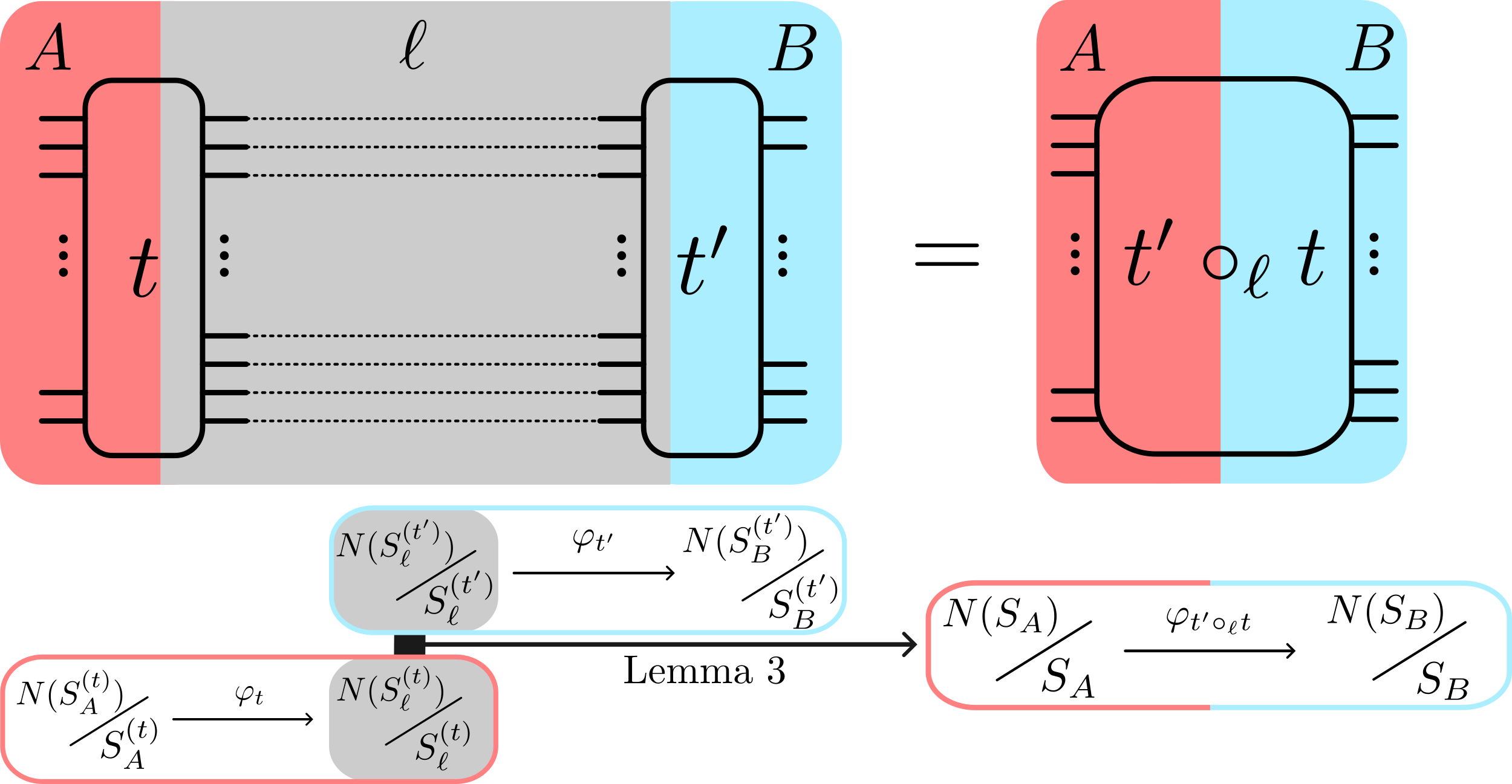}
    \caption{An alternative perspective on Lemma\,\ref{app:lem:composition} and the construction in our proof of it.
    Given two linear Clifford maps $t: A\to \ell$ and $t': \ell\to B$ each of which acts like an isomorphism $\varphi_{t}$ and $\varphi_{t'}$ on an encoded logical Pauli group (possibly a different one for $t$ and $t'$), we can construct the isomorphism applied by the contracted network, a linear map from $A\to B$ explicitly.
    This is done by finding a subgroup of Pauli operators on which the isomorphism can be composed and deriving a resulting stabilizer group on $A\otimes B$ from it.
    Using the equivalence in Lem.\,\ref{app:lem:stab-bipartition_isomorphism}, this stabilizer group can be associated to a logical isomorphism of the contracted network.
    Note that, in a sense, the global stabilizer group on $A\otimes B$ is more fundamental since it fully describes the entire network.
    Lem.\,\ref{app:lem:stab-bipartition_isomorphism} tells us that for any bipartition $(A,B)$ we can infer the induced logical isomorphism from a logical Pauli group on $A$ to a logical Pauli group on $B$.
    For a detailed description of the construction of the composite isomorphism we refer to the proof of Lem.\,\ref{app:lem:composition}.}
    \label{app:fig:join_stabs}
\end{figure}

We illustrate the ``joining operation'' for two stabilizer tensor networks and an alternative interpretation of Lem.\,\ref{app:lem:composition} in Fig.~\ref{app:fig:join_stabs}.
The idea behind the construction of the isomorphism, respectively the stabilizer group, of the contracted network is that $t'$ effectively acts like measuring the stabilizer group $S^{(t')}_\ell$ on a code stabilized by $S^{(t)}_\ell$.\footnote{Equivalently, $t$ can be interpreted as measuring $S^{(t)}_\ell$ on a code stabilized by $S^{(t')}_\ell$. The symmetry becomes apparent in the proof.}
How to treat the action of such a measurement on a set of logical operators is part of the usual stabilizer formalism and captured well by many references, e.g. Refs.\,\cite{gottesman1997stabilizer, Vuillot2019code, Teague2023Floquetifying}.
Our proof builds on that to find the right set of operators along which a composition of the isomorphism of $\varphi_t$ and $\varphi_{t'}$ make sense and shows that this indeed leads to a composite isomorphism, respectively a contracted tensor network with a full-rank stabilizer group.

\begin{proof}
    Let $A$ be the vector space associated to the set of open legs of $t$ in the complement of $\ell$ and $B$ the vector space associated to the set of open legs of $t'$ in the complement of $\ell'$.
    In this proof, we aim to construct a stabilizer group of maximal rank on $A\otimes B$ from $S^{(t)}$ and $S^{(t')}$.
    We denote their maximal subgroups supported on the contracted legs as $S^{(t)}_\ell$ and $S^{(t')}_\ell$ and their restricted subgroups by $S^{(t)}|_\ell$ and $S^{(t')}|_\ell$ (cf. notation from the proof of Lem.\,\ref{app:lem:stab-bipartition_isomorphism}).
    Since there is a unique 1-1 mapping from $\ell$ to $\ell'$, we use $\ell$ in all expressions.
    Additionally, we use the same symbol $\ell$ for the vector space associated to the contracted legs.
    The vector spaces considered in this proof are $A$, $\ell$, and $B$.

    Recall from Lem.\,\ref{app:lem:stab-bipartition_isomorphism} that we can think of $S^{(t)}|_\ell$ and $S^{(t')}|_\ell$ as generating a complete set of (real) logical operators including corresponding stabilizer groups $S^{(t)}_\ell$ and $S^{(t')}_\ell$.
    Moreover, by Lem.\,\ref{app:lem:stab-bipartition_isomorphism} the problem of identifying a stabilizer group of maximal rank on $A\otimes B$ is equivalent to constructing an isomorphism between the logical Pauli group of a (stabilizer) code in $A$ and the logical Pauli group of a (stabilizer) code in $B$, both defined by a stabilizer group $S_A$ and $S_B$.
    In this proof, we will start with the isomorphisms $\varphi_t$ and $\varphi_{t'}$, associated to $t$ and $t'$ with the given bipartition of the space associated to the open legs, $A\otimes \ell$ and $\ell\otimes B$.
    We denote the the number of logical qubits on which $\varphi_t$ acts by $k$ and the number of logical qubits on which $\varphi_{t'}$ acts by $k'$.

    Consider the maximal joint subgroup of (potentially trivial) logical operators (generated by logical $X$ and $Z$ representatives),
    \begin{align}\label{app:proof:eq:intersec}
        P_\ell = S^{(t)}|_\ell \cap S^{(t')}|_\ell.
    \end{align}
    Each element in $P_\ell$ can be identified with a coset of $S^{(t)}_\ell$ and $S^{(t')}_\ell$ by the unique homomorphism that given a normal subgroup $H\leq G$ implements the map $g\mapsto gH$.
    Moreover, $P_\ell$ forms a subgroup of both $S^{(t)}|_\ell$ and $S^{(t')}|_\ell$. 
    Taken together, $P_\ell$ defines both a subgroup of cosets with respect to $S^{(t)}_\ell$ and $S^{(t')}_\ell$.
    
    We are now set to construct the maximal stabilizer group of $t'\circ_\ell t$.
    We start with the (probably non-maximal) stabilizer group $S = S^{(t)}_A\otimes S^{(t')}_B = \{s_a\otimes s_b\;|\; s_a\in S^{(t)}_A, s_b \in S^{(t')}_B\}$ and add commuting terms to get a stabilizer group of maximal rank.
    This is done by composing the isomorphisms of $t$ and $t'$ along a well-defined intersection.
    We can summarize the construction as follows:
    For each $p\in P_\ell$, we add a representative of $\varphi_t^{-1}(pS^{(t)}_\ell)\otimes \varphi_{t'}(pS^{(t')}_\ell)$ to $S$.
    This gives a systematic mapping of elements in $P_\ell$, operators acting on $\ell$, to stabilizers on $A\otimes B$ such that they encode the isomorphism applied by the contracted network, $t'\circ_\ell t$.
    Let us elaborate on the individual steps of the construction.
    It can be understood with the following commutative diagram,
    \begin{align}
        \raisebox{-0.4\height}{\includegraphics[width=0.5\linewidth]{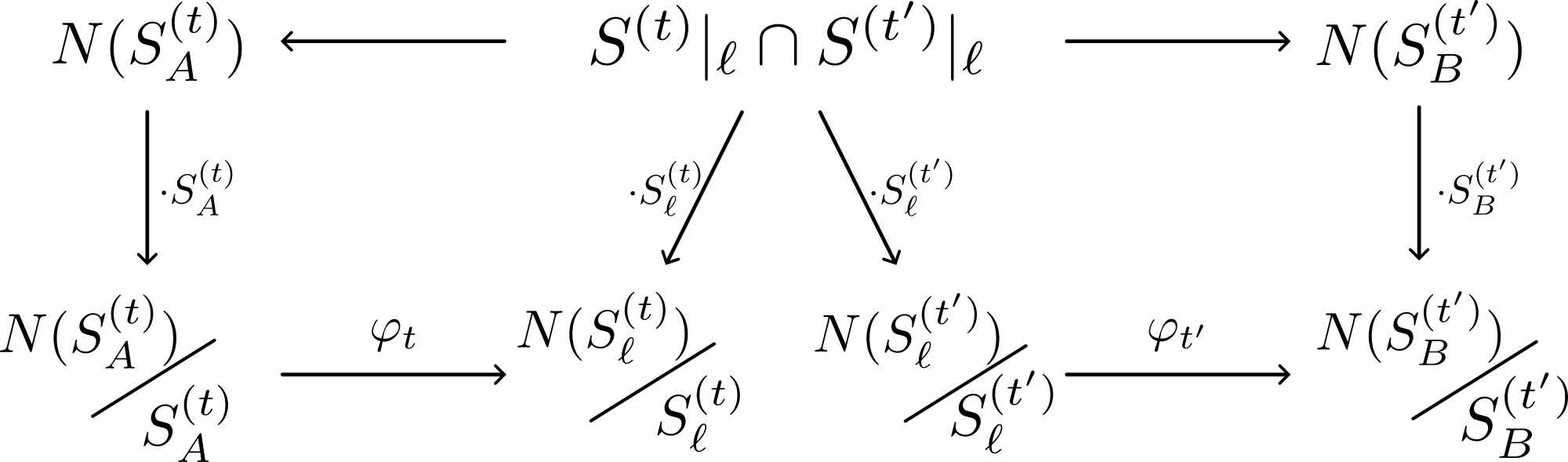}},
    \end{align}
    where in the top row we have sets composed of physical operators on $A$, $\ell$ and $B$ respectively and on the bottom row we have sets of cosets over different stabilizer groups.
    One should interpret this diagram as defining the maps associated to the arrows in the top row.\footnote{Note that the definition of the maps in the top row are only defined up to choosing a representative of a coset $pS^{(t)}_A$. This gives a different element added to the stabilizer group for that specific $p$. The full stabilizer group constructed will be independent of that choice, however, since $S^{(t)}_A \otimes \mathds{1}$ and $\mathds{1} \otimes S^{(t')}_B$ are included in $S$ and is hence uniquely given by the diagram and the associated construction.}
    Each step in the construction outlined above can be understood as an arrow in this diagram. 
    Specifically, we go from top to bottom by identifying an operator with a coset with respect to a given stabilizer group.
    The ``inverse'' operation, going from bottom to top is ``taking a representative of a coset''.
    The horizontal arrows on the bottom represent isomorphisms so admit proper inverses.
    In the procedure described above the stabilizer added to $S$ is obtained by taking an element $p$ in the intersection set in the top center and mapping it to an operator $p_A$ on $A$ by going in clockwise direction in the left loop and then to an operator $p_B$ on $B$ by applying maps in counter clockwise direction in the right loop.
    The operator added to $S$ is given by $p_A\otimes p_B$.
    This is done for each element in $P_\ell$, the intersection set.

    It remains to be shown that $S$ is now a stabilizer group of maximal rank.
    Let $n_A$ be the number of qubits in $A$ and $n_B$ be the number of qubits in $B$.
    We know that $S^{(t)}_A$ has rank $n_A-k$ and $S^{(t')}_B$ has rank $n_B-k'$.
    Hence, we need to show that the procedure of adding elements to $S$ described above increases the rank by exactly $k+k'$.
    Since the construction starts with a stabilizer group that contains the stabilizers in $S_A^{(t)}$ and $S_B^{(t')}$ we only need to consider elements in $P_\ell$ in non-trivial logical cosets with respect to $S^{(t)}_\ell$ and $S^{(t')}_\ell$.
    Additionally, due to the group structure on the cosets, we only need to consider a suitable generating set for the group of cosets.
    For each non-trivial generating coset for which there is at least one $p\in P_\ell$, the rank is increased by one.
    
    To prove that this construction leads to a stabilizer group of maximal rank, we employ techniques introduced in Ref.~\cite{Vuillot2019code} for code deformations.
    Consider the subsystem code with the gauge group 
    \begin{equation}
    G = \langle S^{(t)}_\ell, S^{(t')}_\ell \rangle
    \end{equation}
    generated by the two stabilizer groups on $\ell$ given by $t$ and $t'$.
    We denote the center of that gauge group by $S_G = Z(G)$.
    Up to signs, it corresponds to the stabilizer group associated to $G$.
    We can interpret the sets in the construction of $S$ in light of that subsystem code.
    Let \begin{equation}\cL_G = C(S_G)/G 
    \end{equation}
    be the group of inequivalent sets of logical Pauli operators of this subsystem code and $\kappa = \log_4(\abs{\cL_G}/4)$ the number of logical qubits in that subsystem code.
    
    We can identify both $S^{(t)}_\ell$ and $S^{(t')}_\ell$ as stabilizing a (partly) gauge-fixed subspace since they are Abelian subgroups of $G$. Each $p\in P_\ell$ that is a non-trivial logical with respect to $G$, is hence also a non-trivial logical with respect to $S^{(t)}_\ell$ and $S^{(t')}_\ell$.
    Recalling that $P_\ell$ is exactly the real subgroup of $C(S_G)$, we find that it includes all non-trivial logicals and hence the rank of $S$ is increased by at least $2\kappa$ when following the construction described above.
    Additionally, $P_\ell$ includes logical operators of the gauge qubits of the code, operators that are non-trivial logicals in $S^{(t)}_\ell$ but not in $S^{(t')}_\ell$ and vice versa. 
    In fact, $P_\ell$ contains exactly the gauge logical operators that are not fixed in $S^{(t)}_\ell$ but in $S^{(t')}_\ell$ and vice versa.
    Note that here, $P_\ell$ only contains one operator per logical (gauge) qubit.
    In $S^{(t)}_\ell$ there are $k - \kappa$ gauge qubits unfixed and in $S^{(t)}_\ell$ there are $k' -\kappa$ gauge qubits unfixed.
    Following the construction above we find that for each of these unfixed gauge qubits the rank of $S$ is increased by 1 when following the construction above.

    Taken together, we find that the procedure above increases the rank of $S$ by $2\kappa + k -\kappa + k'-\kappa = k + k'$.
    This leads to a full rank of $S$ of $n-k+m-k + k + k' = n+m$ which is the maximal rank and hence completes our proof.

\end{proof}

This leaves us with all the necessary ingredients to prove the main proposition of this section.

\begin{proof}[Proof of Prop.\,\ref{app:prop:clifford}]
Given Lem.\,\ref{app:lem:stab-bipartition_isomorphism} and Lem.\,\ref{app:lem:composition} the theorem follows by induction.
Any RGB TN can be obtained by a sequence of contractions of the form in Lem.\,\ref{app:lem:composition} and with that admits a stabilizer group of maximal rank.
By Lem.\,\ref{app:lem:stab-bipartition_isomorphism}, this leads to the isomorphism in Prop.\,\ref{app:prop:clifford}.
\end{proof}

In the proof of Lem.\,\ref{app:lem:composition} we used an auxiliary subsystem code to prove that the stabilizer group constructed had maximal rank.
It is defined as being generated by the stabilizer groups that determine the codes on which $t$ and $t'$ act on like an isomorphism.
We found that all non-trivial logical operators of the subsystem code survive the composition.
They determine the size of the logical Pauli group on which the contracted network acts isomorphically.
The other logicals that survive the composition are gauge logicals of the subsystem code and correspond to logical qubits of either $t$ or $t'$ and are ``measured'' within $t'$ or $t$, respectively.

\subsubsection{Basics of Pauli flows}\label{app:sec:flow_intro}

Given this fact, we proceed in defining the concept of a Pauli flow based on the projective symmetries of the RGB tensors and associated colorings defined in the main text and here, and elaborate on their physical meaning.
It is not surprising that they fully characterize the network since its logical action in Prop.\,\ref{app:prop:clifford} was obtained from the symmetries leading to Pauli flows.
Additionally to the logical action, we first show that detector flows give rise to a linear code in the space of measurement outcomes.
Based on this ``redundancy'' within the network, we construct a group homomorphism between (cosets of) Pauli flows and a stabilizer group on the Hilbert space associated to the in- and output legs that captures the logical isomorphism applied by the circuit.

\begin{definition}[Highlights and Pauli flows]
    Let $T$ be an RGB tensor network with a set of tensors $V$ and edges $E$ connecting them.
    Each of the red, green or blue tensors in the network might carry a sign, labeled $s$, that has a \textit{value} $\hat{s}\in\bZ_2=\{0,1\}$.
    We denote the set of all signs in $T$ with $M(T)$.
    Given any any subset $m\subseteq M(T)$ we can define its value by adding the values of its elements,
    \begin{align}
        \widehat{m} := \sum_{j\in m}\hat{j} \mod 2.
    \end{align}
    We define a Pauli \textit{highlight} to be a coloring of edges and signs where each edge and sign is highlighted in one of four colors $\{$trivial, red, green, blue$\}$.
    A \textit{Pauli flow} $F(T)$ is a highlight that for each tensor individually is \emph{valid}, i.e., either it is trivial (all edges connected to $v$ as well as the sign highlighted trivially) or is of the form of one of the  highlights in Eqs.\,\eqref{eq:blue_pauliflows} and \eqref{eq:spider_flow}. 
\end{definition}

\begin{definition}[Detector flows]
    We call a Pauli flow in which only internal edges in an RGB tensor network are highlighted non-trivially a \textit{detector flow}. We denote the set of detector flows of a given RGB TN by $D(T)$.
\end{definition}
For clarity, here we review in a more formal setting the necessary definitions made informally in the main text, and derive the main theorem from these.

Note that we distinguish between the sign of a tensor as an abstract object and its value.
In some cases these values might be fixed (e.g. when applying a Pauli operator on an edge in the network) or obtained in an experiment (e.g. for measurement outcomes).
In most cases one can think of a sign as being analogous to its value (see Def.\,\ref{def:signspace}) but this distinction simplifies upcoming definitions.

\begin{lemma}
    The set of all Pauli flows $F(T)$ of an RGB tensor network $T$ 
    forms an Abelian group under the group operation inherited from the group operation of the highlights on each edge, defined in Eq.\,\eqref{eq:adding_highlights}.
    We denote this group operation by $\oplus_F$.
    The set of detector flows $D(T)$ forms a subgroup of $F(T)$.
\end{lemma}

Since the group operation $\oplus_F$ is Abelian, we refer to it as an addition.

\begin{proof}
First, we define an ``addition'' on highlights by lifting the addition defined for each edge individually to highlights on the network by applying the rules of Eq.\,\eqref{eq:adding_highlights} to each edge individually.
Additionally, we impose the same rules of combining highlights of signs.
Clearly, this fulfills all the conditions of a group multiplication where the trivial highlight on all edges and signs plays the role of the (unique) unit and the group operation is by definition commutative and any flow is its own inverse.
This renders the Pauli flows a group isomorphic to $\bZ_2^{\times d_F}$ for some integer $d_F$.

Given this notion of summing with the obvious choice of identity, it remains to be shown that Pauli flows are closed under that operation.
Looking at Eqs.\,\eqref{eq:blue_pauliflows} and\eqref{eq:spider_flow} we can see that the sum of two valid highlights again is a valid highlight and hence the product of two Pauli flows $f_1$ and $f_2$ is again a Pauli flow $f_1\oplus_F f_2$.

If two flows highlight an edge $e$ trivially, then their sum clearly also highlights $e$ trivially.
From this it follows directly that the set of detector flows is closed under $\oplus_F$, i.e., a subgroup of 
$F(T)$.
\end{proof}

The abstract group formed by $F(T)$ can be viewed as a subgroup of $\bZ_2^{\times 2\abs{E}}$, where $\abs{E}$ is the number of edges in $T$.
Addtionally, we can include the highlight of the signs carried by the tensors into the definition of Pauli flows by incorporating additional $\bZ_2$ variables for each tensor that can be charged with respect to a flow.

\begin{definition}[Pauli looping of a flow]
For any Pauli flow $f\in F(T)$ on an RGB tensor network $T$ we associate the following two-step transformation to the network
\begin{enumerate}
    \item Apply the symmetries associated to the flow, for each tensor in the network.
    \item Combine the remaining Pauli tensors on each contacted leg to a global phase.
\end{enumerate}
We call any such transformation \textit{(Pauli) looping} of the flow $f$.
\end{definition}
The first step can be understood as iteratively moving and transforming Paui operators through the network as we show for an exemplary detector flow in a TN representing a repetition code circuit in Fig.~\ref{fig:detector_proof}.
Note that the last step can always be performed by the definition of the Pauli flow.

\begin{definition}[Sign of a flow]\label{app:def:chargedsigns}
    Let $f\in F(T)$.
    Applying the looping of $f$ results in a network that differs by the starting network only by a $\pm1$ prefactor $c_f$ and Pauli tensors on the open legs.
    Consider a fixed orientation of the green tensors that remain on the open legs, e.g. pointing into the network.
    For this orientation, we define the \textit{sign of the flow $f$} as $\sgn(f) = (1+c_f)/2$.
    It takes values in $\{0,1\}$.
\end{definition}
For the purposes discussed in this paper the choice of fixed ``reference'' orientation on the open legs does not matter, as long as it is fixed.

\subsubsection{Detector flows define a classical linear code}\label{app:detectors}

In this section, we prove that the group of detector flows gives rise to a classical linear code in the space of all possible sign configurations of the network.
This proof works for any RGB TN, including those not necessarily coming directly from a circuit\footnote{Note, however, that any RGB TN is equivalent to a network that can directly be translated back to a circuit using the rewrite rules discussed in Sec.~\ref{app:sec:rewriterules}, potentially acting on additional qubits.}, but for example from a measurement-based protocol, where the tensor network represents the amplitudes associated to measurement outcomes obtained from single-qubit measurements on a ``resource state'' represented as an RGB TN.

\begin{lemma}\label{app:lem:detector}
Let $d$ be a detector flow.
It holds that $\sgn(d)=0$.
\end{lemma}
\begin{proof}
    By definition of a detector flow, it highlights all open legs trivially.
    As such, there is no choice for a reference orientation on the open legs when definting its sign $\sgn(d)$.
    By defintion of $\sgn(d)$, the network before and after applying the Pauli looping of $d$ is exactly the same, up to a prefactor $(-1)^{\sgn(d)}$.
    This prefactor has to equal $1$ for the network to give non-zero values, i.e. $\sgn(d)=0$.
    If this is not fulfilled the network evaluates to zero which says that the probability of observing a particular sign configuration of elementary tensors that leads to $\sgn(d)=1$ is zero.
    
\end{proof}

\begin{figure*}
    \centering
    \includegraphics[width=0.8\linewidth]{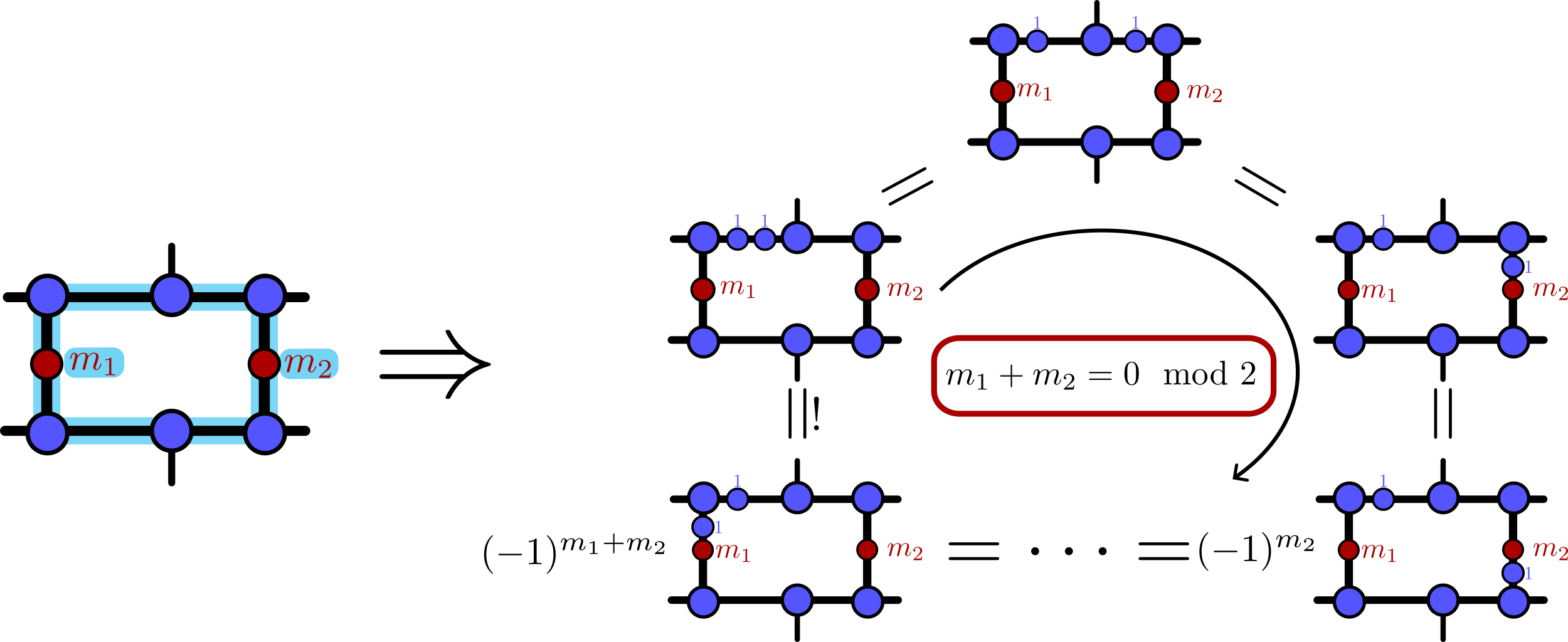}
    \caption{We show an exemplary detector flow associated to two repeated $ZZ$ measurements as happening for example in a repetition code circuit.
    Any such detector flow defines a constraint on the values of the signs charged with respect to that flow.
    In the case shown, $m_1+m_2=0\mod 2$ which can be seen explicitly by constructing a sequence of transformations that leave the contracted network invariant (following in the arrow in the figure).
    We first add two (blue) $s=1$ tensors on one edge of the (blue) highlight and iteratively push one of the tensor through the network using the symmetries of the individual tensors.
    For the shown detector flow, the network acquires a phase of $(-1)^{m_1+m_2}$ through that sequence which has to equal $1$ if the network gives a non-zero value.}
    \label{fig:detector_proof}
\end{figure*}

We illustrate the proof for a detector flow of a repetition code circuit in Fig.~\ref{fig:detector_proof}.

\begin{lemma}\label{app:lem:sgn_homomorphism}
    Let $f_1$ and $f_2$ be two Pauli flows of the same RGB TN.
    In that case it holds that $\sgn(f_1)\oplus \sgn(f_2) = \sgn(f_1\oplus_F f_2)$, where $\oplus_F$ denotes the addition of Pauli flows and $\oplus$ addition modulo 2.
    Phrased differently, $\sgn$ is group homomorphism. 
\end{lemma}

\begin{definition}[Sign space]\label{def:signspace}
    Let $M(T)$ be the set of all signs of an RGB tensor network.
    We define the \textit{sign space} $\widehat{M}_T$ as the formal $\bZ_2$ span over $M(T)$, span$_{\bZ_2}(M(T))$.
    This equips $\widehat{M}_T$ with a canonical basis in which each basis element corresponds to a specific sign label.
    We denote the basis element associated to sign $s\in M(T)$ by $\vb{e}_s\in\widehat{M}_T$.

    Each Pauli flow $f$ defines a projection $p_f:\widehat{M}_T\to \widehat{M}_T$,
    \begin{align}
        \vb{m} = \sum_{s\in M} m_s \vb{e}_s \mapsto p_f(\vb{m}) = \sum_{s\in s(f)}m_s\vb{e}_s.
    \end{align}

    We call the image of $p_f$ the \textit{subspace charged with respect to $f$}.
\end{definition}

\begin{thm}\label{app:thm_detector_linearcode}
Let $D(T)$ be the group of detector flows of an RGB tensor network with a generating set of flows $G_D$.
It defines a linear code $C_D\subseteq \widehat{M}_T$, defined as the kernel of a parity check matrix $H_D$ whose rank (as a linear operator) equals the rank of $D(T)$ (as a finite Abelian group).
\end{thm}

Physically, the space $C_D$ corresponds to the space spanned by all sign values that fulfil Lemma~\ref{app:lem:detector} for every detector flow of the network.
This space is equivalent to the \textit{outcome code} introduced in Ref.~\cite{delfosse2023spacetime} up to some subtleties in how the space of all signs is set up.
Specifically, we include all signs in the network, not only measurement outcomes.
This includes Pauli operators that would potentially flip some stabilizers and thereby render the resulting outcome code (in the language of Ref.~\cite{delfosse2023spacetime}) an \textit{affine} linear code.
Since we include the signs attached to tensors that represent Pauli operators, we obtain a linear code without any further modifications.

We use the code $C_D$ to infer errors occurring while performing a QEC protocol.
Specifically, we associate a vector in $\widehat{M}_T$ to the recorded signs during the QEC protocol and obtain a syndrome by applying $H_D$ to it.

\begin{proof}
Let $d\in D$ be a detector flow and define the vector $\vb{h}_d = \sum_{s\in s(d)}\vb{e}_s\in \Im(p_d)$ lying in the subspace charged with respect to $d$.
In fact, we can define a (linear) subspace $C_d=\ker(\vb{h}_d^T)$, where $\vb{h}_d^T: \widehat{M}\to\bZ_2$ is understood as a linear map that acts via taking the dot product with the vector $\vb{h}_d$.
The kernel of $C_d$ is the subspace spanned by sign values that fulfill $\sgn(d)=0$.

Now consider two independent detector flows $d_1$ and $d_2$ with the associated subspaces $C_{d_1}$ and $C_{d_2}$.
By linearity of the dot product and transposition and Lem.\,\ref{app:lem:sgn_homomorphism} it follows that
\begin{align}
    C_{d_1}\cap C_{d_2} = C_{d_1}\cap C_{d_2} \cap C_{d_1\oplus_F d_2}.
\end{align}
Hence, it suffices to consider a generating set of detector flows, $G_D$, when defining the space
\begin{align}
    C_D = \bigcap_{d\in D(T)} C_d = \bigcap_{g\in G_D} C_g = \ker(H_D),
\end{align}
where $H_D$ is the matrix whose rows are $\{\vb{h}_g^T\;|\; g\in G_D\}$.
\end{proof}

\subsubsection{The logical isomorphism from Pauli flows}\label{app:sec:flows_logical_iso}

Given an RGB tensor network $T$ and its Pauli flow group $F(T)$ as well as its detector subgroup $D(T)$, we can construct a quotient group of $F(T)/D(T)\simeq\cS$ and an isomorphism to a stabilizer group on the in- and output legs of $T$ that we associate to the network.
This stabilizer group is exactly the stabilizer group that captures the logical isomorphism applied by $T$ in Prop.\,\ref{app:prop:clifford}.

Consider an RGB tensor network with $n+m$ open edges. We interpret the first $n$ legs as ``input'' legs and the remaining $m$ legs as ``output'' legs.
In the following, we define a group homomorphism from $F(T)$ onto $\hat{\cP}_{n+m} = \cP_{n+m}/\langle i\rangle$, the \textit{quasi-Pauli group} on the vector spaces associated to the open legs in the network.
\begin{definition}
    Let $f\in F(T)$ be a flow of an RGB tensor network $T$.
    Consider its restriction onto the in- and output legs and let $\text{col}_f(i)$ be the color of the highlight of edge $i\in\{1,2,...,n+m\}$.
    Let $\rho:\{\text{trivial, red, green, blue}\}\to\hat{\cP}_1$ be the inverse map of the map defined in Eq.\,\eqref{eq:pauliflowmap}.
    We define $\hat{p}: F(T)\to \hat{\cP}_{n+m}$
    \begin{align}
        \hat{p}(f) = \bigotimes_{i=1}^{n+m}\rho(\text{col}_f(i)),
    \end{align}
    to map onto Pauli operators modulo phases on the in- and output spaces, where the labeling of the open edges is chosen that the first $n$ tensor factors correspond to the ``input'' space and the remaining $m$ tensor factors to the ``output'' space.
\end{definition}

\begin{lemma}\label{app:lem:hatp_homomorphism}
$\hat{p}$ is a group homomorphism.
\end{lemma}

\begin{proof}
First, we note that $F(T)$, as a group, is isomorphic to a subgroup of $\hat{\cP}_{E}$, the quasi Pauli group on all edges of $T$, contracted or open, which itself is isomorphic to $\bZ_2^{\times 2 \abs{E}}$.

Having established the isomorphism between a highlight and a quasi Pauli group, it remains to be shown that $\hat{p}{(f_1)}\hat{p}({f_2}) = \hat{p}(f_1\oplus_{F}f_2)$.
Noting that the group law on the highlights directly resembles how their images under $\rho$ multiply, i.e., $\rho$ is a group homomorphism onto $\hat{\cP}_1$ and $\hat{p}(f)$ is a simple tensor product over images of $\rho$ we find that this is indeed the case.
\end{proof}

\begin{cor}
    $\ker(\hat{p}) = D(T)$
\end{cor}
\begin{proof}
    By definition, elements in $D(T)$ have trivial highlights on the in- and output edges.
    This shows that $D(T)\subseteq\ker(\hat{p})$.
    Moreover, $\rho$ is injective, i.e., has trivial kernel.
    Hence, $\ker(\hat{p})\subseteq D(T)$.
\end{proof}

As $D$ is the kernel of $\hat{p}$, we can unambiguously define $\hat{p}$ to act on $F(T)/D(T)$ by taking an arbitrary representative in $fD(T)$ as input into $\hat{p}$.
Since $D(T)$ is a normal subgroup ($F(T)$ is Abelian), the induced map $\hat{p}_D: F(T)/D(T) \to \hat{\cP}_{n+m}$ is still a group homomorphism.
Moreover, it is clearly injective.
\begin{definition}
    We define $\iota: \hat{\cP} \to \cP:$
    \begin{align}
    \begin{split}
        [\mathds{1}] \mapsto \mathds{1},\;
        [X] \mapsto X,\;
        [Y] \mapsto Y,\;
        [Z] \mapsto Z,
    \end{split}
    \end{align}
    that maps each coset to its representative, with $+1$ sign.
    This map can be straightforwardly defined on $\hat{\cP}_{n+m}$ by taking its tensor product $\iota^{\otimes (n+m)}$
    We will sometimes omit the superscript if the domain and with that the superscript is clear from context.
\end{definition}
The map $\iota$ allows us to map images from $\hat{p}_D$ to actual Pauli operators in in- and output spaces.
It is defined such that the image in the Pauli group is well-behaved, in the sense of the following key lemma.
\begin{lemma}\label{app:lem:ptilde_homomorphism}
    $\Tilde{p} = \iota\circ \hat{p}_D$ is an injective group homomorphism.
\end{lemma}
\begin{proof}
Since $\hat{p}_D$ and $\iota$ are both injective, $\Tilde{p}$ is injective.
It remains to be shown that $\Tilde{p}$ is a group homomorphism.
Specifically, we need to show that two Pauli operators in $\Im(\Tilde{p})$ commute since it could only fail to respect the (Abelian) group multiplication up to signs that were modded out in the definition of $\hat{p}$.

We can see this directly from Prop.\,\ref{app:prop:clifford}.
In fact, as common in the stabilizer formalism, any Pauli $^\ast$-homomorphism $\cP_n\to\cP_m$ can be understood as a group homomorphism from cosets with respect to an ``input'' stabilizer group to cosets with respect to an ``output'' stabilizer group.
As such, it can be understood as a collection of pairs $(p_i,p_j)$, for some $p_i\in \cP_n,p_j\in\cP_m$, indicating that (a coset containing) $p_i$ gets mapped to (a coset containing) $p_j$.
Importantly, not every operator in $\cP_n$, respectively $\cP_m$, appears in these pairs.
Moreover, there can be many $p_i$'s for the same $p_j$ and many $p_j$s for the same $p_i$ which can be understood as stabilizer equivalences on $\cP_n$, respectively $\cP_m$ induced by some sort of projection (measurement) in the circuit.

Following the construction of $\Tilde{p}$ we find that the image of $\Tilde{p}$ is exactly the minimal set of these pairs of Pauli operators that defines the homomorphism but without any signs.
To show that omitting these signs does not alter the group structure, we need to show that all elements in $\Im(\Tilde{p})$ commute.
To show this, consider two different elements $p_i\otimes p_j, p_i'\otimes p_j' \in \Im(\Tilde{p})$.
By the homomorphism it is guaranteed that $[p_i,p_i'] = [p_j,p_j']$, where $[a,b] = ab -ba$ is the algebra commutator.
Hence, $[p_i\otimes p_j, p_i'\otimes p_j'] = 0$ for all $p_i\otimes p_j, p_i'\otimes p_j'\in \Im(\Tilde{p})$ which completes the proof.
\end{proof}

We have just constructed an isomorphism of (cosets of) flows to the stabilizer groups associated to the RGB tensor network that capture the logical encoding and non-trivial Clifford applied by it, namely $\Im(\Tilde{p})$.
There is still some freedom, however, in choosing the signs of elements in $\Im(\Tilde{p})$ consistently.
On the level of the logical action $\Tilde{p}$ determines the Clifford action up to Paulis since these are exactly the operators that might change the sign of a Pauli operator by conjugation.

We can resolve this ambiguity within the Pauli flow formalism though, by including the signs of the flows explicitly.
Note that this is the first time in this section where we actually need to consider how a given flow extends into the network, not only its restriction on the open legs.

\begin{thm}
    The map $p: F(T)/D(T)\to\cP_{n+m}$ defined as
    \begin{align}
        fD(T) \mapsto (-1)^{\sgn(f)}\Tilde{p}(f),
    \end{align}
    where $f$ is an arbitrary element in $fD(T)$, is well-defined and $\cS = \Im(p)$ is a stabilizer group.
\end{thm}
\begin{proof}
$p$ only differs from $\Tilde{p}$ (which is well-defined) by adding a factor $(-1)^{\sgn(f)}$.
By Lemma\,\ref{app:lem:detector}, this factor is independent of the choice of representative in $fD$ and hence $p$ is well-defined.

Lemma\,\ref{app:lem:ptilde_homomorphism} says that $\Im(\Tilde{p})$ is a stabilizer group.
By Lemma \ref{app:lem:sgn_homomorphism}, The added factor $(-1)^{\sgn(f)}$ also multiplies like the elements in $F(T)$, following the multiplication of the flows.
Combined, the additional sign factor doesn't alter the group structure and by Lemma\,\ref{app:lem:detector} the identity coset is (the only coset that) gets mapped to the identity operator, $p(D(T))=\Tilde{p}(D(T)) = \mathds{1}$, rendering it a stabilizer group.
In particular, $-\mathds{1}\neq\Im(p)$.
\end{proof}

Often, we might abuse the notation slightly and use $p$ for a map that directly maps flows, rather than cosets of flows, to Pauli operators. 
Since the step of identifying a given flow with a coset is unambiguous.

\begin{definition}
As a subgroup of $\cP_{n+m} = \cP_n\otimes\cP_m$, we can decompose $\cS=\Im(p)$ into three parts
\begin{align}
    \cS = \cS_{\text{in}}\times \cS_{\text{out}}\times\cS_{L},
\end{align}
where $\cS_{\text{in}}$ is the maximal subgroup only generated by Paulis of the form $P\otimes\mathds{1}_m$ (only acting on the input space), $\cS_{\text{out}}$ the maximal subgroup generated by Paulis of the form $\mathds{1}_n\otimes P'$ and $\cS_\ell$ the subgroup that cannot be generated by operators that act trivially on one side of the tensor product.

We call $\cS_{\text{in}}$ the \textit{input stabilizer group}, $\cS_{\text{out}}$ the \textit{output stabilizer group} and $\cS_{\ell}$ the \textit{logical stabilizer group}.
Analogously, we call their preimages under $p$ \textit{in-/output stabilizer flows} and \textit{logical flows}.
\end{definition}

In fact, $\cS$ is exactly the stabilizer group obtained when following the procedure described in Sec.~\ref{app:sec:RGB_log_iso} based on the symmetries of the elementary building block of an RGB TN.
As constructed, Pauli flows capture exactly these symmetries and the signs of the flows track how Pauli operators acquire a sign when propagating through the network.
    We can also track the composition described in the proof of Lem.\,\ref{app:lem:composition} graphically by matching up Pauli flows.
This makes this tool very useful when analyzing arbitrary Clifford protocols.

\subsection{Comments on Pauli flows when applying rewrite rules to RGB network}\label{app:sec:flows_rewrite}
We have seen that Pauli flows capture the full action of a given RGB network and give rise to the error-correcting capabilities of the associated protocol via its detector flows.
Since a given tensor network can be used to construct a whole equivalence class of concrete protocols implementing the same logical action via the rewrite rules discussed in Sec.~\ref{app:sec:rewriterules}, it is natural to ask how Pauli flows change under this equivalence relation.
We will see that the flows of two equivalent networks $t$ and $t'$ are in a sense isomorphic: Given $t$ and its flows, the flows of $t'$ are uniquely determined by the sequence of rewrites we need to map $t$ to $t'$.

To understand this more concretely, let us look at the individual rewrite rules.
In fact, we can use the stabilizer group of each individual constituent to argue about how the flows change.
There are different types of rewrite rules.
The simplest one being the splitting rules presented in Eqs.\,\eqref{app:eq:blue_split} and \eqref{app:eq:redgreen_split}.
For these it is easy to see that the flows after the splitting operation are uniquely determined by the flows before.
For example,
\begin{align}
    \raisebox{-0.4\height}{\includegraphics[width=0.275\linewidth]{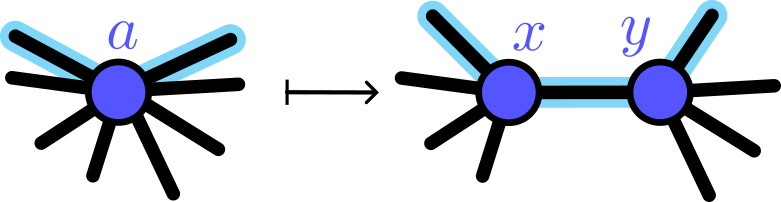}} \qq{and} \raisebox{-0.4\height}{\includegraphics[width=0.275\linewidth]{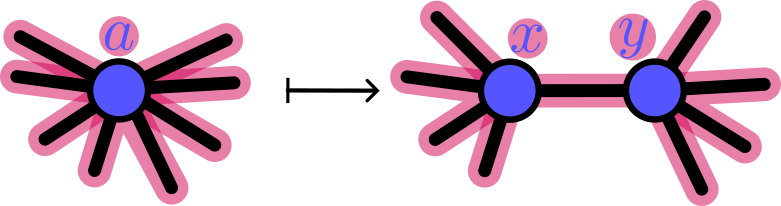}},
\end{align}
where $x+y = a \mod 2$.
Similar relations are obtained when splitting/fusing red and green tensors.
Note that the number of signs charged with respect to a given flow can change when applying the rewrite rules.
As a consequence, the isomorphism class of the linear code obtained from the detector flows (see Thm.\,\ref{app:thm_detector_linearcode}) is not an invariant under rewritings.
This is the first indication that QEC performance can differ within one family of equivalent protocols.
One should be aware, though, that the performance of a protocol highly depends on the error model and less about the details of the classical code on the signs in the network.
Here we find a 1-1 correspondence of flows before and after the local rewriting.
The same applies when cancelling loops between differently colored tensors, e.g.,
\begin{align}
    \raisebox{-0.4\height}{\includegraphics[width=0.125\linewidth]{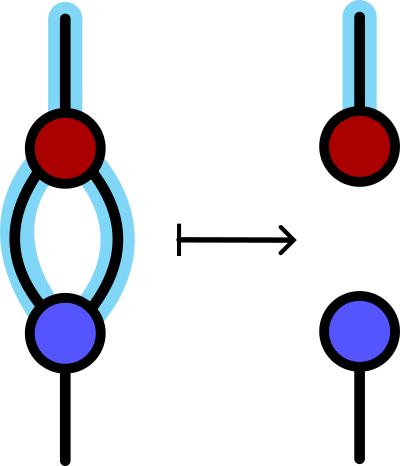}} \qq{and} \raisebox{-0.4\height}{\includegraphics[width=0.125\linewidth]{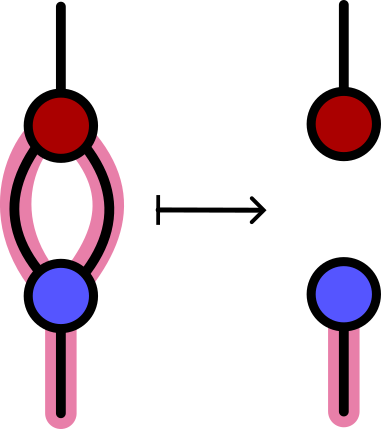}}
\end{align}
and when applying the generalized bialgebra rule, e.g.
\begin{align}
    \raisebox{-0.4\height}{\includegraphics[width=0.225\linewidth]{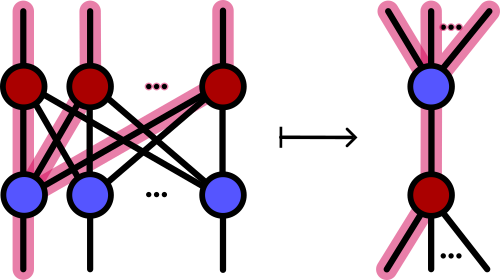}} \qq{and} \raisebox{-0.4\height}{\includegraphics[width=0.225\linewidth]{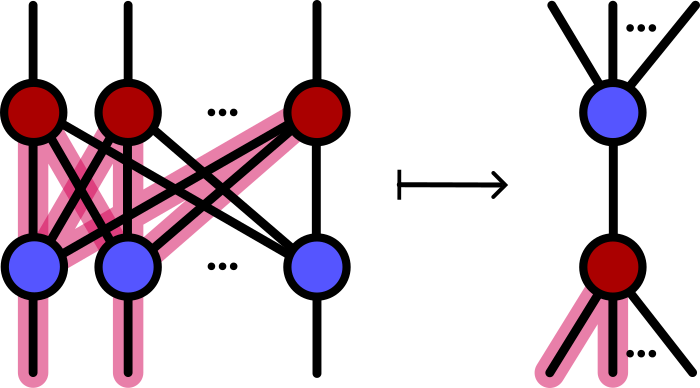}}.
\end{align}
The blue flows are mapped analogously by symmetry of the diagram.

For the rules discussed so far there is a 1-1 mapping between the flows before and after rewriting which makes the statement ``flows are preserved'' hold exactly.
There is a remaining rule that is important to include, namely the rule that allows one to remove loops among tensors of the same color.
Each of these loops can carry a detector flow which has to be added when ``spawning'' such a loop or removed when cancelling such a loop.
Given the explicit rewrite, we can still uniquely determine the flows after the transformation. 
Namely, we add or remove one generator from a set of generating flows,
\begin{align}
    \raisebox{-0.4\height}{\includegraphics[width=0.15\linewidth]{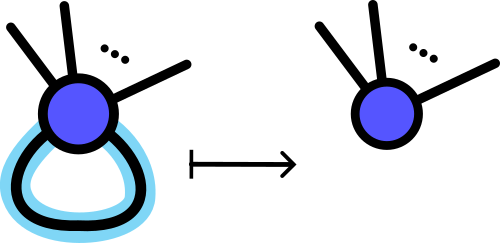}} \qq{and} \raisebox{-0.4\height}{\includegraphics[width=0.15\linewidth]{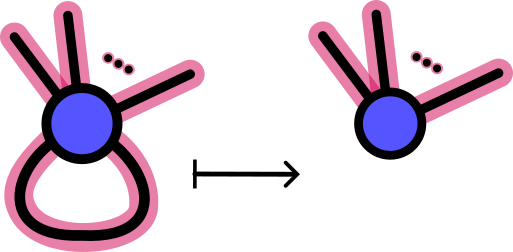}}.
\end{align}
This holds analogously for the red and the green tensor but for the flows of a different color.
Since the resulting map from the group of all flows before to the flows after the rewrite is invertible (even though the flows are not mapped onto each other 1-1), we say that the flows give rise to an invariant under the rewrite rules.
We leave a more rigorous mathematical study of the mathematical structure of that invariant to future work.
Since adding a loop is the only way how to increase the rank of the group of flows $F(T)$ this is the rewrite rule that adds non-trivial constraints (detectors) on the signs of the network.

\section{Dynamical boundaries in the XYZ ruby code}\label{app:boundaries}

\begin{figure}
    \centering
    \includegraphics[width=\linewidth]{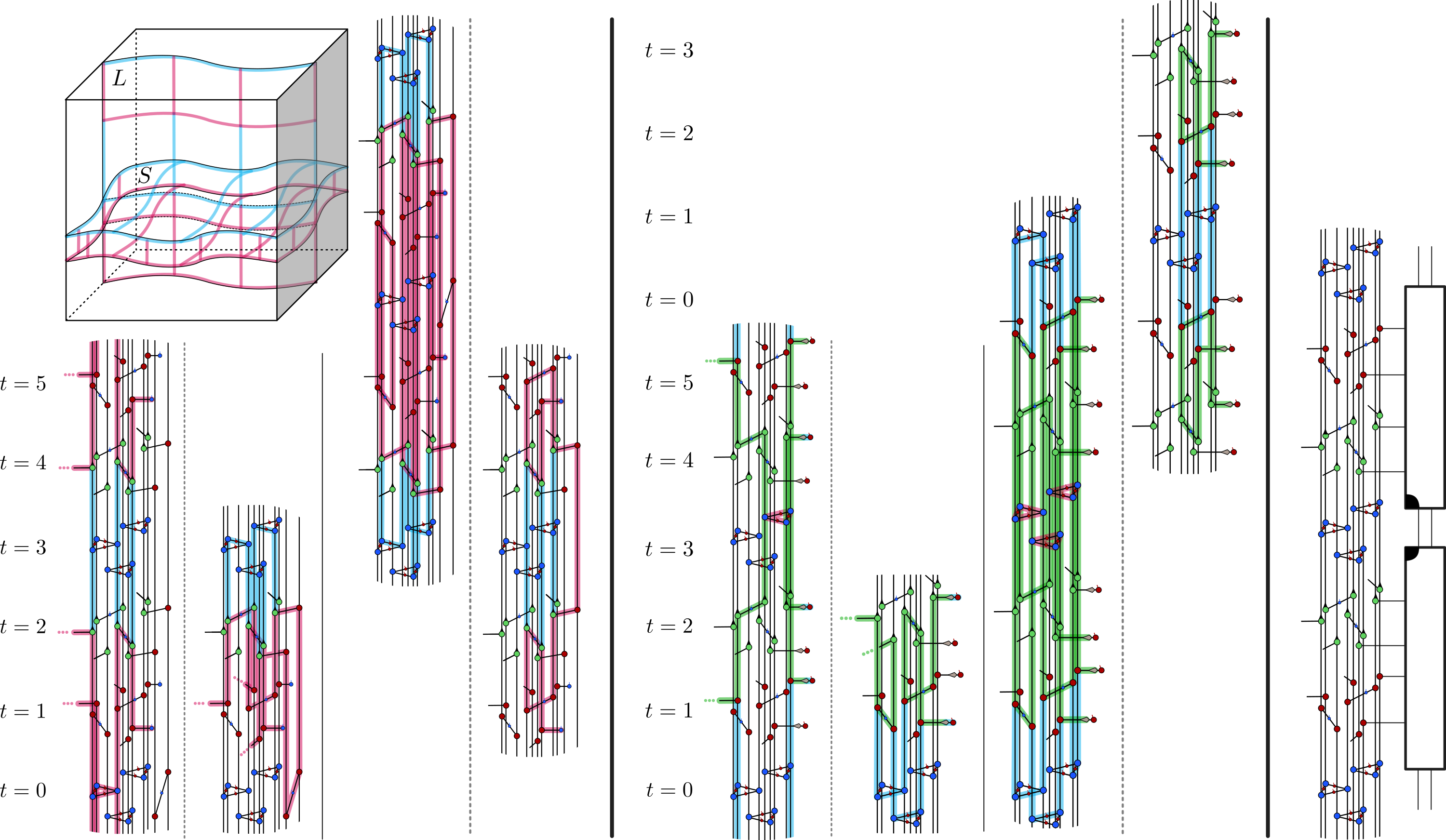}
    \caption{Three dynamical boundaries for the XYZ ruby code in the rewinding schedule (cf. Sec.~\ref{sec:rubycode_definition}). The timesteps are expressed modulo $6$. 
    We construct the boundaries by cutting the network along a plane in spacetime and terminating the open legs by attaching a specific two-dimensional tensor network.
    The stabilizers of the attached tensor network determine which flows are preserved in that procedure.
    There are choices for which the preserved flows give rise to a topological boundary.
    Here we show three inequivalent ways to terminate the boundary topologically, two explicitly resolved in terms of an RGB TN and the third one implicitly via an elementary cell of the boundary tensor network, which is defined by its stabilizer flows in Eq.\,\eqref{app:eq:bdry3_PEPsstabilizer}.
    At each boundary a different set of non-local flows can end, illustrated by $L$ and $S$ on the top left.
    Additionally, the local (detector) flows that deform them in the bulk can end at the boundary.
    For example, for the boundary on the left we show representatives of two inequivalent logical flows (probed in memory and stability experiments) that end at the boundary.
    They are both deformed by the same set of local (detector) flows, shown next to them.
    Analogously, the other boundaries shown allow for inequivalent logical flows to end.
    Importantly, the logical operators associated to the non-local flows that can end at two different boundaries of the three above anti-commute.
    In fact, the boundaries correspond to the three Pauli boundaries of the underlying color code phase and can be used to define a code on a triangle encoding a single logical qubit, as discussed in Ref.~\ref{sec:planar}.
    The color boundaries can be constructed similarly but one might want to consider a different cut to simplify the construction.
    }
    \label{fig:dynamical_pauli_bdrys}
\end{figure}

In this section, we present three distinct boundaries to the XYZ ruby code with the rewinding schedule.
Each boundary is given in terms of its representation as an RGB tensor network and is obtained from cutting the bulk network along a plane that is parallel to the time direction.
This cut can be represented on the spatial lattice of qubits.
In Fig.~\ref{fig:dynamical_pauli_bdrys}, we present boundaries defined along the following cut
\begin{align}
    \raisebox{-0.4\height}{\includegraphics[width=0.3\linewidth]{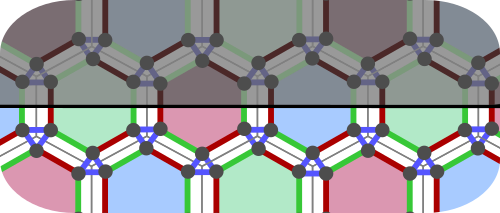}},
\end{align}
where we think of disregarding the network associated to the half-space depicted in gray, above the black line.
After cutting, we obtain a (three-dimensional) network with open legs at the boundary.
The boundaries along this cut are defined by a specific two-dimensional stabilizer tensor network with which we terminate the open legs along the cut.
The stabilizer group of the network with which we terminate the open legs determines the Pauli flows close to the boundary (cf. Lem.\,\ref{app:lem:composition}) and with that the QEC properties, such as the emerging logical flows and the fault distance.

Given the (topological) algebraic data associated to the non-local flows, respectively the logical operators they represent along the spacetime boundary, we can impose conditions on the stabilizer state with which we terminate the boundary.
In Fig.~\ref{fig:dynamical_pauli_bdrys}, we present three distinct boundaries associated to the Pauli boundaries of the underlying color code phase.
From the perspective of the ISGs the boundary type changes according to the discussion in Sec.~\ref{sec:planar}.
In spacetime, however, each boundary should be regarded as a topological boundary of a unique type.
This can be seen by looking at the non-local flows in the system with boundaries.
We find that when introducing a boundary only a subset of the non-local flows survive.
In particular, the non-local flows that are preserved along the boundary are associated to a Lagrangian subgroup of the underlying color code anyon theory.
As a consequence the non-local flows admit a topological symmetry in the bulk of the network, i.e., they can be freely deformed by a subgroup of local detector flows.
The detector flows that deform the non-local flows close to the boundaries are also preserved at the boundary in our construction as can be seen in Fig.~\ref{fig:dynamical_pauli_bdrys}.

Since the microscopic realization and associated circuits have to be constructed separately using rewrite rules of the RGB TN representing the boundary in spacetime, we only show an explicit realization of the boundaries for two of the three types.
The third boundary is defined by the elementary tensor from which we construct the two-dimensional TN with which we terminate the boundary.
It is designed such that the following holds:
If we interpret the non-local flows that can end at the first boundary as $X$-type logicals and the ones that can end at the second boundary as $Z$-type logicals then the third boundary allows their product, $Y$-like logicals, to end.
The boundary itself is defined by an 8-legged tensor with the following stabilizer flows:
\begin{align}\label{app:eq:bdry3_PEPsstabilizer}
\raisebox{-0.4\height}{\includegraphics[width=0.7\linewidth]{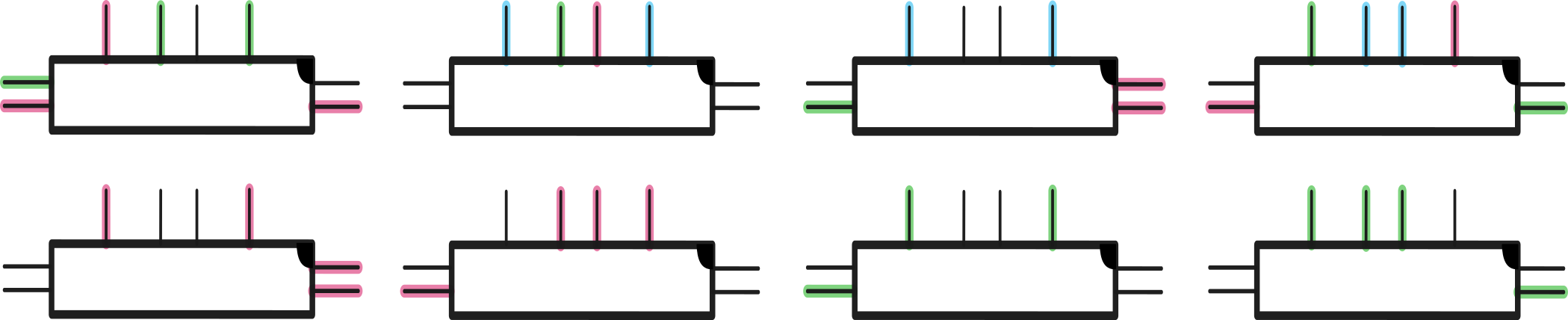}}.
\end{align}
The associated stabilizers (cf. construction in Sec.~\ref{app:sec:flows_logical_iso}) uniquely determine the network up to signs of the stabilizers.

We leave the detailed construction of the other three types of color boundaries and corners between them to future work.

\section{Fault tolerance of elementary circuit building blocks} \label{app:noise}
\begin{figure*}
    \centering
    \includegraphics[]{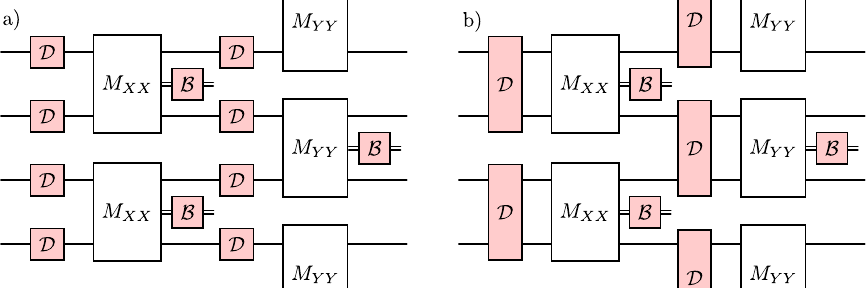}
    \includegraphics[]{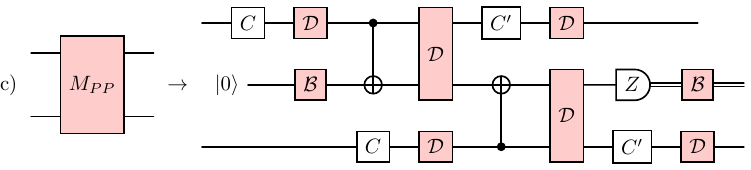}
    \caption{Noise models for direct measurements. a) Phenomenological noise model with single-qubit depolarizing noise in between every noisy two-body measurement. b) EM3 noise model with correlated two-qubit depolarizing noise before the noisy two-body measurement. c) Circuit implementation of the (noisy) two-body Pauli measurement.}
    \label{fig:noise_models}
\end{figure*}

Here, we show fault-tolerance properties of the three noise and circuit models we use. Fig.~\ref{fig:noise_models} shows the \textbf{a)} phenomenological, \textbf{b)} EM3 noise model for direct measurements and \textbf{c)} circuit-level noise model.
We use a uniform noise model with parameter $p$ and depolarizing noise channels
\begin{align}
    \mathcal{D}^{\otimes 1} (\rho) = (1-p) \rho + \frac{p}{3} \sum_{E \in \{X, Y, Z\}} E \rho E
\end{align}
and 
\begin{align}
    \mathcal{D}^{\otimes 2} (\rho) = (1-p) \rho + \frac{p}{15}  \sum_{EE' \in \{I, X, Y, Z\}^{\otimes 2}  \setminus II} EE' \rho E'E.
\end{align}
Qubit initializations and measurements are followed by bit flip channels,
\begin{align}
    \mathcal{B}(\rho) = (1-p) \rho + p X \rho X.
\end{align}
For the phenomenological noise model, any two-qubit error is an event $\mc{O}(p^2)$, whereas for the EM3 noise model, two-qubit Pauli errors are already first order $\mc{O}(p)$. This explains the halved fault distance of the direct measurement circuit with EM3 noise.

\begin{figure}
    \centering
    \includegraphics[width=0.8\linewidth]{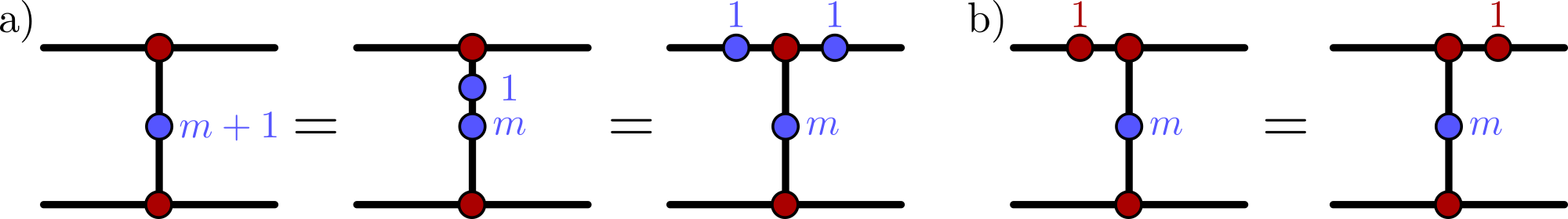}
    \caption{Equivalences of error mechanism to derive the effective noise model of single qubit $X$-, $Y$- and $Z$ faults after $XX$-, $YY$- and $ZZ$-measurements. a) A measurement error, here of an $XX$-measurement, is represented as a flip of the measurement spider sign. Using the rewrite rules\,\eqref{app:eq:blue_split}, we can pull out the flip of the measurement as a two-legged $s=1$ spider. This can be further rewritten to a blue spider just before, and after the measurement. b) Pauli faults in the same basis as the measurement can be pushed through the measurement using the splitting rule for the red spider (Eq.\,\eqref{app:eq:redgreen_split}).}
    \label{fig:noise_model_reduction}
\end{figure}

The implementation of the two-qubit Pauli measurement with an auxiliary qubit, single-qubit rotations and $\CNOT$s keeps detrimental two-qubit faults at order $\mc{O}(p^2)$.
This is due to the fact that all two-qubit errors on the data qubits that are $\mc{O}(p)$ are stabilizer-equivalent to a single-qubit error. These are two-qubit depolarizing errors after the first $\CNOT$, where the Pauli on the auxiliary qubit has part $Z$, i.e., a fault of the form $PZ$ with $P \in \{X,Y,Z\}$. This part can propagate to the lower data qubit, such that (before the single-qubit rotation), the effective Pauli error on the data qubits is $PZ$ with $P \in \{X,Y,Z\}$. Before rotating, any of the circuits has implemented a $ZZ$-measurement, which is then part of the stabilizer group. The effective Pauli error is therefore stabilizer-equivalent to a single-qubit error, $PZ \sim PI$. This shows that the chosen implementation of two-body measurements is fault tolerant.
Fig.~\ref{fig:noise_model_reduction} shows equivalences to derive the effective error model we use for the argumentation on fault-tolerance in the main text.

\section{Belief propagation and postprocessing parameters} \label{app:bp}

\begin{figure}
    \centering
    \includegraphics[width=\linewidth]{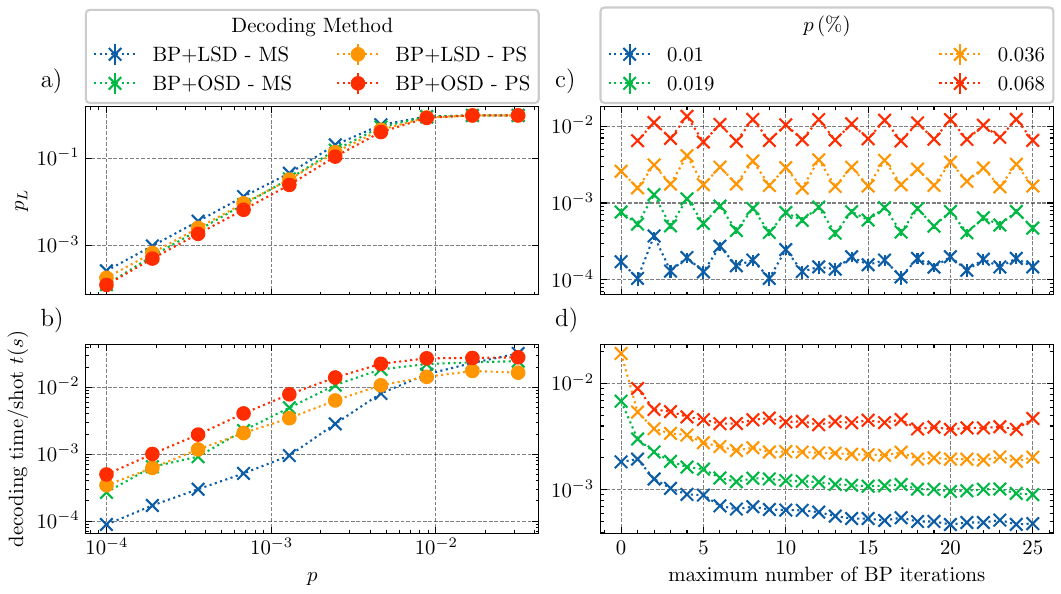}
    \caption{Comparison of different decoding methods for $\mathcal{L}_Y$ memory experiments for the distance $4$ XYZ ruby code using the phenomenological noise model and $4$ rounds. In a), we fix the OSD order to $20$ and the BP Iterations to $19$. In general, BP+OSD with the \emph{product sum} (PS) algorithm gives the lowest logical error rate at the cost of the highest decoding time per shot, shown in b). BP+LSD with the \emph{minimum sum} (MS) algorithm offers almost an order of magnitude faster decoding per shot, at the cost of a lower accuracy. c) Different number of BP iterations show oscillatory behavior in the logical error rate using OSD postprocessing. An odd number of iterations gives lower logical error rates. In d) we show the overall decoding time per shot which decreases with increasing number of BP iterations.}
    \label{fig:bp_params}
\end{figure}

We decode the syndrome by converting the detector error model of the noisy circuits to a detector matrix, a logical matrix and a prior vector, see main text. We use the open source library \texttt{bposd}\,\cite{roffe2022ldpc}. It offers a variety of belief propagation and post-processing methods. As postprocessing methods, we use the well established ordered statistics decoding (OSD)\,\cite{panteleev2021degenerate}, and the recently introduced \emph{localized statistics decoding} (LSD), that offers a better speed-accuracy tradeoff enabling us to perform Monte Carlo sampling for larger distances, at the cost of a higher logical error rate.
For the OSD post-processing and an \emph{OSD order} set to $20$, we observe a similar logical error rate across all available BP methods, with the \emph{product sum} method giving a good trade-off between decoding time and logical error rate, see Fig.~\ref{fig:bp_params} a).
The \emph{maximum number of BP iterations} shows oscillatory behavior: An odd number show significantly lower logical error rates compares to an even number of maximum iterations. 
We do not investigate this behavior any further here and fix the number of 
maximum iterations to $19$ in all of our simulations.

\section{Finite size scaling analysis of thresholds} \label{app:fss}

\begin{figure}
    \centering
    \includegraphics[width=\linewidth]{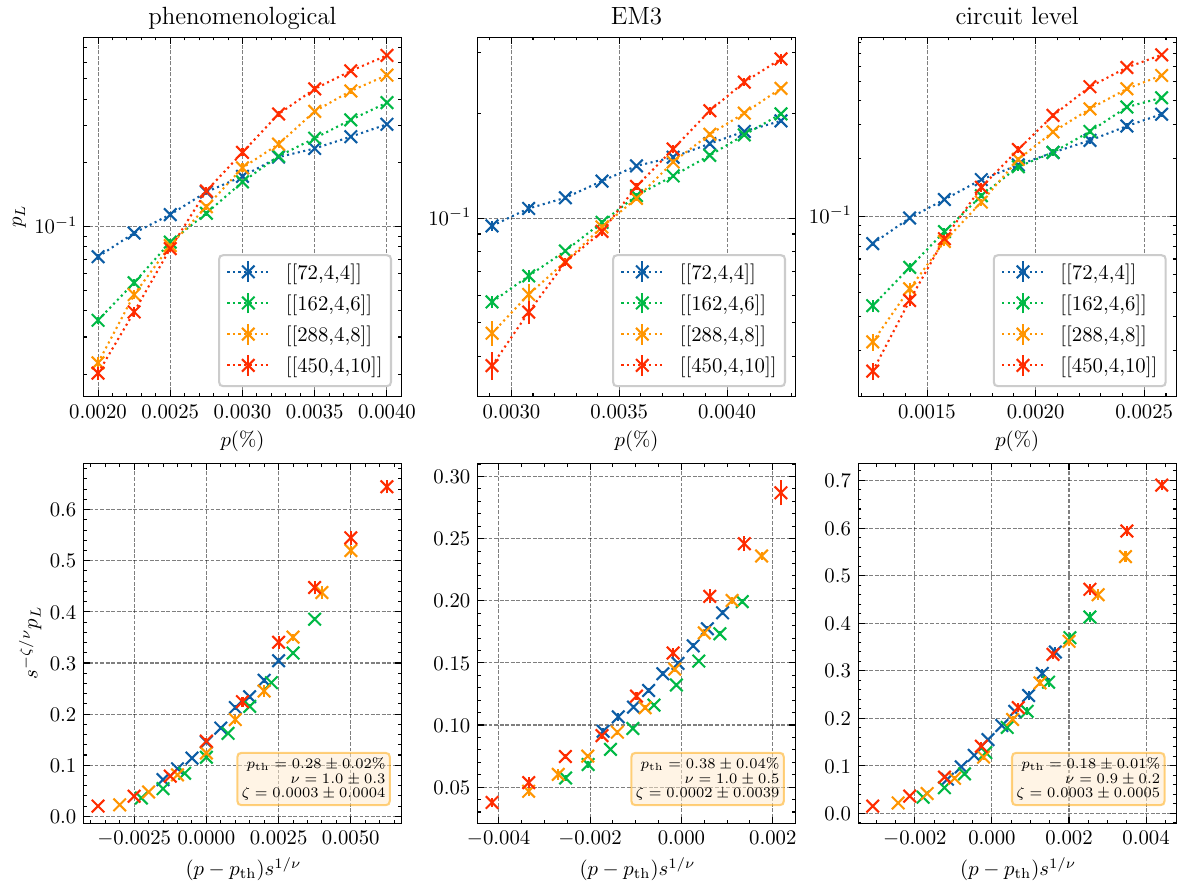}
    \caption{Finite size scaling analysis of thresholds, for $\mathcal{L}_Y$ memory experiments with $d$ cycles and the three circuit/noise model defined in Sec.\ref{sec:circuits}. We make the Ansatz $p_L(p) = s^{\zeta/\nu} f(s^{1/\nu} (p - p_{\mathrm{th}}))$, where $s$ is the linear system size $s = 2d$. The \texttt{pyfssa} package then determines the critical exponents and threshold that gives the best data collapse.}
    \label{fig:plot_th_d_fss}
\end{figure}

We perform finite size scaling analysis on the logical error rate of $\mathcal{L}_Y$ memory experiments using the open source package \texttt{pyfssa}\,\cite{sorge2015pyfssa}. 
We make the Ansatz for a scaling in the form of $p_L(p) = s^{\zeta/\nu} f(s^{1/\nu} (p - p_{\mathrm{th}}))$, where $s$ is the linear system size $s = 2d$ and $f(x)$ a linear dimensionless function. The \texttt{pyfssa} package then determines the critical exponents and threshold that gives the best data collapse.
The results are shown in Fig.~\ref{fig:plot_th_d_fss}. We extract critical exponents of $\approx 1$ thresholds for the phenomenological noise model of 
\begin{align}
    p^{\mathrm{ph}}_{\mathrm{th}} \approx 0.28 \pm 0.02 \%.
\end{align}
For the EM3 noise model we find
\begin{align}
    p^{\mathrm{em3}}_{\mathrm{th}} \approx 0.38 \pm 0.04 \%,
\end{align}
and for the circuit-level noise model
\begin{align}
    p^{\mathrm{cl}}_{\mathrm{th}} \approx 0.18 \pm 0.01 \%.
\end{align}

\NEW{Note that because BP+post-processing is not an MLD decoder, these numbers don't represent fundamental (i.e. optimal) thresholds of the code. 
The collapse is evidently not perfect, which we attribute to the suboptimality of the decoder.
We, however, still use this method to get a reasonable estimate, including uncertainty, for the crossing point of the logical error curves.}

\section{$\cL_X$ experiments}
\begin{figure}
    \centering
    \includegraphics[width=\linewidth]{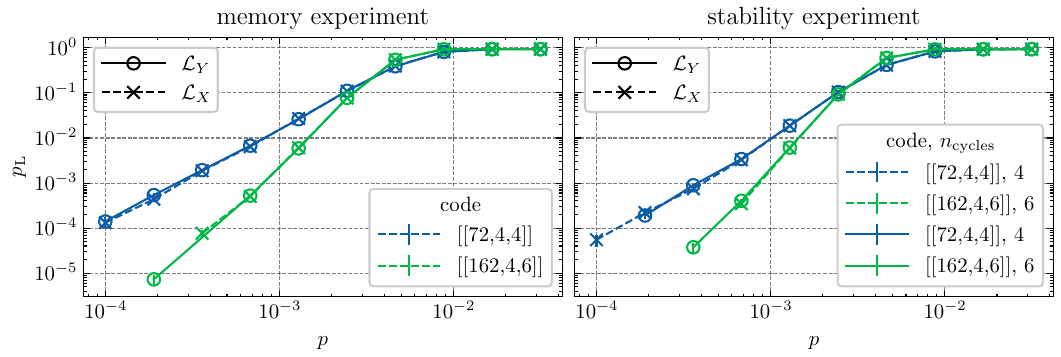}
    \caption{Memory (left) and stability (right) experiments for $\mathcal{L}_X$ observables compared to $\mathcal{L}_Y$ observables shown in the main text in Sec.~\ref{sec:results} show that within error bars, the performance coincides.}
    \label{fig:plot_pL_X_vs_Y}
\end{figure}

In Fig.~\ref{fig:plot_pL_X_vs_Y} we show that the $\mathcal{L}_X$ memory and stability observables performs within error bars the same as the $\mathcal{L}_Y$ observables shown in the main text Sec.~\ref{sec:results}. This is expected due to the symmetry of the code and protocols.

\end{appendix}

\clearpage
\renewcommand*{\bibfont}{\footnotesize} 
\bibliography{references.bib} 

\begin{thebibliography}{98}%
\makeatletter
\providecommand \@ifxundefined [1]{%
 \@ifx{#1\undefined}
}%
\providecommand \@ifnum [1]{%
 \ifnum #1\expandafter \@firstoftwo
 \else \expandafter \@secondoftwo
 \fi
}%
\providecommand \@ifx [1]{%
 \ifx #1\expandafter \@firstoftwo
 \else \expandafter \@secondoftwo
 \fi
}%
\providecommand \natexlab [1]{#1}%
\providecommand \enquote  [1]{``#1''}%
\providecommand \bibnamefont  [1]{#1}%
\providecommand \bibfnamefont [1]{#1}%
\providecommand \citenamefont [1]{#1}%
\providecommand \href@noop [0]{\@secondoftwo}%
\providecommand \href [0]{\begingroup \@sanitize@url \@href}%
\providecommand \@href[1]{\@@startlink{#1}\@@href}%
\providecommand \@@href[1]{\endgroup#1\@@endlink}%
\providecommand \@sanitize@url [0]{\catcode `\\12\catcode `\$12\catcode `\&12\catcode `\#12\catcode `\^12\catcode `\_12\catcode `\%12\relax}%
\providecommand \@@startlink[1]{}%
\providecommand \@@endlink[0]{}%
\providecommand \url  [0]{\begingroup\@sanitize@url \@url }%
\providecommand \@url [1]{\endgroup\@href {#1}{\urlprefix }}%
\providecommand \urlprefix  [0]{URL }%
\providecommand \Eprint [0]{\href }%
\providecommand \doibase [0]{https://doi.org/}%
\providecommand \selectlanguage [0]{\@gobble}%
\providecommand \bibinfo  [0]{\@secondoftwo}%
\providecommand \bibfield  [0]{\@secondoftwo}%
\providecommand \translation [1]{[#1]}%
\providecommand \BibitemOpen [0]{}%
\providecommand \bibitemStop [0]{}%
\providecommand \bibitemNoStop [0]{.\EOS\space}%
\providecommand \EOS [0]{\spacefactor3000\relax}%
\providecommand \BibitemShut  [1]{\csname bibitem#1\endcsname}%
\let\auto@bib@innerbib\@empty
\bibitem [{\citenamefont {Terhal}(2015)}]{RevModPhys.87.307}%
  \BibitemOpen
  \bibfield  {author} {\bibinfo {author} {\bibfnamefont {B.~M.}\ \bibnamefont {Terhal}},\ }\bibfield  {title} {\bibinfo {title} {Quantum error correction for quantum memories},\ }\href {https://doi.org/10.1103/RevModPhys.87.307} {\bibfield  {journal} {\bibinfo  {journal} {Rev. Mod. Phys.}\ }\textbf {\bibinfo {volume} {87}},\ \bibinfo {pages} {307} (\bibinfo {year} {2015})}\BibitemShut {NoStop}%
\bibitem [{\citenamefont {Steane}(1996)}]{PhysRevLett.77.793}%
  \BibitemOpen
  \bibfield  {author} {\bibinfo {author} {\bibfnamefont {A.~M.}\ \bibnamefont {Steane}},\ }\bibfield  {title} {\bibinfo {title} {Error correcting codes in quantum theory},\ }\href {https://doi.org/10.1103/PhysRevLett.77.793} {\bibfield  {journal} {\bibinfo  {journal} {Phys. Rev. Lett.}\ }\textbf {\bibinfo {volume} {77}},\ \bibinfo {pages} {793} (\bibinfo {year} {1996})}\BibitemShut {NoStop}%
\bibitem [{\citenamefont {Takita}\ \emph {et~al.}(2017)\citenamefont {Takita}, \citenamefont {Cross}, \citenamefont {C{\ifmmode\acute{o}\else\'{o}\fi}rcoles}, \citenamefont {Chow},\ and\ \citenamefont {Gambetta}}]{takita2017experimental}%
  \BibitemOpen
  \bibfield  {author} {\bibinfo {author} {\bibfnamefont {M.}~\bibnamefont {Takita}}, \bibinfo {author} {\bibfnamefont {A.~W.}\ \bibnamefont {Cross}}, \bibinfo {author} {\bibfnamefont {A.~D.}\ \bibnamefont {C{\ifmmode\acute{o}\else\'{o}\fi}rcoles}}, \bibinfo {author} {\bibfnamefont {J.~M.}\ \bibnamefont {Chow}},\ and\ \bibinfo {author} {\bibfnamefont {J.~M.}\ \bibnamefont {Gambetta}},\ }\bibfield  {title} {\bibinfo {title} {{Experimental demonstration of fault-tolerant state preparation with superconducting qubits}},\ }\href {https://doi.org/10.1103/PhysRevLett.119.180501} {\bibfield  {journal} {\bibinfo  {journal} {Phys. Rev. Lett.}\ }\textbf {\bibinfo {volume} {119}},\ \bibinfo {pages} {180501} (\bibinfo {year} {2017})}\BibitemShut {NoStop}%
\bibitem [{\citenamefont {Campagne-Ibarcq}\ \emph {et~al.}(2020)\citenamefont {Campagne-Ibarcq}, \citenamefont {Eickbusch}, \citenamefont {Touzard}, \citenamefont {Zalys-Geller}, \citenamefont {Frattini}, \citenamefont {Sivak}, \citenamefont {Reinhold}, \citenamefont {Puri}, \citenamefont {Shankar}, \citenamefont {Schoelkopf}, \citenamefont {Frunzio}, \citenamefont {Mirrahimi},\ and\ \citenamefont {Devoret}}]{GKPExperiment}%
  \BibitemOpen
  \bibfield  {author} {\bibinfo {author} {\bibfnamefont {P.}~\bibnamefont {Campagne-Ibarcq}}, \bibinfo {author} {\bibfnamefont {A.}~\bibnamefont {Eickbusch}}, \bibinfo {author} {\bibfnamefont {S.}~\bibnamefont {Touzard}}, \bibinfo {author} {\bibfnamefont {E.}~\bibnamefont {Zalys-Geller}}, \bibinfo {author} {\bibfnamefont {N.~E.}\ \bibnamefont {Frattini}}, \bibinfo {author} {\bibfnamefont {V.~V.}\ \bibnamefont {Sivak}}, \bibinfo {author} {\bibfnamefont {P.}~\bibnamefont {Reinhold}}, \bibinfo {author} {\bibfnamefont {S.}~\bibnamefont {Puri}}, \bibinfo {author} {\bibfnamefont {S.}~\bibnamefont {Shankar}}, \bibinfo {author} {\bibfnamefont {R.~J.}\ \bibnamefont {Schoelkopf}}, \bibinfo {author} {\bibfnamefont {L.}~\bibnamefont {Frunzio}}, \bibinfo {author} {\bibfnamefont {M.}~\bibnamefont {Mirrahimi}},\ and\ \bibinfo {author} {\bibfnamefont {M.~H.}\ \bibnamefont {Devoret}},\ }\bibfield  {title} {\bibinfo {title} {Quantum error correction of a qubit encoded in grid states of an oscillator},\ }\href {https://doi.org/10.1038/s41586-020-2603-3} {\bibfield  {journal} {\bibinfo  {journal} {Nature}\ }\textbf {\bibinfo {volume} {584}},\ \bibinfo {pages} {368} (\bibinfo {year} {2020})}\BibitemShut {NoStop}%
\bibitem [{\citenamefont {Krinner}\ \emph {et~al.}(2022)\citenamefont {Krinner}, \citenamefont {Lacroix}, \citenamefont {Remm}, \citenamefont {Di~Paolo}, \citenamefont {Genois}, \citenamefont {Leroux}, \citenamefont {Hellings}, \citenamefont {Lazar}, \citenamefont {Swiadek}, \citenamefont {Herrmann}, \citenamefont {Norris}, \citenamefont {Andersen}, \citenamefont {M{\ifmmode\ddot{u}\else\"{u}\fi}ller}, \citenamefont {Blais}, \citenamefont {Eichler},\ and\ \citenamefont {Wallraff}}]{krinner2022realizing}%
  \BibitemOpen
  \bibfield  {author} {\bibinfo {author} {\bibfnamefont {S.}~\bibnamefont {Krinner}}, \bibinfo {author} {\bibfnamefont {N.}~\bibnamefont {Lacroix}}, \bibinfo {author} {\bibfnamefont {A.}~\bibnamefont {Remm}}, \bibinfo {author} {\bibfnamefont {A.}~\bibnamefont {Di~Paolo}}, \bibinfo {author} {\bibfnamefont {E.}~\bibnamefont {Genois}}, \bibinfo {author} {\bibfnamefont {C.}~\bibnamefont {Leroux}}, \bibinfo {author} {\bibfnamefont {C.}~\bibnamefont {Hellings}}, \bibinfo {author} {\bibfnamefont {S.}~\bibnamefont {Lazar}}, \bibinfo {author} {\bibfnamefont {F.}~\bibnamefont {Swiadek}}, \bibinfo {author} {\bibfnamefont {J.}~\bibnamefont {Herrmann}}, \bibinfo {author} {\bibfnamefont {G.~J.}\ \bibnamefont {Norris}}, \bibinfo {author} {\bibfnamefont {C.~K.}\ \bibnamefont {Andersen}}, \bibinfo {author} {\bibfnamefont {M.}~\bibnamefont {M{\ifmmode\ddot{u}\else\"{u}\fi}ller}}, \bibinfo {author} {\bibfnamefont {A.}~\bibnamefont {Blais}}, \bibinfo {author} {\bibfnamefont {C.}~\bibnamefont {Eichler}},\ and\ \bibinfo {author} {\bibfnamefont {A.}~\bibnamefont {Wallraff}},\ }\bibfield  {title} {\bibinfo {title} {{Realizing repeated quantum error correction in a distance-three surface code}},\ }\href {https://doi.org/10.1038/s41586-022-04566-8} {\bibfield  {journal} {\bibinfo  {journal} {Nature}\ }\textbf {\bibinfo {volume} {605}},\ \bibinfo {pages} {669} (\bibinfo {year} {2022})}\BibitemShut {NoStop}%
\bibitem [{\citenamefont {Zhao}\ \emph {et~al.}(2022)\citenamefont {Zhao}, \citenamefont {Ye}, \citenamefont {Huang}, \citenamefont {Zhang}, \citenamefont {Wu}, \citenamefont {Guan}, \citenamefont {Zhu}, \citenamefont {Wei}, \citenamefont {He}, \citenamefont {Cao}, \citenamefont {Chen}, \citenamefont {Chung}, \citenamefont {Deng}, \citenamefont {Fan}, \citenamefont {Gong}, \citenamefont {Guo}, \citenamefont {Guo}, \citenamefont {Han}, \citenamefont {Li}, \citenamefont {Li}, \citenamefont {Li}, \citenamefont {Liang}, \citenamefont {Lin}, \citenamefont {Qian}, \citenamefont {Rong}, \citenamefont {Su}, \citenamefont {Sun}, \citenamefont {Wang}, \citenamefont {Wu}, \citenamefont {Xu}, \citenamefont {Ying}, \citenamefont {Yu}, \citenamefont {Zha}, \citenamefont {Zhang}, \citenamefont {Huo}, \citenamefont {Lu}, \citenamefont {Peng}, \citenamefont {Zhu},\ and\ \citenamefont {Pan}}]{zhao2022realization}%
  \BibitemOpen
  \bibfield  {author} {\bibinfo {author} {\bibfnamefont {Y.}~\bibnamefont {Zhao}}, \bibinfo {author} {\bibfnamefont {Y.}~\bibnamefont {Ye}}, \bibinfo {author} {\bibfnamefont {H.-L.}\ \bibnamefont {Huang}}, \bibinfo {author} {\bibfnamefont {Y.}~\bibnamefont {Zhang}}, \bibinfo {author} {\bibfnamefont {D.}~\bibnamefont {Wu}}, \bibinfo {author} {\bibfnamefont {H.}~\bibnamefont {Guan}}, \bibinfo {author} {\bibfnamefont {Q.}~\bibnamefont {Zhu}}, \bibinfo {author} {\bibfnamefont {Z.}~\bibnamefont {Wei}}, \bibinfo {author} {\bibfnamefont {T.}~\bibnamefont {He}}, \bibinfo {author} {\bibfnamefont {S.}~\bibnamefont {Cao}}, \bibinfo {author} {\bibfnamefont {F.}~\bibnamefont {Chen}}, \bibinfo {author} {\bibfnamefont {T.-H.}\ \bibnamefont {Chung}}, \bibinfo {author} {\bibfnamefont {H.}~\bibnamefont {Deng}}, \bibinfo {author} {\bibfnamefont {D.}~\bibnamefont {Fan}}, \bibinfo {author} {\bibfnamefont {M.}~\bibnamefont {Gong}}, \bibinfo {author} {\bibfnamefont {C.}~\bibnamefont {Guo}}, \bibinfo {author} {\bibfnamefont {S.}~\bibnamefont {Guo}}, \bibinfo {author} {\bibfnamefont {L.}~\bibnamefont {Han}}, \bibinfo {author} {\bibfnamefont {N.}~\bibnamefont {Li}}, \bibinfo {author} {\bibfnamefont {S.}~\bibnamefont {Li}}, \bibinfo {author} {\bibfnamefont {Y.}~\bibnamefont {Li}}, \bibinfo {author} {\bibfnamefont {F.}~\bibnamefont {Liang}}, \bibinfo {author} {\bibfnamefont {J.}~\bibnamefont {Lin}}, \bibinfo {author} {\bibfnamefont {H.}~\bibnamefont {Qian}}, \bibinfo {author} {\bibfnamefont {H.}~\bibnamefont {Rong}}, \bibinfo {author} {\bibfnamefont {H.}~\bibnamefont {Su}}, \bibinfo {author} {\bibfnamefont {L.}~\bibnamefont {Sun}}, \bibinfo {author} {\bibfnamefont {S.}~\bibnamefont {Wang}}, \bibinfo {author} {\bibfnamefont {Y.}~\bibnamefont {Wu}}, \bibinfo {author} {\bibfnamefont {Y.}~\bibnamefont {Xu}}, \bibinfo {author} {\bibfnamefont {C.}~\bibnamefont {Ying}}, \bibinfo {author} {\bibfnamefont {J.}~\bibnamefont {Yu}}, \bibinfo {author} {\bibfnamefont {C.}~\bibnamefont {Zha}}, \bibinfo {author} {\bibfnamefont {K.}~\bibnamefont {Zhang}}, \bibinfo {author} {\bibfnamefont {Y.-H.}\ \bibnamefont {Huo}}, \bibinfo {author} {\bibfnamefont {C.-Y.}\ \bibnamefont {Lu}}, \bibinfo {author} {\bibfnamefont {C.-Z.}\ \bibnamefont {Peng}}, \bibinfo {author} {\bibfnamefont {X.}~\bibnamefont {Zhu}},\ and\ \bibinfo {author} {\bibfnamefont {J.-W.}\ \bibnamefont {Pan}},\ }\bibfield  {title} {\bibinfo {title} {{Realization of an error-correcting surface code with superconducting qubits}},\ }\href {https://doi.org/10.1103/PhysRevLett.129.030501} {\bibfield  {journal} {\bibinfo  {journal} {Phys. Rev. Lett.}\ }\textbf {\bibinfo {volume} {129}},\ \bibinfo {pages} {030501} (\bibinfo {year} {2022})}\BibitemShut {NoStop}%
\bibitem [{\citenamefont {AI}(2023)}]{GoogleQEC}%
  \BibitemOpen
  \bibfield  {author} {\bibinfo {author} {\bibfnamefont {G.~Q.}\ \bibnamefont {AI}},\ }\bibfield  {title} {\bibinfo {title} {Suppressing quantum errors by scaling a surface code logical qubit},\ }\href {https://doi.org/10.1038/s41586-022-05434-1} {\bibfield  {journal} {\bibinfo  {journal} {Nature}\ }\textbf {\bibinfo {volume} {614}},\ \bibinfo {pages} {676–681} (\bibinfo {year} {2023})}\BibitemShut {NoStop}%
\bibitem [{\citenamefont {Postler}\ \emph {et~al.}(2022)\citenamefont {Postler}, \citenamefont {Heu{\ss}en}, \citenamefont {Pogorelov}, \citenamefont {Rispler}, \citenamefont {Feldker}, \citenamefont {Meth}, \citenamefont {Marciniak}, \citenamefont {Stricker}, \citenamefont {Ringbauer}, \citenamefont {Blatt}, \citenamefont {Schindler}, \citenamefont {M{\"u}ller},\ and\ \citenamefont {Monz}}]{postler2022demonstration}%
  \BibitemOpen
  \bibfield  {author} {\bibinfo {author} {\bibfnamefont {L.}~\bibnamefont {Postler}}, \bibinfo {author} {\bibfnamefont {S.}~\bibnamefont {Heu{\ss}en}}, \bibinfo {author} {\bibfnamefont {I.}~\bibnamefont {Pogorelov}}, \bibinfo {author} {\bibfnamefont {M.}~\bibnamefont {Rispler}}, \bibinfo {author} {\bibfnamefont {T.}~\bibnamefont {Feldker}}, \bibinfo {author} {\bibfnamefont {M.}~\bibnamefont {Meth}}, \bibinfo {author} {\bibfnamefont {C.~D.}\ \bibnamefont {Marciniak}}, \bibinfo {author} {\bibfnamefont {R.}~\bibnamefont {Stricker}}, \bibinfo {author} {\bibfnamefont {M.}~\bibnamefont {Ringbauer}}, \bibinfo {author} {\bibfnamefont {R.}~\bibnamefont {Blatt}}, \bibinfo {author} {\bibfnamefont {P.}~\bibnamefont {Schindler}}, \bibinfo {author} {\bibfnamefont {M.}~\bibnamefont {M{\"u}ller}},\ and\ \bibinfo {author} {\bibfnamefont {T.}~\bibnamefont {Monz}},\ }\bibfield  {title} {\bibinfo {title} {Demonstration of fault-tolerant universal quantum gate operations},\ }\href {https://doi.org/10.1038/s41586-022-04721-1} {\bibfield  {journal} {\bibinfo  {journal} {Nature}\ }\textbf {\bibinfo {volume} {605}},\ \bibinfo {pages} {675} (\bibinfo {year} {2022})}\BibitemShut {NoStop}%
\bibitem [{\citenamefont {Ryan-Anderson}\ \emph {et~al.}(2022)\citenamefont {Ryan-Anderson}, \citenamefont {Brown}, \citenamefont {Allman}, \citenamefont {Arkin}, \citenamefont {Asa-Attuah}, \citenamefont {Baldwin}, \citenamefont {Berg}, \citenamefont {Bohnet}, \citenamefont {Braxton}, \citenamefont {Burdick}, \citenamefont {Campora}, \citenamefont {Chernoguzov}, \citenamefont {Esposito}, \citenamefont {Evans}, \citenamefont {Francois}, \citenamefont {Gaebler}, \citenamefont {Gatterman}, \citenamefont {Gerber}, \citenamefont {Gilmore}, \citenamefont {Gresh}, \citenamefont {Hall}, \citenamefont {Hankin}, \citenamefont {Hostetter}, \citenamefont {Lucchetti}, \citenamefont {Mayer}, \citenamefont {Myers}, \citenamefont {Neyenhuis}, \citenamefont {Santiago}, \citenamefont {Sedlacek}, \citenamefont {Skripka}, \citenamefont {Slattery}, \citenamefont {Stutz}, \citenamefont {Tait}, \citenamefont {Tobey}, \citenamefont {Vittorini}, \citenamefont {Walker},\ and\ \citenamefont {Hayes}}]{ryan2022implementing}%
  \BibitemOpen
  \bibfield  {author} {\bibinfo {author} {\bibfnamefont {C.}~\bibnamefont {Ryan-Anderson}}, \bibinfo {author} {\bibfnamefont {N.~C.}\ \bibnamefont {Brown}}, \bibinfo {author} {\bibfnamefont {M.~S.}\ \bibnamefont {Allman}}, \bibinfo {author} {\bibfnamefont {B.}~\bibnamefont {Arkin}}, \bibinfo {author} {\bibfnamefont {G.}~\bibnamefont {Asa-Attuah}}, \bibinfo {author} {\bibfnamefont {C.}~\bibnamefont {Baldwin}}, \bibinfo {author} {\bibfnamefont {J.}~\bibnamefont {Berg}}, \bibinfo {author} {\bibfnamefont {J.~G.}\ \bibnamefont {Bohnet}}, \bibinfo {author} {\bibfnamefont {S.}~\bibnamefont {Braxton}}, \bibinfo {author} {\bibfnamefont {N.}~\bibnamefont {Burdick}}, \bibinfo {author} {\bibfnamefont {J.~P.}\ \bibnamefont {Campora}}, \bibinfo {author} {\bibfnamefont {A.}~\bibnamefont {Chernoguzov}}, \bibinfo {author} {\bibfnamefont {J.}~\bibnamefont {Esposito}}, \bibinfo {author} {\bibfnamefont {B.}~\bibnamefont {Evans}}, \bibinfo {author} {\bibfnamefont {D.}~\bibnamefont {Francois}}, \bibinfo {author} {\bibfnamefont {J.~P.}\ \bibnamefont {Gaebler}}, \bibinfo {author} {\bibfnamefont {T.~M.}\ \bibnamefont {Gatterman}}, \bibinfo {author} {\bibfnamefont {J.}~\bibnamefont {Gerber}}, \bibinfo {author} {\bibfnamefont {K.}~\bibnamefont {Gilmore}}, \bibinfo {author} {\bibfnamefont {D.}~\bibnamefont {Gresh}}, \bibinfo {author} {\bibfnamefont {A.}~\bibnamefont {Hall}}, \bibinfo {author} {\bibfnamefont {A.}~\bibnamefont {Hankin}}, \bibinfo {author} {\bibfnamefont {J.}~\bibnamefont {Hostetter}}, \bibinfo {author} {\bibfnamefont {D.}~\bibnamefont {Lucchetti}}, \bibinfo {author} {\bibfnamefont {K.}~\bibnamefont {Mayer}}, \bibinfo {author} {\bibfnamefont {J.}~\bibnamefont {Myers}}, \bibinfo {author} {\bibfnamefont {B.}~\bibnamefont {Neyenhuis}}, \bibinfo {author} {\bibfnamefont {J.}~\bibnamefont {Santiago}}, \bibinfo {author} {\bibfnamefont {J.}~\bibnamefont {Sedlacek}}, \bibinfo {author} {\bibfnamefont {T.}~\bibnamefont {Skripka}}, \bibinfo {author} {\bibfnamefont {A.}~\bibnamefont {Slattery}}, \bibinfo {author} {\bibfnamefont {R.~P.}\ \bibnamefont {Stutz}}, \bibinfo {author} {\bibfnamefont {J.}~\bibnamefont {Tait}}, \bibinfo {author} {\bibfnamefont {R.}~\bibnamefont {Tobey}}, \bibinfo {author} {\bibfnamefont {G.}~\bibnamefont {Vittorini}}, \bibinfo {author} {\bibfnamefont {J.}~\bibnamefont {Walker}},\ and\ \bibinfo {author} {\bibfnamefont {D.}~\bibnamefont {Hayes}},\ }\href {https://arxiv.org/abs/2208.01863} {\bibinfo {title} {Implementing fault-tolerant entangling gates on the five-qubit code and the color code}} (\bibinfo {year} {2022}),\ \Eprint {https://arxiv.org/abs/2208.01863} {arXiv:2208.01863 [quant-ph]} \BibitemShut {NoStop}%
\bibitem [{\citenamefont {Bluvstein}\ \emph {et~al.}(2024)\citenamefont {Bluvstein}, \citenamefont {Evered}, \citenamefont {Geim}, \citenamefont {Li}, \citenamefont {Zhou}, \citenamefont {Manovitz}, \citenamefont {Ebadi}, \citenamefont {Cain}, \citenamefont {Kalinowski}, \citenamefont {Hangleiter}, \citenamefont {Bonilla~Ataides}, \citenamefont {Maskara}, \citenamefont {Cong}, \citenamefont {Gao}, \citenamefont {Sales~Rodriguez}, \citenamefont {Karolyshyn}, \citenamefont {Semeghini}, \citenamefont {Gullans}, \citenamefont {Greiner}, \citenamefont {Vuleti{\ifmmode\acute{c}\else\'{c}\fi}},\ and\ \citenamefont {Lukin}}]{bluvstein2024logical}%
  \BibitemOpen
  \bibfield  {author} {\bibinfo {author} {\bibfnamefont {D.}~\bibnamefont {Bluvstein}}, \bibinfo {author} {\bibfnamefont {S.~J.}\ \bibnamefont {Evered}}, \bibinfo {author} {\bibfnamefont {A.~A.}\ \bibnamefont {Geim}}, \bibinfo {author} {\bibfnamefont {S.~H.}\ \bibnamefont {Li}}, \bibinfo {author} {\bibfnamefont {H.}~\bibnamefont {Zhou}}, \bibinfo {author} {\bibfnamefont {T.}~\bibnamefont {Manovitz}}, \bibinfo {author} {\bibfnamefont {S.}~\bibnamefont {Ebadi}}, \bibinfo {author} {\bibfnamefont {M.}~\bibnamefont {Cain}}, \bibinfo {author} {\bibfnamefont {M.}~\bibnamefont {Kalinowski}}, \bibinfo {author} {\bibfnamefont {D.}~\bibnamefont {Hangleiter}}, \bibinfo {author} {\bibfnamefont {J.~P.}\ \bibnamefont {Bonilla~Ataides}}, \bibinfo {author} {\bibfnamefont {N.}~\bibnamefont {Maskara}}, \bibinfo {author} {\bibfnamefont {I.}~\bibnamefont {Cong}}, \bibinfo {author} {\bibfnamefont {X.}~\bibnamefont {Gao}}, \bibinfo {author} {\bibfnamefont {P.}~\bibnamefont {Sales~Rodriguez}}, \bibinfo {author} {\bibfnamefont {T.}~\bibnamefont {Karolyshyn}}, \bibinfo {author} {\bibfnamefont {G.}~\bibnamefont {Semeghini}}, \bibinfo {author} {\bibfnamefont {M.~J.}\ \bibnamefont {Gullans}}, \bibinfo {author} {\bibfnamefont {M.}~\bibnamefont {Greiner}}, \bibinfo {author} {\bibfnamefont {V.}~\bibnamefont {Vuleti{\ifmmode\acute{c}\else\'{c}\fi}}},\ and\ \bibinfo {author} {\bibfnamefont {M.~D.}\ \bibnamefont {Lukin}},\ }\bibfield  {title} {\bibinfo {title} {{Logical quantum processor based on reconfigurable atom arrays}},\ }\href {https://doi.org/10.1038/s41586-023-06927-3} {\bibfield  {journal} {\bibinfo  {journal} {Nature}\ }\textbf {\bibinfo {volume} {626}},\ \bibinfo {pages} {58} (\bibinfo {year} {2024})}\BibitemShut {NoStop}%
\bibitem [{\citenamefont {Huang}\ \emph {et~al.}(2024)\citenamefont {Huang}, \citenamefont {Brown},\ and\ \citenamefont {Cetina}}]{huang2023comparing}%
  \BibitemOpen
  \bibfield  {author} {\bibinfo {author} {\bibfnamefont {S.}~\bibnamefont {Huang}}, \bibinfo {author} {\bibfnamefont {K.~R.}\ \bibnamefont {Brown}},\ and\ \bibinfo {author} {\bibfnamefont {M.}~\bibnamefont {Cetina}},\ }\bibfield  {title} {\bibinfo {title} {Comparing shor and steane error correction using the bacon-shor code},\ }\bibfield  {journal} {\bibinfo  {journal} {Science Advances}\ }\textbf {\bibinfo {volume} {10}},\ \href {https://doi.org/10.1126/sciadv.adp2008} {10.1126/sciadv.adp2008} (\bibinfo {year} {2024}),\ \Eprint {https://arxiv.org/abs/https://www.science.org/doi/pdf/10.1126/sciadv.adp2008} {https://www.science.org/doi/pdf/10.1126/sciadv.adp2008} \BibitemShut {NoStop}%
\bibitem [{\citenamefont {Postler}\ \emph {et~al.}(2024)\citenamefont {Postler}, \citenamefont {Butt}, \citenamefont {Pogorelov}, \citenamefont {Marciniak}, \citenamefont {Heußen}, \citenamefont {Blatt}, \citenamefont {Schindler}, \citenamefont {Rispler}, \citenamefont {Müller},\ and\ \citenamefont {Monz}}]{postler2023demonstration}%
  \BibitemOpen
  \bibfield  {author} {\bibinfo {author} {\bibfnamefont {L.}~\bibnamefont {Postler}}, \bibinfo {author} {\bibfnamefont {F.}~\bibnamefont {Butt}}, \bibinfo {author} {\bibfnamefont {I.}~\bibnamefont {Pogorelov}}, \bibinfo {author} {\bibfnamefont {C.~D.}\ \bibnamefont {Marciniak}}, \bibinfo {author} {\bibfnamefont {S.}~\bibnamefont {Heußen}}, \bibinfo {author} {\bibfnamefont {R.}~\bibnamefont {Blatt}}, \bibinfo {author} {\bibfnamefont {P.}~\bibnamefont {Schindler}}, \bibinfo {author} {\bibfnamefont {M.}~\bibnamefont {Rispler}}, \bibinfo {author} {\bibfnamefont {M.}~\bibnamefont {Müller}},\ and\ \bibinfo {author} {\bibfnamefont {T.}~\bibnamefont {Monz}},\ }\bibfield  {title} {\bibinfo {title} {Demonstration of fault-tolerant steane quantum error correction},\ }\bibfield  {journal} {\bibinfo  {journal} {PRX Quantum}\ }\textbf {\bibinfo {volume} {5}},\ \href {https://doi.org/10.1103/prxquantum.5.030326} {10.1103/prxquantum.5.030326} (\bibinfo {year} {2024})\BibitemShut {NoStop}%
\bibitem [{\citenamefont {Gottesman}(1997)}]{gottesman1997stabilizer}%
  \BibitemOpen
  \bibfield  {author} {\bibinfo {author} {\bibfnamefont {D.}~\bibnamefont {Gottesman}},\ }\href@noop {} {\emph {\bibinfo {title} {Stabilizer codes and quantum error correction}}}\ (\bibinfo  {publisher} {California Institute of Technology},\ \bibinfo {year} {1997})\BibitemShut {NoStop}%
\bibitem [{\citenamefont {Hastings}\ and\ \citenamefont {Haah}(2021)}]{hastings2021dynamically}%
  \BibitemOpen
  \bibfield  {author} {\bibinfo {author} {\bibfnamefont {M.~B.}\ \bibnamefont {Hastings}}\ and\ \bibinfo {author} {\bibfnamefont {J.}~\bibnamefont {Haah}},\ }\bibfield  {title} {\bibinfo {title} {Dynamically generated logical qubits},\ }\href {https://doi.org/10.22331/q-2021-10-19-564} {\bibfield  {journal} {\bibinfo  {journal} {Quantum}\ }\textbf {\bibinfo {volume} {5}},\ \bibinfo {pages} {564} (\bibinfo {year} {2021})}\BibitemShut {NoStop}%
\bibitem [{\citenamefont {Kribs}\ \emph {et~al.}(2005)\citenamefont {Kribs}, \citenamefont {Laflamme},\ and\ \citenamefont {Poulin}}]{PhysRevLett.94.180501}%
  \BibitemOpen
  \bibfield  {author} {\bibinfo {author} {\bibfnamefont {D.}~\bibnamefont {Kribs}}, \bibinfo {author} {\bibfnamefont {R.}~\bibnamefont {Laflamme}},\ and\ \bibinfo {author} {\bibfnamefont {D.}~\bibnamefont {Poulin}},\ }\bibfield  {title} {\bibinfo {title} {Unified and generalized approach to quantum error correction},\ }\href {https://doi.org/10.1103/PhysRevLett.94.180501} {\bibfield  {journal} {\bibinfo  {journal} {Phys. Rev. Lett.}\ }\textbf {\bibinfo {volume} {94}},\ \bibinfo {pages} {180501} (\bibinfo {year} {2005})}\BibitemShut {NoStop}%
\bibitem [{\citenamefont {Vuillot}(2021)}]{vuillot2022planar}%
  \BibitemOpen
  \bibfield  {author} {\bibinfo {author} {\bibfnamefont {C.}~\bibnamefont {Vuillot}},\ }\href {https://arxiv.org/abs/2110.05348} {\bibinfo {title} {Planar floquet codes}} (\bibinfo {year} {2021}),\ \Eprint {https://arxiv.org/abs/2110.05348} {arXiv:2110.05348} \BibitemShut {NoStop}%
\bibitem [{\citenamefont {Townsend-Teague}\ \emph {et~al.}(2023)\citenamefont {Townsend-Teague}, \citenamefont {Magdalena de~la Fuente},\ and\ \citenamefont {Kesselring}}]{Teague2023Floquetifying}%
  \BibitemOpen
  \bibfield  {author} {\bibinfo {author} {\bibfnamefont {A.}~\bibnamefont {Townsend-Teague}}, \bibinfo {author} {\bibfnamefont {J.}~\bibnamefont {Magdalena de~la Fuente}},\ and\ \bibinfo {author} {\bibfnamefont {M.}~\bibnamefont {Kesselring}},\ }\bibfield  {title} {\bibinfo {title} {Floquetifying the colour code},\ }\bibfield  {journal} {\bibinfo  {journal} {Electronic Proceedings in Theoretical Computer Science}\ }\textbf {\bibinfo {volume} {384}},\ \href {https://doi.org/10.4204/eptcs.384.14} {10.4204/eptcs.384.14} (\bibinfo {year} {2023})\BibitemShut {NoStop}%
\bibitem [{\citenamefont {Dennis}\ \emph {et~al.}(2002)\citenamefont {Dennis}, \citenamefont {Kitaev}, \citenamefont {Landahl},\ and\ \citenamefont {Preskill}}]{dennis2002topological}%
  \BibitemOpen
  \bibfield  {author} {\bibinfo {author} {\bibfnamefont {E.}~\bibnamefont {Dennis}}, \bibinfo {author} {\bibfnamefont {A.}~\bibnamefont {Kitaev}}, \bibinfo {author} {\bibfnamefont {A.}~\bibnamefont {Landahl}},\ and\ \bibinfo {author} {\bibfnamefont {J.}~\bibnamefont {Preskill}},\ }\bibfield  {title} {\bibinfo {title} {{Topological quantum memory}},\ }\href {https://doi.org/10.1063/1.1499754} {\bibfield  {journal} {\bibinfo  {journal} {J. Math. Phys.}\ }\textbf {\bibinfo {volume} {43}},\ \bibinfo {pages} {4452} (\bibinfo {year} {2002})}\BibitemShut {NoStop}%
\bibitem [{\citenamefont {Kesselring}\ \emph {et~al.}(2024)\citenamefont {Kesselring}, \citenamefont {de~la Fuente}, \citenamefont {Thomsen}, \citenamefont {Eisert}, \citenamefont {Bartlett},\ and\ \citenamefont {Brown}}]{kesselring2022anyon}%
  \BibitemOpen
  \bibfield  {author} {\bibinfo {author} {\bibfnamefont {M.~S.}\ \bibnamefont {Kesselring}}, \bibinfo {author} {\bibfnamefont {J.~C.~M.}\ \bibnamefont {de~la Fuente}}, \bibinfo {author} {\bibfnamefont {F.}~\bibnamefont {Thomsen}}, \bibinfo {author} {\bibfnamefont {J.}~\bibnamefont {Eisert}}, \bibinfo {author} {\bibfnamefont {S.~D.}\ \bibnamefont {Bartlett}},\ and\ \bibinfo {author} {\bibfnamefont {B.~J.}\ \bibnamefont {Brown}},\ }\bibfield  {title} {\bibinfo {title} {Anyon condensation and the color code},\ }\href {https://doi.org/10.1103/PRXQuantum.5.010342} {\bibfield  {journal} {\bibinfo  {journal} {PRX Quantum}\ }\textbf {\bibinfo {volume} {5}},\ \bibinfo {pages} {010342} (\bibinfo {year} {2024})}\BibitemShut {NoStop}%
\bibitem [{\citenamefont {Davydova}\ \emph {et~al.}(2024)\citenamefont {Davydova}, \citenamefont {Tantivasadakarn}, \citenamefont {Balasubramanian},\ and\ \citenamefont {Aasen}}]{davydova2023quantum}%
  \BibitemOpen
  \bibfield  {author} {\bibinfo {author} {\bibfnamefont {M.}~\bibnamefont {Davydova}}, \bibinfo {author} {\bibfnamefont {N.}~\bibnamefont {Tantivasadakarn}}, \bibinfo {author} {\bibfnamefont {S.}~\bibnamefont {Balasubramanian}},\ and\ \bibinfo {author} {\bibfnamefont {D.}~\bibnamefont {Aasen}},\ }\bibfield  {title} {\bibinfo {title} {Quantum computation from dynamic automorphism codes},\ }\href {https://doi.org/10.22331/q-2024-08-27-1448} {\bibfield  {journal} {\bibinfo  {journal} {{Quantum}}\ }\textbf {\bibinfo {volume} {8}},\ \bibinfo {pages} {1448} (\bibinfo {year} {2024})}\BibitemShut {NoStop}%
\bibitem [{\citenamefont {Ellison}\ \emph {et~al.}(2023{\natexlab{a}})\citenamefont {Ellison}, \citenamefont {Sullivan},\ and\ \citenamefont {Dua}}]{ellison2023floquet}%
  \BibitemOpen
  \bibfield  {author} {\bibinfo {author} {\bibfnamefont {T.~D.}\ \bibnamefont {Ellison}}, \bibinfo {author} {\bibfnamefont {J.}~\bibnamefont {Sullivan}},\ and\ \bibinfo {author} {\bibfnamefont {A.}~\bibnamefont {Dua}},\ }\href@noop {} {\bibinfo {title} {Floquet codes with a twist}} (\bibinfo {year} {2023}{\natexlab{a}}),\ \Eprint {https://arxiv.org/abs/2306.08027} {arXiv:2306.08027} \BibitemShut {NoStop}%
\bibitem [{\citenamefont {Dua}\ \emph {et~al.}(2024)\citenamefont {Dua}, \citenamefont {Tantivasadakarn}, \citenamefont {Sullivan},\ and\ \citenamefont {Ellison}}]{dua2023engineering}%
  \BibitemOpen
  \bibfield  {author} {\bibinfo {author} {\bibfnamefont {A.}~\bibnamefont {Dua}}, \bibinfo {author} {\bibfnamefont {N.}~\bibnamefont {Tantivasadakarn}}, \bibinfo {author} {\bibfnamefont {J.}~\bibnamefont {Sullivan}},\ and\ \bibinfo {author} {\bibfnamefont {T.~D.}\ \bibnamefont {Ellison}},\ }\bibfield  {title} {\bibinfo {title} {{Engineering 3D Floquet codes by rewinding}},\ }\href {https://doi.org/10.1103/PRXQuantum.5.020305} {\bibfield  {journal} {\bibinfo  {journal} {PRX Quantum}\ }\textbf {\bibinfo {volume} {5}},\ \bibinfo {pages} {020305} (\bibinfo {year} {2024})}\BibitemShut {NoStop}%
\bibitem [{\citenamefont {Zhang}\ \emph {et~al.}(2023)\citenamefont {Zhang}, \citenamefont {Aasen},\ and\ \citenamefont {Vijay}}]{zhang2022xcube}%
  \BibitemOpen
  \bibfield  {author} {\bibinfo {author} {\bibfnamefont {Z.}~\bibnamefont {Zhang}}, \bibinfo {author} {\bibfnamefont {D.}~\bibnamefont {Aasen}},\ and\ \bibinfo {author} {\bibfnamefont {S.}~\bibnamefont {Vijay}},\ }\bibfield  {title} {\bibinfo {title} {{$X$-cube Floquet code: A dynamical quantum error correcting code with a subextensive number of logical qubits}},\ }\href {https://doi.org/10.1103/PhysRevB.108.205116} {\bibfield  {journal} {\bibinfo  {journal} {Phys. Rev. B}\ }\textbf {\bibinfo {volume} {108}},\ \bibinfo {pages} {205116} (\bibinfo {year} {2023})}\BibitemShut {NoStop}%
\bibitem [{\citenamefont {Aasen}\ \emph {et~al.}(2022)\citenamefont {Aasen}, \citenamefont {Wang},\ and\ \citenamefont {Hastings}}]{aasen2022adiabatic}%
  \BibitemOpen
  \bibfield  {author} {\bibinfo {author} {\bibfnamefont {D.}~\bibnamefont {Aasen}}, \bibinfo {author} {\bibfnamefont {Z.}~\bibnamefont {Wang}},\ and\ \bibinfo {author} {\bibfnamefont {M.~B.}\ \bibnamefont {Hastings}},\ }\bibfield  {title} {\bibinfo {title} {{Adiabatic paths of Hamiltonians, symmetries of topological order, and automorphism codes}},\ }\href {https://doi.org/10.1103/PhysRevB.106.085122} {\bibfield  {journal} {\bibinfo  {journal} {Phys. Rev. B}\ }\textbf {\bibinfo {volume} {106}},\ \bibinfo {pages} {085122} (\bibinfo {year} {2022})}\BibitemShut {NoStop}%
\bibitem [{\citenamefont {Bauer}(2024{\natexlab{a}})}]{bauer2024topological}%
  \BibitemOpen
  \bibfield  {author} {\bibinfo {author} {\bibfnamefont {A.}~\bibnamefont {Bauer}},\ }\bibfield  {title} {\bibinfo {title} {Topological error correcting processes from fixed-point path integrals},\ }\href {https://doi.org/10.22331/q-2024-03-20-1288} {\bibfield  {journal} {\bibinfo  {journal} {Quantum}\ }\textbf {\bibinfo {volume} {8}},\ \bibinfo {pages} {1288} (\bibinfo {year} {2024}{\natexlab{a}})}\BibitemShut {NoStop}%
\bibitem [{\citenamefont {van~de Wetering}(2020)}]{van2020zx}%
  \BibitemOpen
  \bibfield  {author} {\bibinfo {author} {\bibfnamefont {J.}~\bibnamefont {van~de Wetering}},\ }\bibfield  {title} {\bibinfo {title} {{ZX-calculus for the working quantum computer scientist}},\ }\href@noop {} {\bibfield  {journal} {\bibinfo  {journal} {arxiv}\ } (\bibinfo {year} {2020})},\ \Eprint {https://arxiv.org/abs/2012.13966} {arXiv:2012.13966} \BibitemShut {NoStop}%
\bibitem [{\citenamefont {Bomb{\'\i}n}\ and\ \citenamefont {Martin-Delgado}(2007)}]{Bombin2007exact}%
  \BibitemOpen
  \bibfield  {author} {\bibinfo {author} {\bibfnamefont {H.}~\bibnamefont {Bomb{\'\i}n}}\ and\ \bibinfo {author} {\bibfnamefont {M.~A.}\ \bibnamefont {Martin-Delgado}},\ }\bibfield  {title} {\bibinfo {title} {Exact topological quantum order in $d=3$ and beyond: Branyons and brane-net condensates},\ }\href {https://doi.org/10.1103/PhysRevB.75.075103} {\bibfield  {journal} {\bibinfo  {journal} {Phys. Rev. B}\ }\textbf {\bibinfo {volume} {75}},\ \bibinfo {pages} {075103} (\bibinfo {year} {2007})}\BibitemShut {NoStop}%
\bibitem [{\citenamefont {Bomb{\'\i}n}\ and\ \citenamefont {Martin-Delgado}(2006)}]{bombin2006topological}%
  \BibitemOpen
  \bibfield  {author} {\bibinfo {author} {\bibfnamefont {H.}~\bibnamefont {Bomb{\'\i}n}}\ and\ \bibinfo {author} {\bibfnamefont {M.~A.}\ \bibnamefont {Martin-Delgado}},\ }\bibfield  {title} {\bibinfo {title} {Topological quantum distillation},\ }\href {https://doi.org/10.1103/PhysRevLett.97.180501} {\bibfield  {journal} {\bibinfo  {journal} {Phys. Rev. Lett.}\ }\textbf {\bibinfo {volume} {97}},\ \bibinfo {pages} {180501} (\bibinfo {year} {2006})}\BibitemShut {NoStop}%
\bibitem [{\citenamefont {Roberts}\ and\ \citenamefont {Williamson}(2024)}]{roberts20203}%
  \BibitemOpen
  \bibfield  {author} {\bibinfo {author} {\bibfnamefont {S.}~\bibnamefont {Roberts}}\ and\ \bibinfo {author} {\bibfnamefont {D.~J.}\ \bibnamefont {Williamson}},\ }\bibfield  {title} {\bibinfo {title} {3-fermion topological quantum computation},\ }\href {https://doi.org/10.1103/PRXQuantum.5.010315} {\bibfield  {journal} {\bibinfo  {journal} {PRX Quantum}\ }\textbf {\bibinfo {volume} {5}},\ \bibinfo {pages} {010315} (\bibinfo {year} {2024})}\BibitemShut {NoStop}%
\bibitem [{\citenamefont {Ellison}\ \emph {et~al.}(2023{\natexlab{b}})\citenamefont {Ellison}, \citenamefont {Chen}, \citenamefont {Dua}, \citenamefont {Shirley}, \citenamefont {Tantivasadakarn},\ and\ \citenamefont {Williamson}}]{Ellison2023paulitopological}%
  \BibitemOpen
  \bibfield  {author} {\bibinfo {author} {\bibfnamefont {T.~D.}\ \bibnamefont {Ellison}}, \bibinfo {author} {\bibfnamefont {Y.-A.}\ \bibnamefont {Chen}}, \bibinfo {author} {\bibfnamefont {A.}~\bibnamefont {Dua}}, \bibinfo {author} {\bibfnamefont {W.}~\bibnamefont {Shirley}}, \bibinfo {author} {\bibfnamefont {N.}~\bibnamefont {Tantivasadakarn}},\ and\ \bibinfo {author} {\bibfnamefont {D.~J.}\ \bibnamefont {Williamson}},\ }\bibfield  {title} {\bibinfo {title} {Pauli topological subsystem codes from {A}belian anyon theories},\ }\href {https://doi.org/10.22331/q-2023-10-12-1137} {\bibfield  {journal} {\bibinfo  {journal} {{Quantum}}\ }\textbf {\bibinfo {volume} {7}},\ \bibinfo {pages} {1137} (\bibinfo {year} {2023}{\natexlab{b}})}\BibitemShut {NoStop}%
\bibitem [{\citenamefont {Bombin}\ \emph {et~al.}(2024)\citenamefont {Bombin}, \citenamefont {Litinski}, \citenamefont {Nickerson}, \citenamefont {Pastawski},\ and\ \citenamefont {Roberts}}]{bombin2023unifying}%
  \BibitemOpen
  \bibfield  {author} {\bibinfo {author} {\bibfnamefont {H.}~\bibnamefont {Bombin}}, \bibinfo {author} {\bibfnamefont {D.}~\bibnamefont {Litinski}}, \bibinfo {author} {\bibfnamefont {N.}~\bibnamefont {Nickerson}}, \bibinfo {author} {\bibfnamefont {F.}~\bibnamefont {Pastawski}},\ and\ \bibinfo {author} {\bibfnamefont {S.}~\bibnamefont {Roberts}},\ }\bibfield  {title} {\bibinfo {title} {{Unifying flavors of fault tolerance with the ZX calculus}},\ }\href {https://doi.org/10.22331/q-2024-06-18-1379} {\bibfield  {journal} {\bibinfo  {journal} {Quantum}\ }\textbf {\bibinfo {volume} {8}},\ \bibinfo {pages} {1379} (\bibinfo {year} {2024})}\BibitemShut {NoStop}%
\bibitem [{\citenamefont {Wang}\ and\ \citenamefont {Wang}(2020)}]{wang2020and}%
  \BibitemOpen
  \bibfield  {author} {\bibinfo {author} {\bibfnamefont {L.}~\bibnamefont {Wang}}\ and\ \bibinfo {author} {\bibfnamefont {Z.}~\bibnamefont {Wang}},\ }\bibfield  {title} {\bibinfo {title} {In and around abelian anyon models},\ }\href {https://doi.org/10.1088/1751-8121/abc6c0} {\bibfield  {journal} {\bibinfo  {journal} {J. Phys. A}\ }\textbf {\bibinfo {volume} {53}},\ \bibinfo {pages} {505203} (\bibinfo {year} {2020})}\BibitemShut {NoStop}%
\bibitem [{\citenamefont {Nielsen}\ and\ \citenamefont {Chuang}(2001)}]{nielsen2001quantum}%
  \BibitemOpen
  \bibfield  {author} {\bibinfo {author} {\bibfnamefont {M.~A.}\ \bibnamefont {Nielsen}}\ and\ \bibinfo {author} {\bibfnamefont {I.~L.}\ \bibnamefont {Chuang}},\ }\bibfield  {title} {\bibinfo {title} {Quantum computation and quantum information},\ }\href {https://doi.org/10.1063/1.1428442} {\bibfield  {journal} {\bibinfo  {journal} {Phys. Today}\ }\textbf {\bibinfo {volume} {54}},\ \bibinfo {pages} {60} (\bibinfo {year} {2001})}\BibitemShut {NoStop}%
\bibitem [{\citenamefont {Gottesman}(1998)}]{gottesman1998TheHR}%
  \BibitemOpen
  \bibfield  {author} {\bibinfo {author} {\bibfnamefont {D.}~\bibnamefont {Gottesman}},\ }\href {https://arxiv.org/abs/quant-ph/9807006} {\bibinfo {title} {{The Heisenberg representation of quantum computers}}} (\bibinfo {year} {1998}),\ \Eprint {https://arxiv.org/abs/quant-ph/9807006} {arXiv:quant-ph/9807006} \BibitemShut {NoStop}%
\bibitem [{\citenamefont {Aasen}\ \emph {et~al.}(2023)\citenamefont {Aasen}, \citenamefont {Haah}, \citenamefont {Li},\ and\ \citenamefont {Mong}}]{aasen2023measurement}%
  \BibitemOpen
  \bibfield  {author} {\bibinfo {author} {\bibfnamefont {D.}~\bibnamefont {Aasen}}, \bibinfo {author} {\bibfnamefont {J.}~\bibnamefont {Haah}}, \bibinfo {author} {\bibfnamefont {Z.}~\bibnamefont {Li}},\ and\ \bibinfo {author} {\bibfnamefont {R.~S.~K.}\ \bibnamefont {Mong}},\ }\href {https://arxiv.org/abs/2304.01277} {\bibinfo {title} {{Measurement quantum cellular automata and anomalies in Floquet codes}}} (\bibinfo {year} {2023}),\ \Eprint {https://arxiv.org/abs/2304.01277} {arXiv:2304.01277} \BibitemShut {NoStop}%
\bibitem [{\citenamefont {Kitaev}(2006)}]{Kitaev2006anyons}%
  \BibitemOpen
  \bibfield  {author} {\bibinfo {author} {\bibfnamefont {A.}~\bibnamefont {Kitaev}},\ }\bibfield  {title} {\bibinfo {title} {Anyons in an exactly solved model and beyond},\ }\href {https://doi.org/https://doi.org/10.1016/j.aop.2005.10.005} {\bibfield  {journal} {\bibinfo  {journal} {Ann. Phys.}\ }\textbf {\bibinfo {volume} {321}},\ \bibinfo {pages} {2} (\bibinfo {year} {2006})}\BibitemShut {NoStop}%
\bibitem [{\citenamefont {Bravyi}\ and\ \citenamefont {Hastings}(2011)}]{bravyi2011ashortproof}%
  \BibitemOpen
  \bibfield  {author} {\bibinfo {author} {\bibfnamefont {S.}~\bibnamefont {Bravyi}}\ and\ \bibinfo {author} {\bibfnamefont {M.~B.}\ \bibnamefont {Hastings}},\ }\bibfield  {title} {\bibinfo {title} {A short proof of stability of topological order under local perturbations},\ }\href {https://doi.org/10.1007/s00220-011-1346-2} {\bibfield  {journal} {\bibinfo  {journal} {Commun. Math. Phys.}\ }\textbf {\bibinfo {volume} {307}},\ \bibinfo {pages} {609} (\bibinfo {year} {2011})}\BibitemShut {NoStop}%
\bibitem [{\citenamefont {Bravyi}\ and\ \citenamefont {Terhal}(2009)}]{bravyi2009no}%
  \BibitemOpen
  \bibfield  {author} {\bibinfo {author} {\bibfnamefont {S.}~\bibnamefont {Bravyi}}\ and\ \bibinfo {author} {\bibfnamefont {B.}~\bibnamefont {Terhal}},\ }\bibfield  {title} {\bibinfo {title} {A no-go theorem for a two-dimensional self-correcting quantum memory based on stabilizer codes},\ }\href {https://doi.org/10.1088/1367-2630/11/4/043029} {\bibfield  {journal} {\bibinfo  {journal} {New J. Phys.}\ }\textbf {\bibinfo {volume} {11}},\ \bibinfo {pages} {043029} (\bibinfo {year} {2009})}\BibitemShut {NoStop}%
\bibitem [{\citenamefont {Bomb{\'\i}n}(2014)}]{bombin2014structure}%
  \BibitemOpen
  \bibfield  {author} {\bibinfo {author} {\bibfnamefont {H.}~\bibnamefont {Bomb{\'\i}n}},\ }\bibfield  {title} {\bibinfo {title} {{Structure of 2D topological stabilizer codes}},\ }\href {https://doi.org/10.1007/s00220-014-1893-4} {\bibfield  {journal} {\bibinfo  {journal} {Comm. Math. Phys.}\ }\textbf {\bibinfo {volume} {327}},\ \bibinfo {pages} {387} (\bibinfo {year} {2014})}\BibitemShut {NoStop}%
\bibitem [{\citenamefont {Bomb{\'\i}n}\ \emph {et~al.}(2012)\citenamefont {Bomb{\'\i}n}, \citenamefont {Duclos-Cianci},\ and\ \citenamefont {Poulin}}]{bombin2012universal}%
  \BibitemOpen
  \bibfield  {author} {\bibinfo {author} {\bibfnamefont {H.}~\bibnamefont {Bomb{\'\i}n}}, \bibinfo {author} {\bibfnamefont {G.}~\bibnamefont {Duclos-Cianci}},\ and\ \bibinfo {author} {\bibfnamefont {D.}~\bibnamefont {Poulin}},\ }\bibfield  {title} {\bibinfo {title} {Universal topological phase of two-dimensional stabilizer codes},\ }\href {https://doi.org/10.1088/1367-2630/14/7/073048} {\bibfield  {journal} {\bibinfo  {journal} {New J. Phys.}\ }\textbf {\bibinfo {volume} {14}},\ \bibinfo {pages} {073048} (\bibinfo {year} {2012})}\BibitemShut {NoStop}%
\bibitem [{\citenamefont {Kesselring}\ \emph {et~al.}(2018)\citenamefont {Kesselring}, \citenamefont {Pastawski}, \citenamefont {Eisert},\ and\ \citenamefont {Brown}}]{Kesselring2018boundariestwist}%
  \BibitemOpen
  \bibfield  {author} {\bibinfo {author} {\bibfnamefont {M.~S.}\ \bibnamefont {Kesselring}}, \bibinfo {author} {\bibfnamefont {F.}~\bibnamefont {Pastawski}}, \bibinfo {author} {\bibfnamefont {J.}~\bibnamefont {Eisert}},\ and\ \bibinfo {author} {\bibfnamefont {B.~J.}\ \bibnamefont {Brown}},\ }\bibfield  {title} {\bibinfo {title} {The boundaries and twist defects of the color code and their applications to topological quantum computation},\ }\href {https://doi.org/10.22331/q-2018-10-19-101} {\bibfield  {journal} {\bibinfo  {journal} {{Quantum}}\ }\textbf {\bibinfo {volume} {2}},\ \bibinfo {pages} {101} (\bibinfo {year} {2018})}\BibitemShut {NoStop}%
\bibitem [{\citenamefont {Kitaev}(2003)}]{kitaev2003fault}%
  \BibitemOpen
  \bibfield  {author} {\bibinfo {author} {\bibfnamefont {A.~Y.}\ \bibnamefont {Kitaev}},\ }\bibfield  {title} {\bibinfo {title} {Fault-tolerant quantum computation by anyons},\ }\href {https://doi.org/10.1016/S0003-4916(02)00018-0} {\bibfield  {journal} {\bibinfo  {journal} {Ann. Phys.}\ }\textbf {\bibinfo {volume} {303}},\ \bibinfo {pages} {2} (\bibinfo {year} {2003})}\BibitemShut {NoStop}%
\bibitem [{\citenamefont {Kubica}\ \emph {et~al.}(2015)\citenamefont {Kubica}, \citenamefont {Yoshida},\ and\ \citenamefont {Pastawski}}]{kubica2015unfolding}%
  \BibitemOpen
  \bibfield  {author} {\bibinfo {author} {\bibfnamefont {A.}~\bibnamefont {Kubica}}, \bibinfo {author} {\bibfnamefont {B.}~\bibnamefont {Yoshida}},\ and\ \bibinfo {author} {\bibfnamefont {F.}~\bibnamefont {Pastawski}},\ }\bibfield  {title} {\bibinfo {title} {Unfolding the color code},\ }\href {https://doi.org/10.1088/1367-2630/17/8/083026} {\bibfield  {journal} {\bibinfo  {journal} {New J. Phys.}\ }\textbf {\bibinfo {volume} {17}},\ \bibinfo {pages} {083026} (\bibinfo {year} {2015})}\BibitemShut {NoStop}%
\bibitem [{\citenamefont {Bomb{\'\i}n}(2010)}]{Bombin2010subsystem}%
  \BibitemOpen
  \bibfield  {author} {\bibinfo {author} {\bibfnamefont {H.}~\bibnamefont {Bomb{\'\i}n}},\ }\bibfield  {title} {\bibinfo {title} {Topological subsystem codes},\ }\href {https://doi.org/10.1103/PhysRevA.81.032301} {\bibfield  {journal} {\bibinfo  {journal} {Phys. Rev. A}\ }\textbf {\bibinfo {volume} {81}},\ \bibinfo {pages} {032301} (\bibinfo {year} {2010})}\BibitemShut {NoStop}%
\bibitem [{\citenamefont {Kargarian}\ \emph {et~al.}(2010)\citenamefont {Kargarian}, \citenamefont {Bomb{\'\i}n},\ and\ \citenamefont {Martin-Delgado}}]{kargarian2010topological}%
  \BibitemOpen
  \bibfield  {author} {\bibinfo {author} {\bibfnamefont {M.}~\bibnamefont {Kargarian}}, \bibinfo {author} {\bibfnamefont {H.}~\bibnamefont {Bomb{\'\i}n}},\ and\ \bibinfo {author} {\bibfnamefont {M.~A.}\ \bibnamefont {Martin-Delgado}},\ }\bibfield  {title} {\bibinfo {title} {{Topological color codes and two-body quantum lattice Hamiltonians}},\ }\href {https://doi.org/10.1088/1367-2630/12/2/025018} {\bibfield  {journal} {\bibinfo  {journal} {New J. Phys.}\ }\textbf {\bibinfo {volume} {12}},\ \bibinfo {pages} {025018} (\bibinfo {year} {2010})}\BibitemShut {NoStop}%
\bibitem [{\citenamefont {Haah}\ and\ \citenamefont {Hastings}(2022)}]{haah2022boundaries}%
  \BibitemOpen
  \bibfield  {author} {\bibinfo {author} {\bibfnamefont {J.}~\bibnamefont {Haah}}\ and\ \bibinfo {author} {\bibfnamefont {M.~B.}\ \bibnamefont {Hastings}},\ }\bibfield  {title} {\bibinfo {title} {Boundaries for the honeycomb code},\ }\href {https://doi.org/10.22331/q-2022-04-21-693} {\bibfield  {journal} {\bibinfo  {journal} {Quantum}\ }\textbf {\bibinfo {volume} {6}},\ \bibinfo {pages} {693} (\bibinfo {year} {2022})}\BibitemShut {NoStop}%
\bibitem [{\citenamefont {McEwen}\ \emph {et~al.}(2023)\citenamefont {McEwen}, \citenamefont {Bacon},\ and\ \citenamefont {Gidney}}]{McEwen_2023}%
  \BibitemOpen
  \bibfield  {author} {\bibinfo {author} {\bibfnamefont {M.}~\bibnamefont {McEwen}}, \bibinfo {author} {\bibfnamefont {D.}~\bibnamefont {Bacon}},\ and\ \bibinfo {author} {\bibfnamefont {C.}~\bibnamefont {Gidney}},\ }\bibfield  {title} {\bibinfo {title} {Relaxing hardware requirements for surface code circuits using time-dynamics},\ }\href {https://doi.org/10.22331/q-2023-11-07-1172} {\bibfield  {journal} {\bibinfo  {journal} {Quantum}\ }\textbf {\bibinfo {volume} {7}},\ \bibinfo {pages} {1172} (\bibinfo {year} {2023})}\BibitemShut {NoStop}%
\bibitem [{\citenamefont {Raussendorf}\ \emph {et~al.}(2019)\citenamefont {Raussendorf}, \citenamefont {Okay}, \citenamefont {Wang}, \citenamefont {Stephen},\ and\ \citenamefont {Nautrup}}]{Raussendorf2019Computationally}%
  \BibitemOpen
  \bibfield  {author} {\bibinfo {author} {\bibfnamefont {R.}~\bibnamefont {Raussendorf}}, \bibinfo {author} {\bibfnamefont {C.}~\bibnamefont {Okay}}, \bibinfo {author} {\bibfnamefont {D.-S.}\ \bibnamefont {Wang}}, \bibinfo {author} {\bibfnamefont {D.~T.}\ \bibnamefont {Stephen}},\ and\ \bibinfo {author} {\bibfnamefont {H.~P.}\ \bibnamefont {Nautrup}},\ }\bibfield  {title} {\bibinfo {title} {Computationally universal phase of quantum matter},\ }\href {https://doi.org/10.1103/PhysRevLett.122.090501} {\bibfield  {journal} {\bibinfo  {journal} {Phys. Rev. Lett.}\ }\textbf {\bibinfo {volume} {122}},\ \bibinfo {pages} {090501} (\bibinfo {year} {2019})}\BibitemShut {NoStop}%
\bibitem [{\citenamefont {Cao}\ and\ \citenamefont {Lackey}(2022)}]{Cao2022Lego}%
  \BibitemOpen
  \bibfield  {author} {\bibinfo {author} {\bibfnamefont {C.}~\bibnamefont {Cao}}\ and\ \bibinfo {author} {\bibfnamefont {B.}~\bibnamefont {Lackey}},\ }\bibfield  {title} {\bibinfo {title} {{Quantum Lego: Building quantum error correction codes from tensor networks}},\ }\href {https://doi.org/10.1103/PRXQuantum.3.020332} {\bibfield  {journal} {\bibinfo  {journal} {PRX Quantum}\ }\textbf {\bibinfo {volume} {3}},\ \bibinfo {pages} {020332} (\bibinfo {year} {2022})}\BibitemShut {NoStop}%
\bibitem [{\citenamefont {Yeung}(2020)}]{yeung2020diagrammatic}%
  \BibitemOpen
  \bibfield  {author} {\bibinfo {author} {\bibfnamefont {R.}~\bibnamefont {Yeung}},\ }\href@noop {} {\bibinfo {title} {Diagrammatic design and study of ans\"{a}tze for quantum machine learning}} (\bibinfo {year} {2020}),\ \Eprint {https://arxiv.org/abs/2011.11073} {arXiv:2011.11073} \BibitemShut {NoStop}%
\bibitem [{\citenamefont {Lang}\ and\ \citenamefont {Coecke}(2012)}]{Lang_2012}%
  \BibitemOpen
  \bibfield  {author} {\bibinfo {author} {\bibfnamefont {A.}~\bibnamefont {Lang}}\ and\ \bibinfo {author} {\bibfnamefont {B.}~\bibnamefont {Coecke}},\ }\bibfield  {title} {\bibinfo {title} {Trichromatic open digraphs for understanding qubits},\ }\href {https://doi.org/10.4204/eptcs.95.14} {\bibfield  {journal} {\bibinfo  {journal} {Electronic Proceedings in Theoretical Computer Science}\ }\textbf {\bibinfo {volume} {95}},\ \bibinfo {pages} {193–209} (\bibinfo {year} {2012})}\BibitemShut {NoStop}%
\bibitem [{\citenamefont {Bauer}(2024{\natexlab{b}})}]{bauer2024lowoverhead}%
  \BibitemOpen
  \bibfield  {author} {\bibinfo {author} {\bibfnamefont {A.}~\bibnamefont {Bauer}},\ }\href@noop {} {\bibinfo {title} {{Low-overhead non-Clifford topological fault-tolerant circuits for all non-chiral Abelian topological phases}}} (\bibinfo {year} {2024}{\natexlab{b}}),\ \Eprint {https://arxiv.org/abs/2403.12119} {arXiv:2403.12119} \BibitemShut {NoStop}%
\bibitem [{\citenamefont {Bomb\'{\i}n}\ \emph {et~al.}(2023)\citenamefont {Bomb\'{\i}n}, \citenamefont {Dawson}, \citenamefont {Mishmash}, \citenamefont {Nickerson}, \citenamefont {Pastawski},\ and\ \citenamefont {Roberts}}]{Bombin2023logical}%
  \BibitemOpen
  \bibfield  {author} {\bibinfo {author} {\bibfnamefont {H.}~\bibnamefont {Bomb\'{\i}n}}, \bibinfo {author} {\bibfnamefont {C.}~\bibnamefont {Dawson}}, \bibinfo {author} {\bibfnamefont {R.~V.}\ \bibnamefont {Mishmash}}, \bibinfo {author} {\bibfnamefont {N.}~\bibnamefont {Nickerson}}, \bibinfo {author} {\bibfnamefont {F.}~\bibnamefont {Pastawski}},\ and\ \bibinfo {author} {\bibfnamefont {S.}~\bibnamefont {Roberts}},\ }\bibfield  {title} {\bibinfo {title} {Logical blocks for fault-tolerant topological quantum computation},\ }\href {https://doi.org/10.1103/PRXQuantum.4.020303} {\bibfield  {journal} {\bibinfo  {journal} {PRX Quantum}\ }\textbf {\bibinfo {volume} {4}},\ \bibinfo {pages} {020303} (\bibinfo {year} {2023})}\BibitemShut {NoStop}%
\bibitem [{\citenamefont {Gidney}(2021)}]{gidney2021stim}%
  \BibitemOpen
  \bibfield  {author} {\bibinfo {author} {\bibfnamefont {C.}~\bibnamefont {Gidney}},\ }\bibfield  {title} {\bibinfo {title} {Stim: a fast stabilizer circuit simulator},\ }\href {https://doi.org/10.22331/q-2021-07-06-497} {\bibfield  {journal} {\bibinfo  {journal} {{Quantum}}\ }\textbf {\bibinfo {volume} {5}},\ \bibinfo {pages} {497} (\bibinfo {year} {2021})}\BibitemShut {NoStop}%
\bibitem [{\citenamefont {Derks}\ \emph {et~al.}(2024)\citenamefont {Derks}, \citenamefont {Townsend-Teague}, \citenamefont {Burchards},\ and\ \citenamefont {Eisert}}]{DetectorErrorModels}%
  \BibitemOpen
  \bibfield  {author} {\bibinfo {author} {\bibfnamefont {P.-J. H.~S.}\ \bibnamefont {Derks}}, \bibinfo {author} {\bibfnamefont {A.}~\bibnamefont {Townsend-Teague}}, \bibinfo {author} {\bibfnamefont {A.~G.}\ \bibnamefont {Burchards}},\ and\ \bibinfo {author} {\bibfnamefont {J.}~\bibnamefont {Eisert}},\ }\bibfield  {title} {\bibinfo {title} {Designing fault-tolerant circuits using detector error models},\ }\href@noop {} {\bibfield  {journal} {\bibinfo  {journal} {arXiv}\ } (\bibinfo {year} {2024})},\ \Eprint {https://arxiv.org/abs/2407.13826} {2407.13826} \BibitemShut {NoStop}%
\bibitem [{\citenamefont {Delfosse}\ and\ \citenamefont {Paetznick}(2023)}]{delfosse2023spacetime}%
  \BibitemOpen
  \bibfield  {author} {\bibinfo {author} {\bibfnamefont {N.}~\bibnamefont {Delfosse}}\ and\ \bibinfo {author} {\bibfnamefont {A.}~\bibnamefont {Paetznick}},\ }\bibfield  {title} {\bibinfo {title} {{Spacetime codes of Clifford circuits}},\ }\href@noop {} {\bibfield  {journal} {\bibinfo  {journal} {arXiv}\ } (\bibinfo {year} {2023})},\ \Eprint {https://arxiv.org/abs/2304.05943} {arXiv:2304.05943} \BibitemShut {NoStop}%
\bibitem [{\citenamefont {Gidney}(2022)}]{gidney2022stability}%
  \BibitemOpen
  \bibfield  {author} {\bibinfo {author} {\bibfnamefont {C.}~\bibnamefont {Gidney}},\ }\bibfield  {title} {\bibinfo {title} {{Stability experiments: The overlooked dual of memory experiments}},\ }\href {https://doi.org/10.22331/q-2022-08-24-786} {\bibfield  {journal} {\bibinfo  {journal} {Quantum}\ }\textbf {\bibinfo {volume} {6}},\ \bibinfo {pages} {786} (\bibinfo {year} {2022})}\BibitemShut {NoStop}%
\bibitem [{\citenamefont {Brown}(2023)}]{brown2023conservation}%
  \BibitemOpen
  \bibfield  {author} {\bibinfo {author} {\bibfnamefont {B.~J.}\ \bibnamefont {Brown}},\ }\bibfield  {title} {\bibinfo {title} {Conservation laws and quantum error correction: towards a generalised matching decoder},\ }\href {https://doi.org/10.1109/MBITS.2023.3246025} {\bibfield  {journal} {\bibinfo  {journal} {IEEE Bits Inf. Th. Ma.}\ }\textbf {\bibinfo {volume} {2}},\ \bibinfo {pages} {5} (\bibinfo {year} {2023})}\BibitemShut {NoStop}%
\bibitem [{\citenamefont {Andrist}\ \emph {et~al.}(2011)\citenamefont {Andrist}, \citenamefont {Katzgraber}, \citenamefont {Bombin},\ and\ \citenamefont {Martin-Delgado}}]{andrist2010tricolored}%
  \BibitemOpen
  \bibfield  {author} {\bibinfo {author} {\bibfnamefont {R.~S.}\ \bibnamefont {Andrist}}, \bibinfo {author} {\bibfnamefont {H.~G.}\ \bibnamefont {Katzgraber}}, \bibinfo {author} {\bibfnamefont {H.}~\bibnamefont {Bombin}},\ and\ \bibinfo {author} {\bibfnamefont {M.~A.}\ \bibnamefont {Martin-Delgado}},\ }\bibfield  {title} {\bibinfo {title} {Tricolored lattice gauge theory with randomness: fault tolerance in topological color codes},\ }\href {https://doi.org/10.1088/1367-2630/13/8/083006} {\bibfield  {journal} {\bibinfo  {journal} {New Journal of Physics}\ }\textbf {\bibinfo {volume} {13}},\ \bibinfo {pages} {083006} (\bibinfo {year} {2011})}\BibitemShut {NoStop}%
\bibitem [{\citenamefont {Stephens}(2014)}]{stephens2014efficient}%
  \BibitemOpen
  \bibfield  {author} {\bibinfo {author} {\bibfnamefont {A.~M.}\ \bibnamefont {Stephens}},\ }\href {https://arxiv.org/abs/1402.3037} {\bibinfo {title} {Efficient fault-tolerant decoding of topological color codes}} (\bibinfo {year} {2014}),\ \Eprint {https://arxiv.org/abs/1402.3037} {arXiv:1402.3037 [quant-ph]} \BibitemShut {NoStop}%
\bibitem [{\citenamefont {Gidney}\ and\ \citenamefont {Jones}(2023)}]{gidney2023new}%
  \BibitemOpen
  \bibfield  {author} {\bibinfo {author} {\bibfnamefont {C.}~\bibnamefont {Gidney}}\ and\ \bibinfo {author} {\bibfnamefont {C.}~\bibnamefont {Jones}},\ }\href@noop {} {\bibinfo {title} {{New circuits and an open source decoder for the color code}}} (\bibinfo {year} {2023}),\ \Eprint {https://arxiv.org/abs/2312.08813} {arXiv:2312.08813} \BibitemShut {NoStop}%
\bibitem [{\citenamefont {Lee}\ \emph {et~al.}(2024)\citenamefont {Lee}, \citenamefont {Li},\ and\ \citenamefont {Bartlett}}]{lee2024color}%
  \BibitemOpen
  \bibfield  {author} {\bibinfo {author} {\bibfnamefont {S.-H.}\ \bibnamefont {Lee}}, \bibinfo {author} {\bibfnamefont {A.}~\bibnamefont {Li}},\ and\ \bibinfo {author} {\bibfnamefont {S.~D.}\ \bibnamefont {Bartlett}},\ }\href {https://arxiv.org/abs/2404.07482} {\bibinfo {title} {Color code decoder with improved scaling for correcting circuit-level noise}} (\bibinfo {year} {2024}),\ \Eprint {https://arxiv.org/abs/2404.07482} {arXiv:2404.07482} \BibitemShut {NoStop}%
\bibitem [{\citenamefont {Kovalev}\ and\ \citenamefont {Pryadko}(2013)}]{kovalev2013fault}%
  \BibitemOpen
  \bibfield  {author} {\bibinfo {author} {\bibfnamefont {A.~A.}\ \bibnamefont {Kovalev}}\ and\ \bibinfo {author} {\bibfnamefont {L.~P.}\ \bibnamefont {Pryadko}},\ }\bibfield  {title} {\bibinfo {title} {Fault tolerance of quantum low-density parity check codes with sublinear distance scaling},\ }\href {https://doi.org/10.1103/PhysRevA.87.020304} {\bibfield  {journal} {\bibinfo  {journal} {Phys. Rev. A}\ }\textbf {\bibinfo {volume} {87}},\ \bibinfo {pages} {020304} (\bibinfo {year} {2013})}\BibitemShut {NoStop}%
\bibitem [{\citenamefont {Gidney}\ \emph {et~al.}(2021)\citenamefont {Gidney}, \citenamefont {Newman}, \citenamefont {Fowler},\ and\ \citenamefont {Broughton}}]{gidney2021faulttolerant}%
  \BibitemOpen
  \bibfield  {author} {\bibinfo {author} {\bibfnamefont {C.}~\bibnamefont {Gidney}}, \bibinfo {author} {\bibfnamefont {M.}~\bibnamefont {Newman}}, \bibinfo {author} {\bibfnamefont {A.}~\bibnamefont {Fowler}},\ and\ \bibinfo {author} {\bibfnamefont {M.}~\bibnamefont {Broughton}},\ }\bibfield  {title} {\bibinfo {title} {{A fault-tolerant honeycomb memory}},\ }\href {https://doi.org/10.22331/q-2021-12-20-605} {\bibfield  {journal} {\bibinfo  {journal} {Quantum}\ }\textbf {\bibinfo {volume} {5}},\ \bibinfo {pages} {605} (\bibinfo {year} {2021})},\ \Eprint {https://arxiv.org/abs/2108.10457v2} {2108.10457v2} \BibitemShut {NoStop}%
\bibitem [{\citenamefont {Knapp}\ \emph {et~al.}(2018)\citenamefont {Knapp}, \citenamefont {Beverland}, \citenamefont {Pikulin},\ and\ \citenamefont {Karzig}}]{knapp2018modeling}%
  \BibitemOpen
  \bibfield  {author} {\bibinfo {author} {\bibfnamefont {C.}~\bibnamefont {Knapp}}, \bibinfo {author} {\bibfnamefont {M.}~\bibnamefont {Beverland}}, \bibinfo {author} {\bibfnamefont {D.~I.}\ \bibnamefont {Pikulin}},\ and\ \bibinfo {author} {\bibfnamefont {T.}~\bibnamefont {Karzig}},\ }\bibfield  {title} {\bibinfo {title} {{Modeling noise and error correction for Majorana-based quantum computing}},\ }\href {https://doi.org/10.22331/q-2018-09-03-88} {\bibfield  {journal} {\bibinfo  {journal} {Quantum}\ }\textbf {\bibinfo {volume} {2}},\ \bibinfo {pages} {88} (\bibinfo {year} {2018})}\BibitemShut {NoStop}%
\bibitem [{\citenamefont {Panteleev}\ and\ \citenamefont {Kalachev}(2021)}]{panteleev2021degenerate}%
  \BibitemOpen
  \bibfield  {author} {\bibinfo {author} {\bibfnamefont {P.}~\bibnamefont {Panteleev}}\ and\ \bibinfo {author} {\bibfnamefont {G.}~\bibnamefont {Kalachev}},\ }\bibfield  {title} {\bibinfo {title} {{Degenerate quantum LDPC codes with good finite length performance}},\ }\href {https://doi.org/10.22331/q-2021-11-22-585} {\bibfield  {journal} {\bibinfo  {journal} {Quantum}\ }\textbf {\bibinfo {volume} {5}},\ \bibinfo {pages} {585} (\bibinfo {year} {2021})},\ \Eprint {https://arxiv.org/abs/1904.02703v3} {1904.02703v3} \BibitemShut {NoStop}%
\bibitem [{\citenamefont {Roffe}(2022)}]{roffe2022ldpc}%
  \BibitemOpen
  \bibfield  {author} {\bibinfo {author} {\bibfnamefont {J.}~\bibnamefont {Roffe}},\ }\href {https://pypi.org/project/ldpc/} {\bibinfo {title} {{LDPC: Python tools for low density parity check codes}}} (\bibinfo {year} {2022})\BibitemShut {NoStop}%
\bibitem [{\citenamefont {Hillmann}\ \emph {et~al.}(2024)\citenamefont {Hillmann}, \citenamefont {Berent}, \citenamefont {Quintavalle}, \citenamefont {Eisert}, \citenamefont {Wille},\ and\ \citenamefont {Roffe}}]{hillmann2024localized}%
  \BibitemOpen
  \bibfield  {author} {\bibinfo {author} {\bibfnamefont {T.}~\bibnamefont {Hillmann}}, \bibinfo {author} {\bibfnamefont {L.}~\bibnamefont {Berent}}, \bibinfo {author} {\bibfnamefont {A.~O.}\ \bibnamefont {Quintavalle}}, \bibinfo {author} {\bibfnamefont {J.}~\bibnamefont {Eisert}}, \bibinfo {author} {\bibfnamefont {R.}~\bibnamefont {Wille}},\ and\ \bibinfo {author} {\bibfnamefont {J.}~\bibnamefont {Roffe}},\ }\bibfield  {title} {\bibinfo {title} {{Localized statistics decoding: A parallel decoding algorithm for quantum low-density parity-check codes}},\ }\bibfield  {journal} {\bibinfo  {journal} {arXiv}\ }\href {https://doi.org/10.48550/arXiv.2406.18655} {10.48550/arXiv.2406.18655} (\bibinfo {year} {2024}),\ \Eprint {https://arxiv.org/abs/2406.18655} {2406.18655} \BibitemShut {NoStop}%
\bibitem [{\citenamefont {Gidney}\ \emph {et~al.}(2022)\citenamefont {Gidney}, \citenamefont {Newman},\ and\ \citenamefont {McEwen}}]{gidney2022benchmarking}%
  \BibitemOpen
  \bibfield  {author} {\bibinfo {author} {\bibfnamefont {C.}~\bibnamefont {Gidney}}, \bibinfo {author} {\bibfnamefont {M.}~\bibnamefont {Newman}},\ and\ \bibinfo {author} {\bibfnamefont {M.}~\bibnamefont {McEwen}},\ }\bibfield  {title} {\bibinfo {title} {{Benchmarking the planar honeycomb code}},\ }\href {https://doi.org/10.22331/q-2022-09-21-813} {\bibfield  {journal} {\bibinfo  {journal} {Quantum}\ }\textbf {\bibinfo {volume} {6}},\ \bibinfo {pages} {813} (\bibinfo {year} {2022})},\ \Eprint {https://arxiv.org/abs/2202.11845v3} {2202.11845v3} \BibitemShut {NoStop}%
\bibitem [{\citenamefont {Wang}\ \emph {et~al.}(2010)\citenamefont {Wang}, \citenamefont {Fowler}, \citenamefont {Hill},\ and\ \citenamefont {Hollenberg}}]{wang2010graphical}%
  \BibitemOpen
  \bibfield  {author} {\bibinfo {author} {\bibfnamefont {D.~S.}\ \bibnamefont {Wang}}, \bibinfo {author} {\bibfnamefont {A.~G.}\ \bibnamefont {Fowler}}, \bibinfo {author} {\bibfnamefont {C.~D.}\ \bibnamefont {Hill}},\ and\ \bibinfo {author} {\bibfnamefont {L.~C.}\ \bibnamefont {Hollenberg}},\ }\bibfield  {title} {\bibinfo {title} {Graphical algorithms and threshold error rates for the 2d color code},\ }\href@noop {} {\bibfield  {journal} {\bibinfo  {journal} {Quantum Information \& Computation}\ }\textbf {\bibinfo {volume} {10}},\ \bibinfo {pages} {780} (\bibinfo {year} {2010})}\BibitemShut {NoStop}%
\bibitem [{\citenamefont {Chamberland}\ \emph {et~al.}(2020)\citenamefont {Chamberland}, \citenamefont {Kubica}, \citenamefont {Yoder},\ and\ \citenamefont {Zhu}}]{chamberland2020triangular}%
  \BibitemOpen
  \bibfield  {author} {\bibinfo {author} {\bibfnamefont {C.}~\bibnamefont {Chamberland}}, \bibinfo {author} {\bibfnamefont {A.}~\bibnamefont {Kubica}}, \bibinfo {author} {\bibfnamefont {T.~J.}\ \bibnamefont {Yoder}},\ and\ \bibinfo {author} {\bibfnamefont {G.}~\bibnamefont {Zhu}},\ }\bibfield  {title} {\bibinfo {title} {{Triangular color codes on trivalent graphs with flag qubits}},\ }\href {https://doi.org/10.1088/1367-2630/ab68fd} {\bibfield  {journal} {\bibinfo  {journal} {New J. Phys.}\ }\textbf {\bibinfo {volume} {22}},\ \bibinfo {pages} {023019} (\bibinfo {year} {2020})}\BibitemShut {NoStop}%
\bibitem [{\citenamefont {Higgott}\ and\ \citenamefont {Breuckmann}(2023)}]{higgott2023improved}%
  \BibitemOpen
  \bibfield  {author} {\bibinfo {author} {\bibfnamefont {O.}~\bibnamefont {Higgott}}\ and\ \bibinfo {author} {\bibfnamefont {N.~P.}\ \bibnamefont {Breuckmann}},\ }\bibfield  {title} {\bibinfo {title} {{Improved single-shot decoding of higher-dimensional hypergraph-product codes}},\ }\href {https://doi.org/10.1103/PRXQuantum.4.020332} {\bibfield  {journal} {\bibinfo  {journal} {PRX Quantum}\ }\textbf {\bibinfo {volume} {4}},\ \bibinfo {pages} {020332} (\bibinfo {year} {2023})}\BibitemShut {NoStop}%
\bibitem [{\citenamefont {Tan}\ \emph {et~al.}(2023)\citenamefont {Tan}, \citenamefont {Zhang}, \citenamefont {Chao}, \citenamefont {Shi},\ and\ \citenamefont {Chen}}]{tan2023scalable}%
  \BibitemOpen
  \bibfield  {author} {\bibinfo {author} {\bibfnamefont {X.}~\bibnamefont {Tan}}, \bibinfo {author} {\bibfnamefont {F.}~\bibnamefont {Zhang}}, \bibinfo {author} {\bibfnamefont {R.}~\bibnamefont {Chao}}, \bibinfo {author} {\bibfnamefont {Y.}~\bibnamefont {Shi}},\ and\ \bibinfo {author} {\bibfnamefont {J.}~\bibnamefont {Chen}},\ }\bibfield  {title} {\bibinfo {title} {{Scalable surface-code decoders with parallelization in time}},\ }\href {https://doi.org/10.1103/PRXQuantum.4.040344} {\bibfield  {journal} {\bibinfo  {journal} {PRX Quantum}\ }\textbf {\bibinfo {volume} {4}},\ \bibinfo {pages} {040344} (\bibinfo {year} {2023})}\BibitemShut {NoStop}%
\bibitem [{\citenamefont {Skoric}\ \emph {et~al.}(2023)\citenamefont {Skoric}, \citenamefont {Browne}, \citenamefont {Barnes}, \citenamefont {Gillespie},\ and\ \citenamefont {Campbell}}]{ParallelWindow}%
  \BibitemOpen
  \bibfield  {author} {\bibinfo {author} {\bibfnamefont {L.}~\bibnamefont {Skoric}}, \bibinfo {author} {\bibfnamefont {D.~E.}\ \bibnamefont {Browne}}, \bibinfo {author} {\bibfnamefont {K.~M.}\ \bibnamefont {Barnes}}, \bibinfo {author} {\bibfnamefont {N.~I.}\ \bibnamefont {Gillespie}},\ and\ \bibinfo {author} {\bibfnamefont {E.~T.}\ \bibnamefont {Campbell}},\ }\bibfield  {title} {\bibinfo {title} {Parallel window decoding enables scalable fault tolerant quantum computation},\ }\href {https://doi.org/10.1038/s41467-023-42482-1} {\bibfield  {journal} {\bibinfo  {journal} {Nature Comm.}\ }\textbf {\bibinfo {volume} {14}},\ \bibinfo {pages} {7040} (\bibinfo {year} {2023})}\BibitemShut {NoStop}%
\bibitem [{\citenamefont {Meinerz}\ \emph {et~al.}(2022)\citenamefont {Meinerz}, \citenamefont {Park},\ and\ \citenamefont {Trebst}}]{PhysRevLett.128.080505}%
  \BibitemOpen
  \bibfield  {author} {\bibinfo {author} {\bibfnamefont {K.}~\bibnamefont {Meinerz}}, \bibinfo {author} {\bibfnamefont {C.-Y.}\ \bibnamefont {Park}},\ and\ \bibinfo {author} {\bibfnamefont {S.}~\bibnamefont {Trebst}},\ }\bibfield  {title} {\bibinfo {title} {Scalable neural decoder for topological surface codes},\ }\href {https://doi.org/10.1103/PhysRevLett.128.080505} {\bibfield  {journal} {\bibinfo  {journal} {Phys. Rev. Lett.}\ }\textbf {\bibinfo {volume} {128}},\ \bibinfo {pages} {080505} (\bibinfo {year} {2022})}\BibitemShut {NoStop}%
\bibitem [{\citenamefont {Sahay}\ and\ \citenamefont {Brown}(2022)}]{sahay2022decoder}%
  \BibitemOpen
  \bibfield  {author} {\bibinfo {author} {\bibfnamefont {K.}~\bibnamefont {Sahay}}\ and\ \bibinfo {author} {\bibfnamefont {B.~J.}\ \bibnamefont {Brown}},\ }\bibfield  {title} {\bibinfo {title} {{Decoder for the triangular color code by matching on a Möbius strip}},\ }\href {https://doi.org/10.1103/PRXQuantum.3.010310} {\bibfield  {journal} {\bibinfo  {journal} {PRX Quantum}\ }\textbf {\bibinfo {volume} {3}},\ \bibinfo {pages} {010310} (\bibinfo {year} {2022})}\BibitemShut {NoStop}%
\bibitem [{\citenamefont {Horsman}\ \emph {et~al.}(2012)\citenamefont {Horsman}, \citenamefont {Fowler}, \citenamefont {Devitt},\ and\ \citenamefont {Van~Meter}}]{horsman2012surface}%
  \BibitemOpen
  \bibfield  {author} {\bibinfo {author} {\bibfnamefont {D.}~\bibnamefont {Horsman}}, \bibinfo {author} {\bibfnamefont {A.~G.}\ \bibnamefont {Fowler}}, \bibinfo {author} {\bibfnamefont {S.}~\bibnamefont {Devitt}},\ and\ \bibinfo {author} {\bibfnamefont {R.}~\bibnamefont {Van~Meter}},\ }\bibfield  {title} {\bibinfo {title} {{Surface code quantum computing by lattice surgery}},\ }\href {https://doi.org/10.1088/1367-2630/14/12/123011} {\bibfield  {journal} {\bibinfo  {journal} {New J. Phys.}\ }\textbf {\bibinfo {volume} {14}},\ \bibinfo {pages} {123011} (\bibinfo {year} {2012})}\BibitemShut {NoStop}%
\bibitem [{\citenamefont {Kapustin}\ and\ \citenamefont {Saulina}(2011)}]{kapustin2011topological}%
  \BibitemOpen
  \bibfield  {author} {\bibinfo {author} {\bibfnamefont {A.}~\bibnamefont {Kapustin}}\ and\ \bibinfo {author} {\bibfnamefont {N.}~\bibnamefont {Saulina}},\ }\bibfield  {title} {\bibinfo {title} {{Topological boundary conditions in Abelian Chern--Simons theory}},\ }\href {https://doi.org/10.1016/j.nuclphysb.2010.12.017} {\bibfield  {journal} {\bibinfo  {journal} {Nucl. Phys. B}\ }\textbf {\bibinfo {volume} {845}},\ \bibinfo {pages} {393} (\bibinfo {year} {2011})}\BibitemShut {NoStop}%
\bibitem [{\citenamefont {Levin}(2013)}]{Levin2013Protected}%
  \BibitemOpen
  \bibfield  {author} {\bibinfo {author} {\bibfnamefont {M.}~\bibnamefont {Levin}},\ }\bibfield  {title} {\bibinfo {title} {Protected edge modes without symmetry},\ }\href {https://doi.org/10.1103/PhysRevX.3.021009} {\bibfield  {journal} {\bibinfo  {journal} {Phys. Rev. X}\ }\textbf {\bibinfo {volume} {3}},\ \bibinfo {pages} {021009} (\bibinfo {year} {2013})}\BibitemShut {NoStop}%
\bibitem [{\citenamefont {Haah}(2021)}]{haah2021classification}%
  \BibitemOpen
  \bibfield  {author} {\bibinfo {author} {\bibfnamefont {J.}~\bibnamefont {Haah}},\ }\bibfield  {title} {\bibinfo {title} {{Classification of translation invariant topological Pauli stabilizer codes for prime dimensional qudits on two-dimensional lattices}},\ }\href {https://doi.org/10.1063/5.0021068} {\bibfield  {journal} {\bibinfo  {journal} {J. Math. Phys.}\ }\textbf {\bibinfo {volume} {62}},\ \bibinfo {pages} {012201} (\bibinfo {year} {2021})}\BibitemShut {NoStop}%
\bibitem [{\citenamefont {Kapustin}\ and\ \citenamefont {Fidkowski}(2019)}]{Kapustin2019}%
  \BibitemOpen
  \bibfield  {author} {\bibinfo {author} {\bibfnamefont {A.}~\bibnamefont {Kapustin}}\ and\ \bibinfo {author} {\bibfnamefont {L.}~\bibnamefont {Fidkowski}},\ }\bibfield  {title} {\bibinfo {title} {{Local commuting projector Hamiltonians and the quantum Hall effect}},\ }\href {https://doi.org/10.1007/s00220-019-03444-1} {\bibfield  {journal} {\bibinfo  {journal} {Commun, Math. Phys.}\ }\textbf {\bibinfo {volume} {373}},\ \bibinfo {pages} {763–769} (\bibinfo {year} {2019})}\BibitemShut {NoStop}%
\bibitem [{\citenamefont {Davydova}\ \emph {et~al.}(2023)\citenamefont {Davydova}, \citenamefont {Tantivasadakarn},\ and\ \citenamefont {Balasubramanian}}]{Davydova2023Floquet}%
  \BibitemOpen
  \bibfield  {author} {\bibinfo {author} {\bibfnamefont {M.}~\bibnamefont {Davydova}}, \bibinfo {author} {\bibfnamefont {N.}~\bibnamefont {Tantivasadakarn}},\ and\ \bibinfo {author} {\bibfnamefont {S.}~\bibnamefont {Balasubramanian}},\ }\bibfield  {title} {\bibinfo {title} {Floquet codes without parent subsystem codes},\ }\href {https://doi.org/10.1103/PRXQuantum.4.020341} {\bibfield  {journal} {\bibinfo  {journal} {PRX Quantum}\ }\textbf {\bibinfo {volume} {4}},\ \bibinfo {pages} {020341} (\bibinfo {year} {2023})}\BibitemShut {NoStop}%
\bibitem [{\citenamefont {Roberts}\ \emph {et~al.}(2017)\citenamefont {Roberts}, \citenamefont {Yoshida}, \citenamefont {Kubica},\ and\ \citenamefont {Bartlett}}]{roberts2017symmetry}%
  \BibitemOpen
  \bibfield  {author} {\bibinfo {author} {\bibfnamefont {S.}~\bibnamefont {Roberts}}, \bibinfo {author} {\bibfnamefont {B.}~\bibnamefont {Yoshida}}, \bibinfo {author} {\bibfnamefont {A.}~\bibnamefont {Kubica}},\ and\ \bibinfo {author} {\bibfnamefont {S.~D.}\ \bibnamefont {Bartlett}},\ }\bibfield  {title} {\bibinfo {title} {Symmetry-protected topological order at nonzero temperature},\ }\href {https://doi.org/10.1103/PhysRevA.96.022306} {\bibfield  {journal} {\bibinfo  {journal} {Phys. Rev. A}\ }\textbf {\bibinfo {volume} {96}},\ \bibinfo {pages} {022306} (\bibinfo {year} {2017})}\BibitemShut {NoStop}%
\bibitem [{\citenamefont {Iverson}\ and\ \citenamefont {Preskill}(2020)}]{Iverson2020Coherence}%
  \BibitemOpen
  \bibfield  {author} {\bibinfo {author} {\bibfnamefont {J.~K.}\ \bibnamefont {Iverson}}\ and\ \bibinfo {author} {\bibfnamefont {J.}~\bibnamefont {Preskill}},\ }\bibfield  {title} {\bibinfo {title} {Coherence in logical quantum channels},\ }\href {https://doi.org/10.1088/1367-2630/ab8e5c} {\bibfield  {journal} {\bibinfo  {journal} {New J. Phys.}\ }\textbf {\bibinfo {volume} {22}},\ \bibinfo {pages} {073066} (\bibinfo {year} {2020})}\BibitemShut {NoStop}%
\bibitem [{\citenamefont {Gidney}\ and\ \citenamefont {Bacon}(2023)}]{gidney2023less}%
  \BibitemOpen
  \bibfield  {author} {\bibinfo {author} {\bibfnamefont {C.}~\bibnamefont {Gidney}}\ and\ \bibinfo {author} {\bibfnamefont {D.}~\bibnamefont {Bacon}},\ }\bibfield  {title} {\bibinfo {title} {{Less bacon more threshold}},\ }\href@noop {} {\bibfield  {journal} {\bibinfo  {journal} {arXiv}\ } (\bibinfo {year} {2023})},\ \Eprint {https://arxiv.org/abs/2305.12046} {arXiv:2305.12046} \BibitemShut {NoStop}%
\bibitem [{\citenamefont {Haah}(2011)}]{Haah2011local}%
  \BibitemOpen
  \bibfield  {author} {\bibinfo {author} {\bibfnamefont {J.}~\bibnamefont {Haah}},\ }\bibfield  {title} {\bibinfo {title} {Local stabilizer codes in three dimensions without string logical operators},\ }\href {https://doi.org/10.1103/PhysRevA.83.042330} {\bibfield  {journal} {\bibinfo  {journal} {Phys. Rev. A}\ }\textbf {\bibinfo {volume} {83}},\ \bibinfo {pages} {042330} (\bibinfo {year} {2011})}\BibitemShut {NoStop}%
\bibitem [{\citenamefont {Brown}\ and\ \citenamefont {Williamson}(2020)}]{Brown2020parallelized}%
  \BibitemOpen
  \bibfield  {author} {\bibinfo {author} {\bibfnamefont {B.~J.}\ \bibnamefont {Brown}}\ and\ \bibinfo {author} {\bibfnamefont {D.~J.}\ \bibnamefont {Williamson}},\ }\bibfield  {title} {\bibinfo {title} {Parallelized quantum error correction with fracton topological codes},\ }\href {https://doi.org/10.1103/PhysRevResearch.2.013303} {\bibfield  {journal} {\bibinfo  {journal} {Phys. Rev. Res.}\ }\textbf {\bibinfo {volume} {2}},\ \bibinfo {pages} {013303} (\bibinfo {year} {2020})}\BibitemShut {NoStop}%
\bibitem [{\citenamefont {Hamma}\ \emph {et~al.}(2005)\citenamefont {Hamma}, \citenamefont {Zanardi},\ and\ \citenamefont {Wen}}]{Hamma2005String}%
  \BibitemOpen
  \bibfield  {author} {\bibinfo {author} {\bibfnamefont {A.}~\bibnamefont {Hamma}}, \bibinfo {author} {\bibfnamefont {P.}~\bibnamefont {Zanardi}},\ and\ \bibinfo {author} {\bibfnamefont {X.-G.}\ \bibnamefont {Wen}},\ }\bibfield  {title} {\bibinfo {title} {String and membrane condensation on three-dimensional lattices},\ }\href {https://doi.org/10.1103/PhysRevB.72.035307} {\bibfield  {journal} {\bibinfo  {journal} {Phys. Rev. B}\ }\textbf {\bibinfo {volume} {72}},\ \bibinfo {pages} {035307} (\bibinfo {year} {2005})}\BibitemShut {NoStop}%
\bibitem [{\citenamefont {Bomb\'{\i}n}(2015)}]{Bombin2015Single}%
  \BibitemOpen
  \bibfield  {author} {\bibinfo {author} {\bibfnamefont {H.}~\bibnamefont {Bomb\'{\i}n}},\ }\bibfield  {title} {\bibinfo {title} {Single-shot fault-tolerant quantum error correction},\ }\href {https://doi.org/10.1103/PhysRevX.5.031043} {\bibfield  {journal} {\bibinfo  {journal} {Phys. Rev. X}\ }\textbf {\bibinfo {volume} {5}},\ \bibinfo {pages} {031043} (\bibinfo {year} {2015})}\BibitemShut {NoStop}%
\bibitem [{\citenamefont {Bridgeman}\ \emph {et~al.}(2024)\citenamefont {Bridgeman}, \citenamefont {Kubica},\ and\ \citenamefont {Vasmer}}]{Bridgeman2024lifting}%
  \BibitemOpen
  \bibfield  {author} {\bibinfo {author} {\bibfnamefont {J.~C.}\ \bibnamefont {Bridgeman}}, \bibinfo {author} {\bibfnamefont {A.}~\bibnamefont {Kubica}},\ and\ \bibinfo {author} {\bibfnamefont {M.}~\bibnamefont {Vasmer}},\ }\bibfield  {title} {\bibinfo {title} {Lifting topological codes: Three-dimensional subsystem codes from two-dimensional anyon models},\ }\href {https://doi.org/10.1103/PRXQuantum.5.020310} {\bibfield  {journal} {\bibinfo  {journal} {PRX Quantum}\ }\textbf {\bibinfo {volume} {5}},\ \bibinfo {pages} {020310} (\bibinfo {year} {2024})}\BibitemShut {NoStop}%
\bibitem [{\citenamefont {Zhou}\ \emph {et~al.}(2024)\citenamefont {Zhou}, \citenamefont {Zhao}, \citenamefont {Cain}, \citenamefont {Bluvstein}, \citenamefont {Duckering}, \citenamefont {Hu}, \citenamefont {Wang}, \citenamefont {Kubica},\ and\ \citenamefont {Lukin}}]{zhou2024algorithmic}%
  \BibitemOpen
  \bibfield  {author} {\bibinfo {author} {\bibfnamefont {H.}~\bibnamefont {Zhou}}, \bibinfo {author} {\bibfnamefont {C.}~\bibnamefont {Zhao}}, \bibinfo {author} {\bibfnamefont {M.}~\bibnamefont {Cain}}, \bibinfo {author} {\bibfnamefont {D.}~\bibnamefont {Bluvstein}}, \bibinfo {author} {\bibfnamefont {C.}~\bibnamefont {Duckering}}, \bibinfo {author} {\bibfnamefont {H.-Y.}\ \bibnamefont {Hu}}, \bibinfo {author} {\bibfnamefont {S.-T.}\ \bibnamefont {Wang}}, \bibinfo {author} {\bibfnamefont {A.}~\bibnamefont {Kubica}},\ and\ \bibinfo {author} {\bibfnamefont {M.~D.}\ \bibnamefont {Lukin}},\ }\bibfield  {title} {\bibinfo {title} {{Algorithmic fault tolerance for fast quantum computing}},\ }\bibfield  {journal} {\bibinfo  {journal} {arXiv}\ }\href {https://doi.org/10.48550/arXiv.2406.17653} {10.48550/arXiv.2406.17653} (\bibinfo {year} {2024}),\ \Eprint {https://arxiv.org/abs/2406.17653} {2406.17653} \BibitemShut {NoStop}%
\bibitem [{\citenamefont {Cohen}\ \emph {et~al.}(2022)\citenamefont {Cohen}, \citenamefont {Kim}, \citenamefont {Bartlett},\ and\ \citenamefont {Brown}}]{cohen2022low}%
  \BibitemOpen
  \bibfield  {author} {\bibinfo {author} {\bibfnamefont {L.~Z.}\ \bibnamefont {Cohen}}, \bibinfo {author} {\bibfnamefont {I.~H.}\ \bibnamefont {Kim}}, \bibinfo {author} {\bibfnamefont {S.~D.}\ \bibnamefont {Bartlett}},\ and\ \bibinfo {author} {\bibfnamefont {B.~J.}\ \bibnamefont {Brown}},\ }\bibfield  {title} {\bibinfo {title} {{Low-overhead fault-tolerant quantum computing using long-range connectivity}},\ }\href {https://doi.org/10.1126/sciadv.abn1717} {\bibfield  {journal} {\bibinfo  {journal} {Sci. Adv.}\ }\textbf {\bibinfo {volume} {8}},\ \bibinfo {pages} {eabn1717} (\bibinfo {year} {2022})}\BibitemShut {NoStop}%
\bibitem [{\citenamefont {Thomsen}\ \emph {et~al.}(2024)\citenamefont {Thomsen}, \citenamefont {Kesselring}, \citenamefont {Bartlett},\ and\ \citenamefont {Brown}}]{thomsen2022low}%
  \BibitemOpen
  \bibfield  {author} {\bibinfo {author} {\bibfnamefont {F.}~\bibnamefont {Thomsen}}, \bibinfo {author} {\bibfnamefont {M.~S.}\ \bibnamefont {Kesselring}}, \bibinfo {author} {\bibfnamefont {S.~D.}\ \bibnamefont {Bartlett}},\ and\ \bibinfo {author} {\bibfnamefont {B.~J.}\ \bibnamefont {Brown}},\ }\bibfield  {title} {\bibinfo {title} {Low-overhead quantum computing with the color code},\ }\bibfield  {journal} {\bibinfo  {journal} {Phys. Rev. Res.}\ }\textbf {\bibinfo {volume} {6}},\ \href {https://doi.org/10.1103/PhysRevResearch.6.043125} {10.1103/PhysRevResearch.6.043125} (\bibinfo {year} {2024})\BibitemShut {NoStop}%
\bibitem [{\citenamefont {Walker}\ and\ \citenamefont {Wang}(2011)}]{Walker2011}%
  \BibitemOpen
  \bibfield  {author} {\bibinfo {author} {\bibfnamefont {K.}~\bibnamefont {Walker}}\ and\ \bibinfo {author} {\bibfnamefont {Z.}~\bibnamefont {Wang}},\ }\bibfield  {title} {\bibinfo {title} {{(3+1)-TQFTs and topological insulators}},\ }\href {https://doi.org/10.1007/s11467-011-0194-z} {\bibfield  {journal} {\bibinfo  {journal} {Frontiers of Physics}\ }\textbf {\bibinfo {volume} {7}},\ \bibinfo {pages} {150–159} (\bibinfo {year} {2011})}\BibitemShut {NoStop}%
\bibitem [{\citenamefont {Williamson}(2021)}]{williamson20211-form}%
  \BibitemOpen
  \bibfield  {author} {\bibinfo {author} {\bibfnamefont {D.~J.}\ \bibnamefont {Williamson}},\ }\href {https://www.youtube.com/watch?v=gT9BP-a90Jo} {\bibinfo {title} {1-form spts \& measurement-based quantum computation}} (\bibinfo {year} {2021}),\ \bibinfo {note} {talk at Harvard CMSA}\BibitemShut {NoStop}%
\bibitem [{\citenamefont {Gorenstein}(2007)}]{gorenstein2007finite}%
  \BibitemOpen
  \bibfield  {author} {\bibinfo {author} {\bibfnamefont {D.}~\bibnamefont {Gorenstein}},\ }\href@noop {} {\emph {\bibinfo {title} {Finite groups}}},\ Vol.\ \bibinfo {volume} {301}\ (\bibinfo  {publisher} {American Mathematical Soc.},\ \bibinfo {year} {2007})\BibitemShut {NoStop}%
\bibitem [{\citenamefont {Vuillot}\ \emph {et~al.}(2019)\citenamefont {Vuillot}, \citenamefont {Lao}, \citenamefont {Criger}, \citenamefont {García~Almudéver}, \citenamefont {Bertels},\ and\ \citenamefont {Terhal}}]{Vuillot2019code}%
  \BibitemOpen
  \bibfield  {author} {\bibinfo {author} {\bibfnamefont {C.}~\bibnamefont {Vuillot}}, \bibinfo {author} {\bibfnamefont {L.}~\bibnamefont {Lao}}, \bibinfo {author} {\bibfnamefont {B.}~\bibnamefont {Criger}}, \bibinfo {author} {\bibfnamefont {C.}~\bibnamefont {García~Almudéver}}, \bibinfo {author} {\bibfnamefont {K.}~\bibnamefont {Bertels}},\ and\ \bibinfo {author} {\bibfnamefont {B.~M.}\ \bibnamefont {Terhal}},\ }\bibfield  {title} {\bibinfo {title} {Code deformation and lattice surgery are gauge fixing},\ }\href {https://doi.org/10.1088/1367-2630/ab0199} {\bibfield  {journal} {\bibinfo  {journal} {New J. Phys.}\ }\textbf {\bibinfo {volume} {21}},\ \bibinfo {pages} {033028} (\bibinfo {year} {2019})}\BibitemShut {NoStop}%
\bibitem [{\citenamefont {Sorge}(2015)}]{sorge2015pyfssa}%
  \BibitemOpen
  \bibfield  {author} {\bibinfo {author} {\bibfnamefont {A.}~\bibnamefont {Sorge}},\ }\href {https://doi.org/10.5281/zenodo.35293} {\bibinfo {title} {pyfssa 0.7.6}} (\bibinfo {year} {2015})\BibitemShut {NoStop}%
\end{thebibliography}%

\end{document}